\documentclass[12pt]{iopart}

\pdfoutput=1

\usepackage{graphicx,epsfig,sidecap}
\usepackage{bm}
\usepackage{braket}
\usepackage{amssymb}
\usepackage{amsfonts}
\usepackage{mathrsfs}
\usepackage{amsmath}
\usepackage{graphicx}
\usepackage{bm}
\usepackage{xcolor}
\usepackage{hyperref}
\usepackage[toc,title]{appendix}

\catcode`@=11 
\renewcommand\tableofcontents{%
  \section*{\contentsname}%
  \@starttoc{toc}%
}
\catcode`@=12

\def\be{\begin{equation}}
\def\ee{\end{equation}}

\def\bea{\begin{eqnarray}}
\def\eea{\end{eqnarray}}

\def\Tr{{\rm Tr}}

\numberwithin{equation}{section}

\begin{document}

\title[On R\'enyi entropies of disjoint intervals in CFT]{On R\'enyi entropies of disjoint intervals\\ in conformal field theory}

\author{Andrea Coser$^1$, Luca Tagliacozzo$^2$ and Erik Tonni$^1$}
\address{$^1$\,SISSA and INFN, via Bonomea 265, 34136 Trieste, Italy. }
\address{$^2$\,ICFO, Av. Carl Friedrich Gauss 3, 08860 Castelldefels (Barcelona), Spain. }

\date{\today}

\begin{abstract}

We study the R\'enyi entropies of $N$ disjoint intervals in the conformal field theories given by the free compactified boson and the Ising model. They are computed as the $2N$ point function of twist fields, by employing the partition function of the model on a particular class of Riemann surfaces. The results are written in terms of Riemann theta functions.
The prediction for the free boson in the decompactification regime is checked against exact results for the harmonic chain. 
For the Ising model, matrix product states computations agree with the conformal field theory result once the finite size corrections have been taken into account.

\end{abstract}

\maketitle

\tableofcontents

\section{Introduction}

The study of the entanglement in extended quantum systems and of its measures has attracted a lot of interest during the last decade (see the reviews \cite{rev}). Given a system in its ground state $| \Psi \rangle$, a very useful measure of entanglement is the entanglement entropy. When the Hilbert space of the full system can be factorized as $\mathcal{H} = \mathcal{H}_A \otimes \mathcal{H}_B$, the $A$'s reduced density matrix reads $\rho_A = \Tr_B \rho$, being $\rho= | \Psi \rangle \langle \Psi |$ the density matrix of the entire system. 
The Von Neumann entropy associated to $\rho_A$ is the entanglement entropy  
\be
\label{SA def}
S_A = - \Tr (\rho_A \log \rho_A)\,.
\ee
Introducing $S_B$ in an analogous way, we have $S_B = S_A$ because $\rho$ describes a pure state.

In quantum field theory the entanglement entropy (\ref{SA def}) is usually computed by employing the replica trick, which consists in two steps: first one computes $\Tr \rho_A^n$ for any integer $n \geqslant 2$ (when $n=1$ the normalization condition $\Tr \rho_A =1$ is recovered) and then analytically  continues the resulting expression to any complex $n$.
This allows to obtain the entanglement entropy as $S_A = - \lim_{n \rightarrow 1} \partial_n \Tr \rho_A^n$. 
The R\'enyi entropies are defined as follows
\be
\label{Renyi entropies def}
S_A^{(n)} = \frac{1}{1-n} \, \log \Tr \rho_A^n \,.
\ee
Given the normalization condition, the replica trick tells us that $S_A = \lim_{n \rightarrow 1} S_A^{(n)} $.

In this paper we consider one dimensional critical systems when $A$ and $B$ correspond to a spatial bipartition.
The simplest and most important example is  the entanglement entropy of an interval $A$ of length $\ell$ in an infinite line, which is given by \cite{Holzhey, cc-04,cc-rev}
\be
\label{SA 1int cft}
S_A = \frac{c}{3} \log \frac{\ell}{\epsilon} + c_1'\,,
\ee
where $c$ is the central charge of the corresponding conformal field theory (CFT), $\epsilon$ is the UV cutoff and $c_1'$ is a non universal constant. The result (\ref{SA 1int cft}) has been rederived in \cite{cc-04} by computing $\Tr \rho_A^n $ for an interval $A =[u,v]$ as the two point function of twist fields, namely
\be
\Tr \rho_A^n  = 
\frac{c_n}{|u-v|^{2\Delta_n}}\,,
\qquad 
\Delta_n = \frac{c}{12} \left( n-\frac{1}{n} \right)\,,
\ee
being $\Delta_n $ the scaling dimension of the twist fields and $c_n$ a non universal constant such that $c_1 =1$, in order to guarantee the normalization condition. 

\begin{figure}[t]
\vspace{.3cm}
\begin{center}
\includegraphics[width=.95\textwidth]{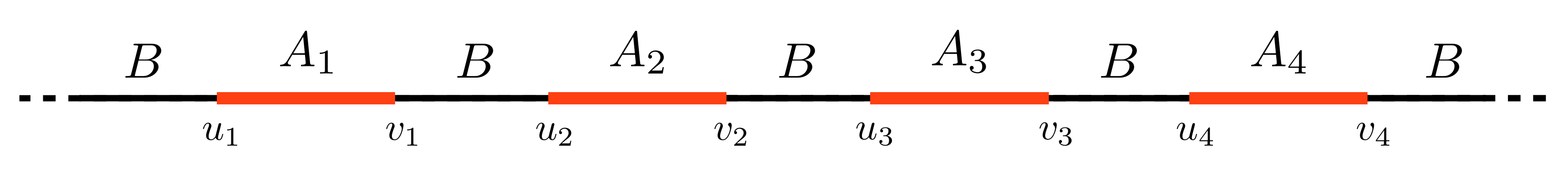}
\end{center}
\vspace{-.5cm}
\caption{A typical configuration of disjoint intervals in the infinite line. We consider the entanglement between $A=\cup_{i=1}^N A_i$ (in this figure $N=4$) and its complement $B$. 
}
\label{fig line intervals}
\end{figure}

When $A$ is a single interval, $\Tr \rho_A^n$ and $S_A$ are sensible only to the central charge of the CFT. 
Instead, when the subsystem $A = \cup_{i=1}^N A_i$ consists of $N \geqslant 2$ disjoint intervals on the infinite line, the R\'enyi entropies encode all the data of the CFT. Denoting by $A_i =[u_i,v_i]$ the $i$-th interval with $i=1,\dots, N$, in Fig. \ref{fig line intervals} we depict a configuration with $N=4$ disjoint intervals.
By employing the method of \cite{cc-04, cc-rev}, $\Tr \rho_A^n$ can be computed as a $2N$ point function of twist fields. 
In CFT, the dependence on the positions in a $2N$ point function of primary operators with $N\geqslant 2$ is not uniquely determined by the global conformal invariance. 
Indeed, we have that \cite{cc-rev}
\be
\label{Tr rhoA N int intro}
\Tr \rho_A^n 
= c_n^N 
\left|\, \frac{\prod_{i<j} (u_j -u_i)(v_j-v_i)}{\prod_{i,j} (v_j - u_i)} \,\right|^{2 \Delta_n}
\mathcal{F}_{N,n}(\boldsymbol{x} )\,,
\ee
where $\mathcal{F}_{N,n}(\boldsymbol{x})$ is a model dependent function of the $2N-3$ independent variables $0 < x_1<  \dots< x_{2N-3}<1$ (indicated by the vector $\boldsymbol{x}$), which are the invariant ratios that can be built with the $2N$ endpoints of the intervals through a conformal map.

For $N=2$ intervals there is only one harmonic ratio $0<x<1$. The function $\mathcal{F}_{2,n}(x)$ has been computed for the free boson compactified on a circle \cite{cct-09} and for the Ising model \cite{cct-11}. A crucial role in the derivation is played by the methods developed in \cite{z-87, dixon,amv, verlinde2-86, amv bosonization, knizhnik-87, BershadskyRadul, dvv} to study CFT on higher genus Riemann surfaces.
The results are expressed in terms of Riemann theta functions \cite{Fay book, Mumford book, Igusa book} and it is still an open problem to compute their analytic continuation in $n$ for the most general case, in order to get the entanglement entropy $S_A$. These CFT predictions are supported by numerical studies performed through various methods \cite{CaraglioGliozzi, pasquier08, atc-10, gt-11, FagottiCalabrese10, atc-11, Fagotti12, rg-12}.

For three or more intervals, few analytic results are available in the literature. For instance, the R\'enyi entropies of $N>2$ disjoint intervals for the Dirac fermion in two dimensions has been computed in \cite{CasiniFoscoHuerta, CasiniHuerta09, CasiniHuerta09a}. This result holds for a specific sector and it is not modular invariant \cite{Headrick-12}.

In this paper we compute $\mathcal{F}_{N,n}(\boldsymbol{x})$ with $N\geqslant 2$  for the free boson compactified on a circle and for the Ising model, by employing the results of \cite{z-87, amv, verlinde2-86,amv bosonization,dvv} and \cite{eg-03}.
The case $n=2$ has been studied in \cite{z-87} and its extension to $n>2$ has been already discussed in \cite{knizhnik-87, BershadskyRadul, cct-09, Headrick-12}. Here we provide explicit expressions for $\mathcal{F}_{N,n}(\boldsymbol{x})$ in terms of Riemann theta functions.
The free boson on the infinite line is obtained as a limiting regime and the corresponding CFT predictions have been checked against exact numerical results for the harmonic chain.
The numerical checks of the CFT formulas for the Ising model have been done by employing the Matrix Product States (MPS) \cite{MPSrev, EvenblyVidal-11}.

We remark that, in the case of several disjoint intervals, the entanglement entropy $S_A$ measures the entanglement of the union of the intervals with the rest of the system $B$. It is not a measure of the entanglement among the intervals, whose union is in a mixed state. In order to address this issue, one needs to consider other quantities which measure the entanglement for mixed states. An interesting example is the negativity \cite{vw-01, Audenaert02}, which has been studied for a two dimensional CFT in \cite{cct-neg-letter,cct-neg-long} by employing the twist fields method (see \cite{Bose-neg-ising, alba13 neg, ctt-neg-ising} for the Ising model).

In the context of the AdS/CFT correspondence, there is a well estabilished prescription to compute $S_A$ in generic spacetime dimensions through the gravitational background in the bulk \cite{RTletter, RTlong, Tak12-rev}, which has been applied also in the case of disjoint regions \cite{HeadrickTakayanagiSSA, HubRang dis, Headrick10, et-11, HaydenHeadrick}. Proposals for the holographic computation of the R\'enyi entropies $S_A^{(n)}$ are also available \cite{MyersCasiniHuerta, Myers-holRenyi, hartman13, faulkner13, maldacena13genEnt}.
The holographic methods hold in the regime of large $c$, while the models that we consider here have $c=1$ and $c=1/2$. 

The layuot of the paper is as follows. 
In \S\ref{sec riemann surf} we describe the relation between $\Tr \rho_A^n $ and the partition functions of two dimensional conformal field theories on the particular class of Riemann surfaces occurring in our problem.
In \S\ref{sec compact boson} we compute the R\'enyi entropies for the free compactified boson in the generic case of $N$ intervals and $n$ sheets, which allows us to write the same quantity also for the Ising model.
In \S\ref{sec 2int case} we discuss how the known case of two intervals is recovered.
In \S\ref{sec HC} we check the CFT predictions for the free boson in the decompactification regime against exact results obtained for the harmonic chain with periodic boundary conditions.
In \S\ref{sec ising} numerical results obtained with MPS for the Ising model with periodic boundary conditions are employed to check the corresponding CFT prediction through a finite size scaling analysis.
In the Appendices, we collect further details and results.

\section{R\'enyi entropies and Riemann surfaces}
\label{sec riemann surf}

Given a two dimensional quantum field theory, let us consider a spatial subsystem $A = \cup_{i=1}^N A_i$ made by  $N$ disjoint intervals $A_1 = [u_1, v_1]$, $\dots$, $A_N = [u_N, v_N]$. 

The path integral representation of $\rho_A$ has been largely discussed in \cite{Holzhey, cc-04, cc-rev}. Tracing over the spatial complement $B$ leaves open cuts, one for each interval, along the line characterized by a fixed value of the Euclidean time. Thus, the path integral giving $\rho_A$ involves fields which live on this sheet with open cuts, whose configurations are fixed on the upper and lower parts of the cuts. 

\begin{figure}[t]
\vspace{.3cm}
\begin{center}
\includegraphics[width=.97\textwidth]{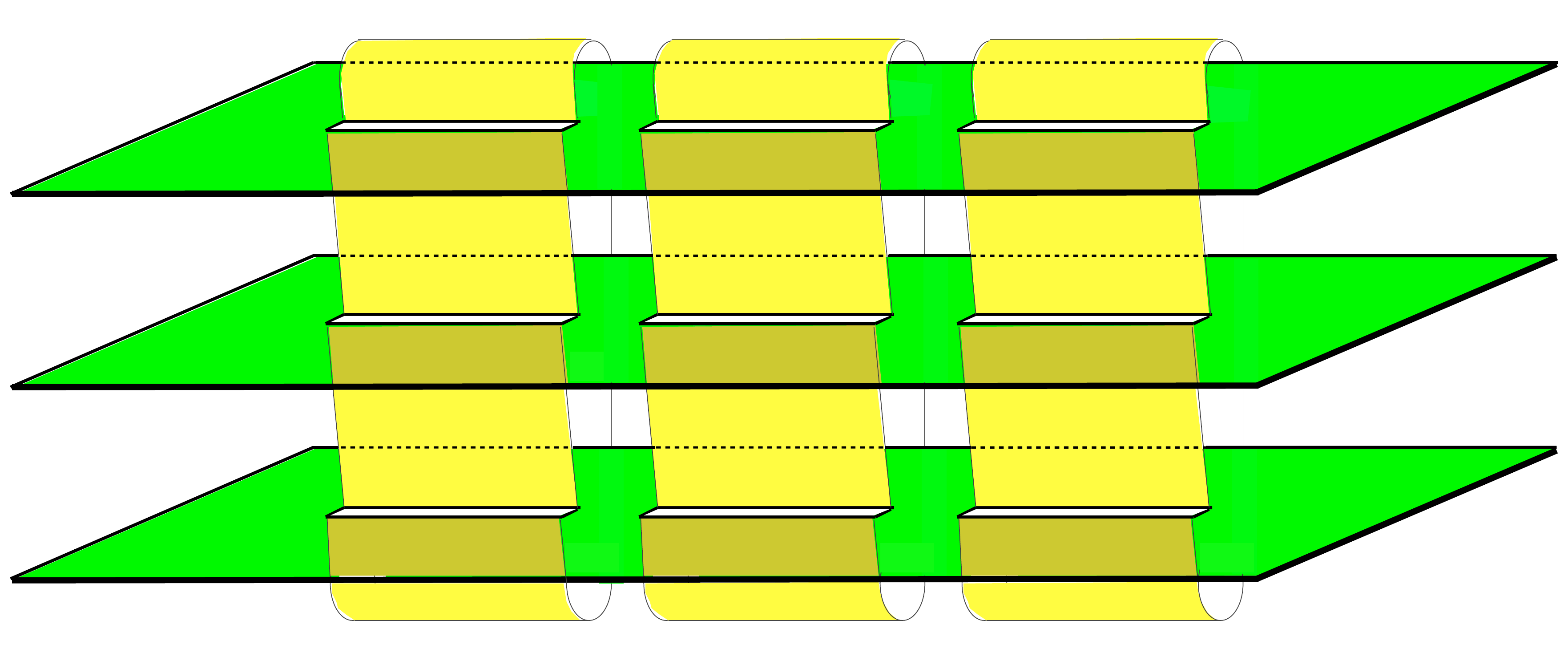}
\end{center}
\vspace{-.2cm}
\caption{The path integral representation of $\textrm{Tr}\rho_A^n$ involves a Riemann surface $\mathscr{R}_{N,n}$, which is shown here for $N=3$ and $n=3$.}
\label{fig sheets}
\end{figure}

To compute $\Tr \rho_A^n$, we take $n$ copies of the path integral representing $\rho_A$ and combine them as briefly explained in the following. For any fixed $x \in A$, we impose that the value of a field on the upper part of the cut on a sheet is equal to the value of the same field on the lower part of the corresponding cut on the sheet right above. This condition is applied in a cyclic way. Then, we integrate over the field configurations along the cuts. 
Correspondingly, the $n$ sheets must be sewed in the same way and this procedure defines the $n$-sheeted Riemann surface $\mathscr{R}_{N,n}$. The endpoints $u_i$ and $v_i$ ($i=1, \dots, N$) are branch points where the $n$ sheets meet.
The Riemann surface $\mathscr{R}_{N,n}$ is depicted in Fig. \ref{fig sheets} for $N=3$ intervals and $n=3$ copies. 
Denoting by $\mathcal{Z}_{N,n}$ the partition function of the model on the Riemann surface $\mathscr{R}_{N,n}$, we can compute $\Tr \rho_A^n$ as \cite{cc-04}
\be
\label{trace rhoA n}
\Tr \rho_A^n = \frac{\mathcal{Z}_{N,n}}{\mathcal{Z}^n}\,,
\ee
where $\mathcal{Z}=\mathcal{Z}_{0,1}$ is the partition function of the model defined on a single copy and without cuts. Notice that (\ref{trace rhoA n}) implies $\Tr \rho_A =1$.
From (\ref{trace rhoA n}), one easily gets the R\'enyi entropies (\ref{Renyi entropies def}). If the analytic continuation of (\ref{trace rhoA n}) to $\textrm{Re} \,n >1$ exists and it is unique, the entanglement entropy is obtained as the replica limit
\be
S_A =  \lim_{n \,\rightarrow\, 1}   S_A^{(n)}  
=
\, - \lim_{n \,\rightarrow\, 1}  \frac{\partial}{\partial n}\, \Tr \rho_A^n\,.
\ee

In order to find the genus of $\mathscr{R}_{N,n}$ \cite{dixon}, let us consider a single sheet and triangulate it through $V$ vertices, $E$ edges and $F$ faces, such that $2N$ vertices are located at the branch points $u_i$ and $v_i$. 
Considering $\mathscr{R}_{N,n}$ constructed as explained above, the replication of the same triangulation on the other sheets generates a triangulation of the Riemann surface $\mathscr{R}_{N,n}$ made by $V'$ vertices, $E'$ edges and $F'$ faces. Notice that, since the branch points belong to all the $n$ sheets, they are not replicated. This observation tells us that $V' = n(V-2N)+2N$, while $E'=nE$ and $F'=nF$ because all the edges and the faces are replicated. 
Then, the genus $g$ of $\mathscr{R}_{N,n}$ is found by plugging these expressions into the relation $V'-E'+F'=2-2g$ and employing the fact that, since each sheet has the topology of the sphere, $V-E+F=2$. The result is
\be
\label{genus}
g=(N-1)(n-1)\,.
\ee
We remark that we are not considering the most general genus $g$ Riemann surface, which is characterized by $3g-3$ complex parameters, but only the subclass of Riemann surfaces obtained through the replication procedure.

Let us consider  a conformal field theory with central charge $c$.
As widely argued in  \cite{cc-04, cc-rev}, in the case of one interval $A=[u,v]$ in an infinite line, $\Tr \rho_A^n$  can be written as the two point function of twist fields on the complex plane plus the point at infinity, i.e.
\be
\label{twist 2point func}
\Tr \rho_A^n  = 
\langle  \mathcal{T}_n(u)  \bar{\mathcal{T}}_n(v)  \rangle =
\frac{c_n}{|u-v|^{2\Delta_n}}\,,
\qquad
\Delta_n = \frac{c}{12} \left( n-\frac{1}{n} \right).
\ee
Both the twist field $\mathcal{T}_n$ and $\bar{\mathcal{T}}_n$, also called branch point twist fields \cite{CardyDoyon}, have the same scaling dimension $\Delta_n$. The constant $c_n$ is non universal and such that $c_1=1$ because of the normalization condition. 

\begin{figure}[t]
\vspace{.3cm}
\begin{center}
\includegraphics[width=.7\textwidth]{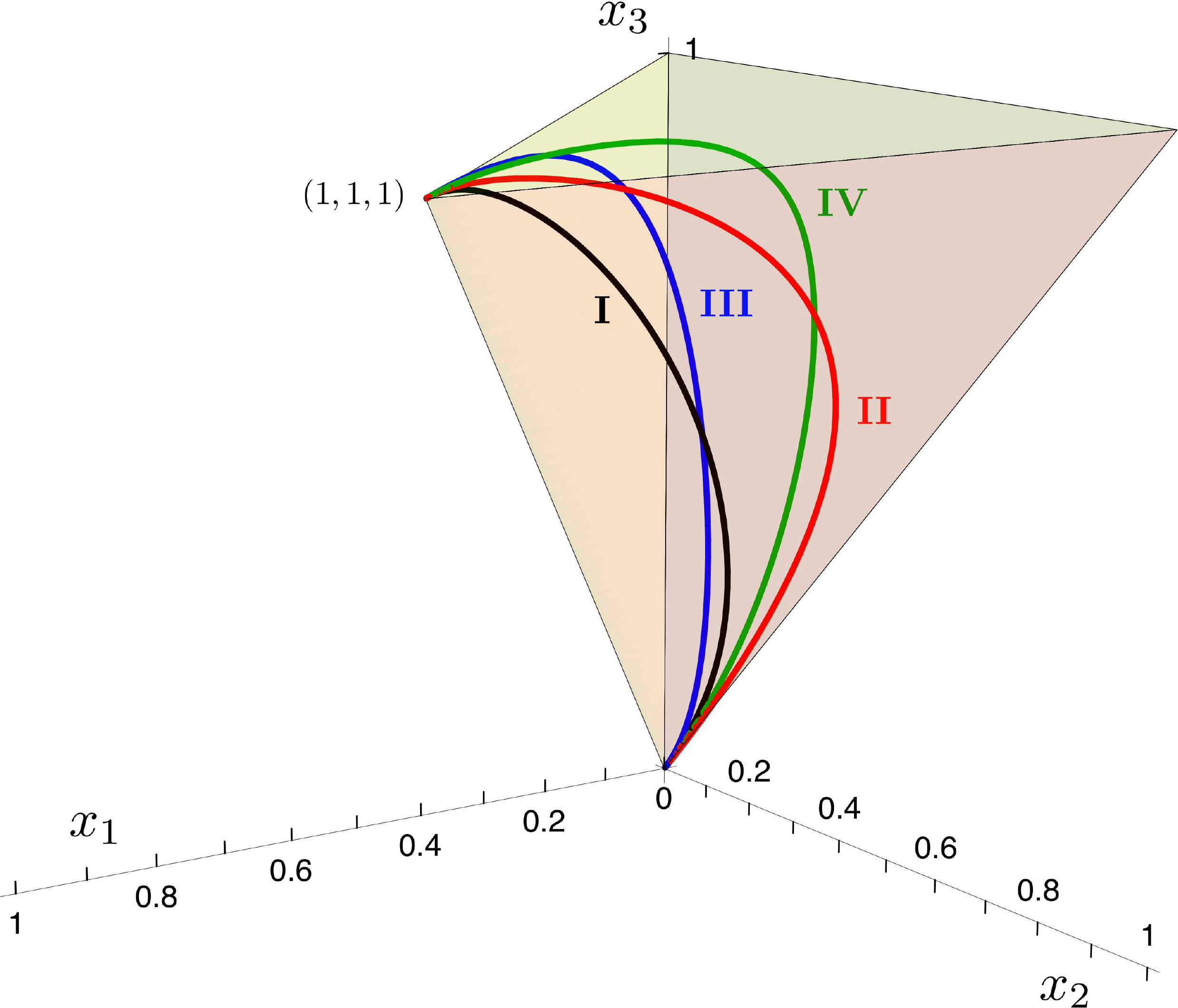}
\end{center}
\vspace{-.1cm}
\caption{The domain $0<x_1<x_2<x_3<1$ of the function $\mathcal{F}_{3,n}(\boldsymbol{x})$. The lines within this domain are the configurations defined in (\ref{N3 configs def}).}
\label{fig configN3}
\end{figure}

Similarly, when $A$ consists of  $N\geqslant 2$ disjoint intervals $A_i =[u_i , v_i]$ with $i=1, \dots, N$, ordered on the infinite line according to $i$, namely $u_1 < v_1 < \dots < u_N < v_N$, we can write $\Tr \rho_A^n$ as the following $2N$ point function of twist fields
\be
\label{renyi 2N point func}
\Tr \rho_A^n 
= \langle \prod_{i\,= 1}^N  \mathcal{T}_n(u_i) \bar{\mathcal{T}}_n(v_i)  \rangle\,.
\ee
In the case of four and higher point correlation functions of primary fields, the global conformal invariance does not fix the precise dependence on $u_i$ and $v_i$ because one can construct invariant ratios involving these points. In particular, let us consider the conformal map such that $u_1 \rightarrow 0$, $u_N \rightarrow 1$ and $v_N \rightarrow \infty$, namely 
\be
\label{map wN}
w_N(z) = \frac{(u_1-z)(u_N-v_N)}{(u_1-u_N)(z-v_N)}\,.
\ee
The remaining $u_i$'s and $v_j$'s are sent into the  $2N-3$ harmonic ratios $x_1 = w_N(v_1)$, $x_2 = w_N(u_2)$, $x_3 = w_N(v_2)$, $\dots$, $x_{2N-3} = w_N(v_{N-1})$ which are invariant under $SL(2,\mathbb{C})$ transformations. The map (\ref{map wN}) preserves the ordering: $0 <x_1<x_2<\dots<x_{2N-3}<1$. 
We denote by $\boldsymbol{x}$ the vector whose elements are the harmonic ratios $x_1, \dots, x_{2N-3}$.\\
Global conformal invariance allows to write the $2N$ point function (\ref{renyi 2N point func}) as \cite{cc-rev}
\be
\label{FNn def general cft}
\Tr \rho_A^n 
= c_n^N 
\left|\, \frac{\prod_{i<j} (u_j -u_i)(v_j-v_i)}{\prod_{i,j} (v_j - u_i)} \,\right|^{2 \Delta_n}
\mathcal{F}_{N,n}(\boldsymbol{x} )\,,
\ee
where $i,j=1, \dots, N$. The function $\mathcal{F}_{N,n}(\boldsymbol{x} )$ encodes the full operator content of the model and therefore it must be computed through its dynamical details. 
Since $\Tr \rho_A =1$, we have $\mathcal{F}_{N,1}(\boldsymbol{x} ) = 1$.
In the case of two intervals, $\mathcal{F}_{2,n}(\boldsymbol{x} )$ has been computed for the free compactified boson \cite{cct-09} and for the Ising model \cite{cct-11}. We remark that the domain of $\mathcal{F}_{N,n}(\boldsymbol{x} )$ is $0 <x_1<\dots<x_{2N-3}<1$ (see Fig. \ref{fig configN3} for $N=3$).

The expression (\ref{FNn def general cft}) is UV divergent. Such divergence is introduced dividing any length occurring in the formula ($u_j-u_i$, $v_j-u_i$, etc.) by the UV cutoff $\epsilon$. Since the ratios $\boldsymbol{x}$ are left unchanged, the whole dependence on $\epsilon$ of (\ref{FNn def general cft}) comes from the ratio of lengths within the absolute value, which gives  $\epsilon^{2N\Delta_n}$.

It is useful to introduce some quantities which are independent of the UV cutoff. 
For $N=2$, we can construct a combination of R\'enyi entropies having this property as follows
\be
\label{mutual Renyi}
I^{(n)}_{A_1, A_2} \equiv S^{(n)}_{A_1} + S^{(n)}_{A_2} - S^{(n)}_{A_1 \cup A_2} 
= \frac{1}{n-1}\,\log \bigg(\frac{\Tr \rho^n_{A_1 \cup A_2} }{\Tr \rho^n_{A_1}  \Tr \rho^n_{A_2} } \bigg).
\ee
The limit $n \rightarrow 1$ of this quantity defines the mutual information $I_{A_1, A_2} $
\be
\label{mutual information N=2}
I_{A_1, A_2} \equiv S_{A_1} + S_{A_2} - S_{A_1 \cup A_2}
= \lim_{n \, \rightarrow\,1} I^{(n)}_{A_1, A_2} \,,
\ee
which is independent of the UV cutoff as well.
The subadditivity of the entanglement entropy tells us that $I_{A_1, A_2} \geqslant 0$, while the strong subadditivity implies that it increases when one of the intervals is enlarged.

For $N>2$ we can find easily two ways to construct quantities such that the short distance divergence cancels.
Let us consider first the following ratio
\be
\label{RNn def}
R_{N,n} \equiv
\prod_{p\,=\,1}^{N} \; \prod_{\sigma_{N,p}} 
\big(\Tr \rho^n_{\sigma_{N,p}}\big)^{(-1)^{N-p}},
\ee
where we denoted by $\sigma_{N,p}$ a generic choice of $1 \leqslant p \leqslant N$ intervals among the $N$ ones we are dealing with. Since $\Tr \rho^n_{\sigma_{N,p}}$ goes like $\epsilon^{2p\Delta_n}$,  one finds that (\ref{RNn def}) is independent of $\epsilon$ by employing that $\sum_{p=1}^N (-1)^{N-p} \binom{N}{p} p =0$.
In the simplest cases of $N=2$ and $N=3$, the ratio (\ref{RNn def}) reads
\be\fl
R_{2,n} = \frac{\Tr \rho^n_{\{1,2\}}}{\Tr \rho^n_{\{1\}} \, \Tr \rho^n_{\{2\}}}\,,
\qquad
R_{3,n} = \frac{\Tr \rho^n_{\{1,2,3\}} \,\big( \Tr \rho^n_{\{1\}} \, \Tr \rho^n_{\{2\}} \, \Tr \rho^n_{\{3\}} \big)}{
\Tr \rho^n_{\{1,2\}} \, \Tr \rho^n_{\{1,3\}} \, \Tr \rho^n_{\{2,3\}} }\,,
\qquad
\dots
\ee
In order to generalize (\ref{mutual Renyi}) for $N\geqslant 2$, one introduces
\be
\label{N renyi mutual info def}
I_{A_1 , \dots , A_N}^{(n)} \equiv \frac{(-1)^N}{n-1} \log R_{N,n} \,,
\ee 
and its limit $n \rightarrow 1$, as done in (\ref{mutual information N=2}) for $N=2$, i.e.
\be
\label{N mutual info def}
I_{A_1 , \dots , A_N} \equiv \lim_{n \rightarrow 1} I_{A_1 , \dots , A_N}^{(n)} \,.
\ee 
For the simplest cases of $N=3$ and $N=4$, one finds respectively
\begin{eqnarray}\fl
I_{A_1 , A_2 , A_3} &=&
S_{A_1} + S_{A_2} + S_{A_3} 
- S_{A_1 \cup A_2} - S_{A_1 \cup A_3} - S_{A_2 \cup A_3}
+ S_{A_1 \cup A_2 \cup A_3}\,, \\ 
\rule{0pt}{.8cm}
\fl
I_{A_1 , A_2 , A_3, A_4} &=& 
\sum_{i=1}^4 S_{A_i} 
- \sum_{\substack{i,j=1 \\ i\,<\,j }}^4 S_{A_i \cup A_j} 
+ \sum_{\substack{i,j,k=1 \\ i<j <k }}^4 S_{A_i \cup A_j \cup A_k} 
- S_{A_1 \cup A_2 \cup A_3 \cup A_4}\,.
\end{eqnarray}
The quantity $I_{A_1, A_2, A_3}$ is called tripartite information \cite{CasiniHuerta09} and it provides a way to establish whether the mutual information is extensive ($I_{A_1, A_2, A_3} =0$) or not. In a general quantum field theory there is no definite sign for $I_{A_1, A_2, A_3}$, but for theories with a holographic dual it has been shown that $I_{A_1, A_2, A_3} \leqslant 0$ \cite{HaydenHeadrick}.

Another cutoff independent ratio is given by
\be
\label{tildeRNn def}
\tilde{R}_{N,n} \equiv
\frac{\Tr \rho_A^n }{ \prod_{i=1}^N \Tr \rho_{A_i}^n}\,.
\ee
When $N=2$ we have $R_{2,n} = \tilde{R}_{2,n} $ but (\ref{RNn def}) and (\ref{tildeRNn def}) are different for $N>2$.\\
From the definitions (\ref{RNn def}) and (\ref{tildeRNn def}), we observe that, when one of the intervals collapses to the empty set, i.e. $A_k \rightarrow \emptyset$ for some $k \in \{1, \dots, N\}$, we have that $R_{N,n} \rightarrow 1$ and $\tilde{R}_{N,n} \rightarrow \tilde{R}_{N-1,n}$, where $\tilde{R}_{N-1,n}$ is defined through $A \setminus A_k$.

For two dimensional conformal field theories at zero temperature we can write $R_{N,n} $ and $\tilde{R}_{N,n}$ more explicitly. In particular, plugging (\ref{twist 2point func}) and (\ref{FNn def general cft}) into (\ref{tildeRNn def}), it is straightforward to observe that $c_n$ simplifies and we are left with
\be\fl
\label{tildeRNn cft}
\tilde{R}_{N,n}(\boldsymbol{x})  =  
\bigg| \prod_{i<j} \frac{(u_i-u_j)(v_i-v_j)}{(u_i - v_j)(u_j -v_i)} \,\bigg|^{2\Delta_n} 
\mathcal{F}_{N,n}(\boldsymbol{x})
\equiv 
| p_N(\boldsymbol{x}) |^{2\Delta_n} \,\mathcal{F}_{N,n}(\boldsymbol{x})\,,
\ee
where the product within the absolute value, that we denote by $p_N$, can be written in terms of $\boldsymbol{x}$. Thus, (\ref{tildeRNn cft}) tells us that $\mathcal{F}_{N,n}(\boldsymbol{x})$ can be easily obtained from $\tilde{R}_{N,n}(\boldsymbol{x})$.
When $N=2$ we have $p_2(x) = -1/(1-x)$, while for $N=3$ we find
\begin{equation}
p_3(\boldsymbol{x}) \equiv
-\frac{(x_3-x_1)(1-x_2)\,x_2}{(x_2-x_1)(1-x_1)  (1-x_3)\, x_3}\,.
\end{equation}
For higher values of $N$, the expression of $p_N(\boldsymbol{x})$ is more complicated.

As for $R_{N,n}$ in (\ref{RNn def}), considering the choice of intervals given by $\sigma_{N,p}$, we have
\be
\label{Tr Np}
\Tr \rho_{\sigma_{N,p}}^n = c_n^p \, \big| P_p(\sigma_{N,p}) \big|^{2\Delta_n} 
\,\mathcal{F}_{p,n}(\boldsymbol{x}^{\sigma_{N,p}})\,,
\ee
where
\be
\label{Ppsigma def}
P_p(\sigma_{N,p}) \equiv 
\left. \frac{\prod_{i<j} (u_j -u_i)(v_j-v_i)}{\prod_{i,j} (v_j - u_i)}\, \right|_{\,i,j \,\in\, \sigma_{N,p}}\,,
\ee
and $\boldsymbol{x}^{\sigma_{N,p}}$ denotes the $2p-3$ harmonic ratios that can be constructed through the $2p$ endpoints of the intervals of $A$ specified by $\sigma_{N,p}$.  Notice that (\ref{Tr Np}) becomes (\ref{FNn def general cft}) when $p=N$ and (\ref{twist 2point func}) for $p=1$ because $\mathcal{F}_{N,1}=1$ by definition and $P_1(\sigma_{N,1}) =1/(v_j - u_j) $, being $j$ the interval specified by $\sigma_{N,1}$. 
Moreover, since (\ref{Ppsigma def}) can be written in terms of the $2N-3$ elements of $\boldsymbol{x}$, we have that $R_{N,n} = R_{N,n}(\boldsymbol{x})$ (see Appendix \ref{app x dependence} for more details).
Plugging (\ref{Tr Np}) into (\ref{RNn def}), one finds that for $N>2$ all the factors $P_p(\sigma_{N,p}) $ cancel (this simplification is explained in Appendix \ref{app x dependence}) and therefore we have
\be
\label{RNn cft}
R_{N,n}(\boldsymbol{x}) = 
\prod_{p\,=\,2}^{N}\, \prod_{\sigma_{N,p}} 
\big[  \mathcal{F}_{p,n}(\boldsymbol{x}^{\sigma_{N,p}})\big]^{(-1)^{N-p}} .
\ee

In order to cancel those parameters which occur only through multiplicative factors, we find it useful to normalize the quantities we introduced by themselves computed for a fixed configuration. 
Thus, for (\ref{RNn def}) and (\ref{N mutual info def}) we have respectively 
\be
\label{RNn norm}
R^{\textrm{\tiny{norm}}}_{N,n} \equiv \frac{R_{N,n}}{
R_{N,n}\big|_{\textrm{\tiny{fixed configuration}}}}\,,
\qquad
I^{\textrm{\tiny{sub}}}_{N} \equiv 
I_{N}  -  I_{N}\big|_{\textrm{\tiny{fixed configuration}}}\,,
\ee
where we adopted the shorthand notation $I_{N} \equiv I_{A_1 , \dots , A_N} $.
In conformal field theories, for the scale invariant quantities depending on the harmonic ratios $\boldsymbol{x}$, the fixed configuration is characterized by fixed values $\boldsymbol{x}_{\textrm{\tiny{fixed}}}$. For instance, we have
\be
\label{RNn cft norm}
R^{\textrm{\tiny{norm}}}_{N,n}(\boldsymbol{x}) = \frac{R_{N,n}(\boldsymbol{x})}{
R_{N,n}(\boldsymbol{x}_{\textrm{\tiny{fixed}}})}\,,
\qquad
\mathcal{F}^{\textrm{\tiny{norm}}}_{N,n}(\boldsymbol{x}) = \frac{\mathcal{F}_{N,n}(\boldsymbol{x})}{
\mathcal{F}_{N,n}(\boldsymbol{x}_{\textrm{\tiny{fixed}}})}\,.
\ee
In \S\ref{sec HC} this normalization is adopted to study the free boson on the infinite line.

\section{Free compactified boson}
\label{sec compact boson}

In this section we consider the real free boson $\phi(z, \bar{z})$ on the Riemann surface $\mathscr{R}_{N,n}$ and compactified on a circle of radius $R$. Its action reads
\be
\label{action free boson RNn}
S[\phi] \propto \int_{\mathscr{R}_{N,n}} \partial_z \phi \,\partial_{\bar{z}} \phi \,  d^2z\,.
\ee
The worldsheet is $\mathscr{R}_{N,n}$ and the target space is $\mathbb{R} / (2\pi R\, \mathbb{Z})$.
This model has $c=1$ and its partition function for a generic compact Riemann surface of genus $g$ has been largely discussed in the literature (see e.g. \cite{z-87, amv, verlinde2-86, dvv, knizhnik-87, BershadskyRadul}).

Instead of working with a single field $\phi$ on $\mathscr{R}_{N,n}$, one could equivalently consider $n$ independent copies of the model with a field $\phi_j$ on the $j$-th sheet \cite{CasiniFoscoHuerta, CardyDoyon}. These $n$ fields are coupled through their boundary conditions along the cuts $A_i$ on the real axis in a cyclic way (see Fig. \ref{fig EGmultisheets})
\be\fl
\label{replica boundary conditions}
\phi_j(x, 0^+) = \phi_{j+1}(x, 0^-)\,,
\qquad
x \in A\,,
\qquad
j \in \{1, \dots, n\}\,,
\qquad
n+1 \equiv 1\,.
\ee
This approach has been adopted in \cite{cct-09} for the $N=2$ case, employing the results of \cite{dixon}. In principle one should properly generalize the construction of \cite{cct-09} to $N>2$. 
For $n=2$ this computation has been done in \cite{z-87}.
Here, instead, in order to address the case $n>2$, we compute (\ref{FNn def general cft}) for the model (\ref{action free boson RNn}) more directly, borrowing heavily from the literature about the free compactified boson on higher genus Riemann surfaces, whose partition function has been constructed in terms of the period matrix of the underlying Riemann surface.

\subsection{The period matrix}
\label{subsec period matrix}

The $n$-sheeted Riemann surface $\mathcal{R}_{N,n}$ obtained by considering $N$ intervals $A_i =[u_i,v_i]$ ($i=1, \dots, N$) is defined by the following complex curve in $\mathbb{C}^2$ \cite{eg-03}
\be\fl
\label{eq curve}
y^n = u(z)v(z)^{n-1}\,,
\qquad
u(z)=\prod_{\gamma\,=\,1}^{N}  (z-x_{2\gamma-2})\,,
\qquad
v(z)=\prod_{\gamma\,=\,1}^{N-1}  (z-x_{2\gamma-1})\,.
\ee
The  complex coordinates $y$ and $z$ parameterize $\mathbb{C}^2$ and in $u(z)$ we introduced $x_0 \equiv 0$ and $x_{2N-2} \equiv 1$ for notational convenience.
For $n=2$, the curve (\ref{eq curve}) is hyperelliptic.
The genus of $\mathcal{R}_{N,n}$ is (\ref{genus}) and it can be found also by applying the Riemann-Hurwitz formula for the curve (\ref{eq curve}).

The period matrix of the curve (\ref{eq curve}) has been computed in \cite{eg-03} by considering the following non normalized basis of holomorphic differentials 
\be\fl
\label{one forms}
\omega_{\alpha,j} = \frac{z^{\alpha-1} \,v(z)^{j-1}}{y^j}\, dz\,,
\qquad
\alpha= 1, \dots , N-1\,,
\qquad
j=1, \dots , n-1\,,
\ee
where $y = y(z)$ through (\ref{eq curve}). The set of one forms defined in (\ref{one forms}) contains $g$ elements.\\
In (\ref{one forms})  we employed a double index notation: a greek index for the intervals and a latin one for the sheets. We make this choice to facilitate the comparison with \cite{cct-09}, slightly changing the notation with respect to the previous section.
These two indices can be combined either as  $r=\alpha+(N-1)(j-1)$ \cite{eg-03} or $r=j+(n-1)(\alpha-1)$ \cite{Headrick-12} in order to have an index $r=1, \dots , g$. Hereafter we assume the first choice. Notice that for the cases of $(N,n)=(2,n)$ and $(N,n)=(N,2)$ we do not need to introduce this distinction.

\begin{figure}[t]
\vspace{.5cm}
\begin{center}
\includegraphics[width=.85\textwidth]{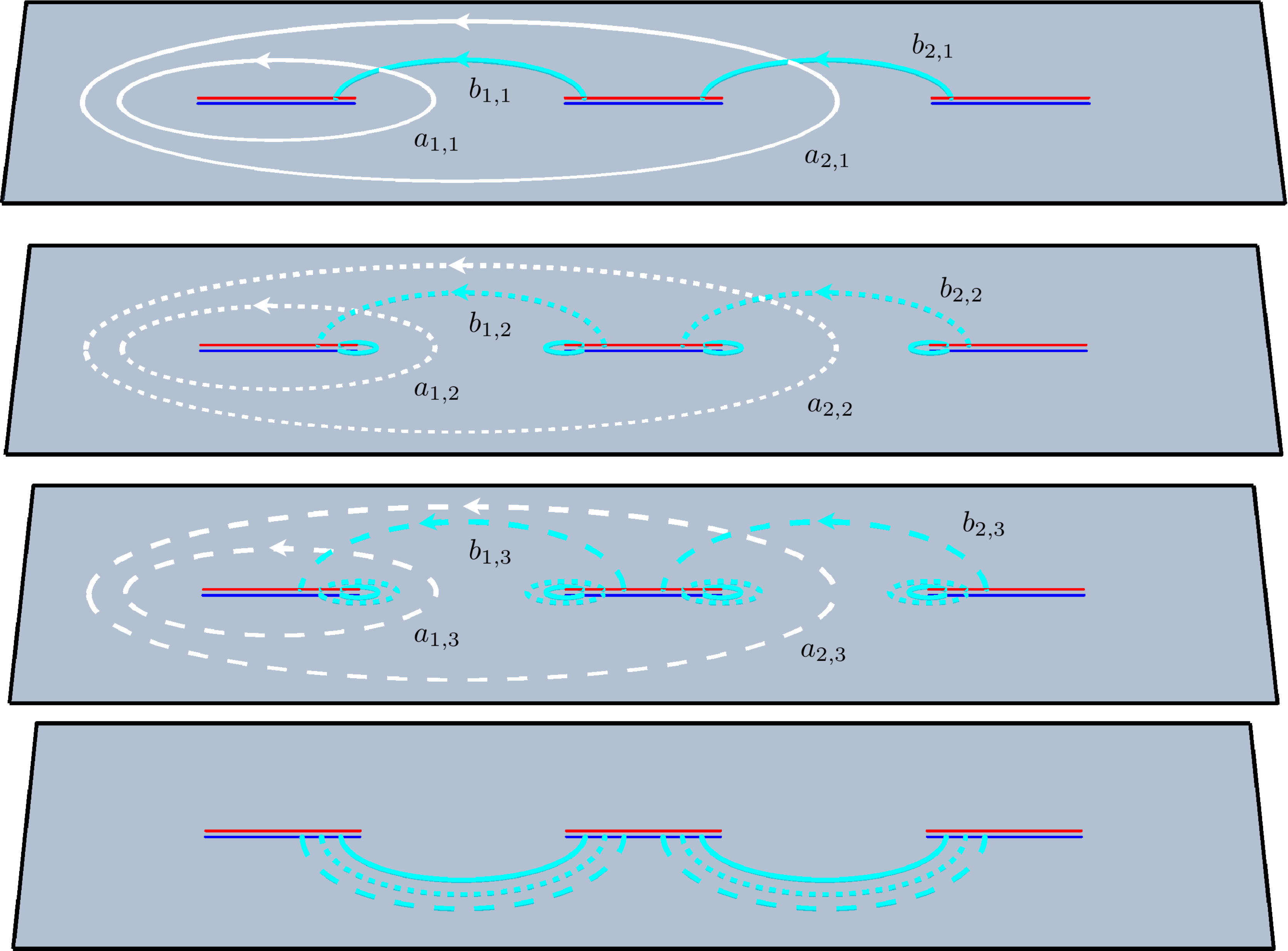}
\end{center}
\caption{The canonical homology basis $\{ a_{\alpha,j}, b_{\alpha,j} \}$ for $N=3$ intervals of equal length and $n=4$ sheets. The sheets are ordered starting from the top. For each cut, the upper part (red) is identified with the lower part (blue) of the corresponding cut on the next sheet in a cyclic way, according to (\ref{replica boundary conditions}).}
\label{fig EGmultisheets}
\end{figure}

\begin{figure}[t]
\vspace{.4cm}
\begin{center}
\includegraphics[width=.75\textwidth]{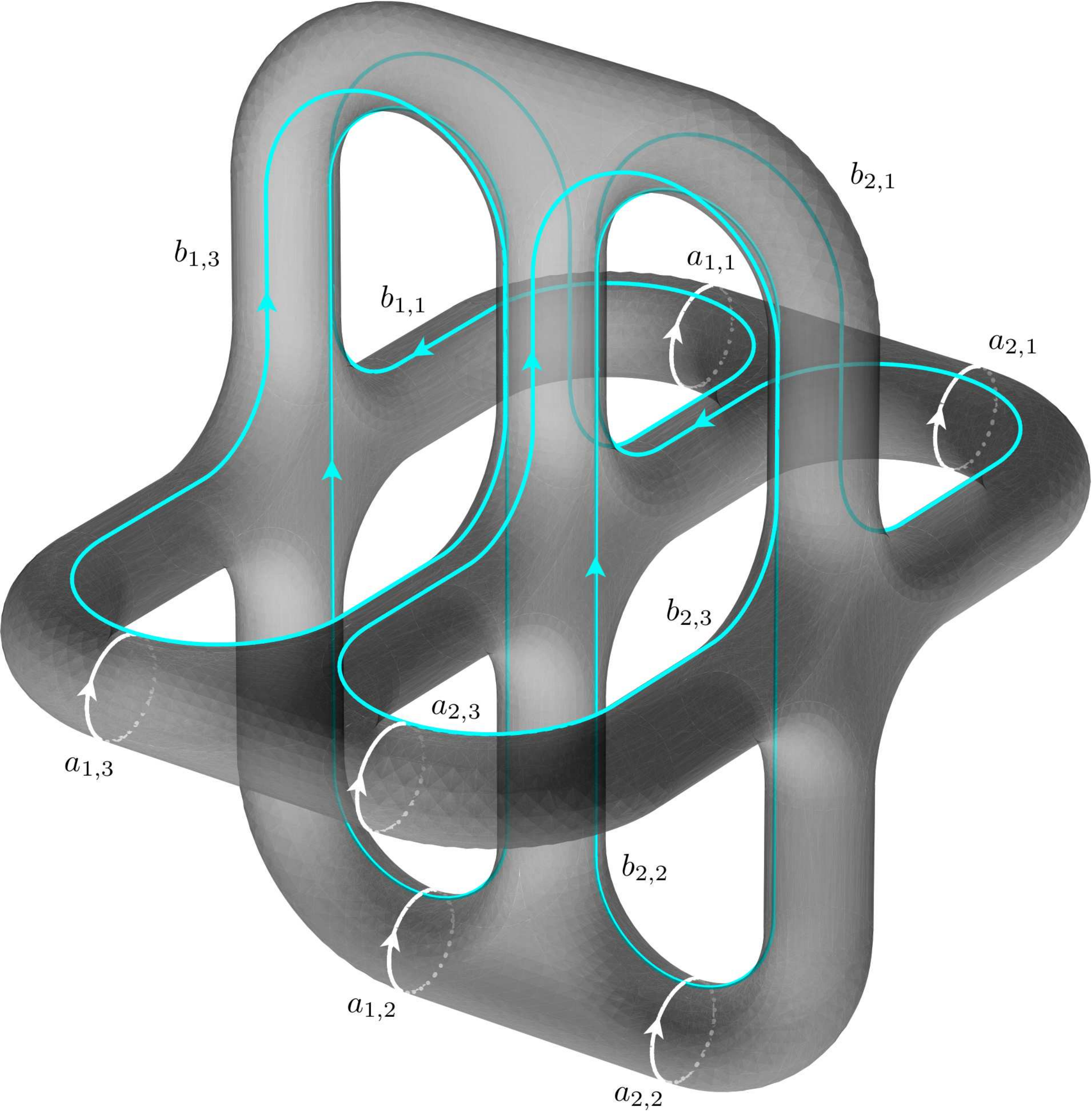}
\end{center}
\vspace{-.1cm}
\caption{The Riemann surface $\mathscr{R}_{3,4}$ with the canonical homology basis $\{ a_{\alpha,j}, b_{\alpha,j} \}$, represented also in Fig. \ref{fig EGmultisheets}.}
\label{fig EGcage}
\end{figure}

The period matrix of the Riemann surface is defined in terms of a canonical homology basis, namely a set of $2g$ closed oriented curves $\{ a_r, b_r\}$ which cannot be contracted to a point  and whose intersections satisfy certain simple relations. 
In particular, defining the intersection number $h \circ \tilde{h}$ between two oriented curves $h$ and $\tilde{h}$ on the Riemann surface as the number of intersection points, with the orientation taken into account (through the tangent vectors at the intersection point and the right hand rule), for a canonical homology basis we have $a_r \circ a_s = b_r \circ b_s =0$ and $a_r\circ b_s =- \,b_r\circ a_s =\delta_{rs}$. By employing the double index notation mentioned above, we choose the canonical homology basis $\{ a_{\alpha,j}, b_{\alpha,j} \}$ adopted in \cite{eg-03}, which is depicted in Fig. \ref{fig EGmultisheets} and in Fig. \ref{fig EGcage} for the special case of $N=3$ intervals and $n=4$ sheets.

Once the canonical homology basis has been chosen, we introduce the $g\times g$ matrices 
\be
\label{ABmat def}
\mathcal{A}_{k,j}^{\beta,\alpha} = \oint_{a_{\alpha,j} } \omega_{\beta,k}\,,
\qquad
\mathcal{B}_{k,j}^{\beta,\alpha} = \oint_{b_{\alpha,j} } \omega_{\beta,k}\,,
\ee
where latin and greek indices run as in (\ref{one forms}). Given the convention adopted above, 
$\mathcal{A}_{k,j}^{\beta,\alpha} $ provides the element $\mathcal{A}_{rs}$ of the $g\times g$ matrix $\mathcal{A}$ by setting $r=\beta+(N-1)(k-1)$ and $s=\alpha+(N-1)(j-1)$ (similarly for $\mathcal{B}$), namely the row index is determined by the one form and the column index by the cycle.
This connection among indices is important because the matrices $\mathcal{A}$ and $\mathcal{B}$ are not symmetric.

From the one forms (\ref{one forms}) and the matrix $\mathcal{A}$ in (\ref{ABmat def}), one constructs the normalized basis of one forms $\nu_r  = \sum_{s=1}^g\mathcal{A}^{-1}_{rs} \omega_s$, which provides the {\it period matrix} $\tau$ as follows
\be
\label{tau def}
\oint_{a_r} \nu_s = \delta_{rs}\,,
\qquad
\oint_{b_r} \nu_s = \tau_{rs}\,,
\qquad
r,s=1, \dots, g\,.
\ee
The period matrix $\tau$ is a $g \times g$ complex and symmetric matrix with positive definite imaginary part, i.e. it belongs to the Siegel upper half space.
Substituting the expression of $\nu_s$ into the definition of $\tau$ in (\ref{tau def}) and employing the definition of the matrix $\mathcal{B}$ in (\ref{ABmat def}), it is straightforward to observe that
\be
\label{period matrix def}
\tau \,= \,\mathcal{A}^{-1} \cdot \mathcal{B} \,\equiv\, \mathcal{R} + {\rm i} \,\mathcal{I} \,,
\ee
where $\mathcal{R}$ and $\mathcal{I}$ are respectively the real and the imaginary part of the period matrix.

In order to compute the period matrix (\ref{period matrix def}), let us introduce the set of auxiliary cycles $\{ a^{\textrm{\tiny aux}}_{\alpha,j}, b^{\textrm{\tiny aux}}_{\alpha,j}  \}$, which is represented in Figs. \ref{fig AUXmultisheets} and \ref{fig AUXcage}. 
It is clear that this set  is not a canonical homology basis. Indeed, some cycles intersect more than one cycle. Nevertheless, we can use them to decompose the cycles of the basis $\{ a_{\alpha,j}, b_{\alpha,j} \}$ as 
\be
\label{eg cycles from aux}
a_{\alpha,j} = \sum_{\gamma \,=\, 1}^\alpha  a^{\textrm{\tiny aux}}_{\gamma,j}\,,
\qquad
b_{\alpha,j} = \sum_{l \,=\, j}^{n-1}  b^{\textrm{\tiny aux}}_{\alpha,l}\,.
\ee
Integrating the one forms (\ref{one forms}) along the auxiliary cycles as shown in (\ref{ABmat def}) for the basis $\{ a_{\alpha,j}, b_{\alpha,j}  \}$, one defines the matrices $\mathcal{A}^{\textrm{\tiny aux}}$ and $\mathcal{B}^{\textrm{\tiny aux}}$.
The advantage of the auxiliary cycles is that the integrals $(\mathcal{A}^{\textrm{\tiny aux}})^{\beta,\alpha}_{k,j}$ and $(\mathcal{B}^{\textrm{\tiny aux}})^{\beta,\alpha}_{k,j}$ on the $j$-th sheet are obtained multiplying the corresponding ones on the first sheet by a phase \cite{dixon}
\be\fl
\label{ABmat aux}
(\mathcal{A}^{\textrm{\tiny aux}})^{\beta,\alpha}_{k,j}
=
\rho_n^{k(j-1)} (\mathcal{A}^{\textrm{\tiny aux}})^{\beta,\alpha}_{k,1}\,, 
\qquad
(\mathcal{B}^{\textrm{\tiny aux}})^{\beta,\alpha}_{k,j}
=
\rho_n^{k(j-1)} (\mathcal{B}^{\textrm{\tiny aux}})^{\beta,\alpha}_{k,1}\,,
\qquad
\rho_n \equiv e^{2\pi \textrm{i}/n}\,.
\ee

Because of the relation (\ref{eg cycles from aux}) among the cycles of the canonical homology basis and the auxiliary ones, the matrices $\mathcal{A}$ and $\mathcal{B}$ in (\ref{ABmat def}) are related to $\mathcal{A}^{\textrm{\tiny aux}}$ and $\mathcal{B}^{\textrm{\tiny aux}}$ as
\begin{eqnarray}
\label{Amat from Aaux}
& & \hspace{-1.5cm}
\mathcal{A}^{\beta,\alpha}_{k,j} =
\sum_{\gamma \,=\, 1}^\alpha  (\mathcal{A}^{\textrm{\tiny aux}})^{\beta,\gamma}_{k,j}=
\rho_n^{k(j-1)} \sum_{\gamma \,=\, 1}^\alpha  
(\mathcal{A}^{\textrm{\tiny aux}})^{\beta,\gamma}_{k,1}\,,
\\
\label{Bmat from Baux}
& & \hspace{-1.5cm}
\mathcal{B}^{\beta,\alpha}_{k,j} =
\sum_{l \,=\, j}^{n-1}  (\mathcal{B}^{\textrm{\tiny aux}})^{\beta,\alpha}_{k,l} =
\sum_{l \,=\, j}^{n-1}  \rho_n^{k(l-1)} (\mathcal{B}^{\textrm{\tiny aux}})^{\beta,\alpha}_{k,1}
= \frac{\rho_n^{k j}-1}{\rho_n^{k} (1-\rho_n^k)}\,
(\mathcal{B}^{\textrm{\tiny aux}})^{\beta,\alpha}_{k,1}\,,
\end{eqnarray}
where the relations (\ref{ABmat aux}) have been used.
Thus, from (\ref{Amat from Aaux}) and (\ref{Bmat from Baux}) we learn that
we just need  $(\mathcal{A}^{\textrm{\tiny aux}})^{\beta,\alpha}_{k,1}$ and $(\mathcal{B}^{\textrm{\tiny aux}})^{\beta,\alpha}_{k,1}$ 
to construct the matrices $\mathcal{A}$ and $\mathcal{B}$.

By carefully considering the phases in the integrand along the cycles, we find
\begin{eqnarray}
\label{Aaux 1k}
(\mathcal{A}^{\textrm{\tiny aux}})^{\beta,\alpha}_{k,1}
=
\oint_{a^{\textrm{\tiny aux}}_{\alpha,1}} \omega_{\beta,k} &=& 
(-1)^{N-\alpha} (\rho_n^{-k} -1)\, 
\mathscr{I}_{\beta,k} \big|_{x_{2\alpha -2}}^{x_{2\alpha -1}}\,,
\\
\rule{0pt}{.7cm}
\label{Baux 1k}
(\mathcal{B}^{\textrm{\tiny aux}})^{\beta,\alpha}_{k,1}
=
\oint_{b^{\textrm{\tiny aux}}_{\alpha,1}} \omega_{\beta,k} &=& 
(-1)^{N-\alpha} \rho_n^{k/2}  (\rho_n^{-k} -1)\, 
\mathscr{I}_{\beta,k}\big|_{x_{2\alpha -1}}^{x_{2\alpha}}\,,
\end{eqnarray}
where we introduced the following integral
\be\fl
\label{integral}
\mathscr{I}_{\beta,k}\big|_a^b \equiv
\int_0^1 
\frac{(b-a) \, \big[ (b-a) t +a \big]^{\beta-1-k/n} \;dt}{
 \prod_{\gamma=2}^N \big| (b-a) t -(x_{2\gamma-2} -a)  \big|^{k/n} 
 \prod_{\gamma=1}^{N-1} \big| (b-a) t -(x_{2\gamma-1} -a)  \big|^{1-k/n}
}.
\ee
We numerically evaluate the integrals needed to get the $g \times g$ matrices $\mathcal{A}$ and $\mathcal{B}$ as explained above and then construct the period matrix $\tau=\mathcal{A}^{-1} \cdot \mathcal{B}$, as in (\ref{period matrix def}).

In Appendix \ref{app lauricella} we write the integrals occurring in (\ref{Aaux 1k}) and (\ref{Baux 1k}) in terms of Lauricella functions, which are generalizations of the hypergeometric functions \cite{exton}.
As a check of our expressions, we employed the formulas for the number of real components of the period matrix found in \cite{Headrick-12}.

Both the matrices in (\ref{Amat from Aaux}) and (\ref{Bmat from Baux}) share the following structure
\be
\label{Hmat case}
\mathcal{H}^{\beta,\alpha}_{k,j} = 
h(k,j)
(\mathcal{H}_k)_{\beta\alpha} \,,
\qquad
(\mathcal{H}_k)_{\beta\alpha} \equiv \mathcal{H}^{\beta,\alpha}_{k,1}\,,
\ee
where we denoted by $\mathcal{H}$ a $g\times g$ matrix whose indices run as explained in the beginning of this subsection, $h$ is a generic function and we also introduced the $(N-1)\times (N-1)$ matrices $\mathcal{H}_k$ labelled by $k=1, \dots, n-1$.
Considering the block diagonal matrix made by the $\mathcal{H}_k$'s, one finds that (\ref{Hmat case}) can be written as
\be\fl
\label{Hmat tensor prod}
\mathcal{H} = \mathcal{H}_{\textrm{d}} \cdot 
(\mathscr{M}_\mathcal{H} \otimes \mathbb{I}_{N-1})\,,
\qquad
\mathcal{H}_{\textrm{d}} \equiv  \textrm{diag}(\dots , \mathcal{H}_k\, , \dots) \,,
\qquad
(\mathscr{M}_\mathcal{H} )_{kj} \equiv h(k,j)\,,
\ee
where we denote by $\mathbb{I}_{p}$ the $p \times p$ identity matrix. For the determinant of (\ref{Hmat tensor prod}), we find 
\be
\label{Hmat tensor prod det}
\textrm{det}(\mathcal{H} ) = \big( \textrm{det}(\mathscr{M}_\mathcal{H} ) \big)^{n-1}
\prod_{k\,=\,1}^{n-1}  \textrm{det}(\mathcal{H}_k)\,.
\ee
Thus, (\ref{Amat from Aaux}) and (\ref{Bmat from Baux}) can be expressed as in (\ref{Hmat tensor prod}) with 
\begin{eqnarray}
\label{Akmat ba}
\fl
 (\mathscr{M}_\mathcal{A})_{kj} & \equiv&  \rho_n^{k(j-1)} \,,
\qquad 
\hspace{1cm}(\mathcal{A}_k)_{\beta\alpha} \equiv  \mathcal{A}^{\beta,\alpha}_{k,1}
= 
(\rho_n^{-k} -1) \sum_{\gamma \,=\, 1}^\alpha  
(-1)^{N-\gamma} 
\mathscr{I}_{\beta,k} \big|_{x_{2\gamma -2}}^{x_{2\gamma -1}}
\,,
\\
\fl
(\mathscr{M}_\mathcal{B})_{kj} &  \equiv& 
\frac{\rho_n^{k j}-1}{\rho_n^{k} (1-\rho_n^{-k})}\,,
\qquad (\mathcal{B}_k)_{\beta\alpha} \equiv  \mathcal{B}^{\beta,\alpha}_{k,1}
= 
(-1)^{N-\alpha} \rho_n^{-k/2}  (1-\rho_n^{-k} )\, 
\mathscr{I}_{\beta,k}\big|_{x_{2\alpha -1}}^{x_{2\alpha}}
\,,
\end{eqnarray}
where (\ref{Aaux 1k}) and (\ref{Baux 1k})  have been employed. 
The period matrix (\ref{period matrix def}) becomes \cite{eg-03}
\be\fl
\label{tau eg tensor}
\tau  = (\mathscr{M}_\mathcal{A} \otimes \mathbb{I}_{N-1})^{-1} \cdot 
\textrm{diag}( \mathcal{A}^{-1}_1 \cdot \mathcal{B}_1 , 
 \mathcal{A}^{-1}_2 \cdot \mathcal{B}_2 , \dots ,
  \mathcal{A}^{-1}_{n-1} \cdot \mathcal{B}_{n-1})
\cdot (\mathscr{M}_\mathcal{B} \otimes \mathbb{I}_{N-1})\,.
\ee
Notice that $\textrm{det} (\mathscr{M}_\mathcal{A}) = \textrm{det} (\mathscr{M}_\mathcal{B}) $ and this implies
\be
 \textrm{det} (\tau) 
  = 
  \textrm{det} \big(
  \textrm{diag}( \mathcal{A}^{-1}_1 \cdot \mathcal{B}_1 , \dots , 
  \mathcal{A}^{-1}_{n-1} \cdot \mathcal{B}_{n-1})\big)
  =
  \prod_{k=1}^{n-1} \frac{\textrm{det} (\mathcal{B}_k) }{\textrm{det} (\mathcal{A}_k) } \,.
\ee
Moreover, since $\textrm{det} (\mathscr{M}_\mathcal{A}) \neq 1$, from the relation (\ref{Hmat tensor prod det}) we have $\textrm{det} (\mathcal{A}) \neq \prod_{k=1}^{n-1} \textrm{det} (\mathcal{A}_k) $ and $\textrm{det} (\mathcal{B}) \neq \prod_{k=1}^{n-1} \textrm{det} (\mathcal{B}_k)$.

\subsection{The partition function}
\label{sucsec partition function}

In order to write the partition function of the free boson on $\mathscr{R}_{N,n}$, we need to introduce the Riemann theta function, which is defined as follows \cite{Fay book, Mumford book}
\be
\label{theta def}
\Theta(\boldsymbol{0} | \Omega ) = \sum_{\boldsymbol{m} \,\in\, \mathbb{Z}^p}
\exp( \textrm{i} \pi\, \boldsymbol{m}^{{\rm t}}\cdot \Omega \cdot \boldsymbol{m}  )\,,
\ee
where $\Omega $ is a $p\times p$ complex, symmetric matrix with positive imaginary part.
Notice that the Riemann theta function $\Theta(\boldsymbol{z}  | \Omega ) $ is defined as a periodic function of a complex vector $\boldsymbol{z} \in \mathbb{C}^p$, but in our problem the special case $\boldsymbol{z} = \boldsymbol{0}$ occurs.

As mentioned at the beginning of this section, we do not explicitly extend the construction of \cite{z-87, dixon, cct-09} to the case $N\geqslant 2$ and $n \geqslant 2$. Given the form of the result for $N = 2$ intervals and $n \geqslant 2$ sheets \cite{pasquier08,cct-09}, we assume its straightforward generalization to $N>2$. 
Let us recall that $\mathcal{F}_{2,n}(x)$ can be obtained as the properly normalized partition function of the model (\ref{action free boson RNn}) on $\mathscr{R}_{2,n}$, once the four endpoints of the two intervals have been mapped to $0$, $x$, $1$ and $\infty$ ($0<x<1$) \cite{cct-09}. Thus, for $N>2$ we compute $\mathcal{F}_{N,n}(\boldsymbol{x})$ in (\ref{FNn def general cft}) as the normalized partition function of (\ref{action free boson RNn}) on $\mathscr{R}_{N,n}$, once (\ref{map wN}) has been applied.

By employing the results of \cite{z-87, amv, verlinde2-86, knizhnik-87, BershadskyRadul, dvv}, for the free compactified boson we can write $\mathcal{F}_{N,n}(\boldsymbol{x}) = \mathcal{F}_{N,n}^{\textrm{\tiny qu}} \, \mathcal{F}_{N,n}^{\textrm{\tiny cl}} (\eta)$, where this splitting comes from the separation of the field as the sum of a classical solution and the quantum fluctuation around it. The classical part is made by the sum over all possible windings around the circular target space and therefore it encodes its compactified nature. This tells us that $\mathcal{F}_{N,n}^{\textrm{\tiny cl}} $ contains all the dependence on the compactification radius through the parameter $\eta \propto R^2$.
We refer the reader to the explicit constructions of \cite{z-87,dixon, cct-09} for the details.

\begin{figure}[t]
\begin{center}
\includegraphics[width=1\textwidth]{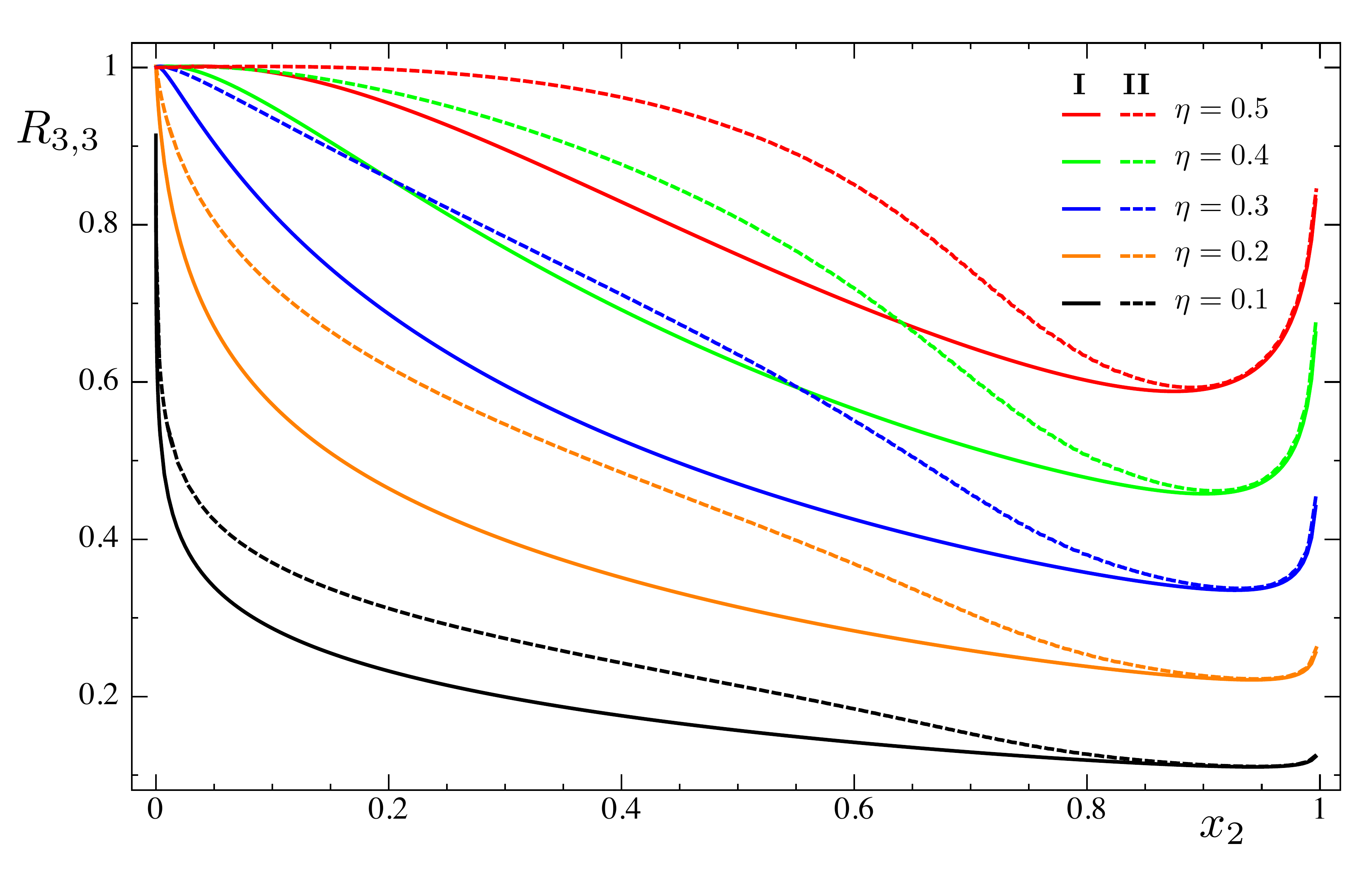}
\end{center}
\vspace{-.5cm}
\caption{The function $R_{3,3}(\boldsymbol{x})$ for the free compactified boson, obtained from (\ref{RNn cft}) and (\ref{FNn final}), computed for two configurations of intervals defined in \S\ref{sec HC} (see Fig. \ref{fig configN3}).}
\label{fig RN3n3eta}
\end{figure}
Given the period matrix $\tau$ for $\mathscr{R}_{N,n}$ computed in \S\ref{subsec period matrix}, the quantum and the classical part in $\mathcal{F}_{N,n}(\boldsymbol{x}) = \mathcal{F}_{N,n}^{\textrm{\tiny qu}} \, \mathcal{F}_{N,n}^{\textrm{\tiny cl}} (\eta)$ read \cite{z-87, amv, verlinde2-86, dvv}
\be\fl
\label{FquFcl}
\mathcal{F}_{N,n}^{\textrm{\tiny qu}}  = 
\frac{1}{| \Theta (\boldsymbol{0} | \tau ) |^2}\,,
\qquad
\mathcal{F}_{N,n}^{\textrm{\tiny cl}} (\eta)
=
\sum_{\boldsymbol{p},\tilde{\boldsymbol{p}}} \exp[
\textrm{i} \pi(\boldsymbol{p}^{{\rm t}}\cdot \tau \cdot \boldsymbol{p} - \tilde{\boldsymbol{p}}^{{\rm t}}\cdot \tau \cdot \tilde{\boldsymbol{p}}) ]\,,
\ee
where
\be
\boldsymbol{p} = \frac{\boldsymbol{m}}{\sqrt{2\eta}} + \frac{\boldsymbol{n}\sqrt{2\eta}}{2}\,,
\qquad
\tilde{\boldsymbol{p} } =  \frac{\boldsymbol{m}}{\sqrt{2\eta}} - \frac{\boldsymbol{n}\sqrt{2\eta}}{2} \,,
\qquad
\boldsymbol{m}, \boldsymbol{n} \in \mathbb{Z}^g\,.
\ee
Expanding the argument of the exponential in (\ref{FquFcl}), one finds that the classical part can be written in terms of the Riemann theta function as
\be
\label{Zclassical}
\mathcal{F}_{N,n}^{\textrm{\tiny cl}} (\eta) =  \Theta (\boldsymbol{0}  | T_\eta ) \,,
\ee
where $T_\eta $ is the following $2g \times 2g$ symmetric matrix
\begin{equation}
\label{Teta def}
T_\eta  = \begin{pmatrix}
\,{\rm i}\, \eta\, \mathcal{I} & \mathcal{R} 
\\
\mathcal{R} & {\rm i}  \,\mathcal{I}/\eta\,
\end{pmatrix}.
\end{equation}
Being $\mathcal{I}$ positive definite and $\eta>0$, also the imaginary part of $T_\eta$ is positive definite.
From (\ref{Zclassical}) and (\ref{Teta def}), it is straightforward to observe that $\mathcal{F}^{{\rm cl}}_{N,n}(\eta) = \mathcal{F}^{{\rm cl}}_{N,n}(1/\eta) $. 
Thus, since all the dependence of $\mathcal{F}_{N,n}(\boldsymbol{x})$ on $\eta$ is contained in $\mathcal{F}^{{\rm cl}}_{N,n}$, we find that $\mathcal{F}_{N,n}(\boldsymbol{x})$ is invariant under $\eta \leftrightarrow 1/\eta$.

\begin{figure}[t]
\begin{center}
\includegraphics[width=1\textwidth]{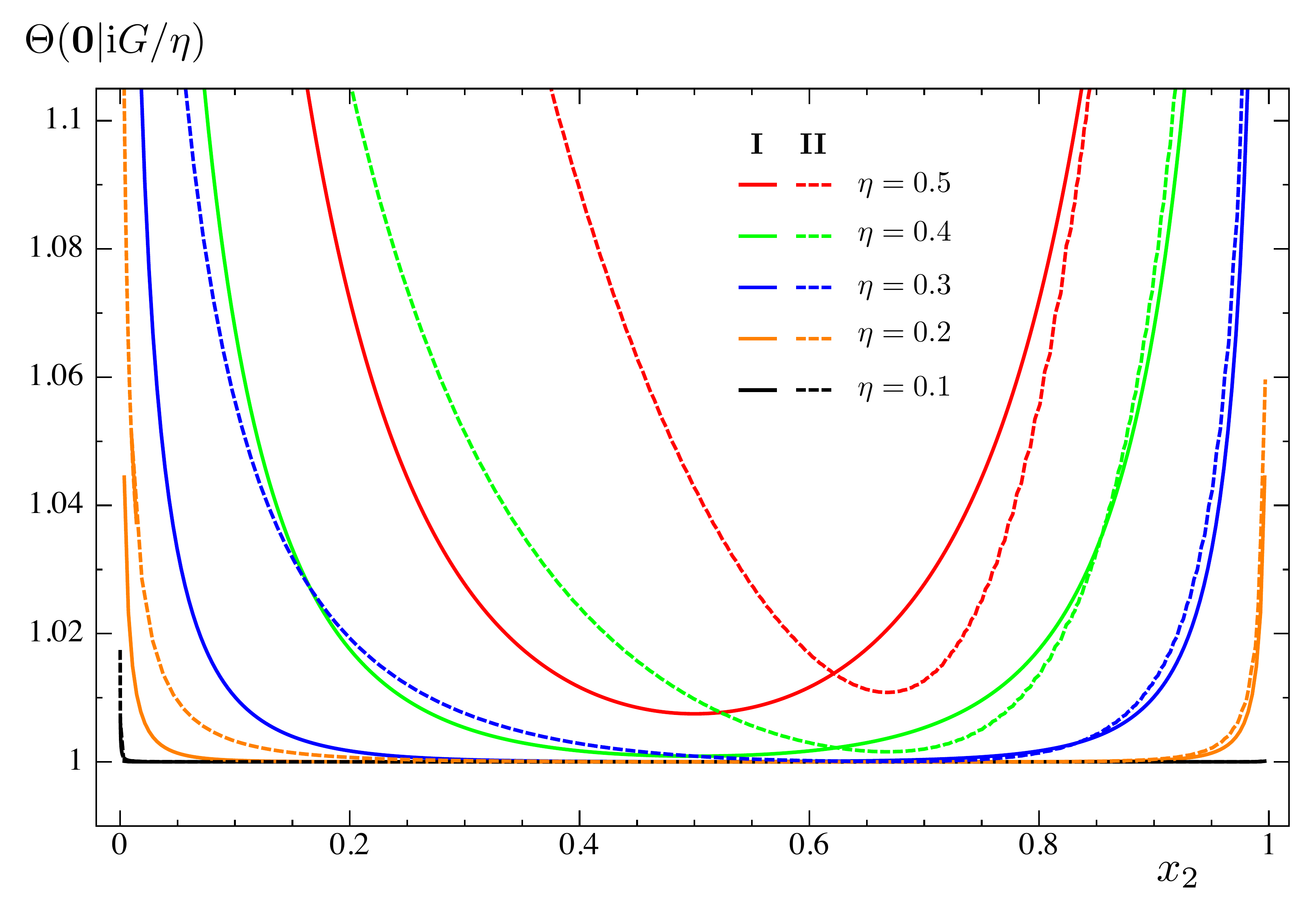}
\end{center}
\vspace{-.5cm}
\caption{The function $\Theta (\boldsymbol{0} | \textrm{i}G/\eta )$ with $N=3$, $n=3$ and for the configurations I and II shown in Fig. \ref{fig configN3}. For small $\eta$ (the decompactification regime) this term can be neglected (see (\ref{Zclassical v2}) and (\ref{FNn final})).}
\label{fig thetadecompact}
\end{figure}

By employing the Poisson summation formula (only for half of the sums), the classical part (\ref{Zclassical}) can be written as 
\be
\label{Zclassical v2}
\mathcal{F}_{N,n}^{\textrm{\tiny cl}} (\eta) 
=  
\eta^{g/2}\,\frac{\Theta (\boldsymbol{0}  | \textrm{i}\eta G ) }{\sqrt{\textrm{det}(\mathcal{I})}}
=
\eta^{-g/2}\,\frac{\Theta (\boldsymbol{0}  | \textrm{i} G/\eta ) }{\sqrt{\textrm{det}(\mathcal{I})}}\,,
\ee
where the $g \times g$ matrix $G$ reads
\be
\label{Gmat def}
G = \begin{pmatrix}
\,\mathcal{I} + \mathcal{R} \cdot \mathcal{I}^{-1} \cdot \mathcal{R} & 
\;\;\mathcal{R}\cdot \mathcal{I}^{-1} \,
\\
\mathcal{I}^{-1} \cdot \mathcal{R} & 
\;\;\mathcal{I}^{-1} 
\end{pmatrix}.
\ee
This matrix is real,  independent of $\eta$ and symmetric, being $\mathcal{R} $ and $\mathcal{I} $ symmetric matrices.\\
Combining (\ref{FquFcl}), (\ref{Zclassical}) and (\ref{Zclassical v2}), we find $\mathcal{F}_{N,n}( \boldsymbol{x}) $ for the free compactified boson 
\begin{equation}\fl
\label{FNn final}
\mathcal{F}_{N,n}( \boldsymbol{x}) 
= 
\frac{\Theta (\boldsymbol{0} | T_\eta )}{| \Theta (\boldsymbol{0} | \tau ) |^2}
=
\frac{\eta^{g/2}\,\Theta (\boldsymbol{0} | \textrm{i}\eta G )}{ \sqrt{\textrm{det}(\mathcal{I})}\, | \Theta (\boldsymbol{0} | \tau ) |^2}
=
\frac{\eta^{-g/2}\,\Theta (\boldsymbol{0} | \textrm{i} G/\eta )}{ \sqrt{\textrm{det}(\mathcal{I})}\, | \Theta (\boldsymbol{0} | \tau ) |^2}\,.
\end{equation}
The term $| \Theta (\boldsymbol{0} | \tau )|$ in the denominator can be rewritten by applying the Thomae type formula for the complex curves (\ref{eq curve}) \cite{eg-03, eg-06} 
\be\fl
\label{thomae formula}
\Theta (\boldsymbol{0} | \tau )^8 
= 
\frac{\prod_{k\,=1}^{n-1} [\textrm{det}(\mathcal{A}_k)]^4}{(2\pi \textrm{i})^{4g}}\,
\Bigg(\,
\prod_{\substack{i,j\,=\,0 \\ i\,<\,j}}^{N-1} 
(x_{2j} - x_{2i})
\prod_{\substack{r,s\,=\,0 \\ r\,<\,s}}^{N-2} 
(x_{2s+1} - x_{2r+1})
\Bigg)^{2(n-1)},
\ee
where the $(N-1)\times (N-1)$ matrices $\mathcal{A}_k$ have been defined in (\ref{Akmat ba}). \\
Plugging (\ref{FNn final}) into (\ref{FNn def general cft}), one finds $\Tr \rho_A^n $ for the free compactified boson in terms of the compactification radius and of the endpoints of the intervals.
Once $\mathcal{F}_{N,n}( \boldsymbol{x}) $ has been found, $\tilde{R}_{N,n}(\boldsymbol{x}) $ and $R_{N,n}(\boldsymbol{x}) $ are obtained through (\ref{tildeRNn cft}) and (\ref{RNn cft}) respectively. 

In \cite{cct-11}  the expansion where all the lengths of the intervals are small with respect to the other characteristic lengths of the systems has been studied.
This means that $x_{2i+1} - x_{2i}$ are small compared to the distances $x_{2j+2} - x_{2j+1}$, where $i,j=0, \dots, N-2$ (we recall that $x_0 = 0$ and $x_{2N-2} = 1$).
Analytic expressions have been found for $N=2$ \cite{cct-11} and one could extend this analysis to $N>2$ by employing (\ref{FNn final}). We leave this analysis for future work.
We checked numerically that $\mathcal{F}_{N,n}( \boldsymbol{0}) = 1$, which generalizes the known result $\mathcal{F}_{2,n}(0) = 1$ \cite{cct-09}.

In  Appendix \ref{sec modular transf} we discuss the invariance of (\ref{FNn final}) under a cyclic change in the ordering of the sheets, an inversion and the exchange $A \leftrightarrow B$, writing explicitly these transformations in terms of symplectic matrices.

\subsection{The decompactification regime}
\label{subsec dec regime}

When $\eta \rightarrow \infty$ the target space of the free boson becomes the infinite line. This regime is important because it can be obtained as the continuum limit of the harmonic chain. Notice that the results of this subsection can be obtained also for $\eta \rightarrow 0$ because of the $\eta \leftrightarrow 1/\eta$ invariance.

Since $\Theta (\boldsymbol{0} | \textrm{i}\eta G ) \rightarrow 1$ when $\eta \rightarrow \infty$ (or equivalently $\Theta (\boldsymbol{0} | \textrm{i} G/\eta ) \rightarrow 1$ when $\eta \rightarrow 0$ as shown in Fig. \ref{fig thetadecompact}), we find that (\ref{FNn final}) becomes 
\begin{equation}
\label{FNn eta large}
\mathcal{F}_{N,n}^{\eta\rightarrow \infty}( \boldsymbol{x}  ) 
=
\frac{\eta^{g/2}}{ \sqrt{\textrm{det}(\mathcal{I})}\, | \Theta (\boldsymbol{0} | \tau ) |^2}
\equiv
\eta^{g/2}\,\widehat{\mathcal{F}}_{N,n}( \boldsymbol{x}  ) \,.
\end{equation}
Writing $| \Theta (\boldsymbol{0} | \tau )|$ through (\ref{thomae formula}), one can improve the numerical evaluation of (\ref{FNn eta large}).\\
Plugging (\ref{FNn eta large}) into (\ref{RNn cft}), we find that in the decompactification regime $R_{N,n}$ becomes
\be
\label{RNn eta large}
R_{N,n}^{\eta \rightarrow \infty}( \boldsymbol{x}  ) = \eta^{(-1)^N(n-1)/2}
\prod_{p\,=\,2}^{N}\, \prod_{\sigma_{N,p}} 
\big[ \widehat{\mathcal{F}}_{p,n}(\boldsymbol{x}^{\sigma_{N,p}}) \big]^{(-1)^{N-p}} \,.
\ee
In this case it is very useful to consider the normalization (\ref{RNn cft norm}) through a fixed configuration of intervals characterized by $\boldsymbol{x}_{\textrm{\tiny{fixed}}}$ because the dependence on $\eta$ simplifies in the ratio. Indeed, from (\ref{RNn eta large}) we find 
\be
\label{rNn dec def}
\lim_{\eta\,\rightarrow\,\infty} R_{N,n}^{\textrm{\tiny{norm}}}( \boldsymbol{x}) =
\frac{R_{N,n}^{\eta \rightarrow \infty}( \boldsymbol{x})}{R_{N,n}^{\eta \rightarrow \infty}( \boldsymbol{x}_{\textrm{\tiny{fixed}}})}
=
\prod_{p\,=\,2}^{N}\, \prod_{\sigma_{N,p}} 
\left[
\frac{\widehat{\mathcal{F}}_{p,n}(\boldsymbol{x}^{\sigma_{N,p}})}{\widehat{\mathcal{F}}_{p,n}(\boldsymbol{x}^{\sigma_{N,p}}_{\textrm{\tiny{fixed}}})}
\right]^{(-1)^{N-p}} ,
\ee
and similarly, from (\ref{FNn eta large}), we have
\be
\label{FNn dec def}
\lim_{\eta\,\rightarrow\,\infty} \mathcal{F}_{N,n}^{\textrm{\tiny{norm}}}( \boldsymbol{x}) 
=
\frac{\mathcal{F}_{N,n}^{\eta \rightarrow \infty}( \boldsymbol{x})}{\mathcal{F}_{N,n}^{\eta \rightarrow \infty}( \boldsymbol{x}_{\textrm{\tiny{fixed}}})}
=
\frac{\widehat{\mathcal{F}}_{N,n}( \boldsymbol{x})}{\widehat{\mathcal{F}}_{N,n}( \boldsymbol{x}_{\textrm{\tiny{fixed}}})}\,.
\ee
In \S\ref{sec HC} we compare (\ref{rNn dec def}) and (\ref{FNn dec def}) to the corresponding results for the harmonic chain with periodic boundary conditions.

\subsection{The Dirac model}
\label{sec dirac}

It is well known that the partition function of the compactified massless free boson describes various systems at criticality. For example, the free Dirac fermion corresponds to the case $\eta=1/2$. Given (\ref{FNn final}), we can write $\Tr \rho_A^n$ for this model by applying the results of \cite{amv, verlinde2-86, amv bosonization, dvv}. Let us introduce the Riemann theta function with characteristic $\boldsymbol{e}^{{\rm t}} = (\boldsymbol{\varepsilon}^{{\rm t}} , \boldsymbol{\delta}^{{\rm t}} )$, namely
\be\fl
\label{theta def spin}
\Theta[\boldsymbol{e} ](\boldsymbol{z} | \Omega ) 
= \sum_{\boldsymbol{m} \,\in\, \mathbb{Z}^p}
\exp\big[ \textrm{i} \pi (\boldsymbol{m} + \boldsymbol{\varepsilon})^{{\rm t}}\cdot \Omega \cdot (\boldsymbol{m} + \boldsymbol{\varepsilon}) 
+ 2\pi \textrm{i}  (\boldsymbol{m} + \boldsymbol{\varepsilon})^{{\rm t}}\cdot 
(\boldsymbol{z} + \boldsymbol{\delta})
 \big]\,,
\ee
where $\boldsymbol{z}  \in \mathbb{C}^p/(\mathbb{Z}^p + \tau\,\mathbb{Z}^p)$ is the independent variable, while $\boldsymbol{\varepsilon}$ and $\boldsymbol{\delta}$ are vectors whose entries are either 0 or $1/2$. 
When $\boldsymbol{\varepsilon} = \boldsymbol{\delta}  = \boldsymbol{0}$ and $\boldsymbol{z} = \boldsymbol{0}$, we recover (\ref{theta def}).
The parity of (\ref{theta def spin})  is the same one of the integer number $4\boldsymbol{\varepsilon} \cdot \boldsymbol{\delta} $; indeed
\be
\Theta[\boldsymbol{e} ]( - \boldsymbol{z} | \Omega ) 
=
(-1)^{4\boldsymbol{\varepsilon} \cdot \boldsymbol{\delta} }\,
\Theta[\boldsymbol{e} ](\boldsymbol{z} | \Omega ) \,.
\ee
The characteristics $\boldsymbol{e}$ are either even or odd, according to the parity of $4\boldsymbol{\varepsilon} \cdot \boldsymbol{\delta} $. It is not difficult to realize that there are $2^{p-1}(2^p+1)$ even characteristics, $2^{p-1}(2^p-1)$ odd ones.

Applying some identities for the Riemann theta functions, from (\ref{FNn final}) one finds 
\be
\label{FNn dirac}
\mathcal{F}^{\textrm{\tiny Dirac}}_{N,n}( \boldsymbol{x}) =
\mathcal{F}_{N,n}( \boldsymbol{x}) \big|_{\eta =1/2}=
\frac{\sum_{\boldsymbol{e}} | \Theta [\boldsymbol{e} ](\boldsymbol{0} | \tau ) |^2}{2^g \,| \Theta (\boldsymbol{0} | \tau ) |^2} \,,
\ee
where the period matrix $\tau$ has been computed in \S\ref{subsec period matrix}. Notice that,
being $\Theta[\boldsymbol{e} ](\boldsymbol{0} | \Omega ) = 0$ when $\boldsymbol{e} $ is odd, in the sum over the characteristics in (\ref{FNn dirac}) only the even ones occur. 
Since (\ref{FNn dirac}) has been obtained as the special case $\eta=1/2$ of (\ref{FNn final}),
$\mathcal{F}^{\textrm{\tiny Dirac}}_{N,n}( \boldsymbol{0}) = 1$.
The result of \cite{CasiniFoscoHuerta} corresponds to keep only $\boldsymbol{e} = \boldsymbol{0} $ in the numerator of (\ref{FNn dirac}) instead of considering the sum over all the sectors of the model.
We refer the reader to \cite{Headrick-12} for a detailed comparison between these two approaches.

\section{Recovering the two intervals case}
\label{sec 2int case}

It is not straightforward to recover the known results for two intervals \cite{cct-09, cct-11}, whose generalization allowed to study the partial transposition and the negativity for a two dimensional CFT \cite{cct-neg-letter,cct-neg-long}. In this section we first review the status of the two intervals case and then we show that the corresponding R\'enyi entropies reduce to a particular case of the expressions discussed in \S\ref{sec compact boson}, as expected.

\subsection{Two disjoint intervals and partial transposition}

The negativity \cite{vw-01} provides a good measure of entanglement for mixed states. Considering a bipartition where $A$ is made by two disjoint intervals, the negativity can be found as a replica limit $n_e \rightarrow 1$ of $ \Tr(\rho_A^{T_{A_2}})^{n_e}$ where $n_e$ is an even number and $\rho_A^{T_{A_2}}$ is obtained by taking $\rho_A$ and partially transpose it with respect to the second interval.
For a two dimensional CFT, it turns out that $\Tr(\rho_A^{T_{A_2}})^n$ is obtained by considering the four point function $\langle \mathcal{T}_n \bar{\mathcal{T}}_n \mathcal{T}_n \bar{\mathcal{T}}_n  \rangle$, and exchanging the twist fields $\mathcal{T}_n$ and $\bar{\mathcal{T}}_n$ at the endpoints af $A_2$. In terms of the harmonic ratio $x$ of the four points, while for the R\'enyi entropies it is enough to consider $x \in (0,1)$, the partial transposition forces us to include also the range $x<0$. 
For generic positions of the twist fields in the complex plane, $x \in \mathbb{C}$ and the corresponding expression of $\langle \mathcal{T}_n \bar{\mathcal{T}}_n \mathcal{T}_n \bar{\mathcal{T}}_n  \rangle$ is given by the r.h.s. of (\ref{FNn def general cft}) with $\mathcal{F}_{2,n} = \mathcal{F}_{2,n}(x,\bar{x})$.

For the free compactified boson, this function reads  \cite{cct-neg-long}
\be\fl
\label{F2 neg}
\mathcal{F}_{2,n}(x,\bar{x}) 
= \frac{\Theta(\boldsymbol{0}|T_{\eta, 2})}{\prod_{k=1}^{n-1} |F_{k/n}(x)|}
= \frac{\Theta(\boldsymbol{0}|T_{\eta, 2})}{|\Theta(\boldsymbol{0}|\tau_2)|^2}\,,
\qquad
F_{k/n}(x) \equiv\, _2F_1(k/n,1-k/n;1;x)\,,
\ee
where $T_{\eta, 2}$ is the $2(n-1) \times 2(n-1)$ symmetric matrix given by
\be
\label{Teta2 def}
T_{\eta, 2} = 
\begin{pmatrix}
\,{\rm i}\, \eta\, \textrm{Im}(\tau_2) &\textrm{Re}(\tau_2)
\\
\textrm{Re}(\tau_2) & {\rm i}  \,\textrm{Im}(\tau_2) /\eta\,
\end{pmatrix},
\ee
defined in terms of the following $(n-1) \times (n-1)$ complex and symmetric matrix
\be
\label{tau2 matrix}
(\tau_2)_{ij} = \frac{2}{n} \sum_{k=1}^{n-1} \sin(\pi k/n) 
\bigg[\, \textrm{i}\, \frac{F_{k/n}(1-x)}{F_{k/n}(x)} \,\bigg]
\cos[2\pi k/n(i-j)]\,.
\ee
The matrix $T_{\eta, 2}$ in (\ref{F2 neg}) is defined as in  (\ref{Teta def}) with $\tau_2$ instead of $\tau$. In the second step of (\ref{F2 neg}) the Thomae formula (\ref{thomae formula}) has been employed. Notice that, because of the sum over $k$ in (\ref{tau2 matrix}), substituting $\cos[2\pi k/n(i-j)]$ with $\rho_n^{k(i-j)}$ the matrix does not change.
The non vanishing of $\textrm{Re}(\tau_2)$ is due to the fact that the term within the square brackets in (\ref{tau2 matrix}) is complex for $x\in \mathbb{C}$.

As briefly explained in \S\ref{sec dirac}, it is straightforward to write the corresponding result for the Dirac model from (\ref{F2 neg}). It reads
\be
\label{F2 neg dirac}
\mathcal{F}^{\textrm{\tiny Dirac}}_{2,n}(x,\bar{x}) 
= 
\frac{\sum_{\boldsymbol{e}} | \Theta [\boldsymbol{e} ](\boldsymbol{0} | \tau_2 ) |^2}{2^{n-1} \, | \Theta (\boldsymbol{0} | \tau_2 ) |^2} \,.
\ee
Given the period matrix (\ref{tau2 matrix}), one can also find $\mathcal{F}^{\textrm{\tiny Ising}}_{2,n}(x,\bar{x}) $ for the Ising model \cite{alba13 neg, ctt-neg-ising}
\be
\label{F2 neg ising}
\mathcal{F}^{\textrm{\tiny Ising}}_{2,n}(x,\bar{x}) 
= 
\frac{\sum_{\boldsymbol{e}} | \Theta [\boldsymbol{e} ](\boldsymbol{0} | \tau_2 ) |}{2^{n-1} \,| \Theta (\boldsymbol{0} | \tau_2 ) |} \,.
\ee

In order to consider the R\'enyi entropies, we must restrict to $x\in (0,1)$. Within this domain, $F_{k/n}(x)$ is real and this leads to a  purely imaginary $\tau_2$. 
Since $\textrm{Re}(\tau_2)$ vanishes identically for $x\in (0,1)$, the matrix $T_{\eta, 2} $ in (\ref{Teta2 def}) becomes block diagonal and therefore $\Theta(\boldsymbol{0}|T_{\eta, 2}) = \Theta(\boldsymbol{0}|\,{\rm i} \eta\, \textrm{Im}(\tau_2) )\, \Theta(\boldsymbol{0}|\,{\rm i}\, \textrm{Im}(\tau_2)/\eta )$ factorizes. Thus, the expressions given in (\ref{F2 neg}) and (\ref{F2 neg ising})  reduce to $\mathcal{F}_{2,n}(x) $ for the free compactified boson \cite{cct-09} and for the Ising model \cite{cct-11} respectively.

\begin{figure}[t]
\vspace{.5cm}
\begin{center}
\includegraphics[width=.85\textwidth]{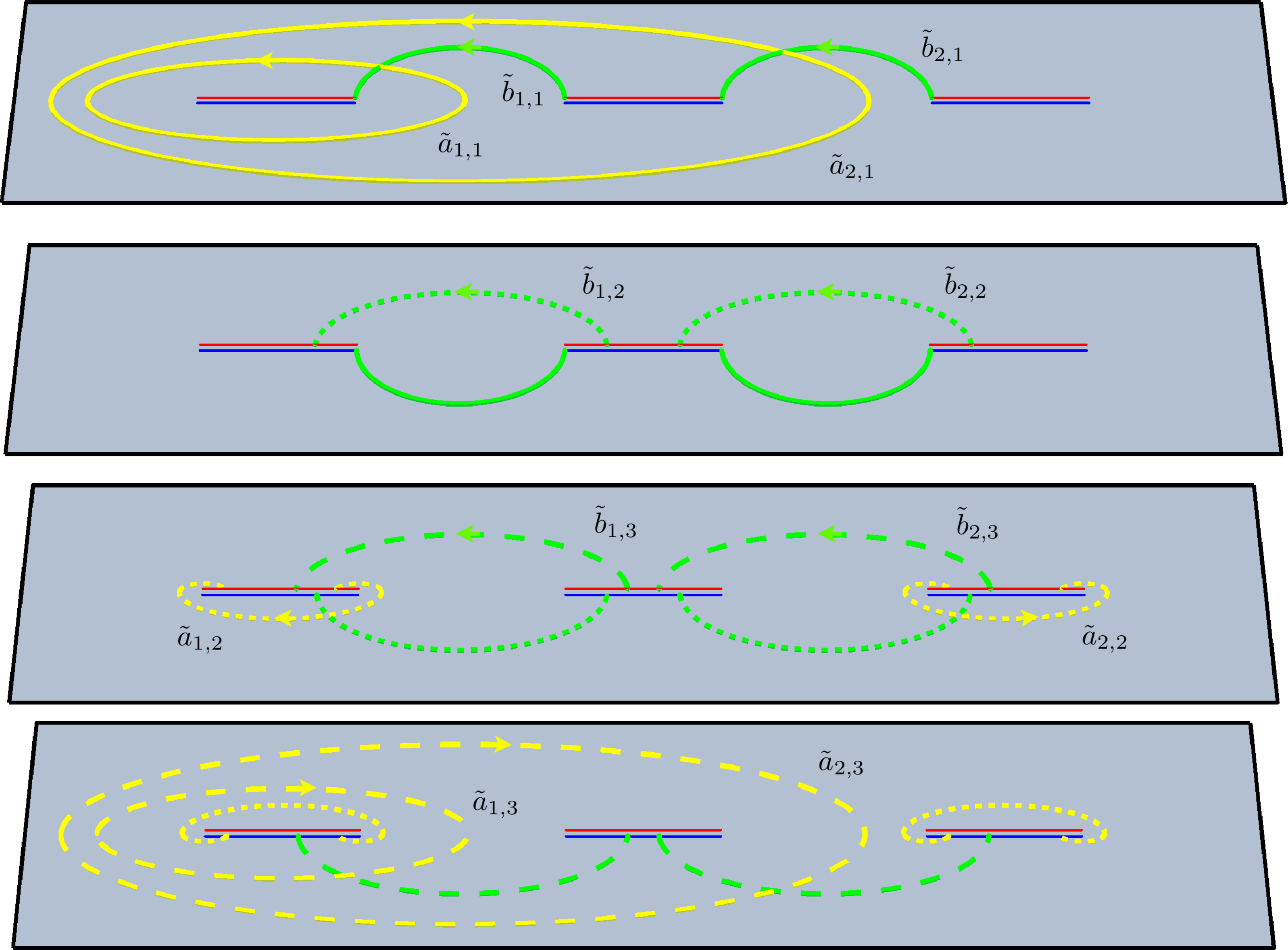}
\end{center}
\caption{The canonical homology basis $\{ \tilde{a}_{\alpha,j}, \tilde{b}_{\alpha,j} \}$ for $N=3$ and $n=4$.}
\label{fig AMVmultisheets}
\end{figure}

\begin{figure}[t]
\vspace{.4cm}
\begin{center}
\includegraphics[width=.75\textwidth]{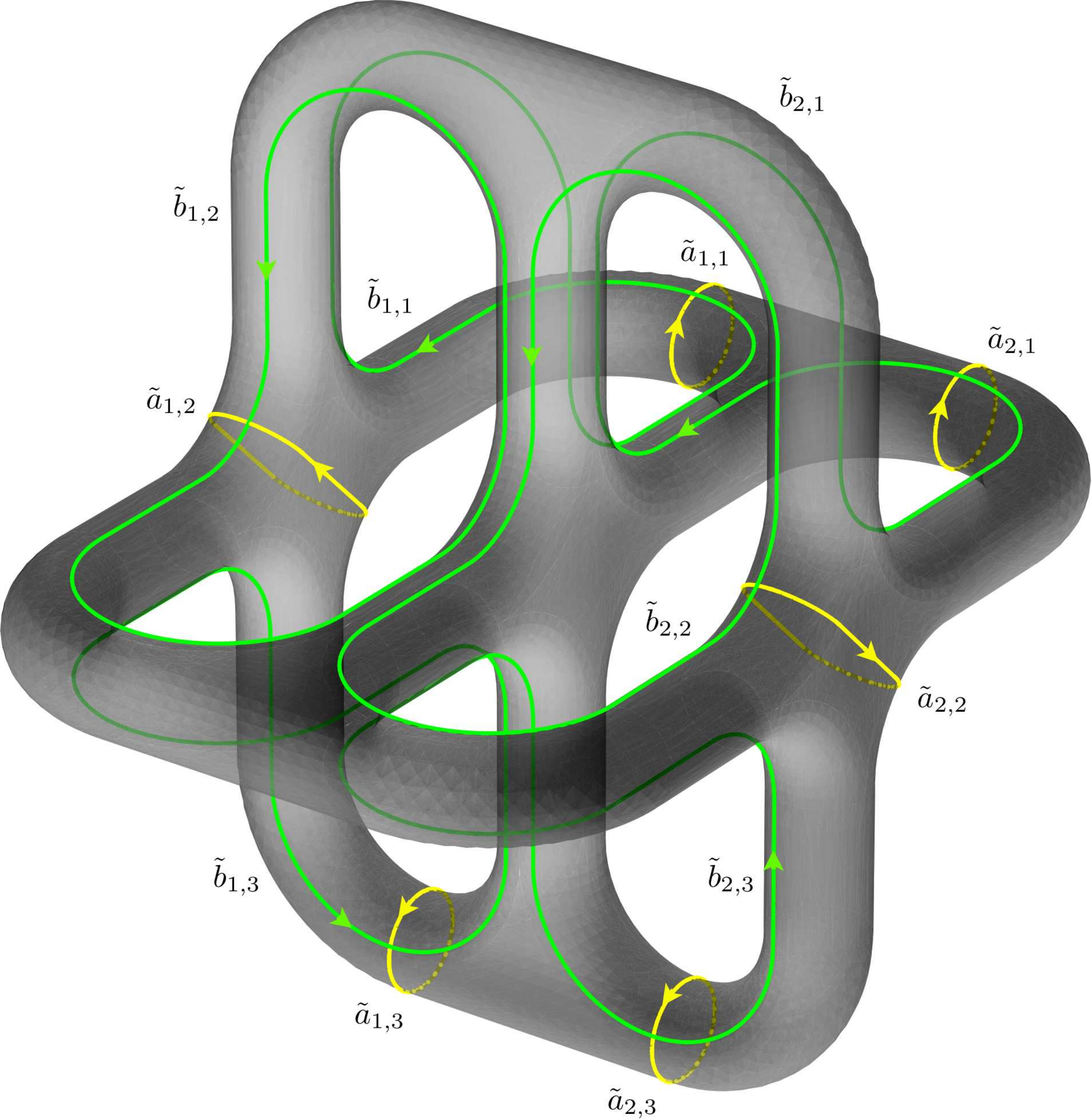}
\end{center}
\vspace{-.1cm}
\caption{The Riemann surface $\mathscr{R}_{3,4}$ with the canonical homology basis $\{ \tilde{a}_{\alpha,j}, \tilde{b}_{\alpha,j} \}$.}
\label{fig AMVcage}
\end{figure}

\subsection{Another canonical homology basis}
\label{subsec amv}

To recover the period matrix (\ref{tau2 matrix}) for $x\in(0,1)$ as the two intervals case of a period matrix characterizing $N \geqslant 2$ intervals, we find it useful to introduce the canonical homology basis $\{ \tilde{a}_{\alpha,j}, \tilde{b}_{\alpha,j} \}$ depicted in Figs. \ref{fig AMVmultisheets} and \ref{fig AMVcage}. This basis is considered very often in the literature on higher genus Riemann surfaces (e.g. see Fig. 1 both in \cite{amv} and \cite{verlinde2-86}).
Integrating the holomorphic differentials (\ref{one forms}) along the cycles $\tilde{\boldsymbol{a}}$ and $\tilde{\boldsymbol{b}}$, as done in (\ref{ABmat def}) for the untilded ones, one gets the matrices $\tilde{\mathcal{A}}$ and $\tilde{\mathcal{B}}$. To evaluate these matrices, we repeat the procedure described in \S\ref{subsec period matrix}. In particular, we first write $\{ \tilde{a}_{\alpha,j}, \tilde{b}_{\alpha,j} \}$ through the auxiliary cycles depicted in Figs.  \ref{fig AUXmultisheets} and \ref{fig AUXcage}, finding that
\be
\label{amv cycles from aux}
\tilde{a}_{\alpha,j} = 
\sum_{\gamma = 1}^\alpha  \sum_{l =1}^j  a^{\textrm{\tiny aux}}_{\gamma,l}\,,
\qquad
\tilde{b}_{\alpha,j} =  b^{\textrm{\tiny aux}}_{\alpha,j}\,.
\ee
Comparing (\ref{eg cycles from aux}) with (\ref{amv cycles from aux}), we observe that for $n=2$ the canonical homology basis introduced here coincides with the one defined in \S\ref{subsec period matrix}.
From (\ref{amv cycles from aux}), one can write the matrices $\tilde{\mathcal{A}}$ and $\tilde{\mathcal{B}}$ as follows
\begin{eqnarray}
\label{tildeAmat}
& & \hspace{-1.5cm}
\tilde{\mathcal{A}}^{\beta,\alpha}_{k,j}  =
\sum_{\gamma = 1}^\alpha  \sum_{l =1}^j 
(\mathcal{A}^{\textrm{\tiny aux}})^{\beta,\gamma}_{k,l}=
\sum_{\gamma = 1}^\alpha  \sum_{l =1}^j 
\rho_n^{k(l-1)} (\mathcal{A}^{\textrm{\tiny aux}})^{\beta,\gamma}_{k,1}
=
\frac{1-\rho_n^{j k}}{1-\rho_n^k} \,
\sum_{\gamma = 1}^\alpha  (\mathcal{A}^{\textrm{\tiny aux}})^{\beta,\gamma}_{k,1}\,,
\\
\rule{0pt}{.5cm}
\label{tildeBmat}
& & \hspace{-1.5cm}
\tilde{\mathcal{B}}^{\beta,\alpha}_{k,j} =
(\mathcal{B}^{\textrm{\tiny aux}})^{\beta,\alpha}_{k,j} = 
\rho_n^{k(j-1)} (\mathcal{B}^{\textrm{\tiny aux}})^{\beta,\alpha}_{k,1}\,,
\end{eqnarray}
where (\ref{ABmat aux}) has been used. Now the elements of $\tilde{\mathcal{A}}$ and $\tilde{\mathcal{B}}$ are expressed in terms of the integrals (\ref{Aaux 1k}) and (\ref{Baux 1k}), which can be numerically evaluated. Once $\tilde{\mathcal{A}}$ and $\tilde{\mathcal{B}}$ have been computed, the period matrix with respect to the basis $\{ \tilde{a}_{\alpha,j}, \tilde{b}_{\alpha,j} \}$  is $\tilde{\tau} = \tilde{\mathcal{A}}^{-1} \cdot \tilde{\mathcal{B}}$.

Since the matrices $\tilde{\mathcal{A}}$ and $\tilde{\mathcal{B}}$ have the structure (\ref{Hmat case}), like $\mathcal{A}$ and $\mathcal{B}$ in \S\ref{subsec period matrix}, we can write them as in (\ref{Hmat tensor prod}) with  
\begin{eqnarray}
\label{Atildekmat ba}
\fl
(\mathscr{M}_{\tilde{\mathcal{A}}})_{kj} &\equiv& \frac{1-\rho_n^{k j}}{1-\rho_n^k}  \,,
\qquad 
\hspace{.8cm}(\tilde{\mathcal{A}}_k)_{\beta\alpha} \equiv  \tilde{\mathcal{A}}^{\beta,\alpha}_{k,1}
= 
(\rho_n^{-k} -1) \sum_{\gamma \,=\, 1}^\alpha  
(-1)^{N-\gamma} 
\mathscr{I}_{\beta,k} \big|_{x_{2\gamma -2}}^{x_{2\gamma -1}}
\,
\\
\label{Btildekmat ba}
\fl
(\mathscr{M}_{\tilde{\mathcal{B}}})_{kj} &\equiv& \rho_n^{k(j-1)}  \,,
\qquad \hspace{1.02cm}
(\tilde{\mathcal{B}}_k)_{\beta\alpha} \equiv  \tilde{\mathcal{B}}^{\beta,\alpha}_{k,1}
= 
(-1)^{N-\alpha} \rho_n^{k/2}  (\rho_n^{-k} -1)\, 
\mathscr{I}_{\beta,k}\big|_{x_{2\alpha -1}}^{x_{2\alpha}}
\,.
\end{eqnarray}
where (\ref{Aaux 1k}) and (\ref{Baux 1k}) have been employed and $\mathscr{I}_{\beta,k}\big|_{b}^{a}$ are the integrals (\ref{integral}).
Notice that $\tilde{\mathcal{A}}_k = \mathcal{A}_k$, while $(\tilde{\mathcal{B}}_k)_{\beta\alpha} = (\mathcal{B}^{\textrm{\tiny aux}})^{\beta,\alpha}_{k,1} = -\rho_n^k (\mathcal{B}_k)_{\beta\alpha} $.
Thus, the period matrix $\tilde{\tau}$ reads
\be\fl
\label{tau amv tensor}
\tilde{\tau}  = (\mathscr{M}_{\tilde{\mathcal{A}}} \otimes \mathbb{I}_{N-1})^{-1} \cdot 
\textrm{diag}( \tilde{\mathcal{A}}^{-1}_1 \cdot \tilde{\mathcal{B}}_1 , 
 \tilde{\mathcal{A}}^{-1}_2 \cdot \tilde{\mathcal{B}}_2 , \dots ,
  \tilde{\mathcal{A}}^{-1}_{n-1} \cdot \tilde{\mathcal{B}}_{n-1})
\cdot (\mathscr{M}_{\tilde{\mathcal{B}}} \otimes \mathbb{I}_{N-1})\,.
\ee

Since (\ref{tau eg tensor}) and (\ref{tau amv tensor}) are the period matrices of the Riemann surface $\mathscr{R}_{N,n}$ with respect to different canonical homology bases, they must be related through a symplectic transformation. The relations (\ref{eg cycles from aux}) and (\ref{amv cycles from aux}) in the matrix form become respectively
\be
\label{auxdep matform}
\left\{ \begin{array}{l}
\boldsymbol{a} =  A \cdot \boldsymbol{a}^{\textrm{\tiny aux}}\\
\boldsymbol{b} =  B \cdot \boldsymbol{b}^{\textrm{\tiny aux}}
\end{array}\right. ,
\qquad
\left\{ \begin{array}{l}
\tilde{\boldsymbol{a}} =  \tilde{A} \cdot \boldsymbol{a}^{\textrm{\tiny aux}}\\
\tilde{\boldsymbol{b}} =  \boldsymbol{b}^{\textrm{\tiny aux}}
\end{array}\right. .
\ee
Introducing the $p\times p$ upper triangular matrix $I_p^{\textrm{\tiny up}}$ made by $1$'s (i.e. $(I_p^{\textrm{\tiny up}})_{ab} = 1$ if $a \leqslant b$ and zero otherwise) and also its transposed $I_p^{\textrm{\tiny low}} \equiv (I_p^{\textrm{\tiny up}})^\textrm{t}$, which is a lower triangular matrix, we can write that $A = \mathbb{I}_{n-1} \otimes I_{N-1}^{\textrm{\tiny low}}$, 
$B = I_{n-1}^{\textrm{\tiny up}} \otimes \mathbb{I}_{N-1} $ and 
$\tilde{A} = I_{n-1}^{\textrm{\tiny low}}  \otimes I_{N-1}^{\textrm{\tiny low}}$.
We remark that the matrices $\textrm{diag}(A, B)$ and $\textrm{diag}(\tilde{A}, \mathbb{I}_g)$ occurring in (\ref{auxdep matform}) are not symplectic matrices because, as already noticed in \S\ref{subsec period matrix}, the auxiliary set of cycles is not a canonical homology basis.
From (\ref{auxdep matform}) it is straightforward to find the relation between the two canonical homology bases, namely
\be
\label{cyc2tildecyc}
\left\{ \begin{array}{l}
\tilde{\boldsymbol{a}} =  \tilde{A} \cdot A^{-1} \cdot \boldsymbol{a}\\
\tilde{\boldsymbol{b}} =  B^{-1} \cdot  \boldsymbol{b}
\end{array}\right.,
\qquad
M\equiv 
\begin{pmatrix}
 \tilde{A} \cdot A^{-1}  & 0_g \\ 0_g & B^{-1} 
\end{pmatrix}
\in Sp(2g, \mathbb{Z})\,,
\ee
which can be constructed by using that $(I_p^{\textrm{\tiny up}})_{ab}^{-1} = \delta_{a,b} - \delta_{a+1,b}$ and the properties of the tensor product, finding $\tilde{A} \cdot A^{-1} = I_{n-1}^{\textrm{\tiny low}} \otimes \mathbb{I}_{N-1} $ and $B^{-1} =   (I_{n-1}^{\textrm{\tiny up}})^{-1}\otimes \mathbb{I}_{N-1} $.
Notice that (\ref{cyc2tildecyc}) belongs to the symplectic modular group, as expected from the fact that it encodes the change between canonical homology bases.

\subsection{The case $N=2$}

Specializing the expressions given in the previous subsection to the $N=2$ case, the greek indices assume only a single value; therefore they can be suppressed. The matrices (\ref{tildeAmat}) and (\ref{tildeBmat}) become respectively
\begin{eqnarray}
\fl
\label{tildeAmat N2}
\tilde{\mathcal{A}}_{kj} 
 & \equiv &
\tilde{\mathcal{A}}^{1,1}_{k,j} =
\frac{1-\rho_n^{j k}}{1-\rho_n^k}  \Big[  (1-\rho_n^{-k}) \mathscr{I}_{1,k} \big|_{0}^{x}\, \Big]
=
\frac{1-\rho_n^{k j}}{1-\rho_n^k} 
\big[2\pi \textrm{i}\, \rho_n^{-k/2} F_{k/n}(x)\big]\,,
\\
\rule{0pt}{.6cm}
\label{tildeBmat N2}
\fl
\tilde{\mathcal{B}}_{kj} 
& \equiv &
\tilde{\mathcal{B}}^{1,1}_{k,j} =
\rho_n^{k(j-1)}   \Big[  \rho_n^{k/2}(1-\rho_n^{-k}) \mathscr{I}_{1,k} \big|_{x}^{1}\, \Big]=
\rho_n^{k(j-1)}  \big[2\pi \textrm{i} \,   F_{k/n}(1-x)\big]\,,
\end{eqnarray}
where $x\in (0,1)$, the indices $j,k\in \{1, \dots, n-1\}$ and the explicit results for $(\mathcal{A}^{\textrm{\tiny aux}})_{k,1}$ and $(\mathcal{B}^{\textrm{\tiny aux}})_{k,1}$, from (\ref{Aaux 1k}) and (\ref{Baux 1k}) respectively, are written within the square brackets (see (4.29) of \cite{dixon} and also (\ref{N2lauricellaA}) and (\ref{N2lauricellaB})).
The matrices (\ref{tildeAmat N2}) and (\ref{tildeBmat N2}) can be written respectively as follow
\begin{eqnarray}
\label{tildeAmat N2 v2}
\tilde{\mathcal{A}}
 & = & \textrm{diag}(\dots , 2\pi \textrm{i}\, \rho_n^{-k/2} F_{k/n}(x), \dots) 
 \cdot \mathscr{M}_{\tilde{\mathcal{A}}}\,,
\\
\rule{0pt}{.5cm}
\label{tildeBmat N2 v2}
\tilde{\mathcal{B}}
& = &
\textrm{diag}(\dots , 2\pi \textrm{i} \,   F_{k/n}(1-x), \dots) 
 \cdot \mathscr{M}_{\tilde{\mathcal{B}}}\,,
\end{eqnarray}
where $\mathscr{M}_{\tilde{\mathcal{A}}}$ and $\mathscr{M}_{\tilde{\mathcal{B}}}$ have been defined in (\ref{Atildekmat ba}) and (\ref{Btildekmat ba}) respectively.
Computing $\mathscr{M}_{\tilde{\mathcal{A}}}^{-1}$, whose elements read $(\mathscr{M}_{\tilde{\mathcal{A}}}^{-1})_{ik} = (\rho_n^k-1)/(n \rho_n^{ik})$, we can easily check that (\ref{tau2 matrix}) becomes
\be
\tau_2 = \tilde{\mathcal{A}}^{-1} \cdot \tilde{\mathcal{B}}
=  \mathscr{M}_{\tilde{\mathcal{A}}}^{-1} \cdot  
\textrm{diag}\bigg(\dots ,\, \rho_n^{k/2}\, \frac{F_{k/n}(1-x)}{F_{k/n}(x)}\, ,\, \dots \bigg) 
\cdot  \mathscr{M}_{\tilde{\mathcal{B}}}\,.
\ee
Thus, the matrix (\ref{tau2 matrix}) for $0<x<1$, found in \cite{cct-09}, is the $N=2$ case of the period matrix $\tilde{\tau}$, written with respect to the canonical homology basis introduced in the section \S\ref{subsec amv}
\be
\tilde{\tau}|_{N=2} = \tau_2\,.
\ee

To conclude this section, let us consider the symplectic transformation (\ref{cyc2tildecyc}), which reduces to $\textrm{diag}(I_{n-1}^{\textrm{\tiny low}} , (I_{n-1}^{\textrm{\tiny up}} )^{-1})$ for $N=2$.
Its inverse reads $\textrm{diag}((I_{n-1}^{\textrm{\tiny low}})^{-1} , I_{n-1}^{\textrm{\tiny up}})$ and it allows us to find the period matrix $\tau_2'$ with respect to the canonical homology basis given by the cycles $\boldsymbol{a}$ and $\boldsymbol{b}$ through (\ref{modular transf tau}), namely
\be
\label{tau2 eg}
\tau_2' =  I_{n-1}^{\textrm{\tiny up}} \cdot \tau_2 \cdot I_{n-1}^{\textrm{\tiny low}}\,.
\ee
Introducing the symmetric matrix $\mathscr{A}_{ij} \equiv 2/n \sum_{k=1}^{n-1} \sin(\pi k/n) \, e^{2\pi \textrm{i} (j-i)}$ (which has been denoted by $A$ in the Appendix C of \cite{cct-09}), after some algebra we find
\be
\label{tau2 Aidentity}
\mathscr{A}\cdot I_{n-1}^{\textrm{\tiny up}} \cdot \tau_2 
\cdot I_{n-1}^{\textrm{\tiny low}} \cdot \mathscr{A}
= \,\tau_2\,.
\ee
Combining (\ref{tau2 eg}) and (\ref{tau2 Aidentity}), we easily get that $\tau_2' = \mathscr{A}^{-1} \cdot \tau_2 \cdot \mathscr{A}^{-1}$. Then, by employing (\ref{theta transf symp}) and the fact that $\textrm{det}(I_{n-1}^{\textrm{\tiny up}}) =1$, we get
\be
\label{missing identity}
\Theta(\boldsymbol{0}|\tau_2') = 
\Theta(\boldsymbol{0}| \mathscr{A}^{-1} \cdot \tau_2 \cdot \mathscr{A}^{-1} ) =
\Theta(\boldsymbol{0}|\tau_2) \,.
\ee
In \cite{cct-09} the second equality in (\ref{missing identity}) has been given as a numerical observation. We have shown that it is a consequence of the relation between the two canonical homology bases considered here.

\section{The harmonic chain}
\label{sec HC}

In this section we consider the R\'enyi entropies and the entanglement entropy for the harmonic chain with periodic boundary conditions, which have been largely studied in the literature \cite{PeschelChung, Peschel03, BoteroReznik, PlenioEisert05, CramerEisert06, PeschelEisler09, ip-09}. We compute new data for  the case of disjoint blocks in order to check the CFT formulas found in \S\ref{sec compact boson} for the decompactification regime.
\begin{figure}[t]
\vspace{.4cm}
\begin{center}
\includegraphics[width=.52\textwidth]{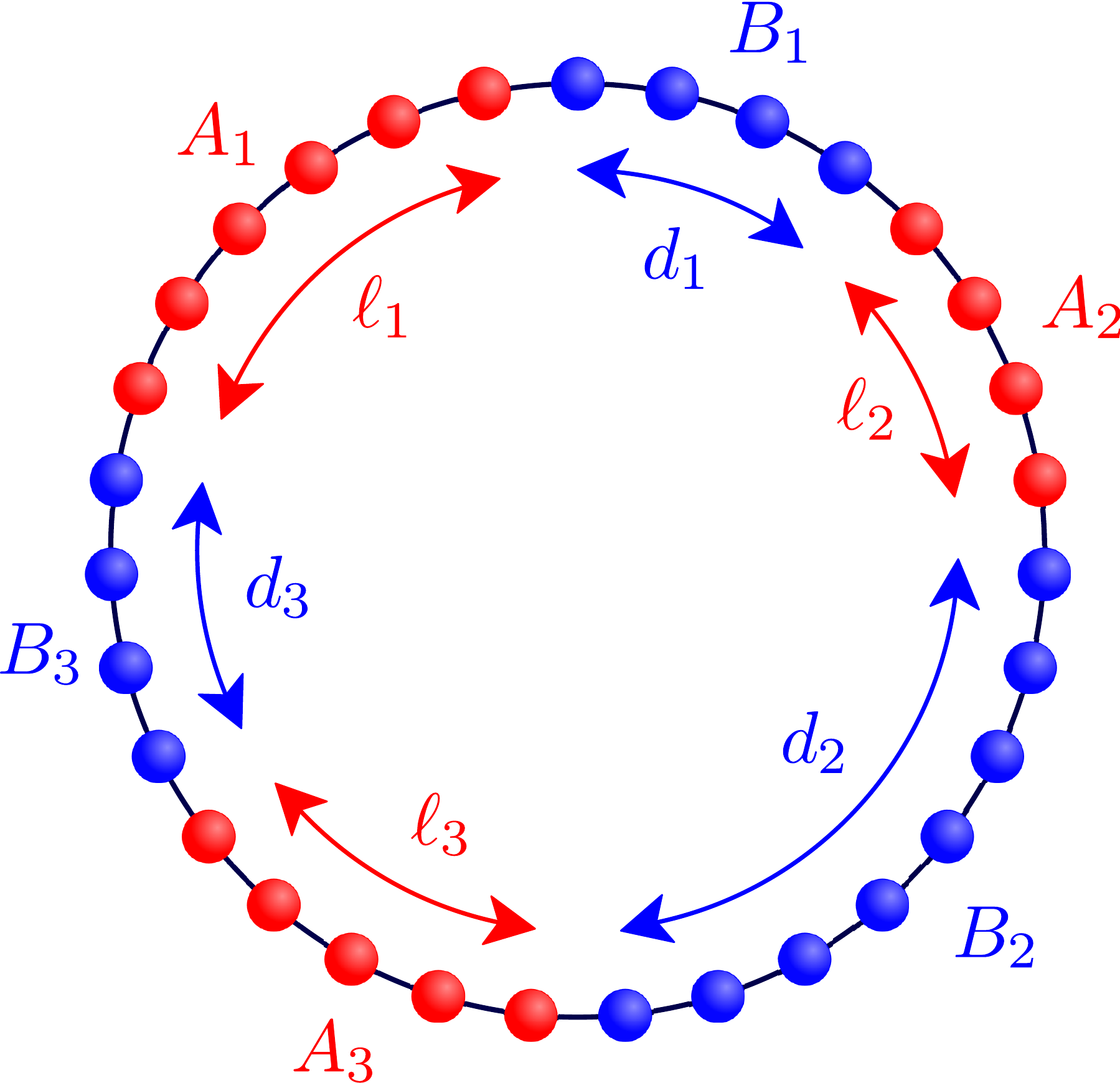}
\end{center}
\vspace{-.1cm}
\caption{A bipartition of the periodic chain where $A$ is made by the union of three disjoint blocks of lattice sites.}
\label{fig hc}
\end{figure} 

The Hamiltonian of the harmonic chain made by $L$ lattice sites and with nearest neighbor interaction reads
\be
\label{HC ham qp}
H = \sum_{n=0}^{L-1} \left(
\frac{1}{2M}\,p_n^2+\frac{M\omega^2}{2}\,q_n^2 +\frac{K}{2}(q_{n+1} -q_n)^2
\right),
\ee
where periodic boundary conditions $q_0 = q_L$ and $p_0 = p_L$ are imposed and the variables $q_n$ and $p_m$ satisfy the commutation relations $[q_n, q_m]=[p_n,p_m]=0$ and $[q_n,p_m] = \textrm{i} \delta_{n,m}$.
The Hamiltonian (\ref{HC ham qp}) contains three parameters $\omega$, $M$, $K$ but, through a canonical rescaling of the variables, it can be written in a form where these parameters occur only in a global factor and in the coupling $\tfrac{2K}{M\omega^2}/(1+\tfrac{2K}{M\omega^2})$ \cite{Audenaert02,BoteroReznik}.
The  Hamiltonian (\ref{HC ham qp}) is the lattice discretization of a free massive boson. When $\omega=0$ the theory is conformal with central charge $c=1$.
Since the bosonic field is not compactified, we must compare the continuum limit of (\ref{HC ham qp}) for $\omega=0$ with the regime $\eta \rightarrow \infty$ of the CFT expressions computed in \S\ref{sec compact boson}, which has been considered in \S\ref{subsec dec regime}.

To diagonalize (\ref{HC ham qp}), first one exploits the translational invariance of the system by Fourier transforming $q_n$ and $p_n$. Then the annihilation and creation operators $a_k$ and $a_k^\dagger$ are introduced, whose algebra is $[a_k, a_{k'}] = [a_k^\dagger, a_{k'}^\dagger]=0$ and $[a_{k},a_{k'}^\dagger]=\textrm{i} \delta_{k,k'}$. The ground state of the system $| 0 \rangle$ is annihilated by all the $a_k$'s and it is a pure Gaussian state.
In terms of the annihilation and creation operators, the Hamiltonian (\ref{HC ham qp}) is diagonal 
\be
\label{HC ham aadag}
H = \sum_{k=0}^{L-1} \omega_k \left( a^\dagger_k a_k +\frac{1}{2} \right),
\ee
where 
\be
\label{hc disp rel periodic}
\omega_k \equiv \sqrt{\omega^2 + \frac{4K}{M} \sin\Big( \frac{\pi k}{L} \Big)^2}\, \geqslant \omega\,,
\qquad k=0,\dots, L-1\,.
\ee
Notice that the lowest value of $\omega_k$ is obtained for $\omega_{0} = \omega$.

\begin{figure}[t]
\vspace{.3cm}
\begin{center}
\includegraphics[width=.92\textwidth]{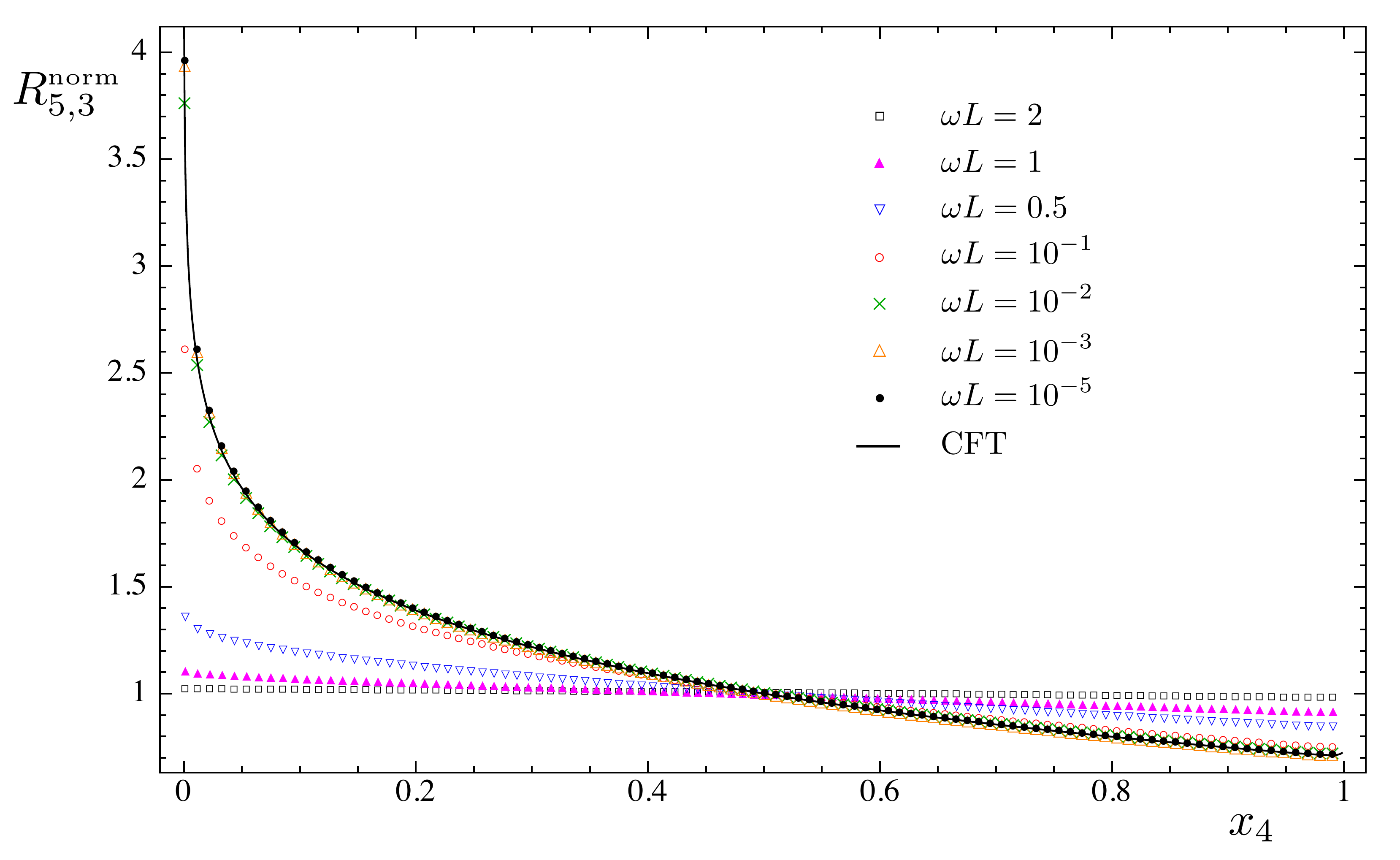}
\\
\hspace{.04cm}
\includegraphics[width=.915\textwidth]{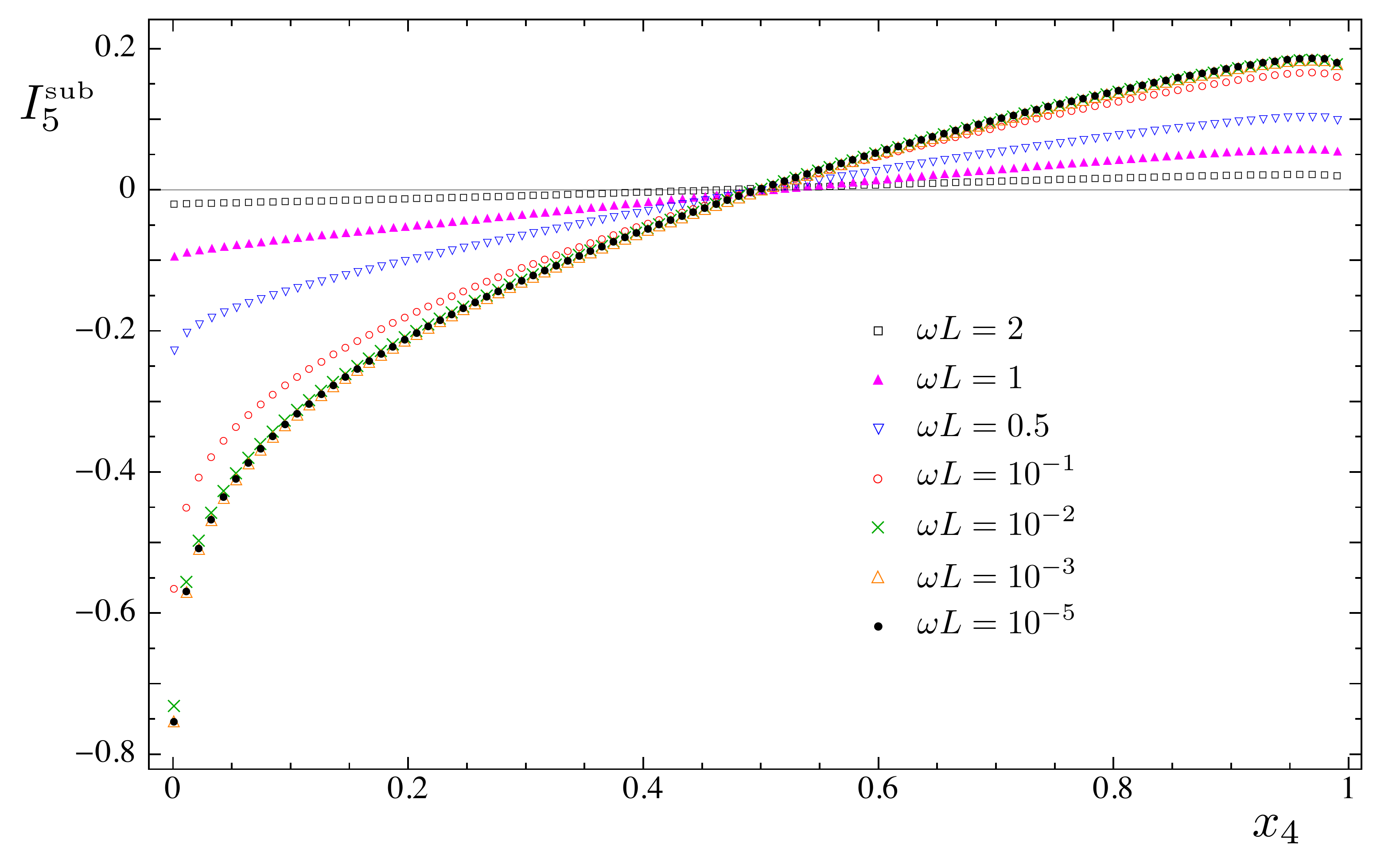}
\end{center}
\vspace{-.5cm}
\caption{The quantities $R^{\textrm{\tiny{norm}}}_{N=5,n=3}$ (top) and $I^{\textrm{\tiny{sub}}}_{N=5}$ (bottom) in (\ref{RNn norm}) computed for the harmonic chain with periodic boundary conditions by employing (\ref{renyi entropies hc}) and (\ref{EE hc}). 
The total length of the chain is $L=5000$. The configuration of the intervals is (\ref{symm config def}) and the fixed one chosen for the normalization is (\ref{fixed config def}).
The continuos curve in the top panel is the CFT prediction (\ref{rNn dec def}) and it agrees with the lattice results for $\omega L \ll 1$. We are not able to compute the CFT prediction for the bottom panel.}
\label{fig HC omegaL}
\end{figure}

The two point functions $\langle 0| q_i q_j|0 \rangle$ and $\langle0| p_i p_j|0\rangle$ are the elements the correlation matrices $\mathbb{Q}_{rs} = \langle 0| q_r q_s |0\rangle $ and $\mathbb{P}_{rs} = \langle 0| p_r p_s|0 \rangle$ respectively. For the harmonic chain with periodic boundary conditions that we are considering, they read
\begin{eqnarray}
\label{qq corr}
\langle 0| q_i q_j|0 \rangle &=& 
\frac{1}{2L} \sum_{k=0}^{L-1} \frac{1}{M\omega_k}  \cos\bigg( \frac{2\pi k (i-j)}{L} \bigg)\,,
\\
\langle 0| p_i p_j|0 \rangle &=& 
\frac{1}{2L} \sum_{k=0}^{L-1} \,M\omega_k  \cos\bigg( \frac{2\pi k (i-j)}{L} \bigg)\,.
\end{eqnarray}
When $i,j = 0, \dots , L-1$ run over the whole chain, then $\mathbb{Q} \cdot \mathbb{P} = \mathbb{I}_L /4$, which is also known as the generalized uncertainty relation. We remark that the limit $\omega \rightarrow 0$ is not well defined because the $k=0$ term in $\langle 0| q_i q_j|0 \rangle $ diverges; therefore we must keep $\omega>0$. Thus, we set $\omega L \ll 1$ in order to stay in the conformal regime. 
As explained above, we can work in units $M=K=1$ without loss of generality.
\\
In \cite{Peschel03, BoteroReznik, PeschelEisler09} it has been discussed that, in order to compute the Renyi entropies and the entanglement entropy of a proper subset $A$ (made by $\tilde{\ell}$ sites) of the harmonic chain, first we have to consider the matrices $\mathbb{Q}_A$ and  $\mathbb{P}_A$, obtained by restricting the indices of the correlation matrices $\mathbb{Q}$ and  $\mathbb{P}$ to the sites belonging to $A$. Then we compute the eigenvalues of the $\tilde{\ell} \times \tilde{\ell} $ matrix $\mathbb{Q}_A \cdot \mathbb{P}_A$. Since they are larger than (or equal to) $1/4$, we can denote them by $\{ \mu_1^2, \dots, \mu_\ell^2\}$. 
Finally, the Renyi entropies are obtained as follows
\be
\label{renyi entropies hc}
\textrm{Tr} \rho_A^n = \prod_{a\,=\,1}^{\tilde{\ell}}
\left[
\bigg( \mu_a+\frac{1}{2} \bigg)^n - \bigg( \mu_a-\frac{1}{2} \bigg)^n\,
\right]^{-1},
\ee
and the entanglement entropy as
\be
\label{EE hc}
S_A = \sum_{a\,=\,1}^{\tilde{\ell}} 
\left[
\bigg( \mu_a+\frac{1}{2} \bigg) \log\bigg( \mu_a+\frac{1}{2} \bigg) 
- \bigg( \mu_a-\frac{1}{2} \bigg) \log\bigg( \mu_a-\frac{1}{2} \bigg) 
\,\right]\,.
\ee
This procedure holds also when $A$ is the union of $N$ disjoint intervals $A_i$ ($i=1, \dots , N$), which is the situation we are interested in.

\begin{figure}[t]
\vspace{.2cm}
\begin{center}
\hspace{0.cm}
\includegraphics[width=.91\textwidth]{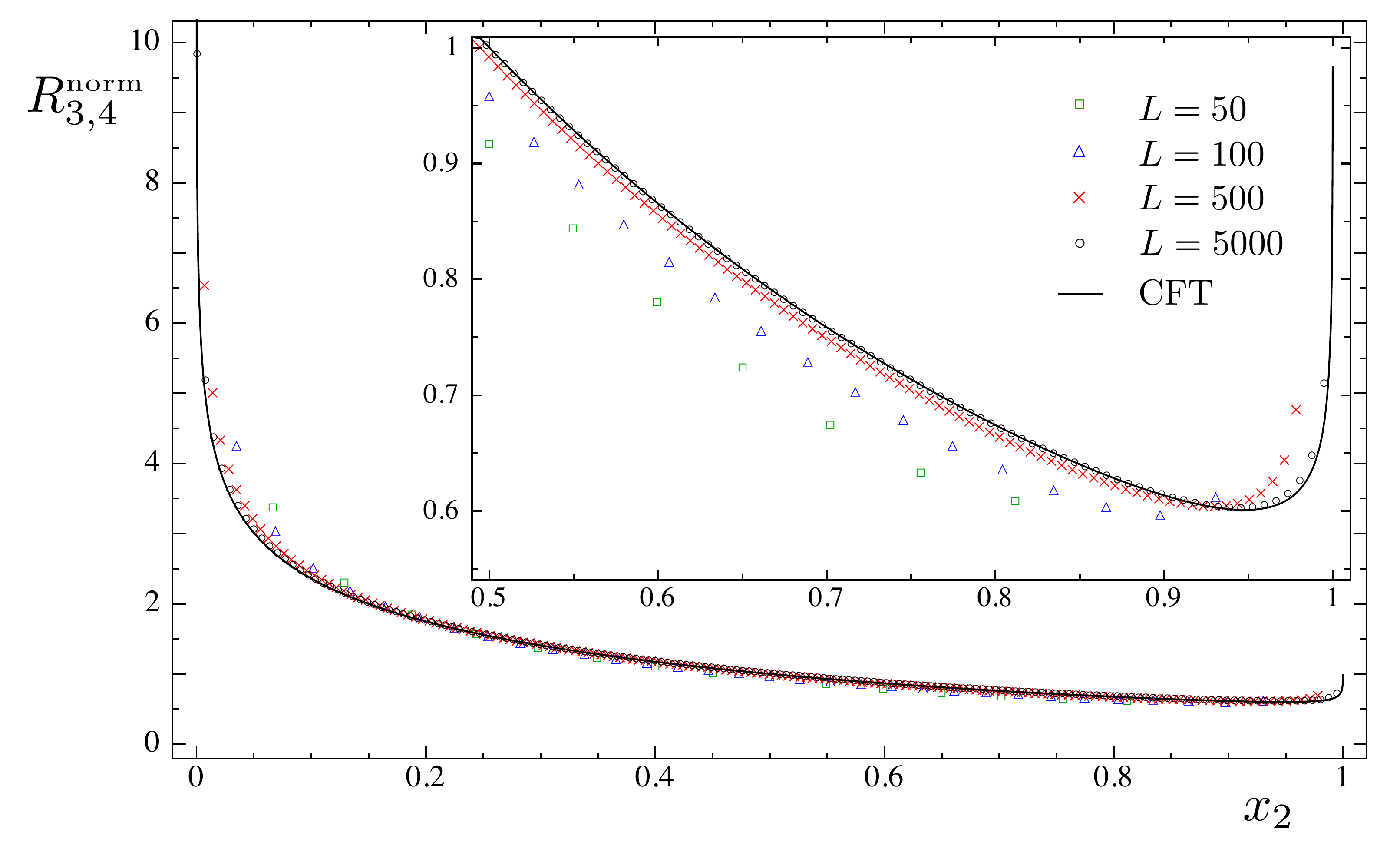}
\\
\hspace{0.05cm}
\includegraphics[width=.906\textwidth]{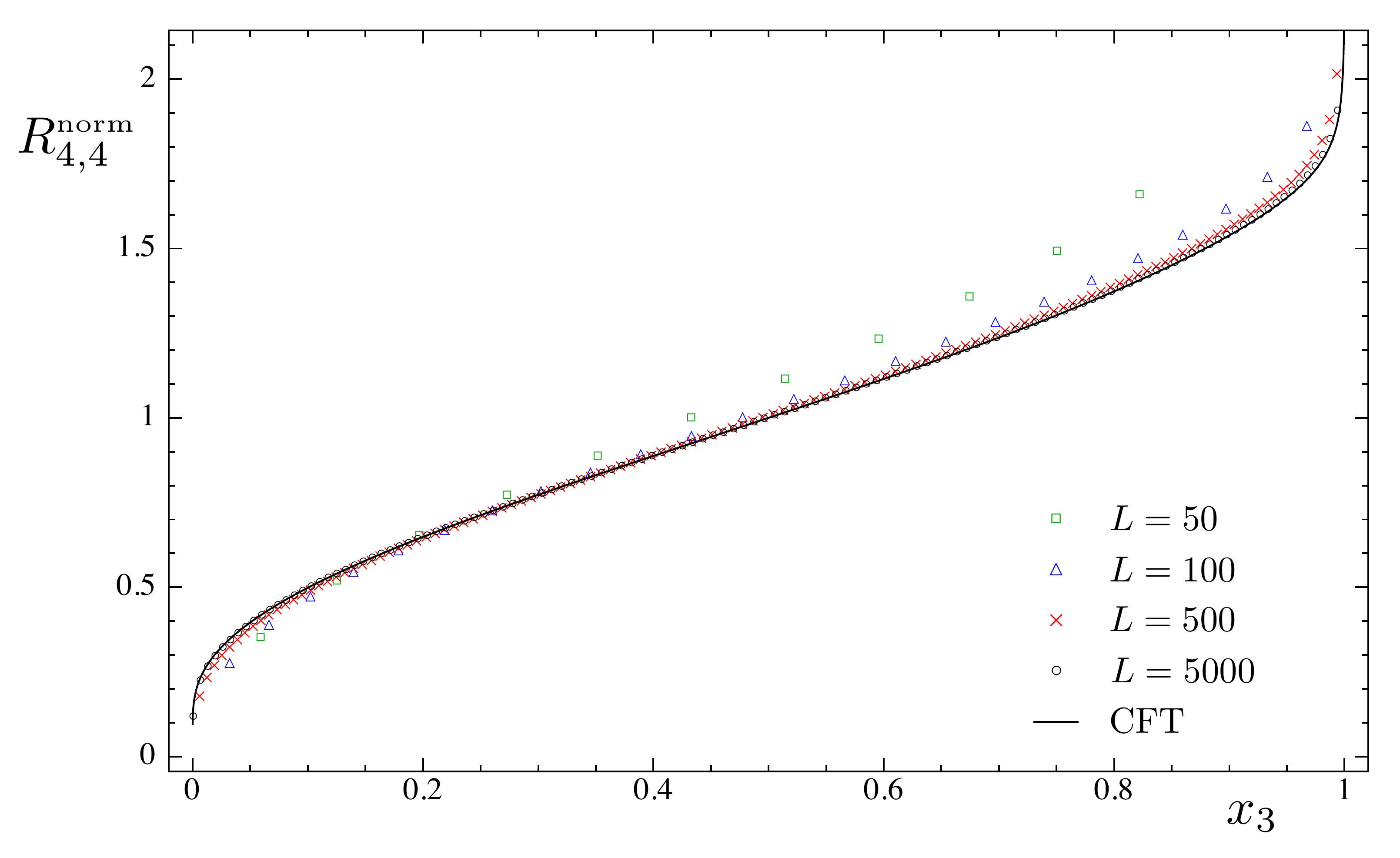}
\end{center}
\vspace{-.5cm}
\caption{The ratio $R^{\textrm{\tiny{norm}}}_{N,n}$ in (\ref{RNn norm}) for the periodic harmonic chain with $\omega L = 10^{-3}$ and the configuration of the intervals given by (\ref{symm config def}), normalized through (\ref{fixed config def}). The continuos curves are the CFT predictions (\ref{rNn dec def}).
Top: $N=3$ and $n=4$ (in the inset we zoom on part of the region $0.5 < x_2<1$).
Bottom: $N=4$ and $n=4$.}
\label{fig HC RNn4}
\end{figure}

\begin{figure}[t]
\vspace{.1cm}
\begin{center}
\hspace{0.015cm}
\includegraphics[width=.92\textwidth]{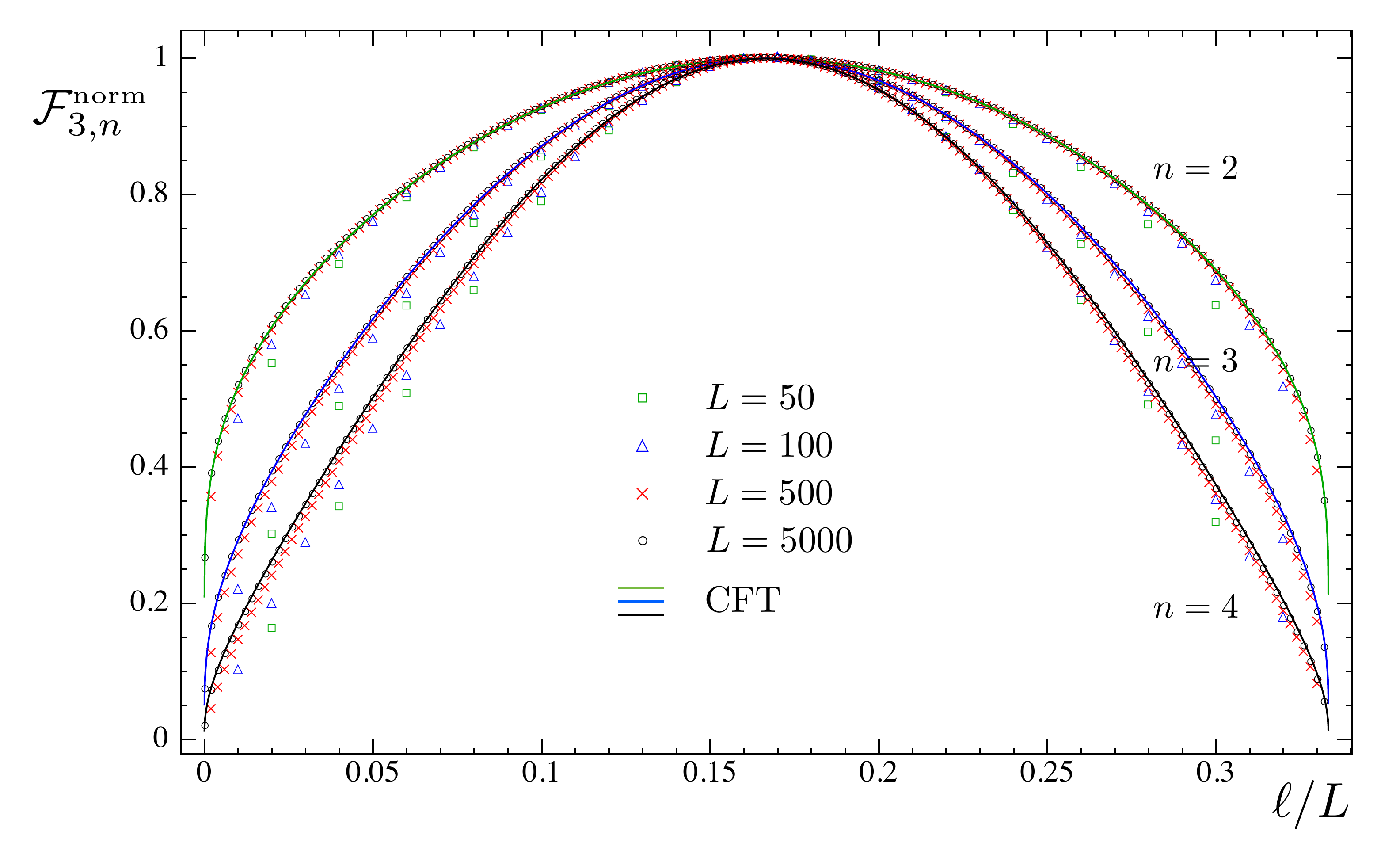}
\\
\hspace{0.cm}
\includegraphics[width=.9017\textwidth]{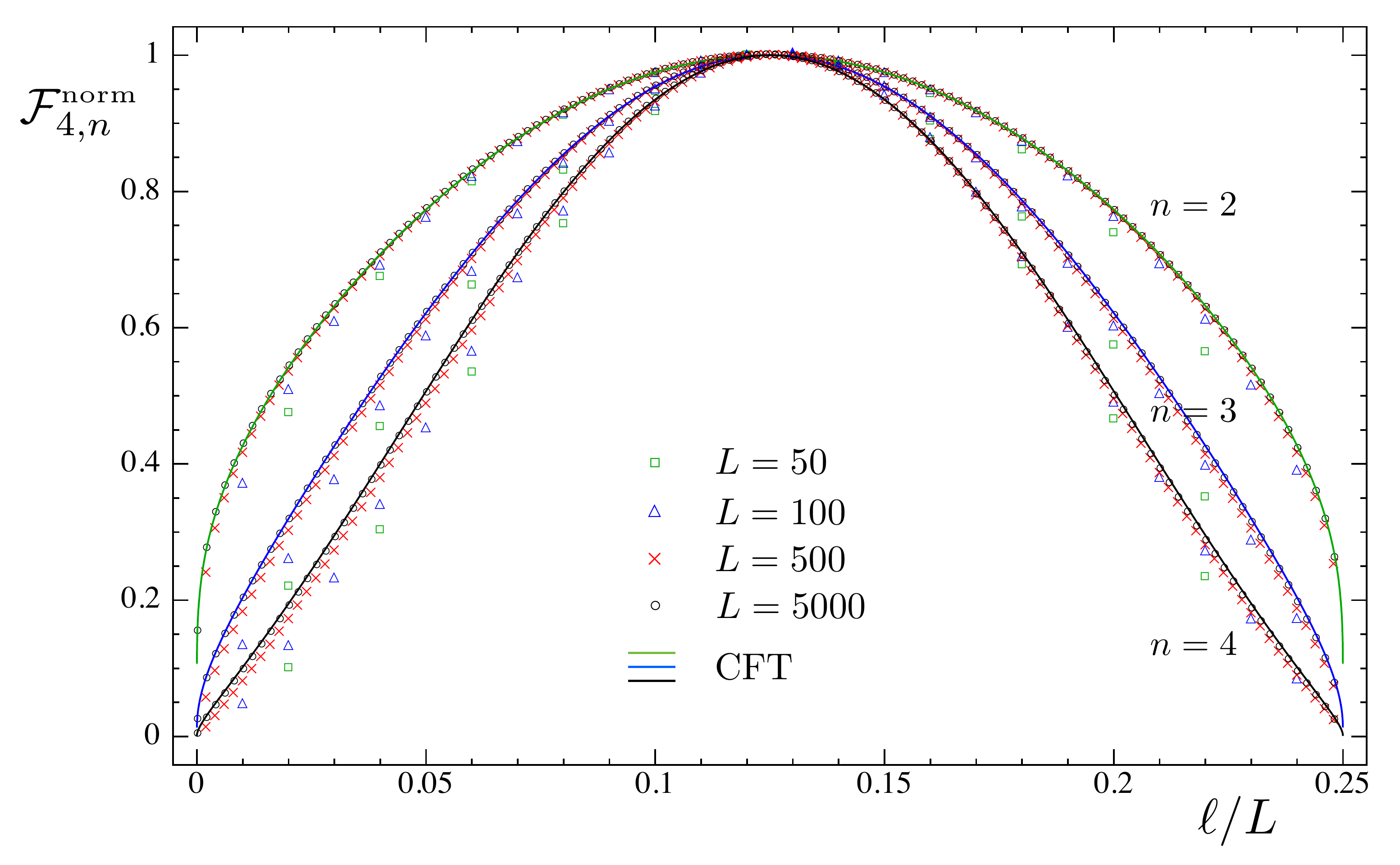}
\end{center}
\vspace{-.5cm}
\caption{The quantity $\mathcal{F}^{\textrm{\tiny{norm}}}_{N,n}$ computed for the periodic harmonic chain with $\omega L = 10^{-3}$ in the configuration of intervals (\ref{symm config def}), normalized through (\ref{fixed config def}). 
The lattice data are obtained by using (\ref{tildeRNn def}), (\ref{tildeRNn cft}), (\ref{renyi entropies hc}) and (\ref{EE hc}).
The continuos curves are given by (\ref{FNn dec def}).
The maximum value on the horizontal axis is $1/N$.
We show the cases of $N=3$ (top) and $N=4$ (bottom) with $n=2,3,4$. 
}
\label{fig HC FN34n}
\end{figure}

Let us denote by $\ell_i$ the number of sites included in $A_i$ and by $d_i$ the number of sites in the separations between $A_i$ and $A_{i+1 \,\textrm{mod}\, N}$, for $i=1, \dots, N$ (see Fig. \ref{fig hc} for $N=3$). Then, we have that $\tilde{\ell} = \sum_{i=1}^N \ell_i$ and the following consistency condition about the total length of the chain must be imposed
\be
\label{Ltot condition}
L = \sum_{i\,=\,1}^N (\ell_i +d_i)\,.
\ee

\begin{figure}[t]
\begin{center}
\hspace{0.02cm}
\includegraphics[width=.91\textwidth]{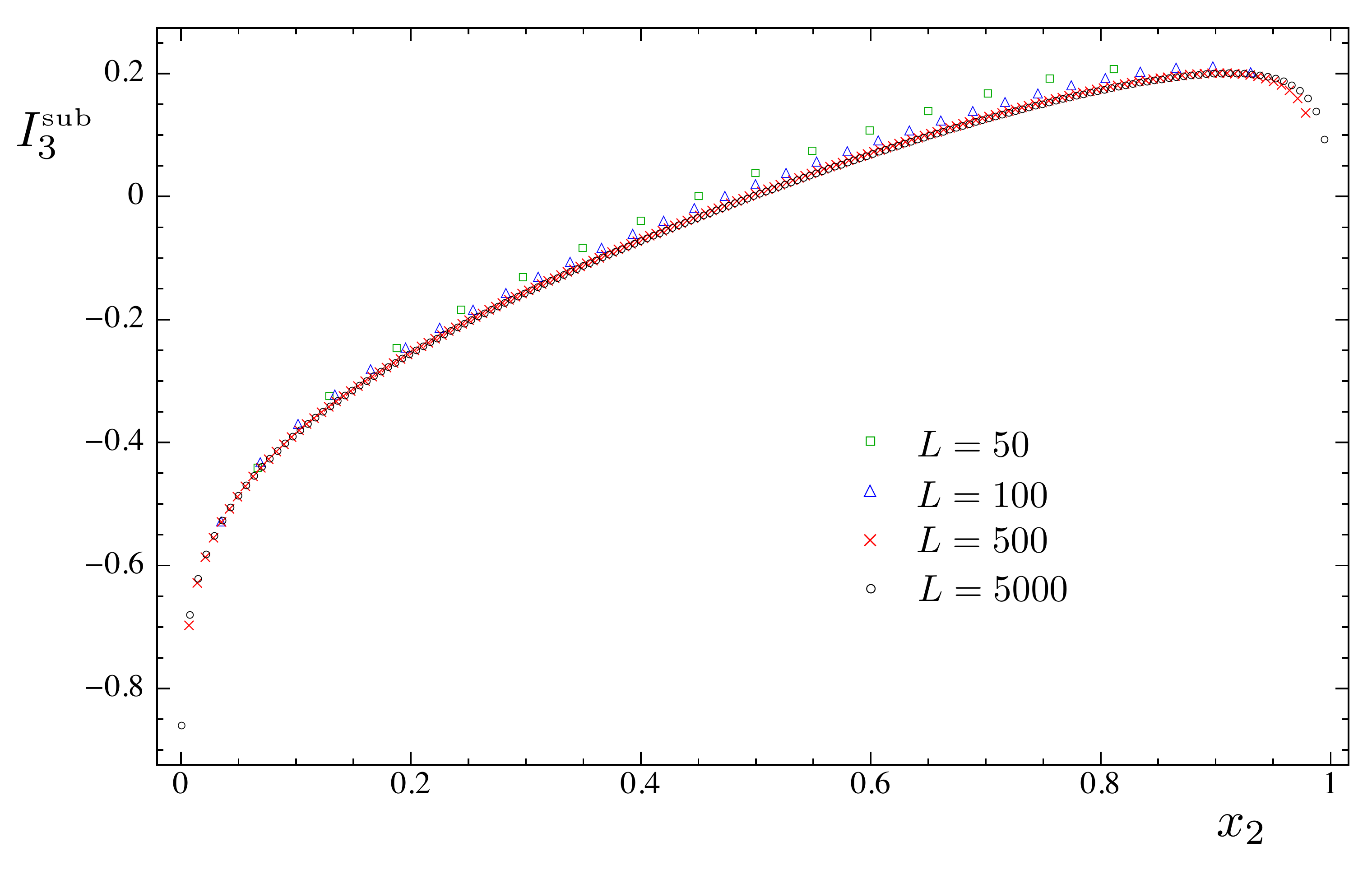}
\\
\hspace{0.cm}
\includegraphics[width=.9028\textwidth]{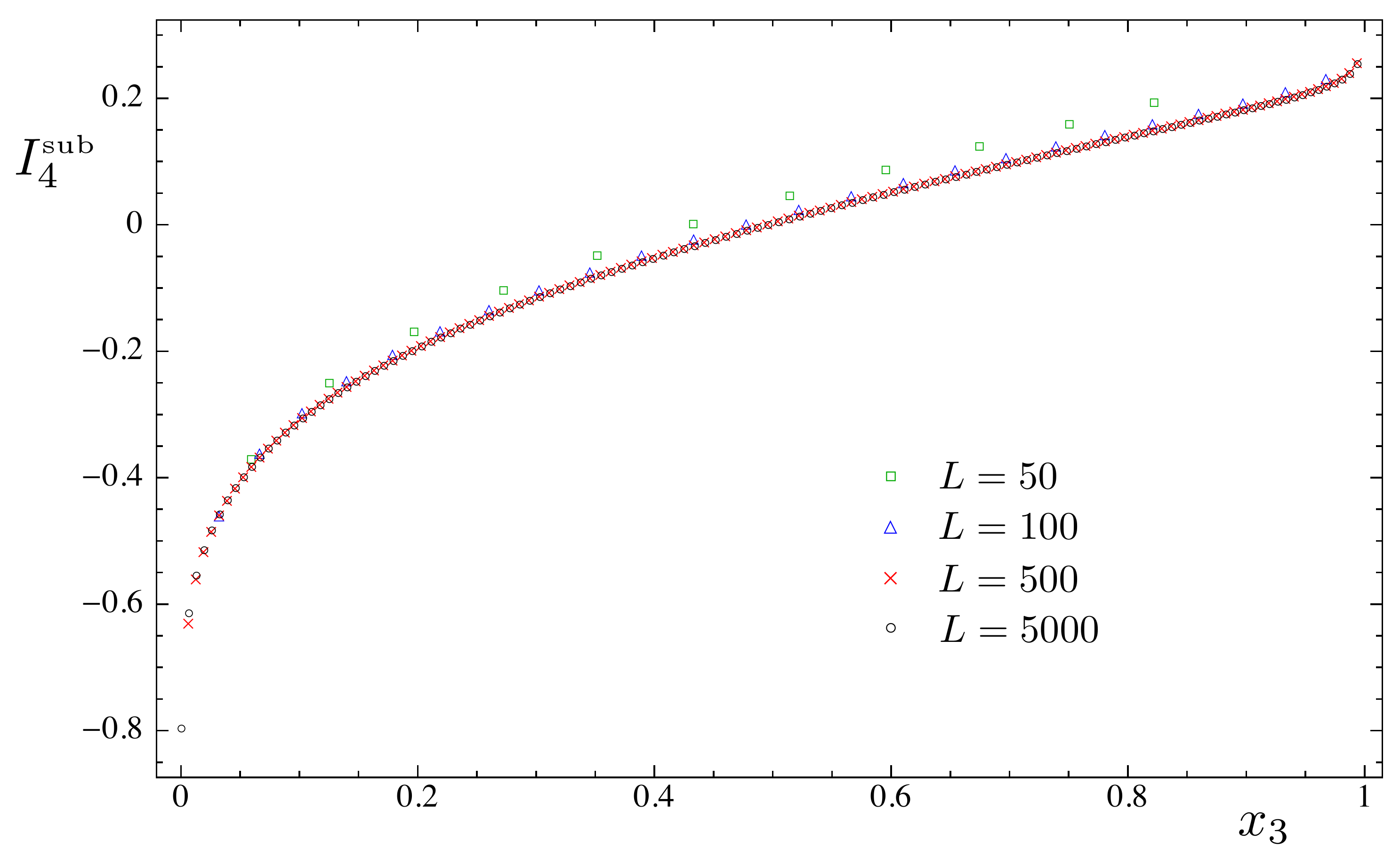}
\end{center}
\vspace{-.5cm}
\caption{The quantity $I^{\textrm{\tiny{sub}}}_{N} $ (see (\ref{RNn norm})) computed for the periodic harmonic chain with $\omega L =10^{-3}$. The configuration of intervals is given by (\ref{symm config def}) and the fixed one by (\ref{fixed config def}). We show $N=3$ (top) and $N=4$ (bottom).}
\label{fig HC IsubN34}
\end{figure}

In order to compare the CFT results found in the previous sections with the ones obtained from the harmonic chain in the continuum limit, we have to generalize the CFT formulas to the case of a finite system of total length $L$ with periodic boundary conditions. This can be done by employing the conformal map from the cylinder to the plane, whose net effect is to replace each length $y$ (e.g. $\ell$, $d$, $2\ell+d$, etc.) with the corresponding chord length $(L/\pi) \sin(\pi y/L)$. Thus, for $x_{2j+1}$ with $j=0, \dots, N-2$ we have
\be
 \label{xratios finiteL odd}\fl
x_{2j+1} \,=\, 
 \frac{
 \sin\big(\pi \big[\sum_{i=1}^j (\ell_i+d_i)+\ell_{j+1}\big] /L\big)
 \, \sin(\pi \ell_N/L)}{
 \sin\big(\pi \sum_{i=1}^{N-1} (\ell_i+d_i)/L\big)
 \,\sin\big(\pi \big[d_{j+1}+\sum_{i=j+2}^{N-1} (\ell_i+d_i)+\ell_{N}\big] /L\big)
 }\,,
 \ee
 while  for the harmonic ratios  $x_{2j}$, where $j=1, \dots, N-2$, we must consider
\be
  \label{xratios finiteL even}\fl
x_{2j} \,=\, 
 \frac{
 \sin\big(\pi \sum_{i=1}^j (\ell_i+d_i)/L\big)
 \, \sin(\pi \ell_N/L)}{
 \sin\big(\pi \sum_{i=1}^{N-1} (\ell_i+d_i)/L\big)
 \,\sin\big(\pi \big[\sum_{i=j+2}^{N-1} (\ell_i+d_i)+\ell_{N}\big] /L\big)
 }\,.
\ee
Notice that $d_N$, which can be obtained from (\ref{Ltot condition}), does not occur in these ratios. Moreover, (\ref{xratios finiteL odd}) and (\ref{xratios finiteL even}) depend only on $\ell_i/L$ and $d_i/L$, with $i=1, \dots, N-1$.

We often consider the configuration where all the intervals have the same length and also the segments separating them have the same size, namely
\be
\label{symm config def}
\ell_1 = \dots = \ell_N \equiv \ell\,,
\qquad
d_1 = \dots = d_N \equiv d\,.
\ee
This configuration is parameterized by $\ell$, once $d$ has been found in terms of $\ell$ through the condition (\ref{Ltot condition}).
As mentioned in \S\ref{sec riemann surf}, in order to eliminate some parameters, it is useful to normalize the results through a fixed configuration of intervals, as done e.g. in \cite{cct-neg-letter, cct-neg-long, ctt-neg-ising}. We choose the following one 
\be\fl
\label{fixed config def}
\textrm{fixed configuration:}
\qquad
\ell_1 = \dots = \ell_N = 
d_1 = \dots = d_{N-1} = \textrm{int}\bigg(\frac{L}{2N}\bigg)\,,
\ee
where $\textrm{int}(\dots)$ denotes the integer part of the number within the brackets and $d_N$ is obtained from (\ref{Ltot condition}).
\\
In Figs. \ref{fig HC omegaL}, \ref{fig HC RNn4}, \ref{fig HC FN34n}  and \ref{fig HC IsubN34} we choose the configuration (\ref{symm config def}) normalized through the fixed one in (\ref{fixed config def}). A chain made by $L=5000$ sites gives us a very good approximation of the continuum case. We also made some checks with $L=10000$ in order to be sure that the results do not change significantly. 
From Fig. \ref{fig HC omegaL} we learn that for $\omega L  \sim 10^{-3}$ we are already in a regime which is suitable to check the CFT prediction of \S\ref{subsec dec regime}, therefore we keep  $\omega L  = 10^{-3}$ for the other plots obtained from the harmonic chain.
In order to compare the lattice results from the periodic chain with the CFT expressions (\ref{FNn eta large}) and (\ref{RNn eta large}), one needs to adjust $\eta$. We find that this value of $\eta$ depends on the product $\omega L \ll 1$.
Nevertheless, as already noticed in \S\ref{subsec dec regime}, normalizing the interesting quantities through a fixed configuration as in (\ref{RNn norm}), we can ignore this important issue because $\eta$ simplifies (see \ref{rNn dec def} and \ref{FNn dec def})).
The Figs. \ref{fig HC RNn4} and \ref{fig HC FN34n} show that the agreement between the exact results from the harmonic chain and the corresponding CFT predictions is very good. 
Instead, for the plots in Fig. \ref{fig HC IsubN34} we do not have a CFT prediction because, ultimately, we are not able to compute $\partial_n \widehat{\mathcal{F}}_{N,n}( \boldsymbol{x}  )$ when $n\rightarrow 1$ for the function defined in (\ref{FNn eta large}).

\begin{figure}[t]
\begin{center}
\vspace{0.cm}
\hspace{-0.04cm}
\includegraphics[width=.704\textwidth]{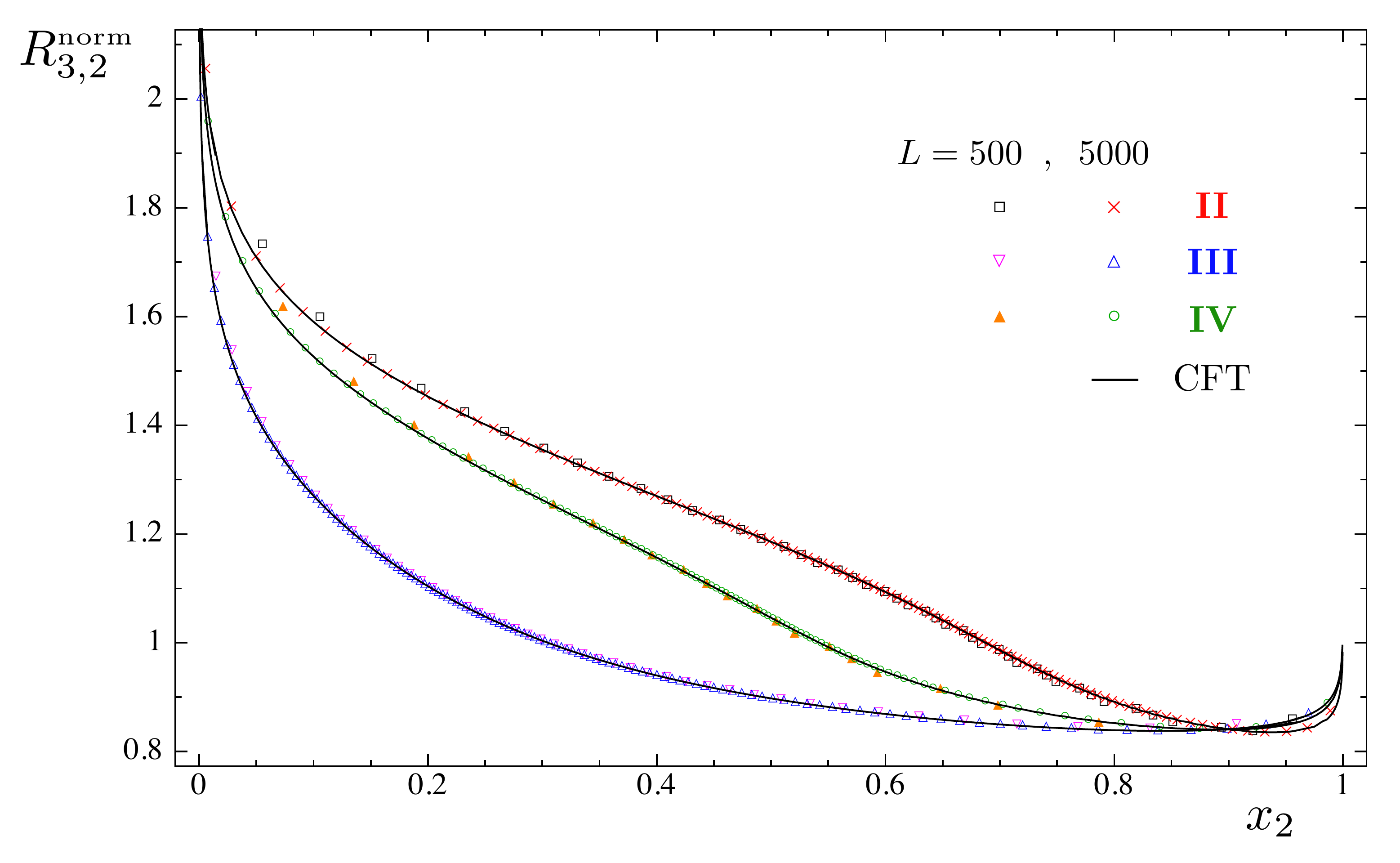}\\
\includegraphics[width=.7\textwidth]{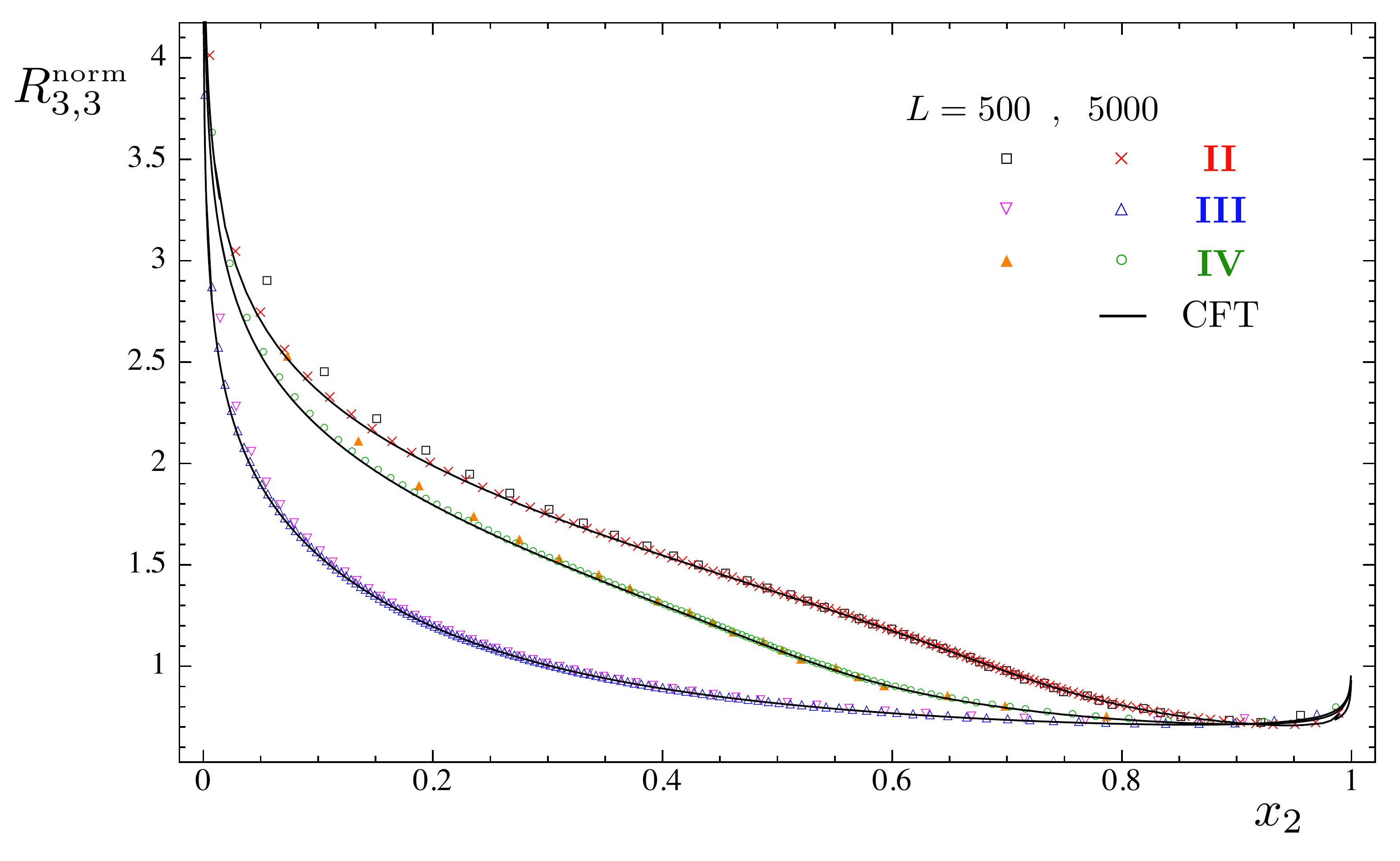}\\
\hspace{.05cm}
\includegraphics[width=.697\textwidth]{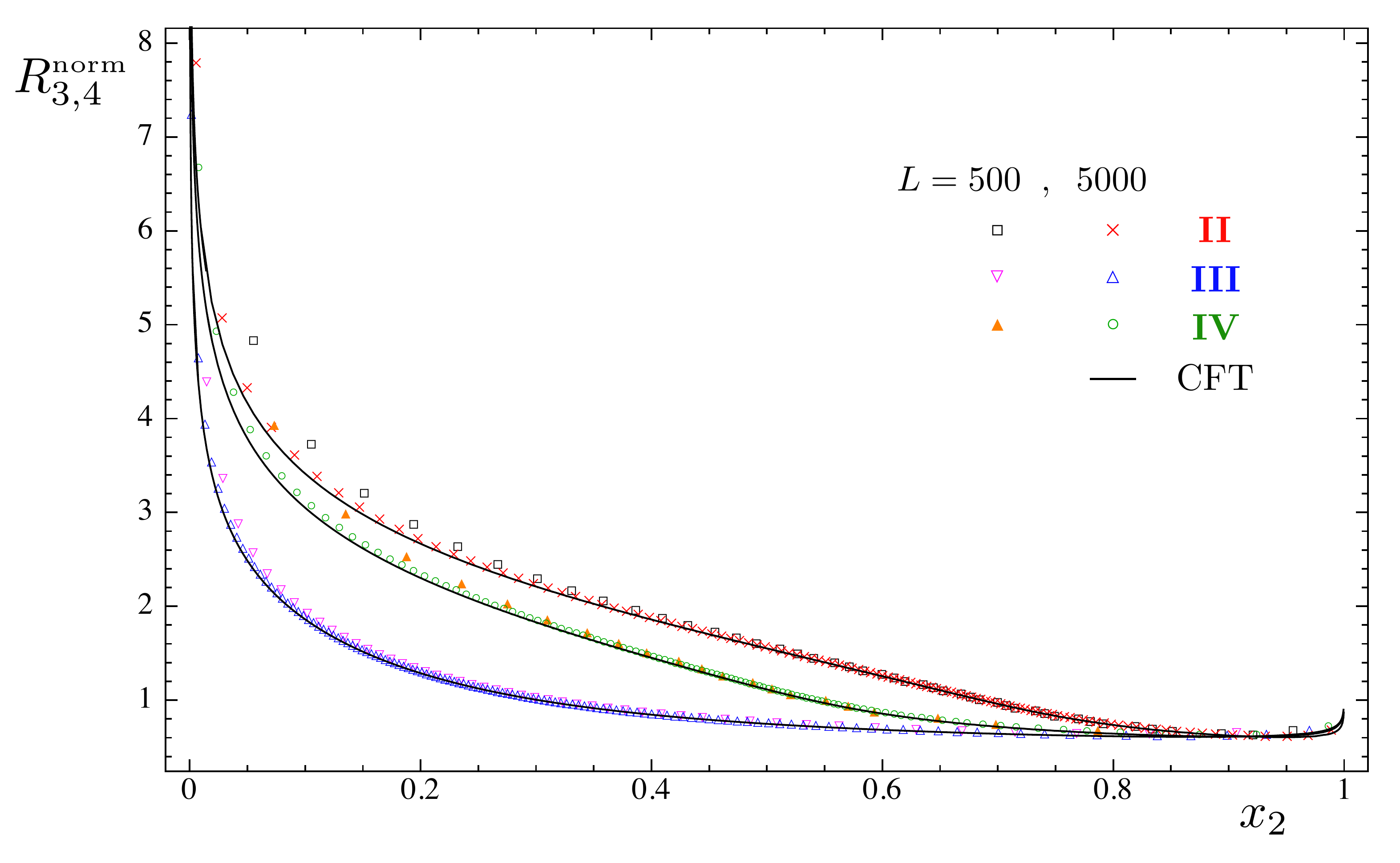}
\end{center}
\vspace{-.5cm}
\caption{
The ratio $R^{\textrm{\tiny{norm}}}_{N,n}$ in (\ref{RNn norm}) for the harmonic chain with $\omega L =10^{-3}$.
The configurations II, III and IV, which are defined in (\ref{N3 configs def}), have been normalized through (\ref{fixed config def}).  The continuos curve is the CFT prediction (\ref{rNn dec def}).
We show $N=3$ and $n=2,3,4$ (top, middle, bottom). 
}
\label{fig HC RN3n234configs}
\end{figure}

\begin{figure}[t]
\begin{center}
\hspace{0.cm}
\includegraphics[width=.9\textwidth]{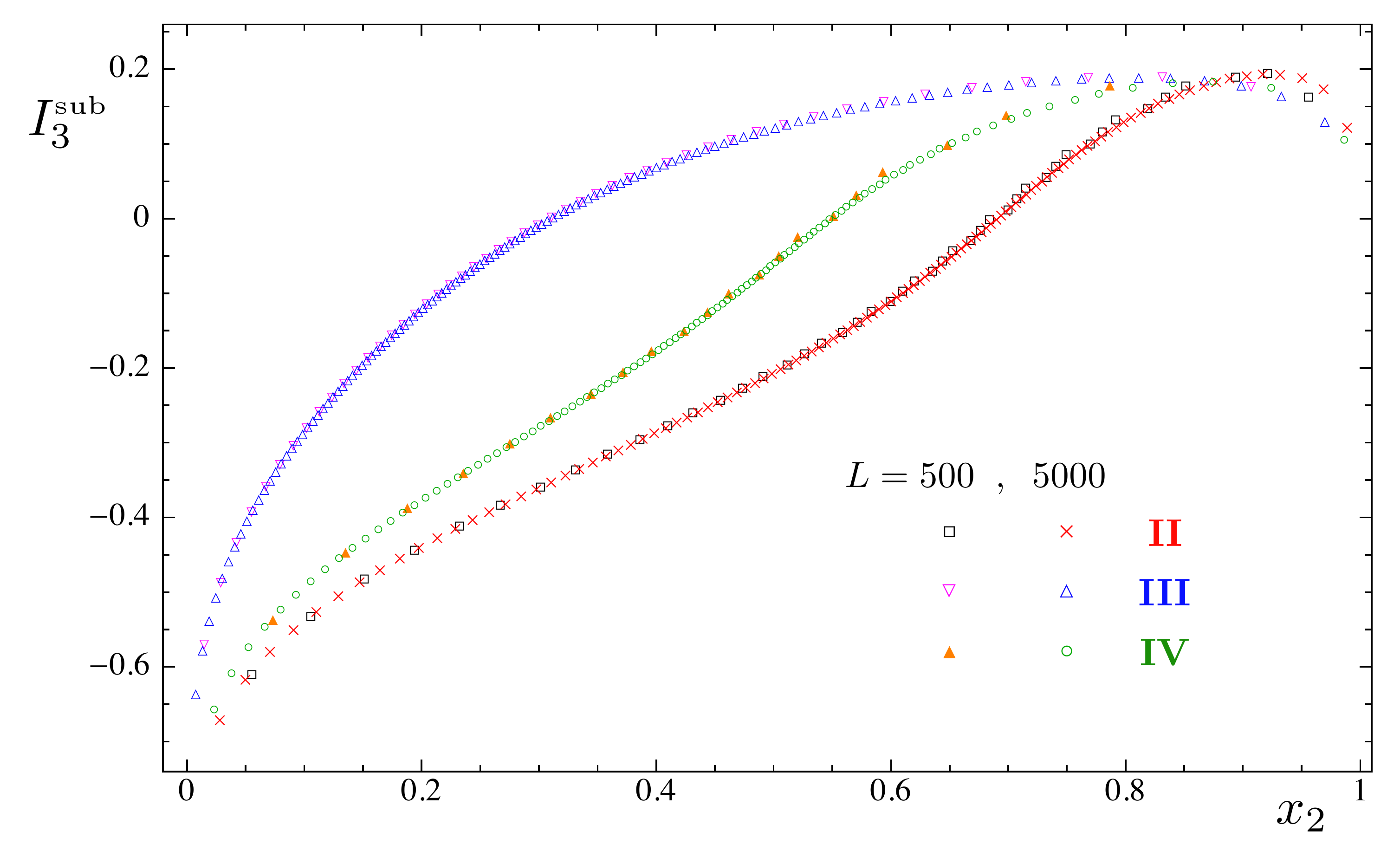}
\end{center}
\vspace{-.5cm}
\caption{$I^{\textrm{\tiny{sub}}}_{N=3} $ in (\ref{RNn norm}) for the periodic harmonic chain with $\omega L =10^{-3}$. 
The configurations are defined in (\ref{N3 configs def}) and the fixed one is given by (\ref{fixed config def}).
}
\label{fig HC IN3configs}
\end{figure}

When $N>2$ we have many possibilities to choose the configuration of the intervals. In principle we should test all of them and not only (\ref{symm config def}), as above.
For simplicity, we consider two other kinds of configurations defined as follows
\be
\label{hc configs}
\rule{0pt}{1.cm}
\begin{array}{c|ccccccccc|}
                & \ell_1 & d_1 & \ell_2 & d_2 & \ell_3 & d_3 & \dots & \ell_N & d_N \\
\hline
\textrm{$\boldsymbol{\lambda}$} & \ell  & d  & \lambda_2 \ell  & d  & \lambda_3 \ell & d  & \dots & \lambda_N \ell  & d     \\
\hline
\textrm{$\boldsymbol{\gamma}$}  & \ell & d  & \gamma_2 \ell  & \gamma_2 d  &  \gamma_3 \ell  & \gamma_3  d  & \dots &\gamma_N \ell  & \gamma_N d     \\
\hline
\end{array}
\ee
where $\lambda_i$ and $\gamma_i$ are integer numbers which can be collected as components of the vectors $\boldsymbol{\lambda}$ and $\boldsymbol{\gamma}$.
Notice that the configuration (\ref{symm config def})  is obtained either with $\lambda_i =1$ or with $\gamma_i =1$, for $i=2, \dots, N$.
Once the ratios $\lambda_i$ or $\gamma_i$ have been chosen in (\ref{hc configs}), we are left with $\ell$ and $d$ as free parameters. As above, $d$ can be found as a function of $\ell$ through the condition (\ref{Ltot condition}) and the maximum value for $\ell$ corresponds to $d=1$.
The configurations in (\ref{hc configs}) depend only on the parameter $\ell$; therefore they provide one dimensional curves in the configurations space, which is $2N-3$ dimensional and parameterized by $0< x_1< x_2 < \dots < x_{2N-3}<1$. \\
When $N=3$, let us consider the configurations (\ref{hc configs}) with the following choices
\be
\label{N3 configs def}
\begin{array}{c|lll|}
\hline
\textrm{ I} & \gamma_1=1 & \gamma_2=1 & \gamma_3=1 
\\
\hline
\textrm{ II}  & \lambda_1=1 & \lambda_2=2 & \lambda_3=8
\\
\hline
\textrm{III}  & \gamma_1=1 & \gamma_2=3 & \gamma_3=6 
\\
\hline
\textrm{IV}  & \lambda_1=1 & \lambda_2=11 & \lambda_3=11
\\
\hline
\end{array}
\ee
where the first one is (\ref{symm config def}) specialized to the case of three intervals.
Plugging these configurations in (\ref{xratios finiteL odd}) and (\ref{xratios finiteL even}) for $N=3$, we can find the corresponding curves within the domain $0<x_1 <x_2<x_3<1$, as shown in Fig. \ref{fig configN3}. These curves can be equivalently parameterized either by $\ell/L$ or by one of the harmonic ratios $x_i$.
In Fig. \ref{fig HC RN3n234configs} we show $R_{3,n}^{\textrm{\tiny{norm}}}$ ($n=2,3,4$), finding a good agreement with the CFT prediction (\ref{rNn dec def}).
In Fig. \ref{fig HC IN3configs} we plot $I_{3}^{\textrm{\tiny{sub}}}$ for the harmonic chain but, as for Fig. \ref{fig HC IsubN34}, we do not have a CFT formula to compare with for the reason mentioned above.

\section{The Ising model}
\label{sec ising}

The Ising model in transverse field provides a simple scenario where we can compute the R\'enyi entropies of several disjoint intervals and compare them with the corresponding predictions obtained through the CFT methods. 
The Hamiltonian is given by
\begin{equation}
 \label{eq:ising}
 H = \sum_{s\,=\,1}^L \Big(\sigma^x_s \sigma^x_{s+1} + h \sigma^z_s\Big) \,,
\end{equation}
where $s$ labels the $L$ sites of a 1D lattice $\mathcal{L}$ and the $\sigma_s^{x,z}$ are the Pauli matrices acting on the spin at site $s$ and periodic boundary conditions are imposed.
The model has two phases, one polarized along $x$ for $\lambda <1$ and another one polarized along $z$ for $\lambda >1$, which are separated by  a second order phase transition at $h =1$. 
\\
 The Ising model in transverse field  can be rewritten as  a model of free fermions \cite{SchultzMttisLieb}. The map underlying this equivalence has been employed in \cite{LatorreRicoVidal} to compute the R\'enyi entropies for one block and in  \cite{FagottiCalabrese10} for two disjoint blocks, where the generalization to $N$ blocks is also discussed. 
 
Our approach is based on the Matrix Product States (MPS), which is completely general and therefore it can be applied for every one dimensional model. 
We choose the MPS because they are the simplest tensor networks (see \S\ref{sec mps defs} for a proper definition).
 The same calculation can be done through other variational ansatz methods, like the Tree Tensor Networks or the MERA 
 \cite{TagliacozzoEvenblyVidal, atc-10, atc-11}).

\subsection{R\'enyi entropies for the Ising CFT}
\label{sec ising cft}

\begin{figure}[t]
\begin{center}
\hspace{0.cm}
\includegraphics[width=.99\textwidth]{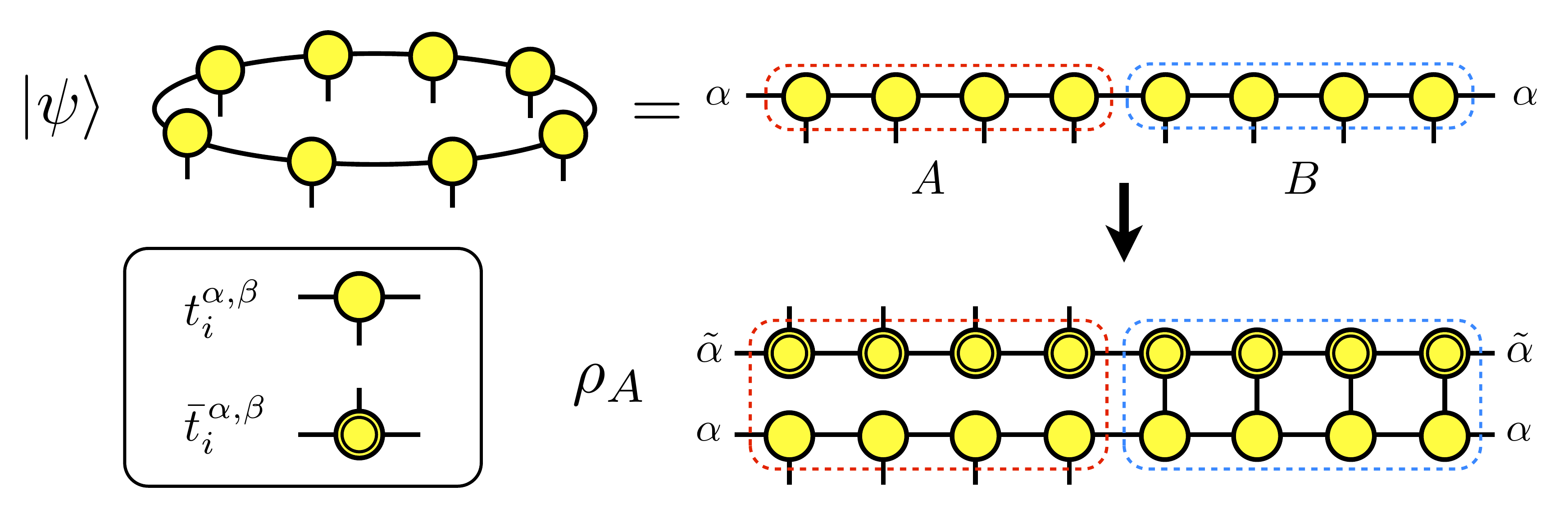}
\end{center}
\vspace{-.3cm}
\caption{\label{fig:MPS}
The contraction giving the MPS state $\ket{\psi}$ of a chain with $L=8$ sites and periodic boundary conditions (points labeled by the same greek index are considered as the same point). The individual tensor $t_i^{\alpha, \beta}$, which defines the MPS state, and its complex conjugate $\bar{t}_i^{\,\alpha, \beta}$ are shown in the box on the left. Considering the bipartition of the chain with $A$ made by 4 contiguous sites, we show the tensor network contraction occurring in the computation of the reduced density matrix $\rho_A$.}
\label{fig:MPS} 
\end{figure}

The continuum limit of the quantum critical point $h =1$ corresponds to a free massless Majorana fermion, which is a CFT with $c=1/2$.

Identifying $\phi$ with $-\,\phi$ in (\ref{action free boson RNn}), the target space becomes $S^1/\mathbb{Z}_2$ and the compactification radius (orbifold radius) parameterizes  the critical line of the Ashkin-Teller model, which can be seen as two Ising models coupled through a four fermion interaction. When the interaction vanishes, the partition function of the Ashkin-Teller model reduces to the square of the partition function of the Ising model.\\
This set of $c=1$ conformal field theories has been studied in \cite{amv, verlinde2-86, amv bosonization, dvv} in the case of a worldsheet given by a generic Riemann surface and the relations found within this context allow us to write $\Tr \rho_A^n$ for the Ising model in terms of Riemann theta functions with characteristic (\ref{theta def spin}).
The peculiar feature of the Ising model with respect to the other points of the Ashkin-Teller line is that we just need the period matrix $\tau$ to find the partition function on the corresponding Riemann surface.

In our case, the Riemann surface is given by (\ref{eq curve}) and its period matrix has been computed in \S\ref{subsec period matrix}. Thus, $\Tr \rho_A^n$ for the Ising model is given by (\ref{Tr rhoA N int intro}) with $c=1/2$ and 
\be
\label{FNn ising}
\mathcal{F}^{\textrm{\tiny Ising}}_{N,n}( \boldsymbol{x}) =
\frac{\sum_{\boldsymbol{e}} | \Theta [\boldsymbol{e} ](\boldsymbol{0} | \tau ) |}{2^g \,| \Theta (\boldsymbol{0} | \tau ) |} \,,
\ee
where the period matrix $\tau$ has been discussed in \S\ref{subsec period matrix}.
As already remarked in \S\ref{sec dirac}, the sum over the characteristics in the numerator of (\ref{FNn ising}) contains only the even ones. We checked numerically that $\mathcal{F}^{\textrm{\tiny Ising}}_{N,n}( \boldsymbol{0}) =1$. 
Moreover, by employing the results of \S\ref{sec 2int case} and of Appendix \ref{sec modular transf}, one finds that, specializing (\ref{FNn ising}) to $N=2$, the expression for $\mathcal{F}^{\textrm{\tiny Ising}}_{2,n}(x)$ found in \cite{cct-11} is recovered. 
In Appendix \ref{sec modular transf} we also discuss the invariance of (\ref{FNn ising}) under a cyclic transformations or an inversion in the ordering of the sheets and under the exchange $A \leftrightarrow B$.

\subsection{Matrix product states: notation and examples}
\label{sec mps defs}

\begin{figure}[t] 
\begin{center}
\hspace{0.cm}
\includegraphics[width=.99\textwidth]{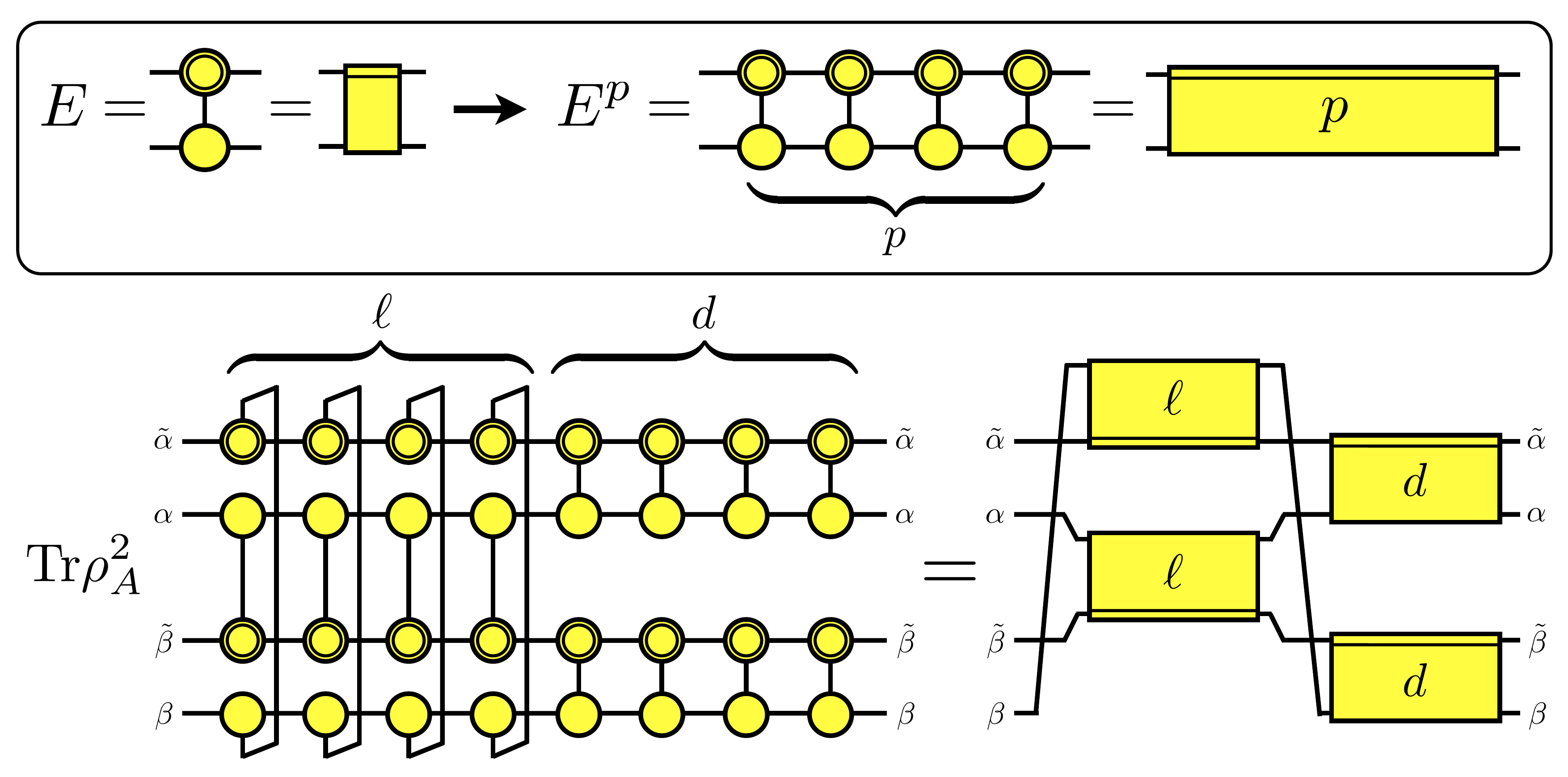}
\end{center}
\vspace{-.3cm}
\caption{\label{fig:mpsrenyi2}
The computation of $\textrm{Tr} \rho_A^2$ for the bipartition of Fig. \ref{fig:MPS}, where $\ell=d=4$. The MPS transfer matrix $E$ and its $p$-th power are shown in the box as yellow rectangles. The pattern for the contractions of the indices is on the right.}
\end{figure}

A pure state $\ket{\Psi} \in \mathbb{V}^{\otimes L}$ defined on the lattice $\mathcal{L}$
can be expanded in the local basis of $\mathbb{V}_s$ given by $\{\ket{1_s}, \ket{2_s}, \cdots, \ket{\delta_s}\}$ as follows
\be
\label{eq:local_expansion}
	|\Psi\rangle \,= 
	\sum_{i_1=1}^{\delta} ~ \sum_{i_2=1}^{\delta} \cdots \sum_{i_L=1}^{\delta} 
	T_{i_1i_2 \cdots i_L} \ket{i_1} \ket{ i_2} \cdots \ket{i_L}\,.
\ee
This means that $\ket{\Psi}$ is encoded in a tensor $T$ with 
$\delta^{L}$ complex components $T_{i_1i_2 \cdots i_L} \in \mathbb{C}$. 
We refer to the index $1 \leqslant i_s \leqslant \delta$, labelling a local basis for site $s$, as the physical index.

The tensor network approach (see e.g. the review \cite{EvenblyVidal-11}) is a powerful way to rewrite the exponentially large tensor $T$ in  (\ref{eq:local_expansion}) as a combination of smaller tensors.
In order to simplify the notation, drawings are employed to represent the various quantities occurring in the computation.
Tensors are represented by geometric shapes (circles or rectangles) having as many legs as the number of indices of the tensor.
The complex conjugate of a tensor is denoted through the same geometric object  
delimited by a double line.
A line shared by two tensors represents the contraction over the pair of  indices joined by it.

The Matrix Product States (MPS) are tensor networks that naturally arise in the context of the Density Matrix Renormalization Group \cite{White, WhiteNoack, OstlundRommel}. 
They are build through a set of tensors $t_i^{\alpha, \beta}$ (one for each lattice site) with three indices (see the box in Fig. \ref{fig:MPS}): $i$ is the physical index mentioned above, while $\alpha$ and $\beta$ are auxiliary indices. The tensors are contracted following the pattern shown in Fig. \ref{fig:MPS}, where the translational invariance of the state is imposed by employing the same elementary tensor for each site. 
The state in Fig. \ref{fig:MPS} has $L=8$ and it is given by
\be
\label{eq:mps}
|\Psi\rangle = 
\sum_{i_1, \dots, i_8=1}^{\delta} \;
\sum_{\alpha_1, \dots, \alpha_8=1}^{\chi} 
t_{i_1}^{\alpha_1 \alpha_2} \,t_{i_2}^{\alpha_2 \alpha_3}  \cdots \, t_{i_8}^{\alpha_8 \alpha_1} 
\ket{i_1} \ket{ i_2} \cdots \ket{i_8}\,,
\ee
where $\chi$ is the rank of the auxiliary indices, which is called bond dimension in this context. Since we are using the same tensor for each site, the state is completely determined by the components of the tensor $t_{i}^{\alpha \beta}$, which are $\delta \chi^2 $ free parameters.
In the MPS approach, the expectation value of local observables can be computed by performing $\mathcal{O}(\delta \chi^3)$ operations. 
The components $t_{i}^{\alpha \beta}$  of the tensor are obtained numerically by minimizing $\langle \Psi | H | \Psi \rangle$ for the Hamiltonian (\ref{eq:ising}).

\begin{figure}[t] 
\begin{center}
\hspace{0.cm}
\includegraphics[width=.99\textwidth]{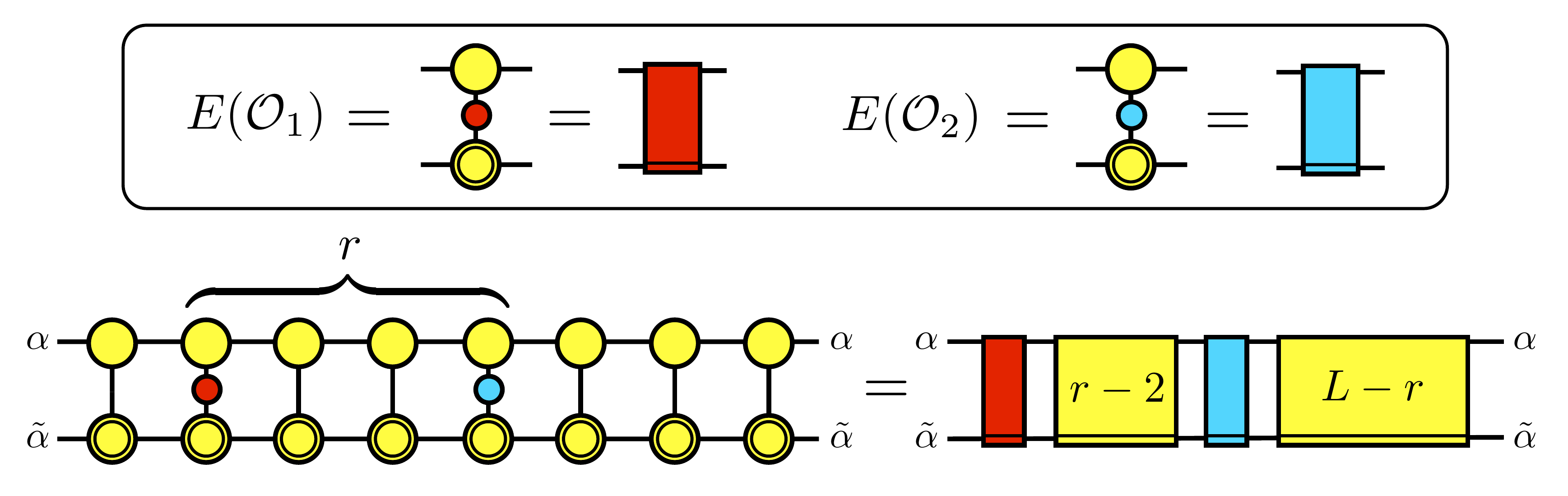}
\end{center}
\vspace{-.3cm}
\caption{\label{fig:mpscorr} 
The two point correlation function $C(r)_{{\cal O}_1, {\cal O}_2}$ of the local operators ${\cal O}_1$ and ${\cal O}_2$.
The corresponding generalized transfer matrices $E({\cal O}_1)$ and $E({\cal O}_2)$, depicted in the box, must be contracted with the proper powers of $E$.}
\end{figure}

The bond dimension $\chi$ controls the accuracy of the results.
Increasing $\chi$,  one can describe an arbitrary state of the Hilbert space \cite{Vidal91}.
In practice, a finite bond dimension which is independent of $L$ allows to describe accurately ground states of gapped local Hamiltonians \cite{Hastings07 area}.
 For gapless Hamiltonians described by a CFT, the  bond dimension has to increase polynomially with the system size \cite{VerstraeteCirac06}, namely $\chi = L ^{1/\kappa}$, where $\kappa $ is an universal exponent \cite{TagliacozzoOliveiraIblisdirLatorre} which depends only on the central charge $c$ as follows: $\kappa= 6/[c(\sqrt{12/c}+1)]$ \cite{PollmannEtAl, PirvuVidalVerstraeteTagliacozzo}. Since the Ising model has $c=1/2$, we have $\kappa \simeq 2$.

In principle, the MPS representation of the ground state allows us to compute several observables. In practice, different computations require a different computational effort.
For instance, considering the bipartition shown in Fig. \ref{fig:MPS}, where $L=8$ and $\ell=4$,
the reduced density matrix $\rho_A$ in a MPS representation  has at most rank $\chi^2$ \cite{EvenblyVidal-11, VerstraeteCiracLatorre-05}, independently on the size of the block. This implies that it  can be computed exactly by performing at most $\mathcal{O}(\delta^3 \chi^6)$ operations.

The case of $N$ disjoint blocks is more challenging. Indeed, the corresponding reduced density matrices in the MPS representation can have rank up to $\chi^{2 N}$, which means that these computations  are exponentially hard in $N$.
Some of these computation can be done by projecting the reduced density matrices on their minimal  rank \cite{atc-10, atc-11}.
Here we describe an alternative approach, which is based on the direct computation of the R\'enyi entropies.

\begin{figure}[t] 
\begin{center}
\vspace{.3cm}
\hspace{0.cm}
\includegraphics[width=.99\textwidth]{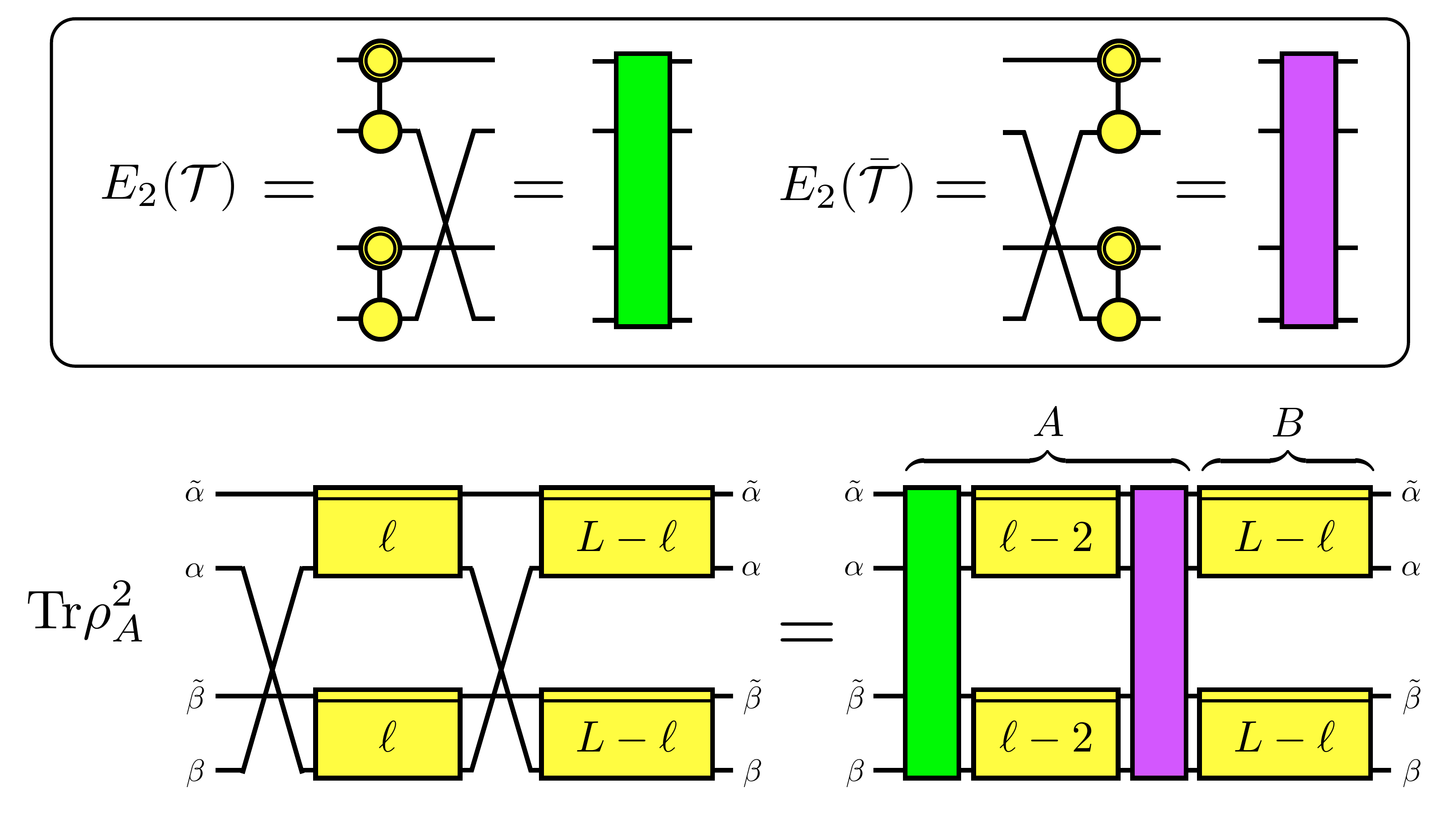}
\end{center}
\vspace{-.3cm}
\caption{\label{fig:twistn2} 
The computation of $\textrm{Tr} \rho_A^2$ of Fig. \ref{fig:mpsrenyi2} as the two point correlation function (see Fig. \ref{fig:mpscorr}) of twist fields in the MPS formalism, i.e. through (\ref{eq:mps_twist}). They are operators acting on the auxiliary degrees of freedom and 
this allows us to define the generalized transfer matrices $E_2({\cal T})$ and $E_2(\bar{{\cal T}})$, which must be contracted with the proper powers of $E_2$.}
\end{figure}

\begin{figure}[t] 
\begin{center}
\vspace{.3cm}
\hspace{0.cm}
\includegraphics[width=.95\textwidth]{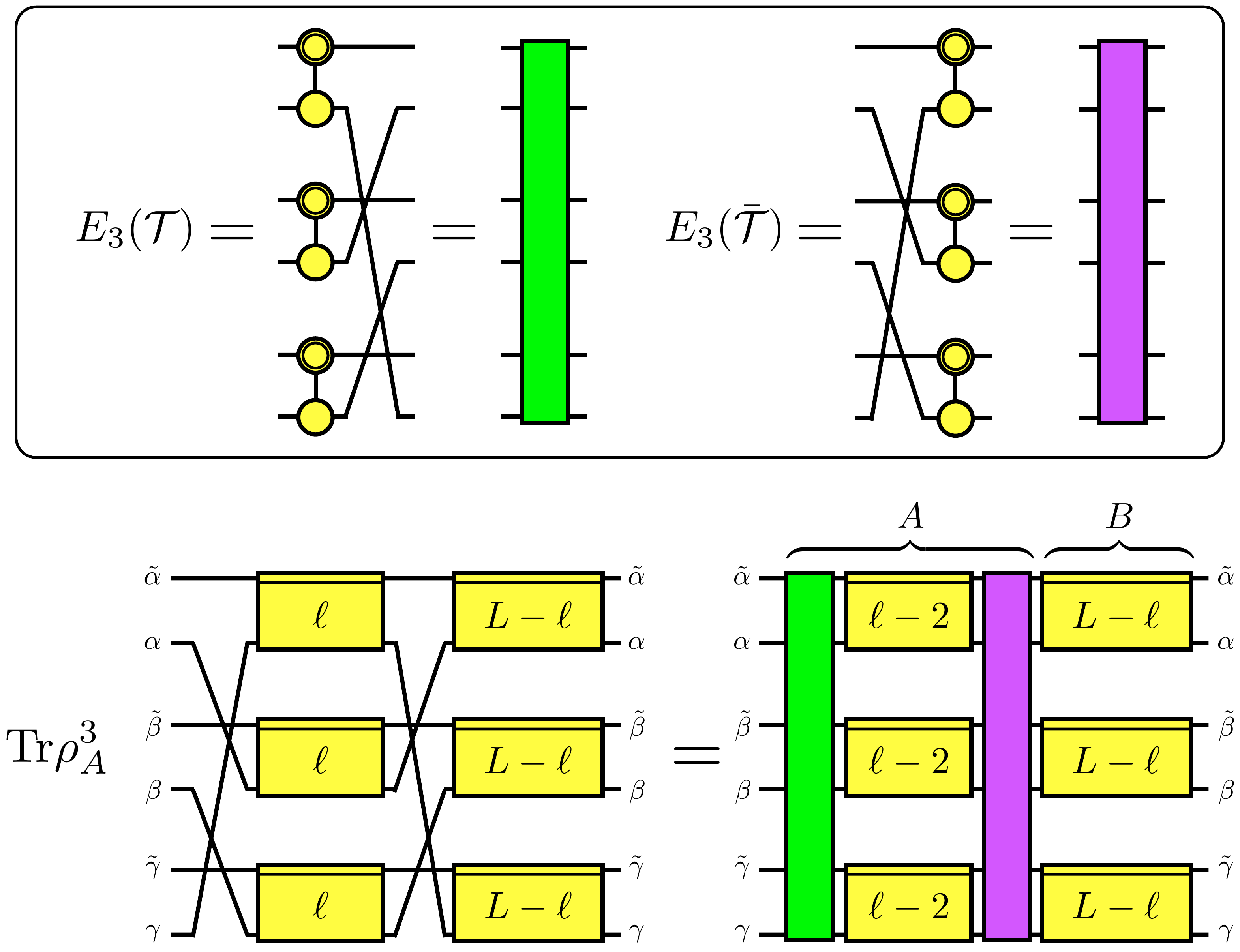}
\end{center}
\vspace{-.3cm}
\caption{\label{fig:twistn3} 
The computation of $\textrm{Tr} \rho_A^3$ of Fig. \ref{fig:mpsrenyi2} as the two point correlation function (see Fig. \ref{fig:mpscorr}) of twist fields (\ref{eq:mps_twist}). In this case the twist fields act on the tensor product of three pairs of virtual indices.
The generalized transfer matrices $E_3({\cal T})$ and $E_3(\bar{{\cal T}})$ are contracted with the proper powers of $E_3 = E \otimes E \otimes E$.}
\end{figure}

\begin{figure}[t] 
\begin{center}
\hspace{-0.3cm}
\includegraphics[width=1\textwidth]{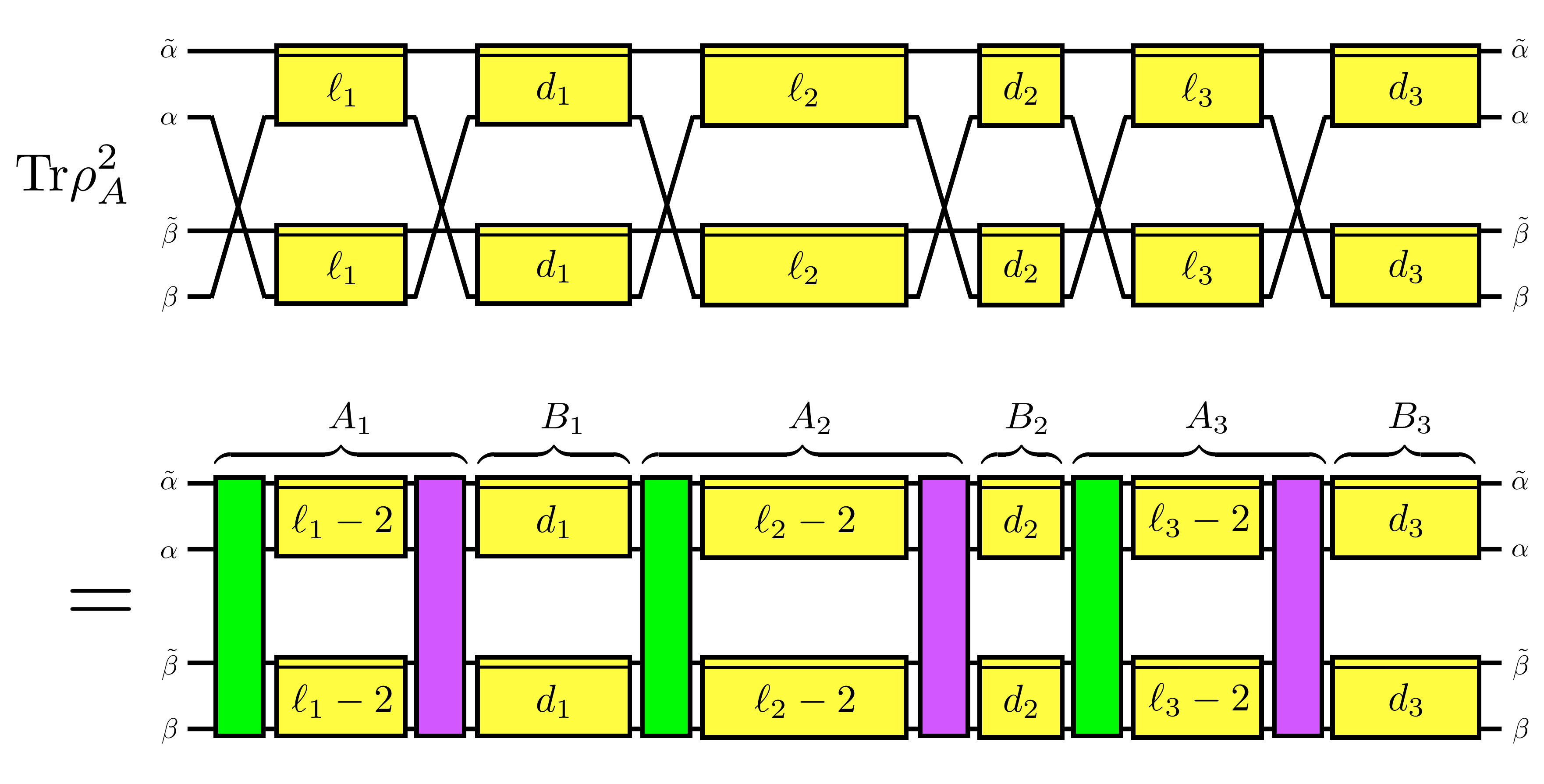}
\end{center}
\vspace{-.3cm}
\caption{\label{fig:mpsRenyiN3} 
The computation of $\textrm{Tr} \rho_A ^n$ through (\ref{eq:mps_twist Nint}) in the case of $N=3$ and $n=2$ as the six point function of twist fields.}
\end{figure}

\subsection{R\'enyi entropies from MPS: correlation functions of twist fields}
\label{sec mps renyi twist}

In the computation of $\textrm{Tr} \rho_A ^n$, which gives the R\'enyi entropies through (\ref{Renyi entropies def}), we need the powers of the MPS transfer matrix $E^{(\alpha, \tilde{\alpha}), (\beta, \tilde{\beta})} \equiv \sum_i{t_{i}^{\alpha, \beta} \bar{t}_i ^{\,\tilde{\alpha}, \tilde{\beta}}}$. Being a mixed tensor involving both $t$ and $\bar{t}$, we represent $E$ as the yellow rectangle in the box of Fig. \ref{fig:mpsrenyi2}, where the double line on one side keeps track of the position of $\bar{t}$. Then, we can straightforwardly construct the $p$-th power $E^p$, which is the key ingredient to obtain $\textrm{Tr} \rho_A ^n$ for a bipartition of the chain. Indeed, when $A$ is made by a block of length $\ell$, it is computed in terms of $E^\ell$ and $E^d$, where $d=L-\ell$.
In Fig. \ref{fig:mpsrenyi2} we represent the computation of $\textrm{Tr} \rho_A ^2$ for the bipartition of Fig. \ref{fig:MPS}. 

Simple manipulations allow us to write the above expression for $\textrm{Tr} \rho_A ^n$ as the two point function of twist fields. 
In order to see this, let us first consider the two point correlation function $C_{{\cal O}_1, {\cal O}_2}(r) \equiv \langle \psi |{\cal O}_1(x) {\cal O}_2(x+r) |\psi \rangle$ of local operators ${\cal O}_1$ and ${\cal O}_2$.
For this computation we introduce the generalized transfer matrix for a generic local operator $\mathcal{O}$ as
\be
\label{Egen def}
E(\mathcal{O})^{(\alpha, \alpha'), (\beta, \beta')} 
\equiv  \sum_{i,j} 
t_{i}^{\alpha, \beta} 
\bar{t}_j^{\,\tilde{\alpha}, \tilde{\beta}}
\mathcal{O}^{i,j} \,,
\ee
whose graphical representation is shown in the box of Fig. \ref{fig:mpscorr}.
Given (\ref{Egen def}), the two point correlation function becomes the following trace of the product of transfer matrices
\begin{equation}
\label{eq:two_point}
C_{{\cal O}_1, {\cal O}_2}(r)
=  
\textrm{Tr} \big(
 E({\cal O}_1)E^{r-2}  E({\cal O}_2) E^{L-r} 
\big) \,,
\end{equation}
which is depicted in Fig. \ref{fig:mpscorr}, where different colors correspond to different operators.

In a similar way, we can write $\textrm{Tr} \rho_A^n$ for the bipartition of Fig. \ref{fig:MPS}
as the two point correlation function of twist fields. This is done by introducing other generalized transfer matrices, namely the tensor product $E_n = E \otimes  \dots \otimes E$ of $n$ transfer matrices and the transfer matrices $E_n({\cal T})$ and $E_n(\bar{{\cal T}})$ associated to the twist fields (see the box in Fig. \ref{fig:twistn2} for $n=2$ and in Fig. \ref{fig:twistn3} for $n=3$). Given these matrices, $\textrm{Tr} \rho_A^n$ reads
\be
\label{eq:mps_twist}
\textrm{Tr} \rho_A^n
= \textrm{Tr} \big(
 E_n({\cal T})E_n^{\ell-2}  E_n(\bar{{\cal T}}) E_n^{L-\ell} 
\big)\,.
\ee
Notice that (\ref{eq:mps_twist}) has the structure of the two point function given in \ref{eq:two_point}, but it is not exactly the same. Indeed, since the twist fields are operators acting on the virtual bonds rather than on the physical bonds, they are not local operators on the original spin chain. In Figs. \ref{fig:twistn2} and \ref{fig:twistn3} we show (\ref{eq:mps_twist}) for $n=2$ and $n=3$ respectively.

It is straightforward to generalize this construction to the case of $N$ disjoint blocks (see Fig. \ref{fig hc} for the notation). In this case $A = \cup_{i=1}^N A_i$ and the generalization of (\ref{eq:mps_twist}) to $N \geqslant 2$ reads
\be
\label{eq:mps_twist Nint}
\textrm{Tr} \rho_A^n
= \textrm{Tr} \big( 
 E_n({\cal T})E_n^{\ell_1-2}  E_n(\bar{{\cal T}}) E_n^{d_1} 
 \cdots 
   E_n({\cal T})E_n^{\ell_N-2}  E_n(\bar{{\cal T}}) E_n^{d_N} 
\big)\,,
\ee
where the dots replace the sequence of terms $E_n({\cal T})E_n^{\ell_j-2}  E_n(\bar{{\cal T}}) E_n^{d_j} $, ordered according to the increasing value of interval index $j=2, \dots, N-1$. 
In Fig. \ref{fig:mpsRenyiN3}, the MPS computation (\ref{eq:mps_twist Nint}) for $N=3$ and $n=2$ is depicted.
It is important to remark that in (\ref{eq:mps_twist Nint}) the computational cost is $O (N \delta \chi^{4 n +1})$, i.e. exponential in $n$ and linear in $N$. Thus, for the simplest cases of $n=2$ and $n=3$ the cost is $\chi^9$ and $\chi^{13}$ respectively.
Because of this, in the remaining part of this section we present numerical results obtained through the exact formula (\ref{eq:mps_twist Nint}) with $n=2$ only, for configurations made by either $N=3$ or $N=4$ disjoint blocks.

The method that we just discussed is very general and, in principle, it can be applied for many lattice models. Nevertheless, the feasibility of the computation strongly depends on the value of the bond dimension $\chi$, which depends on the central charge $c$ as mentioned above. 
Thus, having $c=1/2$, the Ising model is the easiest model that we can deal with. 
A model with $c=1$ would lead to a very high computational cost already for the 
R\'enyi entropy with $n=2$ and this would be a very challenging computation, given the numerical resources at our disposal.
 
As for the approximate calculations of the R\'enyi entropies, a very different scenario arises. 
 In particular, Monte Carlo techniques \cite{CaraglioGliozzi,Hastings10 renyi, WangEtAl, HumeniukRoscilde} look very promising because they allow to obtain an approximate result for $\textrm{Tr} \rho_A^n$  by sampling over the physical indices. Each configuration can be computed with $n \chi^ 3$ operations, but the number of configurations which are necessary to extract a reliable estimation of the R\'enyi entropies in terms of $\chi$ and $n$ is still not understood.
 
\begin{figure}[t]
\vspace{.3cm}
\begin{center}
\includegraphics[width=.7\textwidth]{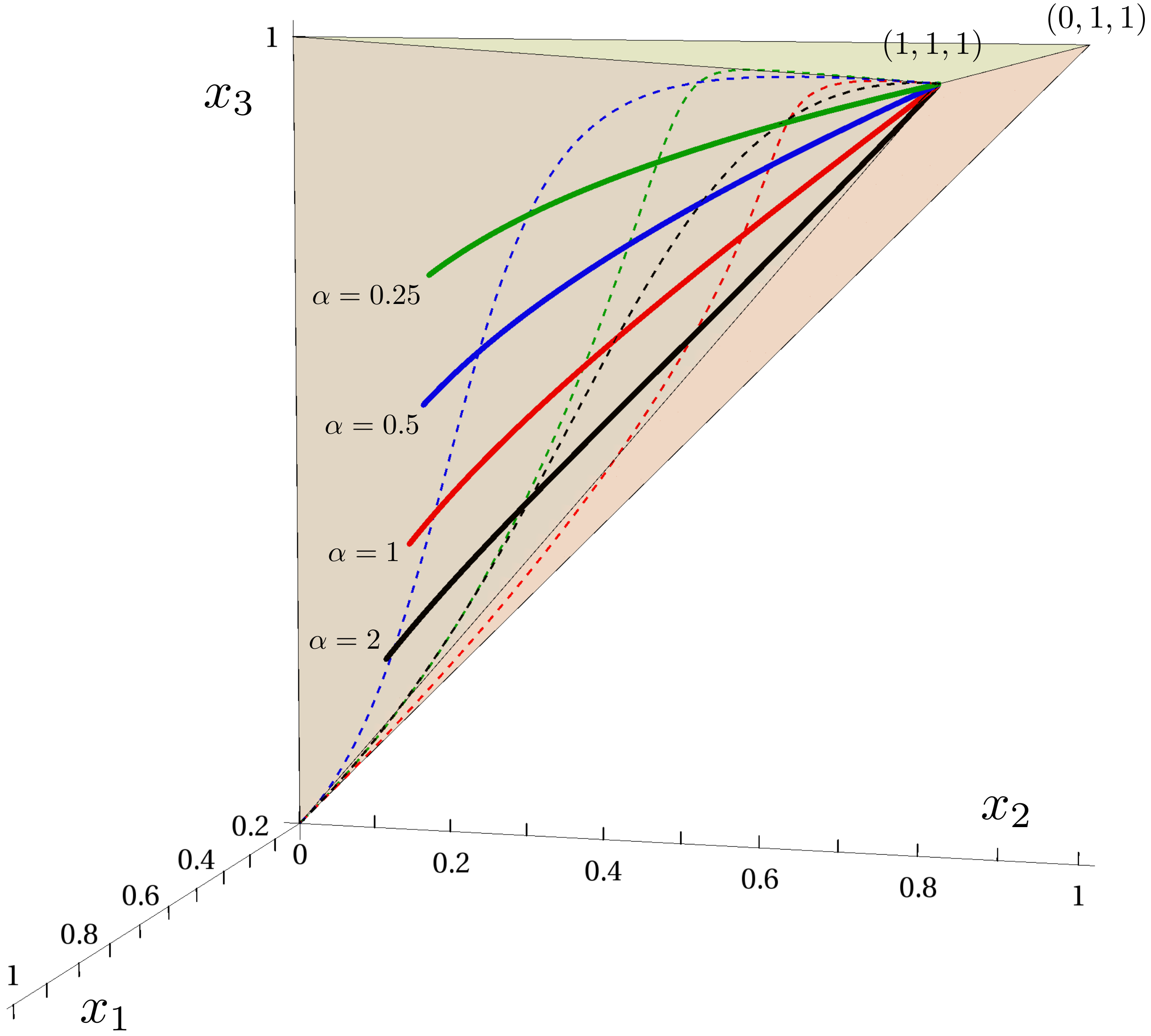}
\end{center}
\vspace{-.4cm}
\caption{The domain $0<x_1<x_2<x_3<1$ for $N=3$, as in Fig. \ref{fig configN3}. The thick lines represent the configurations (\ref{ising configs}) for some choices of $\alpha$. The dashed thin curves are the configurations shown in Fig. \ref{fig configN3}, with the same colors.}
\label{fig isingconfigN3}
\end{figure} 

\begin{figure}[t] 
\begin{center}
\vspace{0.3cm}
\hspace{-.18cm}
\includegraphics[width=.49\textwidth]{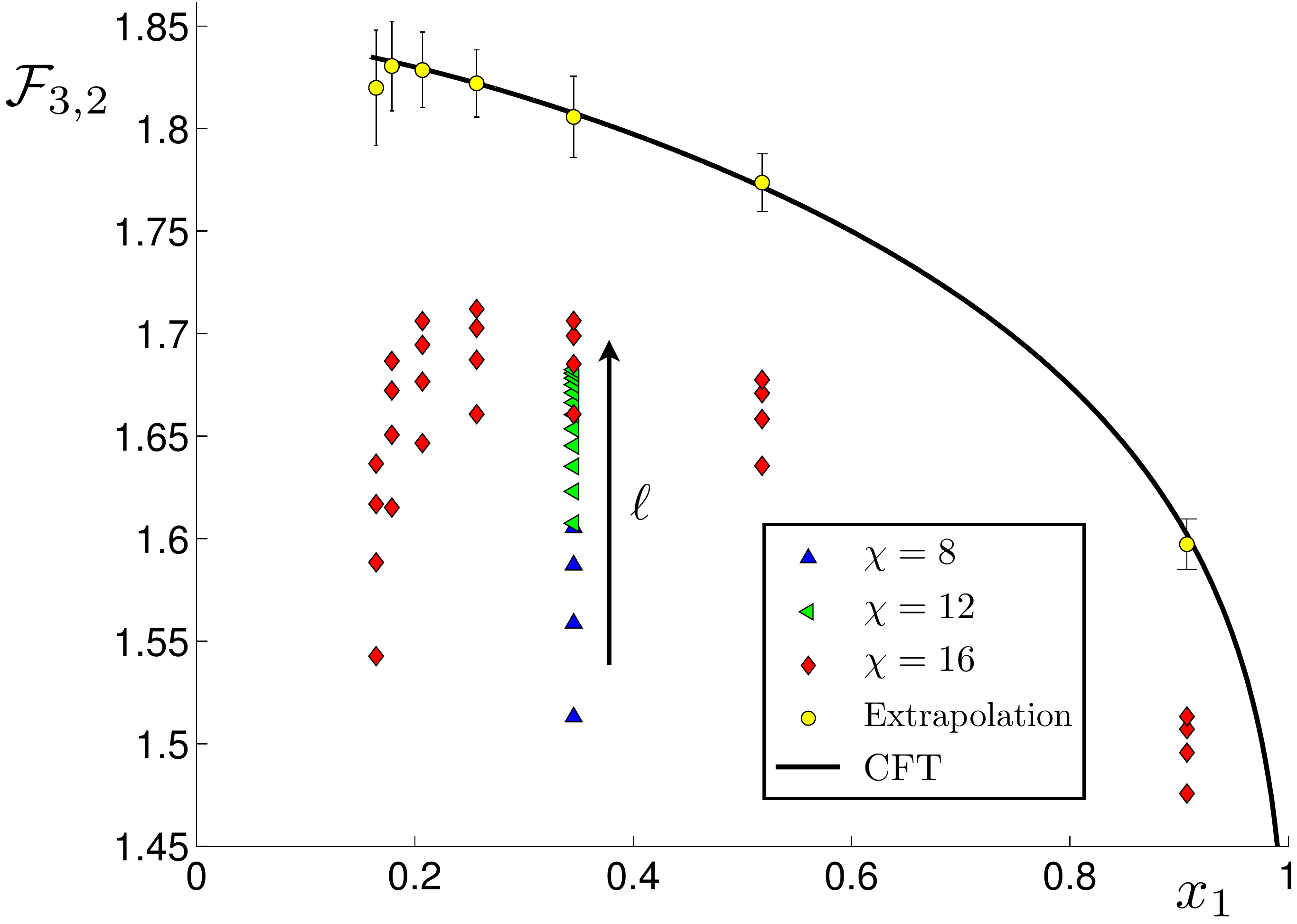}\hspace{.0cm}
\includegraphics[width=.49\textwidth]{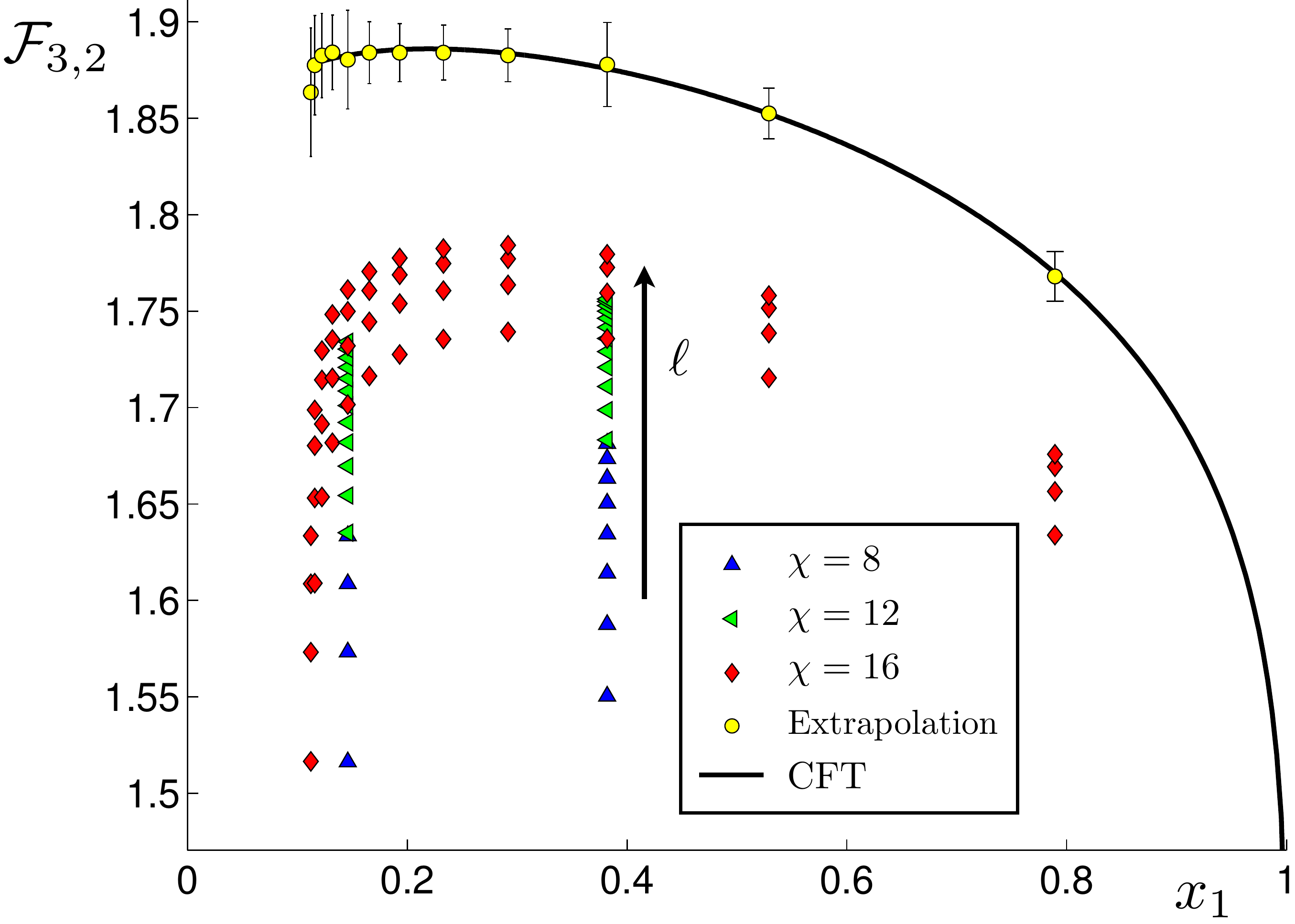}
\\
\vspace{0.5cm}
\includegraphics[width=.49\textwidth]{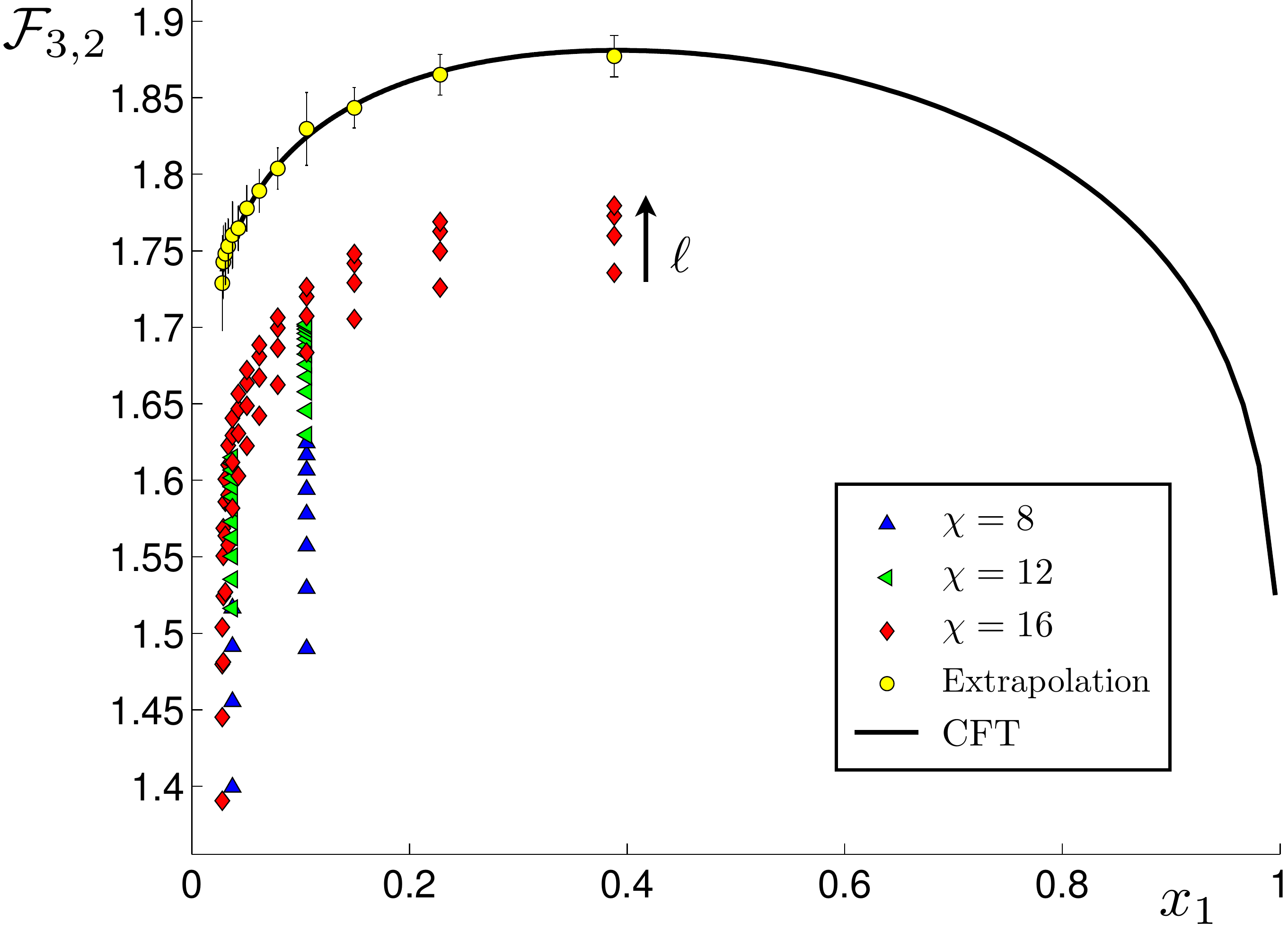}
\includegraphics[width=.49\textwidth]{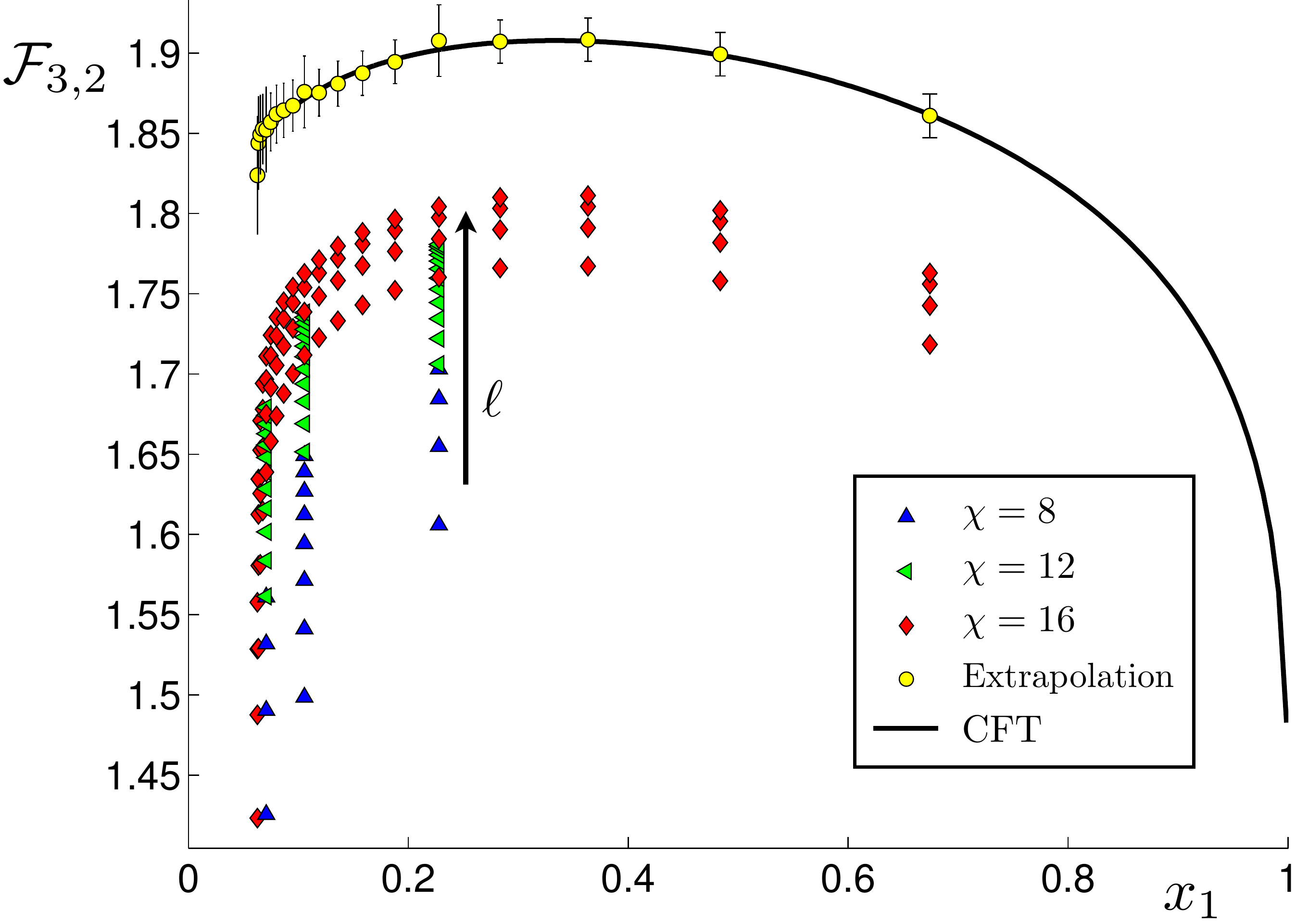}
\end{center}
\vspace{-.3cm}
\caption{\label{fig:isingN3aF}
The results for $\mathcal{F}_{3,2}$ computed through MPS.
The configurations are (\ref{ising configs}) with (from the top left panel, in clockwise direction) $\alpha=0.25$, $\alpha=0.5$, $\alpha=1$ and $\alpha=2$. For a fixed $\boldsymbol{x}$, the length $\ell$ of the blocks increases along the black arrow. The extrapolated points are obtained as explained in \S\ref{sec num ising}.
}
\end{figure}

\begin{figure}[t] 
\begin{center}
\vspace{0.3cm}
\hspace{-0cm}
\includegraphics[width=.488\textwidth]{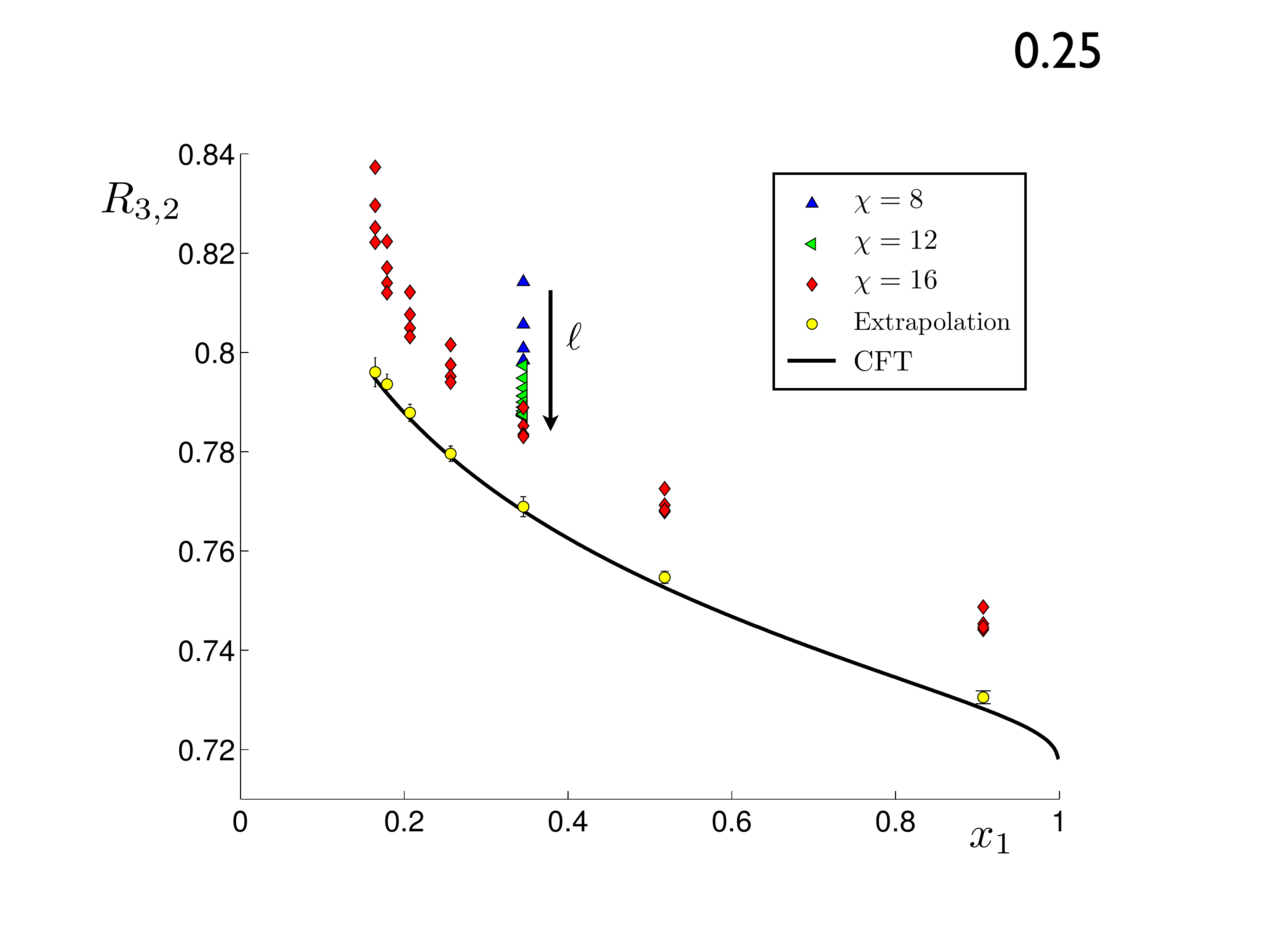}
\includegraphics[width=.488\textwidth]{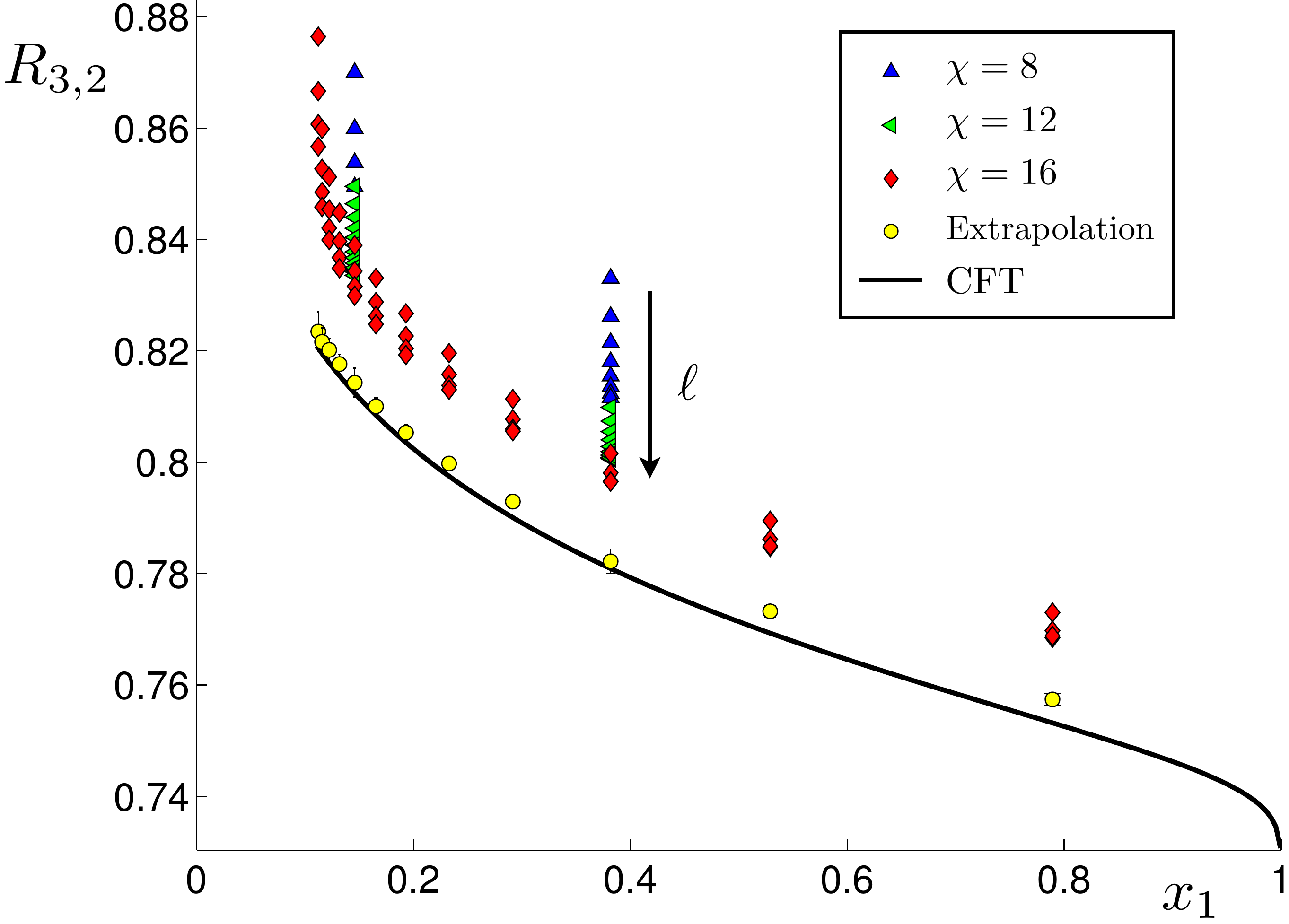}
\\
\vspace{.5cm}
\hspace{-.02cm}
\includegraphics[width=.488\textwidth]{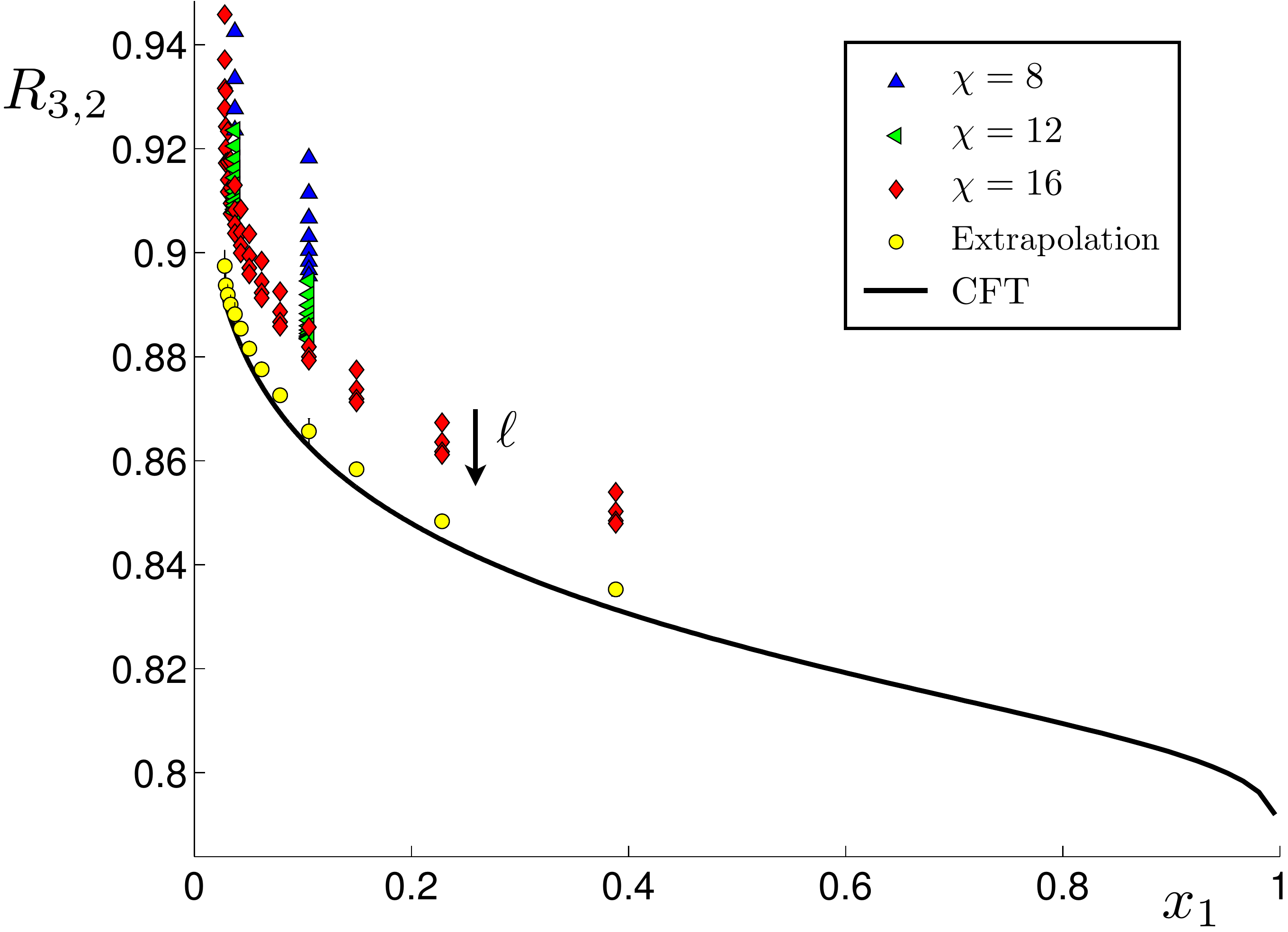}
\hspace{.03cm}
\includegraphics[width=.488\textwidth]{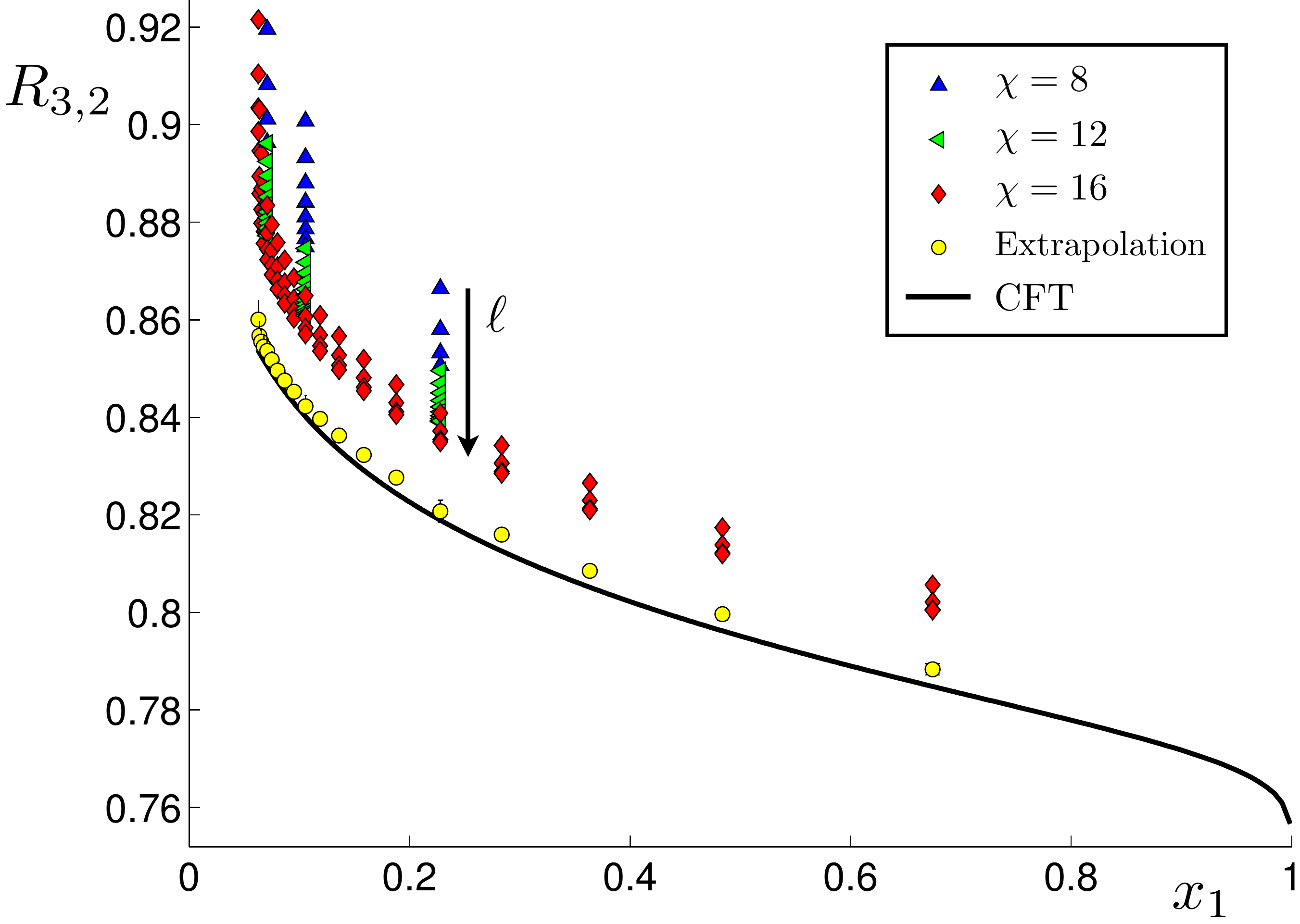}
\end{center}
\vspace{-.3cm}
\caption{\label{fig:isingN3aR}
The results for $R_{3,2}$ computed through MPS.
The configurations are (\ref{ising configs}) with (from the top left panel, in clockwise direction) $\alpha=0.25$, $\alpha=0.5$, $\alpha=1$ and $\alpha=2$. For a fixed $\boldsymbol{x}$, the length $\ell$ of the blocks increases along the black arrow. The extrapolated points are obtained as explained in \S\ref{sec num ising}.
}
\end{figure}

\begin{figure}[t] 
\begin{center}
\vspace{0.3cm}
\hspace{-.1cm}
\includegraphics[width=.488\textwidth]{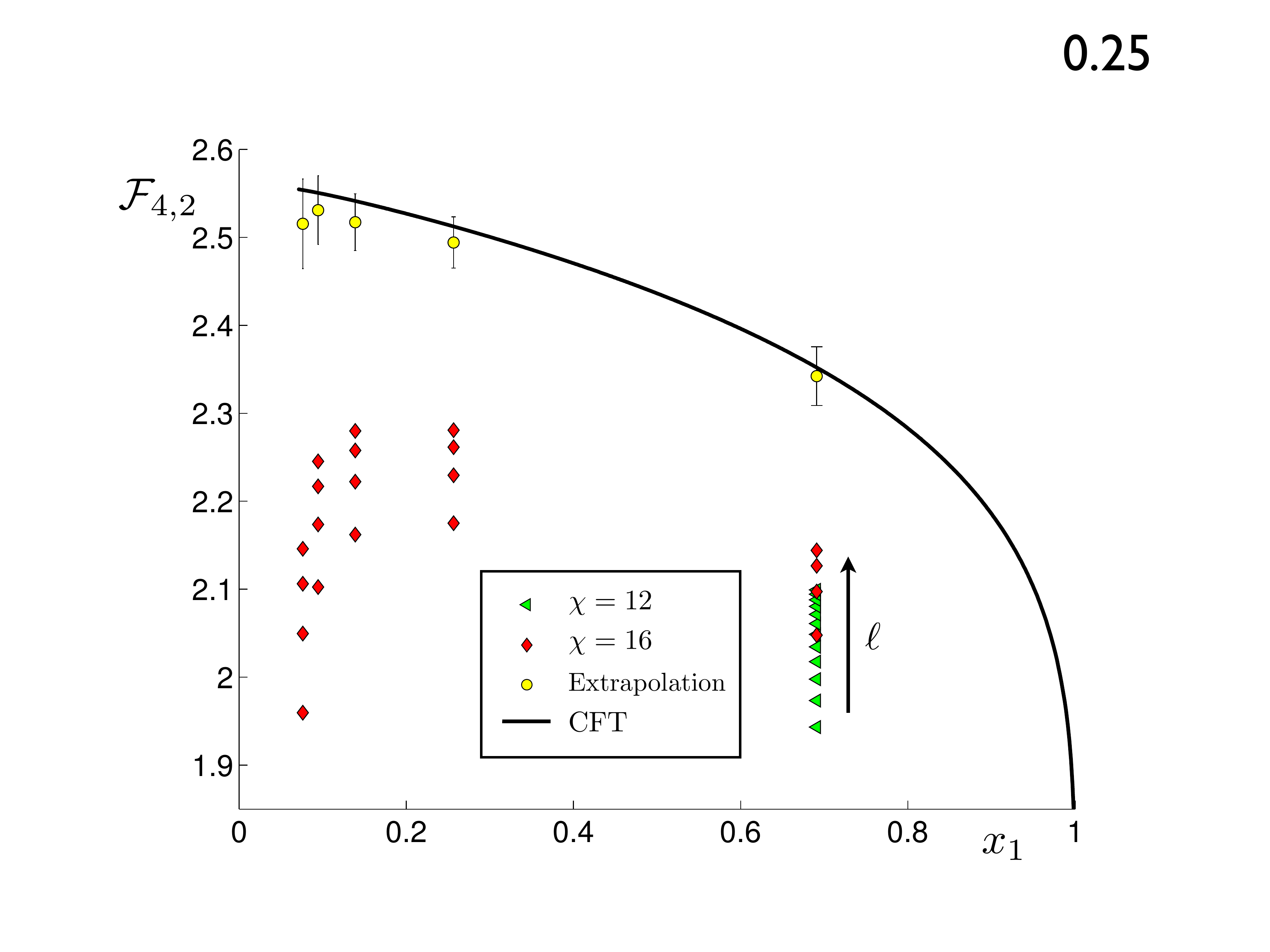}
\hspace{-.2cm}
\includegraphics[width=.488\textwidth]{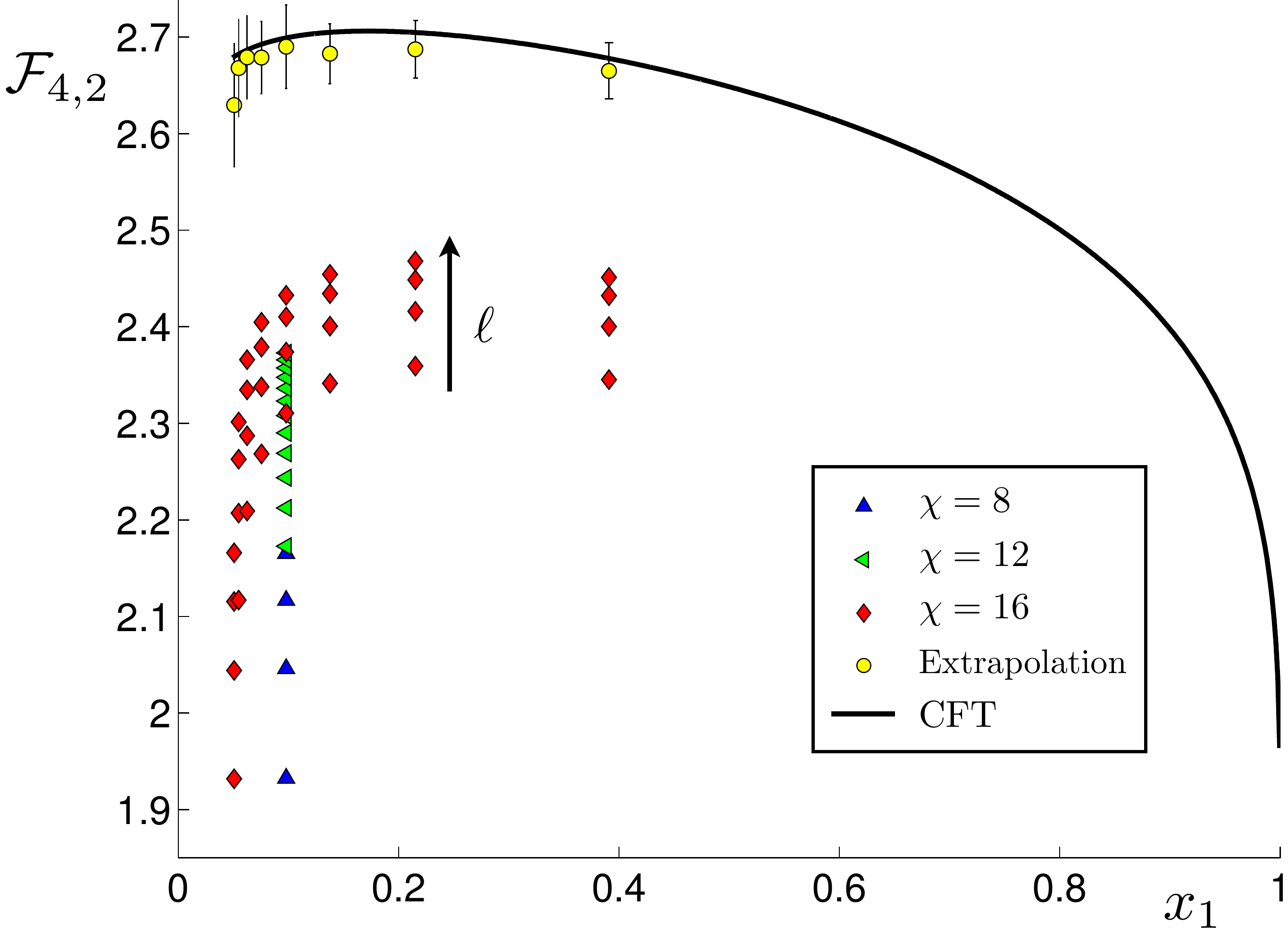}
\\
\vspace{.5cm}
\includegraphics[width=.488\textwidth]{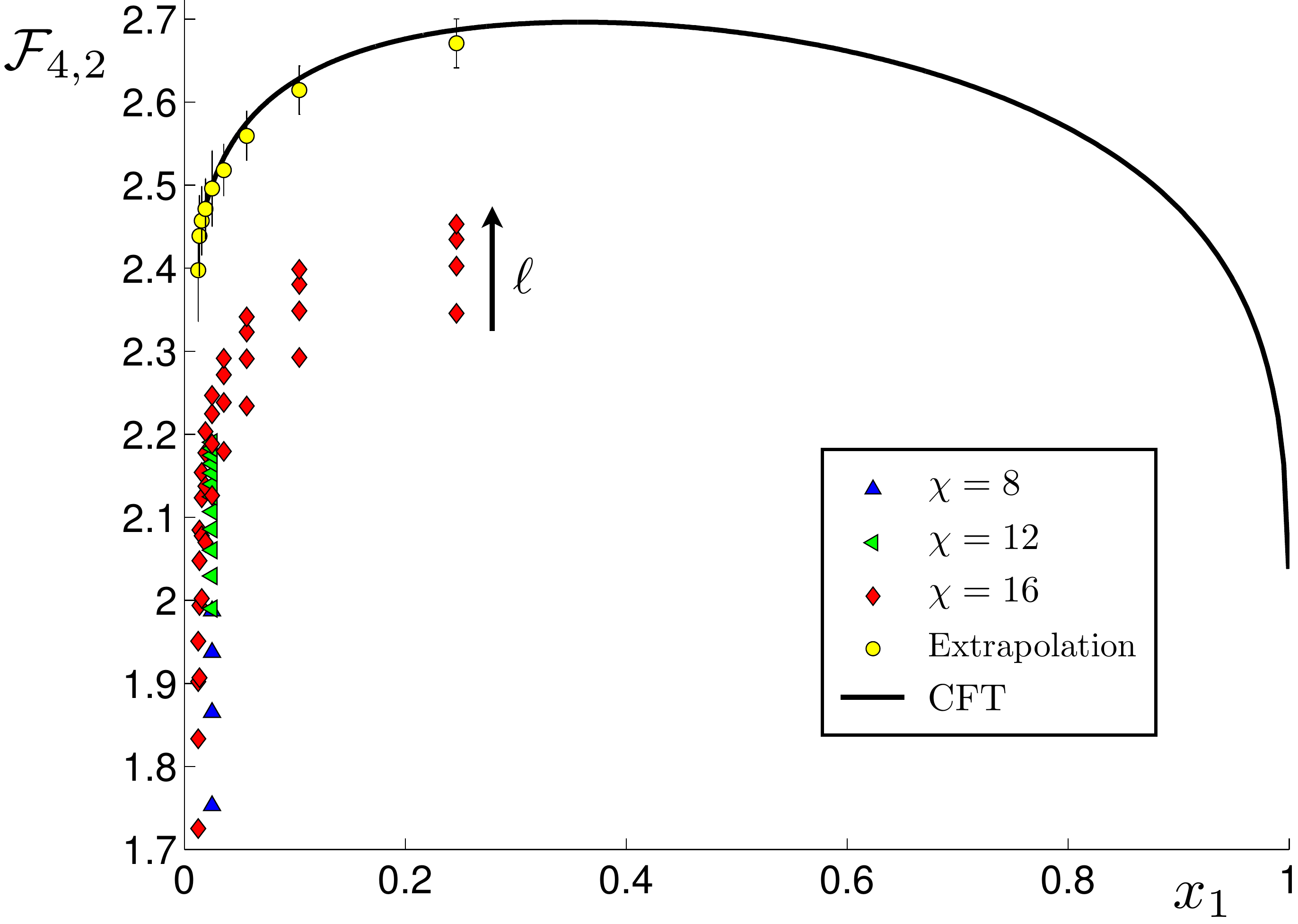}
\hspace{.02cm}
\includegraphics[width=.488\textwidth]{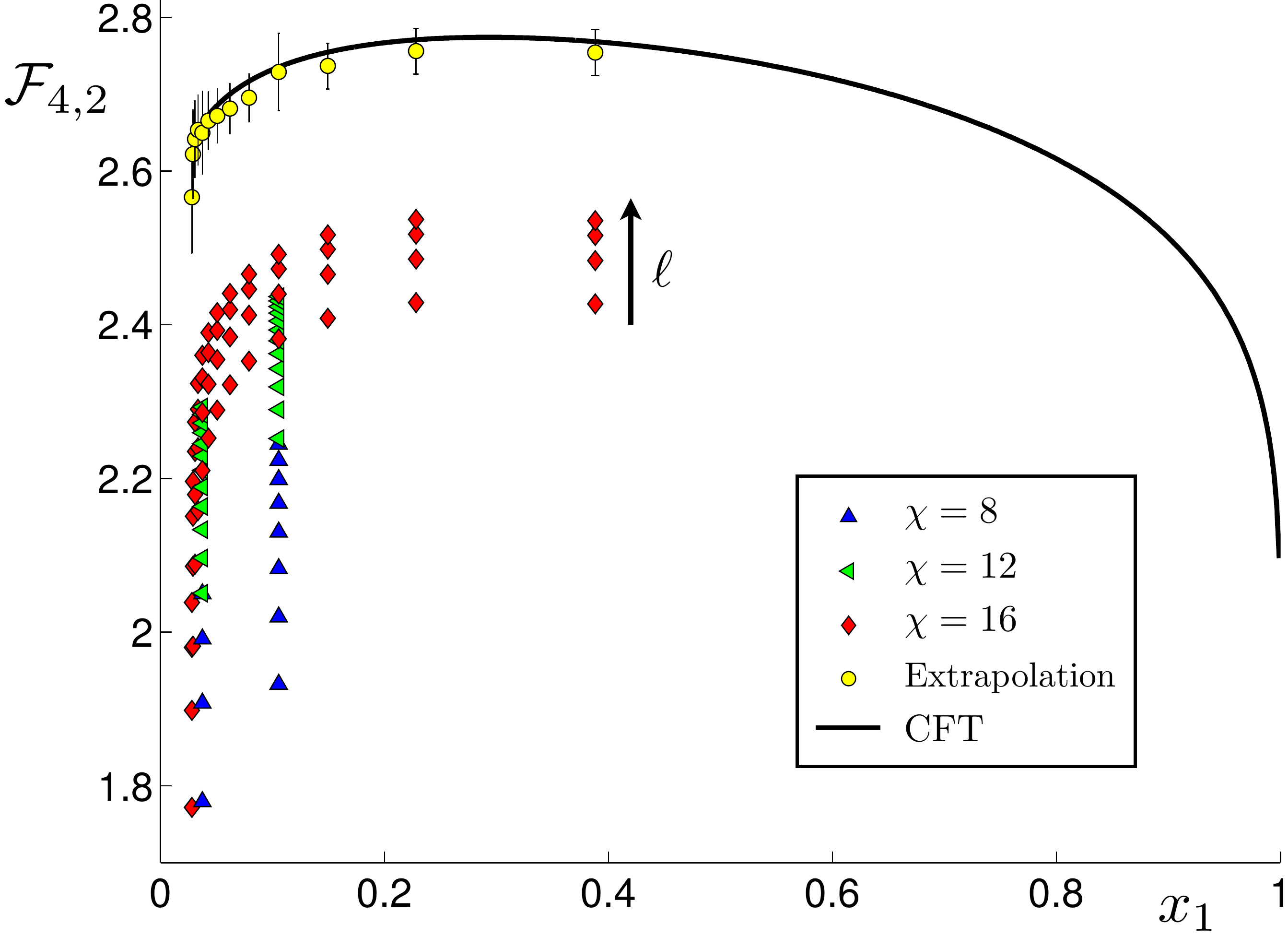}
\end{center}
\vspace{-.3cm}
\caption{\label{fig:isingN4}
The results for $\mathcal{F}_{4,2}$ computed through MPS.
The configurations are (\ref{ising configs}) with (from the top left panel, in clockwise direction) $\alpha=0.25$, $\alpha=0.5$, $\alpha=1$ and $\alpha=2$. For a fixed $\boldsymbol{x}$, the length $\ell$ of the blocks increases along the black arrow. The extrapolated points are obtained as explained in \S\ref{sec num ising}.
}
\end{figure}

 \subsection{Numerical results for $n=2$}
 \label{sec num ising}

Let us discuss the numerical results obtained through the method discussed in \S\ref{sec mps renyi twist} about $\Tr \rho^2_A$  for the Ising model with periodic boundary conditions.
The length $L$ of the chains varies within the range  $30 \leqslant L \leqslant 500$.
The MPS matrices have been computed by employing the variational algorithm described in \cite{PirvuEtAl} (see also the ones in \cite{VerstraetePorrasCirac-04, PippanWhiteEvertz}). 
Moreover, from Fig. 2 of \cite{PirvuVidalVerstraeteTagliacozzo} one observes that, in order to find accurate results for the Ising model in the range of total lengths given above, we need $8 \leqslant \chi \leqslant 16$.

As for the configurations of the $N$ disjoint blocks of sites, denoting by $\ell_i$ the number of sites for the block $A_i$ and by $d_i$ the number of sites separating $A_i$ and $A_{i+1 \,\textrm{mod}\, N}$ with $i =1, \dots, N$ as in \S\ref{sec HC} (see Fig. \ref{fig hc} for the case $N=3$), we find it convenient to choose the following ones
\be
\label{ising configs}
\rule{0pt}{.7cm}
\begin{array}{|ccccccccc|}
                \ell_1 & d_1 & \ell_2 & d_2 & \ell_3 & d_3 & \dots & \ell_N & d_N \\
\hline
\ell  & d  &  \ell  & d  &  \ell & d  & \dots &  \ell  & d_N     \\
\hline
\end{array}
\qquad 
\begin{array}{c}
\\ 
d = \alpha \ell\,,
\end{array}
\ee
where $d_N = L - [N + \alpha(N-1)]\ell$ is fixed by the consistency condition (\ref{Ltot condition}) on the total length of the chain. Thus, each configuration is characterized by the coefficient $\alpha$ and the free parameter is $\ell$. 
In the comparison with the CFT expressions discussed in \S\ref{sec riemann surf} and \S\ref{sec ising cft}, we have taken the finiteness of the system into account  through (\ref{xratios finiteL odd}) and (\ref{xratios finiteL even}), as already done in \S\ref{sec HC} for the harmonic chain. 
Like for (\ref{hc configs}) with the vectors $\boldsymbol{\lambda}$ and $\boldsymbol{\gamma}$ fixed, also for the configurations (\ref{ising configs}) with $\alpha$ fixed 
the harmonic ratios $x_i$ depend only on $\ell/L$, providing one dimensional curves within the $2N-3$ dimensional configuration space $0< x_1< x_2 < \dots < x_{2N-3}<1$.
Nevertheless, notice that in this case the harmonic ratios have a strictly positive lower bound, which can be computed by taking the limit $\ell/L \rightarrow 0$ in the expressions of $x_i$ obtained by specializing (\ref{xratios finiteL odd}) and (\ref{xratios finiteL even}) to (\ref{ising configs}).
For instance, when $N=3$ we have $x_1 = [\sin(\pi \ell/L)/ \sin(2\pi (1+\alpha)\ell/L)]^2$, whose smallest value reads $1/[2(1+\alpha)]^2$. 
Always for $N=3$, in Fig. \ref{fig isingconfigN3} we show the curves corresponding to the configurations (\ref{ising configs}) for the numerical values of $\alpha$ considered in the remaining figures. Each curve can be equivalently parameterized by one of the harmonic ratios and in this section we choose $x_1$ as the independent variable.

Given the configurations (\ref{ising configs}), for any fixed $\alpha$ different values of $\ell$ and $L$ having the same $\ell/L$ provide the same $\boldsymbol{x}$, i.e. the same point in the configurations space. Aligning the numerical data corresponding to the same $\boldsymbol{x}$, one observes that, as $\ell$ increases, they approach the CFT prediction. Nevertheless, the discrepancy is quite large because the chains at our disposal are not long enough. Thus, unlike the case of the harmonic chain discussed in \S\ref{sec HC}, for the Ising model the plots of the data do not immediately confirm the CFT expressions. 

During the last few years many papers have studied the 
corrections to the leading scaling behavior of the R\'enyi entropies \cite{LaflorencieAffleck-05,CalabreseCampostrini09, cc-unusual, atc-10, FagottiCalabrese10, CalabreseEssler-10, CalabreseFagotti-10open, atc-11, ctt-neg-ising, XavierAlcaraz-12}.
When $A$ is a single block made by $\ell$ contiguous lattice sites within a periodic chain of length $L$, the first deviation of $\Tr \rho_A^n$ from the corresponding value obtained through the CFT expression is proportional to $\ell^{-2\Delta/n}$, for some $\Delta <2$. From the field theoretical point of view, this unusual scaling can be understood by assuming that the criticality is locally broken at the branch points and this allows the occurrence of relevant operators with scaling dimension $\Delta < 2$ at those points \cite{cc-unusual}.
For the Ising model the relevant operators must be also parity even and this means that the first correction is proportional  to $\ell^{-2/n}$. 
Instead, when $A$ is made by two disjoint blocks, it has been numerically observed that the leading correction for the Ising model is proportional to $\ell^{-1/n}$ \cite{atc-10, FagottiCalabrese10}, which agrees with $\ell^{-2\Delta/n}$ with $\Delta=1/2$. This could be the contribution of the Majorana fermion introduced by the Jordan-Wigner string between the two blocks \cite{FagottiCalabrese10}.

In the following we consider the case of $A$ made by three and four disjoint blocks, focusing  on $\mathcal{F}_{3,2}$ and $R_{3,2}$ for $N=3$ and on $\mathcal{F}_{4,2}$ for $N=4$.
We studied the configurations (\ref{ising configs}) with $\alpha = p$ and $\alpha=1/p$, where for the integer $p$ we took $1 \leqslant p \leqslant 8$.
Here we show the plots only for $\alpha \in \{0.25, 0.5, 1,2\}$ because the ones for the remaining values of $\alpha$ are very similar.
The results for $N=3$ are reported in Figs. \ref{fig:isingN3aF} and \ref{fig:isingN3aR}, while the ones for $N=4$ are given in Fig. \ref{fig:isingN4}.
Different colored shapes denote numerical data which have been obtained from ground states with different bound dimensions. Moreover, for fixed values of $\boldsymbol{x}$ and $\chi$, the black arrow indicates the direction along which $\ell$ increases.
For a given $\chi$, the maximum value $L_{\textrm{max}}$ of the total size of the chain has been determined according to Fig. 2 of \cite{PirvuVidalVerstraeteTagliacozzo}.
In particular, for $\chi=8$, $\chi=12$ and $\chi=16$ we used respectively $L_{\textrm{max}} = 100$, $L_{\textrm{max}} = 320$ and $L_{\textrm{max}} = 500$.

Notice that larger values of $\chi$ and $\ell$ better approximate the points obtained through the CFT formulas, as expected. Nevertheless, since the discrepancy between our best numerical value and the one predicted by the CFT is quite large, a finite size scaling analysis is necessary, as discussed above. 
For almost every value of $\boldsymbol{x}$ that we are considering, taking the effects of the first correction into account is enough to find reasonable agreement with the CFT predictions.
According to the analysis discussed in Appendix \ref{app:expo}, we find that the first correction is proportional to $\ell^{-\Delta_{\textrm{num}}}$, where $\Delta_{\textrm{num}} = 0.45(5)$ for both $\mathcal{F}_{3,2}$ and $\mathcal{F}_{4,2}$, and  $\Delta_{\textrm{num}} = 0.51(4)$ for $R_{3,2}$. We remark that these exponents have been found just from the numerical data, without assuming the CFT formulas.
The result is compatible with $\Delta=1/2$ found for two disjoint blocks \cite{atc-10, FagottiCalabrese10}. Thus, this result seems to be independent of the number of intervals.

Once the exponents have been determined, we can compare the numerical results with the CFT predictions. This means that, for $N=3$ and $N=4$, we consider
\be
\label{1corr cft}
\mathcal{F}^{\textrm{\tiny \,lat}}_{N,2}(\boldsymbol{x}) 
= \mathcal{F}^{\textrm{\tiny \,ext}}_{N,2}(\boldsymbol{x}) + 
\frac{f_{N}(\boldsymbol{x})}{\ell^{\Delta_{\textrm{num}} }}\,,
\qquad
R^{\textrm{\tiny \,lat}}_{3,2}(\boldsymbol{x}) 
= R^{\textrm{\tiny \,ext}}_{3,2}(\boldsymbol{x}) + 
\frac{r(\boldsymbol{x})}{\ell^{\Delta_{\textrm{num}} }}\,,
\ee
where $\Delta_{\textrm{num}}$ are the exponents given above.
For any fixed $\boldsymbol{x}$, we have two parameters to fit: the coefficient of $\ell^{-\Delta_{\textrm{num}} }$ and the extrapolated value. The latter one must be compared with the corresponding value obtained through the CFT formula. 
Since we have to find only two parameters through this fitting procedure, we can carry out this analysis for all the $\boldsymbol{x}$'s at our disposal, also when few numerical points occur.
Because of the uncertainty on $\Delta_{\textrm{num}}$, for any fixed $\boldsymbol{x}$ we perform the extrapolation for both the maximum and the minimum value of $\Delta_{\textrm{num}}$. This provides the error bars indicated in Figs. \ref{fig:isingN3aF}, \ref{fig:isingN3aR} and \ref{fig:isingN4}, where the yellow circles denote the mean values.

In Appendix \ref{app:fss3corr} we consider more than one correction, keeping the same exponents employed for the case $N=2$ \cite{FagottiCalabrese10, atc-11,ctt-neg-ising}. Unfortunately, this analysis can be performed only for those few values of $\boldsymbol{x}$ at fixed $\alpha$ which have many numerical points (see Figs. \ref{fig:isingN3fss3corrF} and  \ref{fig:isingN3fss3corrR}). 
We typically find that the second correction improves the agreement with the corresponding CFT prediction, as expected, while the third one does not, telling us that, probably, given our numerical data, we cannot catch the third correction.

In Appendix \ref{app:finitechi} we briefly consider the effects due to the finiteness of the bond dimension in our MPS computations. They occur because finite $\chi$ leads to a finite correlation length $\xi_\chi$ and, whenever it is smaller than the relevant length scales
a deviation from the expected power law behavior of the correction is observed
\cite{Fannes92, VerstraeteCirac06, TagliacozzoOliveiraIblisdirLatorre, PirvuVidalVerstraeteTagliacozzo}.

\section{Conclusions}

In this paper we have computed the R\'enyi entropies of $N$ disjoint intervals for the simple conformal field theories given by the free compactified boson and the Ising model.  

For the free boson compactified on a circle of radius $R$, we find that $\Tr \rho_A^n$ for $A=\cup_{i=1}^N A_i$ with $N\geqslant 2$ is given by (\ref{Tr rhoA N int intro}) with $c=1$ and 
\be
\label{Fboson conc}
\mathcal{F}_{N,n}( \boldsymbol{x}) 
= 
\frac{\Theta (\boldsymbol{0} | T_\eta )}{| \Theta (\boldsymbol{0} | \tau ) |^2}\,,
\qquad
T_\eta  = \begin{pmatrix}
\,{\rm i}\, \eta\, \mathcal{I} & \mathcal{R} 
\\
\mathcal{R} & {\rm i}  \,\mathcal{I}/\eta\,
\end{pmatrix} ,
\ee
where $\eta \propto R^2$, the function $\Theta$ is the Riemann theta function (\ref{theta def}) and  $\tau = \mathcal{R}  +{\rm i}\,\mathcal{I} $ is the period matrix of the Riemann surface $\mathscr{R}_{N,n}$ defined by (\ref{eq curve}), which has genus $g=(N-1)(n-1)$ (see e.g. Fig. \ref{fig EGcage}, where $N=3$ and $n=4$).
As for the Ising model, we find that $\Tr \rho_A^n$ is (\ref{Tr rhoA N int intro}) with $c=1/2$ and
\be
\label{Fising conc}
\mathcal{F}^{\textrm{\tiny Ising}}_{N,n}( \boldsymbol{x}) =
\frac{\sum_{\boldsymbol{e}} | \Theta [\boldsymbol{e} ](\boldsymbol{0} | \tau ) |}{
2^{g} \,| \Theta (\boldsymbol{0} | \tau ) |} \,,
\ee
being $\boldsymbol{e} $ the characteristics of the Riemann theta function, defined through (\ref{theta def spin}).
The period matrix of $\mathscr{R}_{N,n}$ \cite{eg-03} has been computed for two different canonical homology bases and, given the relation between them, one can employ either (\ref{tau eg tensor}) or (\ref{tau amv tensor}) in the expressions (\ref{Fboson conc}) and (\ref{Fising conc}).
The peculiar feature of the free compactified boson and of the Ising model is that, in order to write the R\'enyi entropies, we just need the period matrix of $\mathscr{R}_{N,n}$.

We have checked (\ref{Fboson conc}) in the decompactification regime against exact results for the harmonic chain with periodic boundary conditions, finding excellent agreement.
As for the Ising model, we have performed an accurate finite size scaling analysis using Matrix Product States. 
In particular we have identified the twist fields within this formalism, showing that the R\'enyi entropies can be computed as correlation functions of twist fields also in this case. 
Whenever a reliable finite size scaling analysis can be performed, the numerical results confirm  (\ref{Fising conc}).
The results of \cite{cct-09, cct-11} for two disjoint intervals are recovered as special cases of (\ref{Fboson conc}) and (\ref{Fising conc}). 

We have not been able to analytically continue (\ref{Fboson conc}) and (\ref{Fising conc}), in order to find the entanglement entropy. We recall that this is still an open problem in the simplest case of two intervals for the free boson at finite $\eta$ and for the Ising model.
For the boson on the infinite line, we have shown numerical predictions for the tripartite information and for the corresponding quantities in the case of $N>3$.

It is very important to provide further numerical checks of our CFT predictions, in particular for the free boson at finite compactification radius, as done in  \cite{FagottiCalabrese10, atc-11} for two intervals.
Let us mention that it would be extremely interesting to extend the field theoretical computation of the R\'enyi entropies and of the entanglement entropy of disjoint regions to the massive case \cite{Doyon-review} and to higher dimensions \cite{Cardy13}.

\section*{Acknowledgments}

We thank Pasquale Calabrese and Ferdinando Gliozzi for their comments on the draft.
AC and ET are particularly grateful to Tamara Grava for many useful discussions.
LT would like to thank Alessio Celi, Andrew Ferris and Tommaso Roscilde.
ET would like to thank the organizers of the workshop ``Gravity - New perspectives from strings and higher dimensions'', Centro de Ciencias de Benasque, for hospitality during part of this project. LC is supported by FP7-PEOPLE-2010-IIF ENGAGES 273524 and ERC QUAGATUA.

\begin{appendices}

\section*{Appendices}

\section{On the $\boldsymbol{x}$ dependence of $R_{N,n}$}
\label{app x dependence}

In this appendix we give some details about the ratio $R_{N,n}$ defined in (\ref{RNn def}) in the case of two dimensional conformal field theories, when $A = \cup_{i=1}^N A_i$.

In the simplest case of $N=2$ there is only one harmonic ratio $x \in (0,1)$ defined through (\ref{map wN}).
The two quantities (\ref{tildeRNn def}) and (\ref{RNn def}) coincide and one easily finds that
\be
R_{2,n}(x) = \tilde{R}_{2,n}(x) = \frac{\mathcal{F}_{2,n}(x)}{(1-x)^{2\Delta_n}}\,.
\ee
When $N>2$, first we remark that the non universal constant $c_n$ cancels in the ratio (\ref{RNn def}) and this is found by employing the same combinatorial identity occurring for the cutoff independence of $R_{N,n}$, discussed in the section \ref{sec riemann surf}.
Moreover in (\ref{RNn def}) all the factors $P_p(\sigma_{N,p}) $ cancel, namely 
\be
\prod_{p\,=\,1}^{N}\, \prod_{\sigma_{N,p}} \big[  P_p(\sigma_{N,p}) \big]^{(-1)^{N-p}}  = 1\,.
\ee
This result can be obtained by writing the l.h.s. as the product of two factors
\be\fl
\label{prod sep}
\prod_{p\,=\,1}^{N}\, \prod_{\sigma_{N,p}}  \, \prod_{i \, \in\, \sigma_{N,p}} 
\frac{1}{(v_i -u_i)^{(-1)^{N-p}}} \,,
\qquad
\prod_{p\,=\,1}^{N}\, \prod_{\sigma_{N,p}} \;
\prod_{\substack{ i,j \in \sigma_{N,p} \\ i\,<\,j }}
\bigg[\,\frac{(u_j-u_i)(v_j-v_i)}{(v_j-u_i)(v_i-u_j)}\,\bigg]^{(-1)^{N-p}} .
\ee
Then, collecting the different factors, they become respectively 
\be\fl
\prod_{p=1}^{N}\, \prod_{i\,=\,1}^{N}
\frac{1}{(v_i -u_i)^{\xi_p(-1)^{N-p}}} \,,
\qquad \hspace{.9cm}
\prod_{p\,=\,1}^{N}\;
\prod_{\substack{  i,j =1  \\ i\,<\,j }}^N
\bigg[\,\frac{(u_j-u_i)(v_j-v_i)}{(v_j-u_i)(v_i-u_j)}\,\bigg]^{\zeta_p(-1)^{N-p}} ,
\ee
where we denoted by $\xi_p = \binom{N-1}{p-1}$ the number of choices $\sigma_{N,p}$ containing the $i$-th interval and by $\zeta_p= \binom{N-2}{p-2}$ the number of $\sigma_{N,p}$'s containing both the $i$-th and $j$-th interval. By employing the combinatorial identities $\sum_{p=1}^N (-1)^{N-p}\xi_p =0$ and $\sum_{p=2}^N (-1)^{N-p}\zeta_p =0$ respectively, it is straightforward to conclude that  the products in (\ref{prod sep}) are separately equal to 1.
Thus, we have that $R_{N,n}(\boldsymbol{x})$ is given by (\ref{RNn cft}).

As for  the dependence on $\boldsymbol{x}$ of (\ref{RNn cft}), let us consider the choice  $\sigma_{N,p} = \{ i_1 ,\dots , i_p\}$ of $p$ intervals with $1<p \leqslant N$, corresponding to the subregion $A_{i_1} \cup \dots \cup A_{i_p} $ included in $A$. Then one introduces the map
\be
\label{map wsigma}
w_{\sigma_{N,p} }(z) = \frac{(u_{i_1}-z)(u_{i_p}-v_{i_p})}{(u_{i_1}-u_{i_p})(z-v_{i_p})}\,,
\ee
which is constructed to send $u_{i_1} \rightarrow 0$, $u_{i_p} \rightarrow 1$ and $v_{i_p} \rightarrow \infty$. When $p=N$, the map (\ref{map wsigma}) becomes (\ref{map wN}).
The function $\mathcal{F}_{p,n}(\boldsymbol{x}^{\sigma_{N,p}})$ depends on the $2p-3$ harmonic ratios obtained as the images of the remaining endpoints through the map (\ref{map wsigma}), namely
\be
\mathcal{F}_{p,n}(\boldsymbol{x}^{\sigma_{N,p}})= 
\mathcal{F}_{p,n}( w_{\sigma_{N,p} }(v_{i_1}) , \dots ,  w_{\sigma_{N,p} }(v_{i_{p-1}})  )\,.
\ee
Since the ratios $w_{\sigma_{N,s} }(u_{i_r}) $ and $w_{\sigma_{N,s} }(v_{i_r}) $ can be expressed in terms of the harmonic ratios in $\boldsymbol{x}$ by applying (\ref{map wN}), we have that $R_{N,n}=R_{N,n}(\boldsymbol{x}) $.
The final expression can be checked by considering the limits $x_j \rightarrow x_{j\pm 1}$, whose result can be understood by using that the first operator occurring in the OPE of a twist field $\mathcal{T}_n$ with $\bar{\mathcal{T}}_n$ is the identity. 

We find it useful to write explicitly $R_{N,n}(\boldsymbol{x}) $ in the simplest cases. For $N=3$ 
\be
\label{R3n cft explicit exp}
R_{3,n}(\boldsymbol{x}) 
=
\frac{\mathcal{F}_{3,n}(x_1,x_2,x_3)}{
\mathcal{F}_{2,n}(\frac{x_1(x_3-x_2)}{x_2(x_3-x_1)}) 
\, \mathcal{F}_{2,n}(x_1)
\, \mathcal{F}_{2,n}(\frac{x_3-x_2}{1-x_2}) }\,.
\ee
From this expression (we recall that $\mathcal{F}_{2,n}(0)=\mathcal{F}_{2,n}(1)=1$), we can check that $R_{3,n}  \rightarrow 1$ when $x_3  \rightarrow x_2$ (i.e. $A_2 \rightarrow \emptyset$), which is obtained by using $\mathcal{F}_{3,n}(x_1,x_2,x_3) \rightarrow \mathcal{F}_{2,n}(x_1)$, that we checked numerically. In a similar way, we find that $R_{3,n}  \rightarrow 1$ for $x_1  \rightarrow 0$ ($A_1 \rightarrow \emptyset$). Notice that we cannot take $A_3 \rightarrow \emptyset$ in (\ref{R3n cft explicit exp}) because the map (\ref{map wN}) with $N=3$ is not well defined in this limit. 
We can also consider e.g. $x_2 \rightarrow x_1$ , i.e. $B_1 \rightarrow \emptyset$. In this case we verified that $\mathcal{F}_{3,n}(x_1, x_2, x_3) \rightarrow \mathcal{F}_{2,n}(x_3)$, as expected, and this implies that the corresponding limit for $R_{3,n}$ is not 1 identically.
Also when  $B_2 \rightarrow \emptyset$ we find that $R_{3,n}$ does not tend to 1. Indeed, $\mathcal{F}_{3,n}(x_1,x_2,x_3) \rightarrow \mathcal{F}_{2,n}(x_1/x_2)$.

When $N=4$ the elements of $\boldsymbol{x}$ are $x_1$, $\dots$, $x_5$ and $R_{4,n}(\boldsymbol{x}) $ reads
 \be
 \label{R4n explicit}
R_{4,n}(\boldsymbol{x}) 
=
\frac{\mathcal{F}_{4,n}(\boldsymbol{x})\, 
\prod_{i<j} \mathcal{F}_{2,n}(x^{\{i,j\}})}{
\mathcal{F}_{3,n}(\boldsymbol{x}^{\{1,2,3\}})\,
\mathcal{F}_{3,n}(\boldsymbol{x}^{\{1,2,4\}})\,  
\mathcal{F}_{3,n}(\boldsymbol{x}^{\{1,3,4\}})\, 
\mathcal{F}_{3,n}(\boldsymbol{x}^{\{2,3,4\}})} \,,
\ee
where the terms in the denominators are given by 
\be
\begin{array}{l}
\mathcal{F}_{3,n}(\boldsymbol{x}^{\{ 1,2,3 \}}) = 
\mathcal{F}_{3,n}\bigg( \displaystyle
\frac{x_1(x_5-x_4)}{x_4(x_5-x_1)} , \frac{x_2(x_5-x_4)}{x_4(x_5-x_2)} , \frac{x_3(x_5-x_4)}{x_4(x_5-x_3)} \bigg),\\
\rule{0pt}{.6cm}
\mathcal{F}_{3,n}(\boldsymbol{x}^{\{ 1,2,4 \}}) = 
\mathcal{F}_{3,n}(x_1 , x_2 , x_3 ) \,, \\
\rule{0pt}{.6cm}
\mathcal{F}_{3,n}(\boldsymbol{x}^{\{ 1,3,4 \}}) = 
\mathcal{F}_{3,n}(x_1 , x_4 , x_5)\,, \\
\rule{0pt}{.85cm}
\mathcal{F}_{3,n}(\boldsymbol{x}^{\{ 2,3,4 \}}) = 
\mathcal{F}_{3,n}\bigg( \displaystyle
\frac{x_3-x_2}{1-x_2} , \frac{x_4-x_2}{1-x_2}  , \frac{x_5-x_2}{1-x_2} \bigg). \\
\end{array}
\ee
As for the product in the numerator of (\ref{R4n explicit}), the arguments of the $\mathcal{F}_{2,n}$'s are not multicomponent vector and they read
\be\fl
\begin{array}{lll}
\displaystyle
x^{\{ 1,2 \}} = \frac{x_1(x_3-x_2)}{x_2(x_3-x_1)}\,,&\hspace{.6cm}
\displaystyle
x^{\{ 1,3 \}} = \frac{x_1(x_5-x_4)}{x_4(x_5-x_1)}\,,&\hspace{.6cm}
x^{\{ 1,4 \}} = x_1\,,
\\
\rule{0pt}{.7cm}
\displaystyle
x^{\{ 2,3 \}} = \frac{(x_3-x_2)(x_5-x_4)}{(x_4-x_2)(x_5-x_3)}\,,&\hspace{.6cm}
\displaystyle
x^{\{ 2,4 \}} = \frac{x_3-x_2}{1-x_2}\,,&\hspace{.6cm}
\displaystyle
x^{\{ 3,4 \}} =\frac{x_5-x_4}{1-x_4}\,.
\end{array}
\ee
The expression (\ref{R4n explicit}) allows us to check explicitly  that $R_{4,n}  \rightarrow 1$ 
when we send either $x_1  \rightarrow 0$ ($A_1 \rightarrow \emptyset$)  or $x_3  \rightarrow x_2$ ($A_2 \rightarrow \emptyset$) or $x_5  \rightarrow x_4$ ($A_3 \rightarrow \emptyset$).
In a similar way, we observed numerically that $\mathcal{F}_{4,n}(\boldsymbol{x}) \rightarrow \mathcal{F}_{3,n}(x_3, x_4, x_5)$ for $x_2 \rightarrow x_1$ ($B_1 \rightarrow \emptyset$) and that $\mathcal{F}_{4,n}(\boldsymbol{x}) \rightarrow \mathcal{F}_{3,n}(x_1, x_2, x_5)$ for $x_4 \rightarrow x_3$ ($B_2 \rightarrow \emptyset$). Taking the limit $x_5 \rightarrow 1$ ($B_3 \rightarrow \emptyset$), we are joining the last two intervals and we find $\mathcal{F}_{4,n}(\boldsymbol{x}) \rightarrow \mathcal{F}_{3,n}(x_1/x_4, x_2/x_4, x_3/x_4)$, as expected.

For higher $N$, more terms occur to deal with, but it is  always possible to write explicitly $R_{N,n}(\boldsymbol{x})$ in terms of its $2N-3$ independent variables.
The checks given above for the simplest cases of $N=3$ and $N=4$ can be generalized, finding that $R_{N,n}  \rightarrow 1$ when $x_{2k-1} \rightarrow x_{2k-2}$ ($A_k \rightarrow \emptyset$), for some fixed $k \in \{ 1, \dots, N-1 \}$ (we recall that $x_0 =0$). The limit $A_N \rightarrow \emptyset$ (i.e. $u_N \rightarrow v_N$) cannot be considered on $\mathcal{F}_{N,n}(\boldsymbol{x})$ because the map (\ref{map wN}) is not well defined. We have to compute it before applying (\ref{map wN}). As for the limit of joining intervals, for $x_{2l} \rightarrow x_{2l-1}$ ($B_l \rightarrow \emptyset$) with $l \in \{ 1, \dots, N-2 \}$ one finds $\mathcal{F}_{N,n}(\boldsymbol{x}) \rightarrow \mathcal{F}_{N-1,n}(\boldsymbol{x} \setminus \{x_{2l-1},x_{2l}\})$, while for $x_{2N-3} \rightarrow 1$ ($B_{N-1} \rightarrow \emptyset$) we have $\mathcal{F}_{N,n}(\boldsymbol{x}) \rightarrow \mathcal{F}_{N-1,n}(x_1/x_{2N-4}, x_2/x_{2N-4} , \dots , x_{2N-5}/x_{2N-4})$.

\begin{figure}[t]
\vspace{.5cm}
\begin{center}
\includegraphics[width=.85\textwidth]{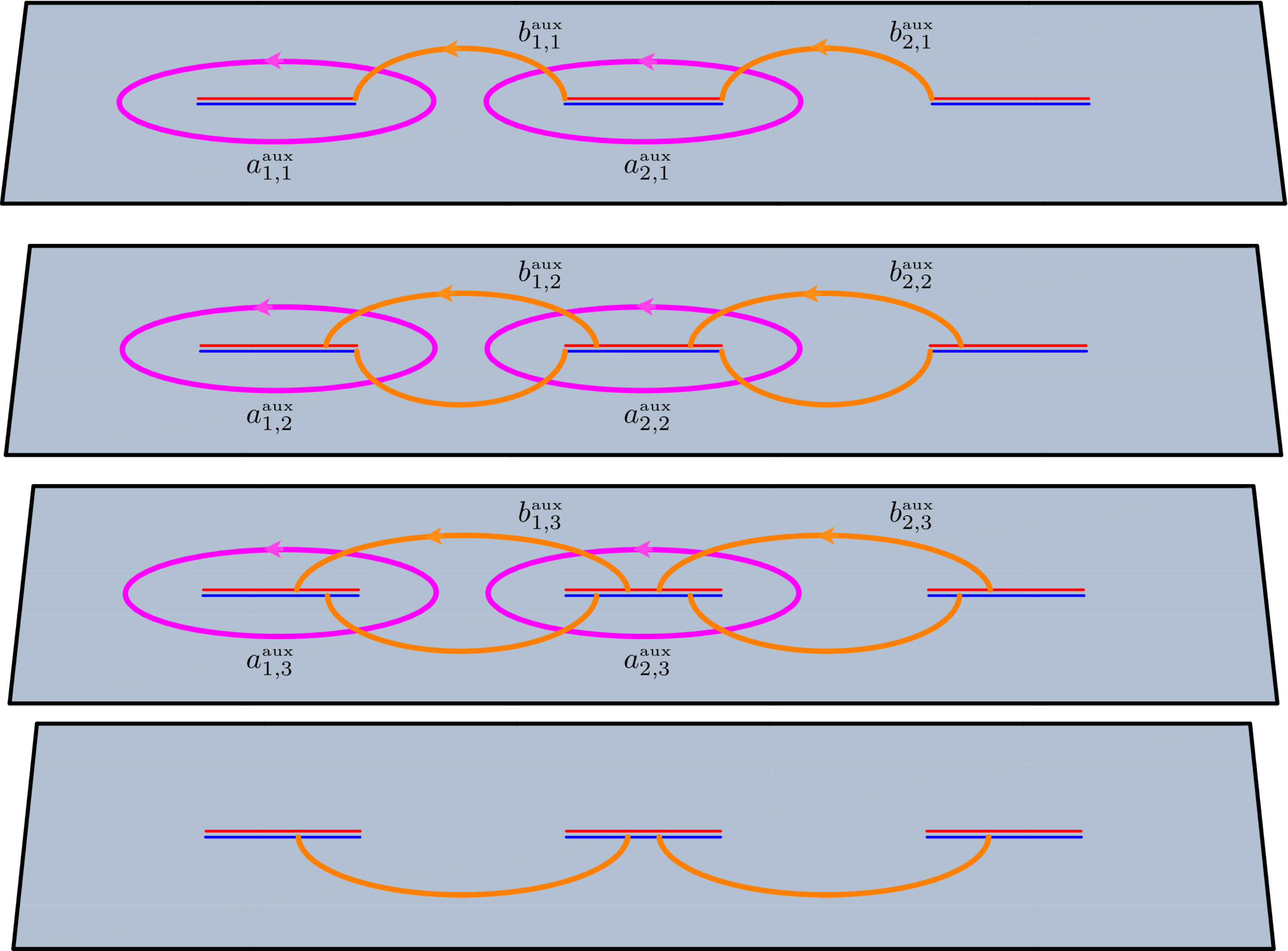}
\end{center}
\caption{The auxiliary cycles $\{ a^{\textrm{\tiny aux}}_{\alpha,j}, b^{\textrm{\tiny aux}}_{\alpha,j} \}$ for $N=3$ and $n=4$.}
\label{fig AUXmultisheets}
\end{figure}

\begin{figure}[t]
\vspace{.4cm}
\begin{center}
\includegraphics[width=.75\textwidth]{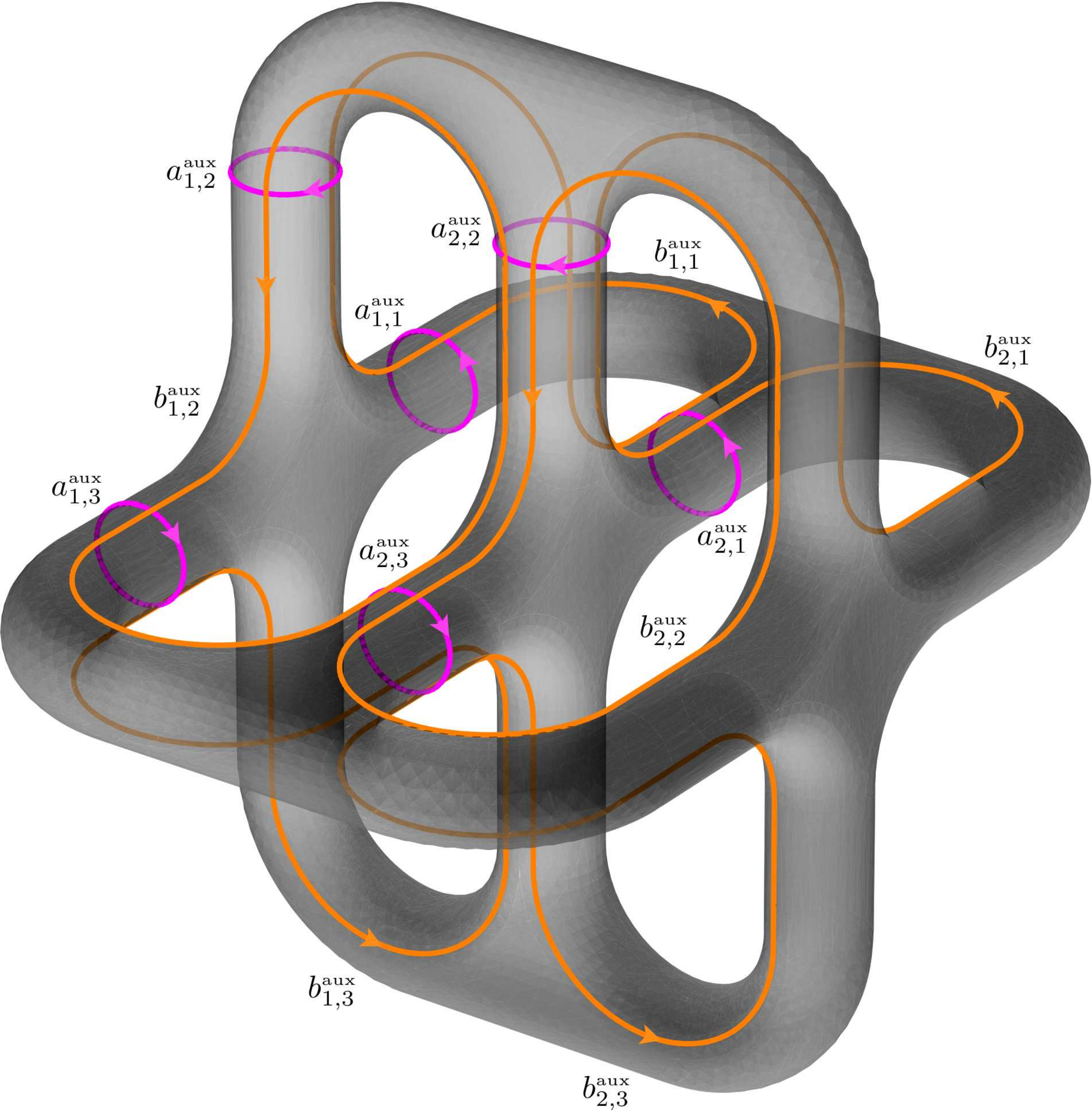}
\end{center}
\caption{The Riemann surface $\mathscr{R}_{3,4}$ with the set of auxiliary cycles $\{ a^{\textrm{\tiny aux}}_{\alpha,j}, b^{\textrm{\tiny aux}}_{\alpha,j} \}$ depicted also in Fig. \ref{fig AUXmultisheets}.}
\label{fig AUXcage}
\end{figure}

\section{Lauricella functions}
\label{app lauricella}

In this appendix we show that the integrals (\ref{Aaux 1k}) and (\ref{Baux 1k}), occurring in \S\ref{subsec period matrix} and \S\ref{subsec amv} for the computation of the period matrices, can be written in terms of the fourth Lauricella function $F_D^{(m)}$ \cite{exton}, which is a generalization of the hypergeometric function $_2F_1$ involving several variables.

The integral representation of $F_D^{(m)}$ for $\textrm{Re}(c)>\textrm{Re}(a)>0$ reads
\be\fl
\label{lauricella def}
\int_0^1  \frac{t^{a-1} (1-t)^{c-a-1}}{\prod_{j=1}^m (1-y_j t)^{b_j}}\, dt
=
\frac{\Gamma(a)\,\Gamma(c-a)}{\Gamma(c)}\, F_D^{(m)}(a, b_1 , \dots , b_m ; c\, ; y_1 , \dots , y_m)\,.
\ee
For $m=1$ the function $F_D^{(m)}$ reduces to the hypergeometric function $_2F_1(a,b_1;c;y_1)$ and for $m=2$ it becomes the Appell function $F_1(a;b_1, b_2;c;y_1, y_2)$. In our problem $m=2N-3$ and therefore $m \geqslant 3$ for $N>2$. 

In terms of the Lauricella function, the integral in (\ref{Aaux 1k}) for $\alpha=1$ reads
\begin{eqnarray}
\label{lauricella acyc alpha1}
& &\fl
\mathscr{I}_{\beta,k} \big|_{0}^{x_1} = \,
\frac{\Gamma(\beta-k/n)\,\Gamma(k/n)}{\Gamma(\beta)} \; 
x_1^{\beta-1} 
\prod_{\gamma=2}^{N-1} x_{2\gamma-2}^{-k/n} \, 
\prod_{\lambda=2}^{N-1} x_{2\lambda-1}^{k/n-1} 
\\
\rule{0pt}{.7cm}
& &\fl \hspace{1.8cm}
\times  F_D^{(2N-3)}\bigg(\beta -\frac{k}{n},  \frac{k}{n} ,  1- \frac{k}{n}, \dots , \frac{k}{n}; \beta\, ; 
\frac{x_1}{x_2} , \frac{x_1}{x_3}, \dots ,  \frac{x_1}{x_{2N-2}} \bigg)\,,
\nonumber
\end{eqnarray}
where we recall that $x_{2N-2}=1$ and $1 \leqslant \beta \leqslant N-1$.
Also the remaining integrals in (\ref{Aaux 1k}), which have $\alpha>1$, can be written through $F_D^{(m)}$
\begin{eqnarray}
\label{lauricella acyc alpha}
& &\fl
\mathscr{I}_{\beta,k} \big|_{x_{2\alpha-2}}^{x_{2\alpha-1}} = \,
\frac{\pi}{\sin(\pi k/n)}\;
x_{2\alpha-2}^{\beta-1-k/n} 
\prod_{\substack{\gamma=2 \\ \gamma \neq \alpha}}^{N} 
 |x_{2\gamma-2} - x_{2\alpha-2} |^{-k/n} 
\prod_{ \substack{ \lambda =1 \\ \lambda \neq \alpha }}^{N-1} 
|x_{2\lambda-1} - x_{2\alpha-2} |^{k/n-1} 
\nonumber
\\
\rule{0pt}{.0cm}
& &\fl \hspace{1.8cm}
\times  F_D^{(2N-3)}\bigg(1 -\frac{k}{n},  \frac{k}{n}+1-\beta ,  1- \frac{k}{n}, \dots , \frac{k}{n} ; 1 ; 
\,\boldsymbol{y}^{(\alpha)}\bigg),
\end{eqnarray}
where $\Gamma(1-k/n)\,\Gamma(k/n) = \pi \csc(\pi k/n)$ has been used and we introduced the $2N-3$ dimensional vector $\boldsymbol{y}^{(\alpha)}$, whose elements read
\be\fl
y^{(\alpha)}_\zeta \equiv \frac{x_{2\alpha-1} - x_{2\alpha-2}}{x_\zeta - x_{2\alpha-2}}\,,
\qquad
\zeta \in \{ 0, 1, \dots, 2N-2\}\setminus \{ 2\alpha -2, 2\alpha -1\}\,.
\ee
As for the integrals in (\ref{Baux 1k}) for $\alpha\geqslant 1$, in terms of Lauricella functions they become
\begin{eqnarray}
\label{lauricella bcyc alpha}
& &\fl
\mathscr{I}_{\beta,k} \big|_{x_{2\alpha-1}}^{x_{2\alpha}} = \,
\frac{\pi}{\sin(\pi k/n)}\;
x_{2\alpha-1}^{\beta-1-k/n} 
\prod_{\substack{ \gamma=1 \\  \gamma \neq \alpha }}^{N} 
 |x_{2\gamma-2} - x_{2\alpha-2} |^{-k/n}
\prod_{\substack{ \gamma=1 \\  \gamma \neq \alpha }}^{N-1} 
|x_{2\gamma-1} - x_{2\alpha-2} |^{k/n-1} 
\nonumber
\\
\rule{0pt}{.0cm}
& &\fl \hspace{1.8cm}
\times  F_D^{(2N-3)}\bigg(\frac{k}{n},  \frac{k}{n}+1-\beta ,  1- \frac{k}{n}, \dots , \frac{k}{n} ; 1 ; 
\,\boldsymbol{w}^{(\alpha)}\bigg),
\end{eqnarray}
where we defined the $2N-3$ dimensional vector $\boldsymbol{w}^{(\alpha)}$, whose elements are
\be\fl
w^{(\alpha)}_\zeta \equiv \frac{x_{2\alpha} - x_{2\alpha-1}}{x_\zeta - x_{2\alpha-1}}\,,
\qquad
\zeta \in \{ 0, 1, \dots, 2N-2\}\setminus \{ 2\alpha -1, 2\alpha \}\,.
\ee
We remark that both in (\ref{lauricella acyc alpha}) and (\ref{lauricella bcyc alpha}) the dots denote the alternating occurrence of $k/n$ and $1-k/n$, like in (\ref{lauricella acyc alpha1}). 
For even $n$, the case $k/n =1/2$ occurs and these expressions slightly simplify.
In order to realize that (\ref{lauricella acyc alpha1}) is (\ref{lauricella acyc alpha}) with $\alpha =1$, it is more convenient to go back to the original integral representation and set $\alpha=1$ there.

For $N=2$ intervals we have only one harmonic ratio $x_1=x \in (0,1)$. Moreover, $\alpha=\beta=1$ and therefore we have to consider only (\ref{lauricella acyc alpha1}) and (\ref{lauricella bcyc alpha}), which reduce respectively to
\begin{eqnarray}\fl
\label{N2lauricellaA}
\mathscr{I}_{1,k} \big|_{0}^{x} &=&  
\frac{\pi}{\sin(\pi k/n)} \, F_{k/n}(x)\,,
\\
\label{N2lauricellaB}
\fl
\mathscr{I}_{1,k} \big|_{x}^{1} &=&  
\frac{\pi}{\sin(\pi k/n)}\;x^{-k/n}\,  _2F_1 \bigg( \,\frac{k}{n} , \frac{k}{n} ;1; \frac{x-1}{x}  \bigg)
=\;
\frac{\pi}{\sin(\pi k/n)}\; F_{k/n}(1-x)\,,
\end{eqnarray}
being $F_{k/n}$ the hypergeometric function defined in (\ref{F2 neg}).
In the last step of (\ref{N2lauricellaB}) we have employed the Kummer's relation
$\, _2F_1(a,b;c;y) = (1-y)^{-a}\, _2F_1(a,c-b;c;y/(y-1))$.

\section{Symmetries of $\mathcal{F}_{N,n}$ as symplectic transformations}
\label{sec modular transf}

In this appendix we discuss some symmetries of $\mathcal{F}_{N,n}$ through the symplectic modular transformations. In Appendix \ref{app symp group} we define the group $Sp(2g,\mathbb{Z})$ and its action on the Riemann theta functions, introducing the subset of transformations we are interested in. In Appendix \ref{app modular invariance} we show that $\mathcal{F}_{N,n}$ is invariant under such class of modular transformations, for both the compactified boson and the Ising model, and in Appendix \ref{app mod transformations} we construct the symplectic matrices implementing the cyclic transformation in the sequence of the sheets, the inversion of their order and the exchange $A \leftrightarrow B$.

\subsection{The symplectic modular group}
\label{app symp group}

Let us consider the group $Sp(2g,\mathbb{Z})$ of the integer symplectic matrices, which is also known as symplectic modular group.  
The generic element $M \in Sp(2g,\mathbb{Z})$ is a $2g \times 2g$ matrix which satisfies 
\be\fl
\label{symplectic matrices}
M = \begin{pmatrix} 
D & C \\
B & A
\end{pmatrix} ,
\qquad
M^{\textrm{t}} \cdot J \cdot M = J\,,
\qquad
J = \begin{pmatrix} 
 0_g & \mathbb{I}_g \\
 - \mathbb{I}_g & 0_g
\end{pmatrix},
\ee
where the $g \times g$ matrices $A$, $B$, $C$ and $D$ are  made of integers, $0_g$ is the $g \times g$ matrix whose elements are all equal to zero and $\mathbb{I}_g$ is the identity matrix.
The condition in (\ref{symplectic matrices}) on $M$ corresponds to require that $D^{\textrm{t}}  \cdot B$ and $C^{\textrm{t}}  \cdot A$ are symmetric matrices and also $D^{\textrm{t}}  \cdot A - B^{\textrm{t}}  \cdot C = \mathbb{I}_g$.

Under a symplectic transformation, the canonical basis of cycles and the normalized basis of the holomorphic one forms  transform respectively as follows
\be
\label{modular trans cycles}
\begin{pmatrix}   \boldsymbol{a}'  \\   \boldsymbol{b}'  \end{pmatrix} 
= M \cdot
\begin{pmatrix}  \boldsymbol{a}  \\   \boldsymbol{b}   \end{pmatrix},
 \qquad
\boldsymbol{\nu}'^{\,\textrm{t}} =  \boldsymbol{\nu}^{\,\textrm{t}}  \cdot (C \cdot \tau + D)^{-1}\,.
\ee
From the first transformation rule, it is straightforward to observe that a canonical homology basis is sent into another canonical homology basis. Moreover, combining the transformation rules  in (\ref{modular trans cycles}), one finds that the period matrix $\tau'$ computed through $\boldsymbol{\nu}'$ and the cycles $\boldsymbol{b}' $ is related to $\tau$ in (\ref{tau def}) as follows
\be
\label{modular transf tau}
\tau' = (A \cdot \tau + B) \cdot  (C \cdot \tau + D)^{-1}\,.
\ee
The transformation rule for the absolute value of the  Riemann theta function with characteristic defined in (\ref{theta def spin}) reads \cite{amv, verlinde2-86, amv bosonization,Fay book, Mumford book} 
\be
\label{theta char transf symp}
\big| \Theta [\boldsymbol{e}' ] (\boldsymbol{0} | \tau' ) \big| 
=
\sqrt{|\textrm{det}(C \cdot \tau + D)|}\; 
\big|\Theta [\boldsymbol{e} ] (\boldsymbol{0} | \tau ) \big|\,,
\ee
where the characteristic $\boldsymbol{e}' $ is given by
\be
\label{mod transf char}
\begin{pmatrix}
 \boldsymbol{\varepsilon}'  \\   \boldsymbol{\delta}'  
\end{pmatrix}
 =
\begin{pmatrix}
D & - C \\
- B & A
\end{pmatrix}
\cdot
\begin{pmatrix} \boldsymbol{\varepsilon} \\   \boldsymbol{\delta} \end{pmatrix}
+\frac{1}{2}\,
\begin{pmatrix}  (C \cdot D^{\textrm{t}})_{\textrm{d}}  \\   
(A \cdot B^{\textrm{t}})_{\textrm{d}} \end{pmatrix},
\ee
where $(\dots)_{\textrm{d}}$ is the vector made by the diagonal of the matrix within the brackets.

Let us consider the subset of $Sp(2g,\mathbb{Z})$ given by the following matrices
\be
\label{classes matrices}
\begin{pmatrix}
D & 0_g \\
0_g & (D^{-1})^{\textrm{t}}
\end{pmatrix},
\qquad
\begin{pmatrix}
0_g & C\, \\
-(C^{-1})^{\textrm{t}} & 0_g
\end{pmatrix}.
\ee
Under the transformations of the first kind, the cycles $\boldsymbol{a}'$ ($\boldsymbol{b}'$) are obtained through $\boldsymbol{a}$ ($\boldsymbol{b}$) cycles only; while applying the transformations of the second kind, the cycles $\boldsymbol{a}'$ ($\boldsymbol{b}'$) are combinations of the cycles $\boldsymbol{b}$ ($\boldsymbol{a}$).
Moreover, for the transformations (\ref{classes matrices}) the relation (\ref{mod transf char}) between the characteristics becomes homogenous. In particular, the zero characteristic is mapped into itself and therefore (\ref{theta char transf symp}) becomes 
\be
\label{theta transf symp}
\big| \Theta(\boldsymbol{0} | \tau' ) \big| 
=
\sqrt{|\textrm{det}(C \cdot \tau + D)|}\; 
\big|\Theta (\boldsymbol{0} | \tau ) \big|\,.
\ee
In the remaining part of this appendix, we will restrict to the transformations (\ref{classes matrices}).

\subsection{Invariance of $\mathcal{F}_{N,n}$}
\label{app modular invariance}

Let us discuss the invariance of $\mathcal{F}_{N,n}(\boldsymbol{x})$ under (\ref{classes matrices}) for the free compactified boson. \\
Considering the two expressions in (\ref{FNn final}) which are not explicitly invariant under $\eta \leftrightarrow 1/\eta$, one finds that $\sqrt{\textrm{det}(\mathcal{I})}\, | \Theta (\boldsymbol{0} | \tau ) |^2$ and $\Theta (\boldsymbol{0} | \textrm{i}\eta G )$ (or $\Theta (\boldsymbol{0} | \textrm{i} G/\eta )$ equivalently)  are separately invariant. 
The invariance of $\sqrt{\textrm{det}(\mathcal{I})}\, | \Theta (\boldsymbol{0} | \tau ) |^2$ is easily obtained combining (\ref{theta transf symp}) and the following relation \cite{amv bosonization}
\be
\label{Imprime inv}
(\mathcal{I}')^{-1} = 
(C\cdot \bar{\tau} +D) \cdot \mathcal{I}^{-1} \cdot (C \cdot \tau +D)^{\textrm{t}}\,,
\ee
which can be verified starting from (\ref{modular transf tau}). This allows us to claim that the expression $\mathcal{F}_{N,n}^{\eta\rightarrow \infty}( \boldsymbol{x}  ) $ in (\ref{FNn eta large}), which characterizes the decompactification regime, is invariant under symplectic transformations.\\
As for the invariance $\Theta (\boldsymbol{0} | \textrm{i}\eta G )$, first we find it convenient to write $G$ in (\ref{Gmat def}) as
\be
\label{Gmat v2}
G =  
\begin{pmatrix}
\tau \cdot \mathcal{I}^{-1} \cdot \bar{\tau}& 
\;\tau \cdot \mathcal{I}^{-1} -\textrm{i} \,\mathbb{I}_g\,
\\
\;\mathcal{I}^{-1} \cdot \bar{\tau} + \textrm{i}  \,\mathbb{I}_g& 
\;\;\;\;\mathcal{I}^{-1} 
\end{pmatrix}.
\ee
The terms $\pm \textrm{i} \, \mathbb{I}_g$ in the off diagonal blocks can be dropped because they cancel each others in the exponent of the general term of the series defining $\Theta (\boldsymbol{0} | \textrm{i}\eta G )$. Then, we can employ the fact that $\Theta (\boldsymbol{0} | \textrm{i}\eta G )$ does not change under simultaneous inversion of the sign for both the off diagonal matrices in $G$. Considering the exponent of the general term of the series, after some algebra one finds that
\begin{eqnarray}
& &
\begin{pmatrix}
 \boldsymbol{m}^{\textrm{t}} &\boldsymbol{n}^{\textrm{t}} 
\end{pmatrix}
\cdot
\begin{pmatrix}
\;\tau' \cdot (\mathcal{I}')^{-1} \cdot \bar{\tau}' & 
\; -\, \tau' \cdot (\mathcal{I}')^{-1}\,
\\
- (\mathcal{I}')^{-1} \cdot \bar{\tau}' & 
\;\;\;(\mathcal{I}')^{-1} 
\end{pmatrix}
\cdot
\begin{pmatrix}
 \boldsymbol{m}\\
 \boldsymbol{n}
\end{pmatrix}
\\
\rule{0pt}{.9cm}
& &
\;=\;
\begin{pmatrix}
 \boldsymbol{m}'^{\,\textrm{t}} &\boldsymbol{n}'^{\,\textrm{t}} 
\end{pmatrix}
\cdot
\begin{pmatrix}
\;\tau \cdot \mathcal{I}^{-1} \cdot \bar{\tau} & 
\; -\, \tau \cdot \mathcal{I}^{-1}\,
\\
- \,\mathcal{I}^{-1} \cdot \bar{\tau} & 
\;\;\;\mathcal{I}^{-1} 
\end{pmatrix}
\cdot
\begin{pmatrix}
 \boldsymbol{m}'\\
 \boldsymbol{n}'
\end{pmatrix},
\nonumber
\end{eqnarray}
where $(\mathcal{I}')^{-1}$ is defined in (\ref{Imprime inv}), $\tau'$ in (\ref{modular transf tau}) and we also introduced
\be
\label{M inverse}
\begin{pmatrix}
 \boldsymbol{m}'\\
 \boldsymbol{n}'
\end{pmatrix}
=
M^{-1}  \cdot
\begin{pmatrix}
 \boldsymbol{m}\\
 \boldsymbol{n}
\end{pmatrix},
\qquad
M^{-1} = 
\begin{pmatrix}
A^{\textrm{t}}  & -\,C^{\textrm{t}}   \\
-B^{\textrm{t}}   & D^{\textrm{t}} 
\end{pmatrix}.
\ee
The vectors $ \boldsymbol{m}'$ and $ \boldsymbol{n}'$ are made of integers and they are related to $ \boldsymbol{m}$ and $ \boldsymbol{n}$ through the inverse $M^{-1}$ of symplectic transformation (\ref{symplectic matrices}), which is also a symplectic matrix.
Since also $( \boldsymbol{m}'^{\,\textrm{t}} ,  \boldsymbol{n}'^{\,\textrm{t}}  )$ cover the whole $\mathbb{Z}^{2g}$, we have that  $\Theta (\boldsymbol{0} | \textrm{i}\eta G )$ is invariant under $Sp(2g,\mathbb{Z})$ for any $\eta$.

For the Ising model, we have that $\mathcal{F}^{\textrm{\tiny Ising}}_{N,n}( \boldsymbol{x}) $ in (\ref{FNn ising}) is invariant under (\ref{classes matrices}). Indeed, from (\ref{theta char transf symp}) and (\ref{theta transf symp}) it is straightforward to conclude that 
\be
\bigg|
\frac{\Theta [\boldsymbol{e}' ] (\boldsymbol{0} | \tau' )}{\Theta(\boldsymbol{0} | \tau' )}
\bigg|
=
\bigg|
\frac{\Theta [\boldsymbol{e}] (\boldsymbol{0} | \tau)}{\Theta(\boldsymbol{0} | \tau )}
\bigg|\,.
\ee
Moreover, each term of the sum over the characteristics in (\ref{FNn ising}) is sent into a different one (except for $\boldsymbol{e}^{\textrm{t}} = (\boldsymbol{0}^{\textrm{t}},\boldsymbol{0}^{\textrm{t}})$) so that the whole sum is invariant because the net effect of (\ref{classes matrices}) is to reshuffle its terms.

\subsection{Some explicit modular transformations}
\label{app mod transformations}

\subsubsection{Cyclic transformation.}
\label{app cyc}
As a concrete example of a symmetry written in terms of a symplectic matrix, we consider first the cyclic change in the ordering of the sheets. 
Indeed, the choice of the first sheet is arbitrary and therefore the period matrix cannot depend on it. This symmetry has been already studied in \cite{Headrick-12}.\\
It is useful to start from the effect of this transformation on the auxiliary cycles of Figs. \ref{fig AUXmultisheets} and \ref{fig AUXcage}: $a^{\textrm{\tiny aux}}_{\alpha,j} \rightarrow a^{\textrm{\tiny aux}}_{\alpha,j+1}$ and $b^{\textrm{\tiny aux}}_{\alpha,j} \rightarrow b^{\textrm{\tiny aux}}_{\alpha,j+1}$. Notice that we introduced the cycles $a^{\textrm{\tiny aux}}_{\alpha,n} \equiv a^{\textrm{\tiny aux}}_{\alpha,0} $ and $b^{\textrm{\tiny aux}}_{\alpha,n} \equiv b^{\textrm{\tiny aux}}_{\alpha,0} $, which are not shown in Figs. \ref{fig AUXmultisheets} and \ref{fig AUXcage}, but, given their indices, it is clear how to place them. In particular, considering this enlarged set of auxiliary cycles, we have that $\sum_{j=1}^n  a^{\textrm{\tiny aux}}_{\alpha,j} = \sum_{j=1}^n  b^{\textrm{\tiny aux}}_{\alpha,j}  =0$, which allow to write $a^{\textrm{\tiny aux}}_{\alpha,n}$ and $b^{\textrm{\tiny aux}}_{\alpha,n}$ in terms of the other ones. From these relations and (\ref{eg cycles from aux}), we find that the canonical homology basis introduced in \S\ref{subsec period matrix} changes as follows
\be\fl
\label{eg basis cyc}
 a_{\alpha, j} \rightarrow a_{\alpha, j+1} \hspace{.5cm} j\neq n-1\,,  
 \qquad 
 a_{\alpha, n-1} \rightarrow -\displaystyle\sum_{k=1}^{n-1}a_{\alpha,k}\,,
 \qquad
  b_{\alpha, j} \rightarrow b_{\alpha, j+1} - b_{\alpha,1}\,.
\ee 
As for the canonical homology basis defined in \S\ref{subsec amv}, from (\ref{amv cycles from aux}) we have
\be\fl
\label{amv basis cyc}
 \tilde a_{\alpha, j} \rightarrow \tilde a_{\alpha, j+1} - \tilde a_{\alpha,1} \,,
 \qquad
 \tilde b_{\alpha, j} \rightarrow \tilde b_{\alpha, j+1} \hspace{.5cm} j\neq n-1\,, 
  \qquad
 b_{\alpha, n-1} \rightarrow -\displaystyle\sum_{k=1}^{n-1}\tilde b_{\alpha,k}\,.
\ee
Since these transformations do not affect the greek index, their rewriting in a matrix form involves $\mathbb{I}_{N-1}$. In particular, (\ref{eg basis cyc}) and (\ref{amv basis cyc})  become respectively
\be\fl
\label{basis cyc mat}
 M_{\text{cyc}} =
\begin{pmatrix}
 D_{\text{cyc}} & 0_{n-1} \\ 0_{n-1} & A_{\text{cyc}}
\end{pmatrix}
\otimes \mathbb{I}_{N-1}\,,
\qquad
 \tilde M_{\text{cyc}} =
\begin{pmatrix}
 \tilde D_{\text{cyc}} & 0_{n-1} \\ 0_{n-1} & \tilde A_{\text{cyc}}
\end{pmatrix}
\otimes \mathbb{I}_{N-1}\,,
\ee
where
\be\fl
\label{ADmat cyc}
\bigg\{
\begin{array}{l}
 (A_{\text{cyc}})_{jk} = \delta_{k-j,1} - \delta_{k,1} 
 \\
 (D_{\text{cyc}})_{jk} = \delta_{k-j,1} - \delta_{j,n-1} 
 \end{array},
 \qquad
\bigg\{
\begin{array}{l}
 (\tilde A_{\text{cyc}} )_{jk} = \delta_{k-j,1} - \delta_{j,n-1} =
 ( D_{\text{cyc}} )_{jk}
 \\
 (\tilde D_{\text{cyc}} )_{jk} = \delta_{k-j,1} - \delta_{k,1}   =
 (A_{\text{cyc}} )_{jk}
 \end{array} .
\ee
Since $A_{\text{cyc}} = (D_{\text{cyc}}^{-1})^{\textrm{t}}$, we have that $ M_{\text{cyc}}$ and $ \tilde M_{\text{cyc}}$ belong to subset of $Sp(2g, \mathbb{Z})$ defined by the first expression in (\ref{classes matrices}). Notice that $(D_{\text{cyc}}^{-1})^{\textrm{t}}$ is the matrix given in Eq. (3.28) of \cite{Headrick-12}.
Moreover, we checked that $ M_{\text{cyc}}^{n} = \tilde{M}_{\text{cyc}}^{n} = \mathbb{I}_{2g} $ and also that $M_{\text{cyc}}  = M^{-1}\cdot \tilde M_{\text{cyc}} \cdot M $, being $M$ the matrix defined in (\ref{cyc2tildecyc}), which relates the two canonical homology bases.
As for the period matrix, by applying (\ref{modular transf tau}) for the transformations (\ref{basis cyc mat}), we numerically checked that $ \tau'_{\text{cyc}}(\boldsymbol{x}) = \tau(\boldsymbol{x})$ and $ \tilde{\tau}'_{\text{cyc}}(\boldsymbol{x}) = \tilde{\tau}(\boldsymbol{x})$, as expected.

\subsubsection{Inversion.}
\label{app inv}

Another symmetry that we can consider is obtained by taking the sheets in the inverse order. As above, we start from the action of this transformation on the auxiliary cycles, which is $a^{\textrm{\tiny aux}}_{\alpha,j} \rightarrow - a^{\textrm{\tiny aux}}_{\alpha,n-j+1}$ and $b^{\textrm{\tiny aux}}_{\alpha,j} \rightarrow b^{\textrm{\tiny aux}}_{\alpha,n-j}$ (we assume the enlarged set of auxiliary cycles introduced in Appendix \ref{app cyc}), where the opposite sign has been introduced to preserve the correct intersection number. 
Then, plugging it into (\ref{eg cycles from aux}), one finds that it acts on the canonical homology basis as follows
\be\fl
\label{eg basis inv}
 a_{\alpha, 1} \rightarrow \sum_{k=1}^{n-1} a_{\alpha, k}\,, 
\qquad
 a_{\alpha, j} \rightarrow -a_{\alpha, n-j+1} \hspace{.5cm} j\neq 1\,,  
 \qquad 
  b_{\alpha, j} \rightarrow b_{\alpha,1} - b_{\alpha, n-j+1} \,,
\ee
while, from (\ref{amv cycles from aux}), we get that the action on the canonical homology basis introduced in \S\ref{subsec amv} is simply $ \tilde{a}_{\alpha, j} \rightarrow \tilde{a}_{\alpha, n-j} $ and $ \tilde{b}_{\alpha, j} \rightarrow \tilde{b}_{\alpha, n-j}$.
The corresponding symplectic matrices $ M_{\text{inv}} $ and $ \tilde{M}_{\text{inv}} $ have the structure of (\ref{basis cyc mat}) with
\be\fl
 (A_{\text{inv}})_{jk} =  (D^{\textrm{t}}_{\text{inv}})_{jk} = \delta_{k,1} - \delta_{j+k-1,n} \,,
 \qquad
 (\tilde A_{\text{inv}} )_{jk} =  (\tilde D_{\text{inv}} )_{jk} = \delta_{j,n-k}\,.
\ee
They are related as $M_{\text{inv}}  = M^{-1}\cdot \tilde M_{\text{inv}} \cdot M $, with $M$ is given by (\ref{cyc2tildecyc}), as expected.
A transformation very close to the one we are considering has been already studied in \cite{Headrick-12}. In particular, their Eq. (3.29) is given $A^{\textrm{t}}_{\text{inv}}$ up to a global minus sign and a cyclic transformation. Since the inversion is involutive, we have $M_{\text{inv}}^2 = \tilde{M}_{\text{inv}}^2 = \mathbb{I}_{2g}$. \\
As for the period matrix, from (\ref{modular transf tau}) we numerically find $ \tau'_{\text{inv}}(\boldsymbol{x}) = - \,\bar{\tau}(\boldsymbol{x})$ and similarly, for the canonical basis of \S\ref{subsec amv}, we have $ \tilde{\tau}'_{\text{inv}}(\boldsymbol{x}) = - \,\bar{\tilde{\tau}}(\boldsymbol{x})$. Since the imaginary part of the period matrix is left invariant, the inversion leaves the period matrix invariant only for $N=2$ or $n=2$ \cite{Headrick-12}.

\subsubsection{Exchange $A \leftrightarrow B$.}
The transformations considered in Appendices \ref{app cyc} and \ref{app inv} do not change the positions of the branch points. This means that $\boldsymbol{x}_{\text{cyc}}  = \boldsymbol{x}_{\text{inv}} = \boldsymbol{x}$. Instead, exchanging $A = \cup_{i=1}^N A_1$ with its complement $B$, we move the intervals and this leads to a change of the harmonic ratios $\boldsymbol{x}$.

A way to implement the transformation $A \leftrightarrow B$ is given by
\be
\label{ABex v1}
\left\{\begin{array}{l}
A_i \,\rightarrow\, B_i \\
B_i \,\rightarrow\, A_{i+1\;\textrm{mod}\;N} 
\end{array}
\right. ,
\qquad
\left\{\begin{array}{l}
u_i \,\rightarrow\, v_i \\
v_i \,\rightarrow\, u_{i+1\;\textrm{mod}\;N} 
\end{array}
\right. ,
\ee
where $i=1, \dots, N$. Applying this transformation twice, $A \rightarrow A$ and $B\rightarrow B$, but their components do not go back to themselves when $N>2$. Indeed, we have $A_i \rightarrow A_{i+2 \,\textrm{mod}\,N} $ and $B_i \rightarrow B_{i+2 \,\textrm{mod}\,N} $.
Moreover, if we give to the intervals $A_i$ and $B_i$ an orientation, the transformation (\ref{ABex v1}) does not change it. Indeed, twist fields $\mathcal{T}_n$ are sent into $\bar{\mathcal{T}}_n$ and viceversa.
Under (\ref{ABex v1}), the components of the vector $\boldsymbol{x}$ change as follows
\be
\label{ABex for x 1}
x_\zeta \;\rightarrow\; 
1-\frac{x_1}{x_{\zeta+1}}\,,
\qquad
\zeta = 1, \dots, 2N-3\,,
\ee
i.e. $\boldsymbol{x} \rightarrow \boldsymbol{x}_{\text{ex,1}}$, where $(\boldsymbol{x}_{\text{ex,1}})_\zeta \equiv 1-x_1/x_{\zeta+1}$ (we recall that $x_{2N-2} \equiv 1$). \\
In order to describe the effect of (\ref{ABex v1}) on the auxiliary cycles of Figs. \ref{fig AUXmultisheets} and \ref{fig AUXcage}, we find it useful to introduce, besides the $a^{\textrm{\tiny aux}}_{\alpha,n}$ and $b^{\textrm{\tiny aux}}_{\alpha,n}$ already defined in Appendix \ref{app cyc}, also the auxiliary cycles $a^{\textrm{\tiny aux}}_{N,j}$ and $b^{\textrm{\tiny aux}}_{N,j}$, so that $\sum_{\alpha=1}^N a^{\textrm{\tiny aux}}_{N,j} = \sum_{\alpha=1}^N b^{\textrm{\tiny aux}}_{N,j}  = 0$, where $j=1, \dots, n$.
Considering this enlarged set of auxiliary cycles $\{a^{\textrm{\tiny aux}}_{\alpha,j} , b^{\textrm{\tiny aux}}_{\alpha,j}  \}$ where $\alpha =1, \dots, N$ and $j=1,\dots, n$, we find that (\ref{ABex v1}) leads to $a^{\textrm{\tiny aux}}_{\alpha,j} \rightarrow b^{\textrm{\tiny aux}}_{\alpha,j}$ and $b^{\textrm{\tiny aux}}_{\alpha,j} \rightarrow -a^{\textrm{\tiny aux}}_{\alpha+1,j+1}$.
By employing these relations in (\ref{eg cycles from aux}) and (\ref{amv cycles from aux}), we find respectively
\be\fl
\Bigg\{\begin{array}{l}
 a_{\alpha,j} \rightarrow \sum_{\gamma=1}^\alpha (b_{\gamma,j} - b_{\gamma,j+1}  ) \\
 \rule{0pt}{.5cm}
 b_{\alpha,j} \rightarrow \sum_{k=1}^j (a_{\alpha+1,k} - a_{\alpha, k}  ) 
\end{array},
\qquad
\hspace{-.3cm}
\Bigg\{\begin{array}{l}
 \tilde{a}_{\alpha,j} \rightarrow 
 \sum_{\gamma=1}^\alpha \sum_{k=1}^j \tilde{b}_{\gamma,k} \\
 \rule{0pt}{.5cm}
  \tilde{b}_{\alpha,j} \rightarrow 
  - \,\tilde{a}_{\alpha+1, j+1} + \tilde{a}_{\alpha+1, j}  + \tilde{a}_{\alpha, j+1}  - \tilde{a}_{\alpha, j}  
\end{array},
\ee
which can be written in matrix form respectively as
\be
 M_{\text{ex,1}} = 
\begin{pmatrix}
0_{g} & -( I_{n-1}^{\textrm{\tiny up}})^{-1} \otimes I_{N-1}^{\textrm{\tiny low}} \\ 
\,I_{n-1}^{\textrm{\tiny low}} \otimes ( I_{N-1}^{\textrm{\tiny up}})^{-1} & 0_{g} 
\end{pmatrix},
\ee
and
\be
\tilde{M}_{\text{ex,1}} = 
\begin{pmatrix}
0_{g} &  I_{n-1}^{\textrm{\tiny low}} \otimes I_{N-1}^{\textrm{\tiny low}} \\ 
\,-(I_{n-1}^{\textrm{\tiny up}})^{-1} \otimes ( I_{N-1}^{\textrm{\tiny up}})^{-1} & 0_{g} 
\end{pmatrix}.
\ee
Applying (\ref{modular transf tau}) for this transformation, we find 
$\tau'_{\text{ex,1}}(\boldsymbol{x}) = - \, \bar{\tau}(\boldsymbol{x}_{\text{ex,1}}) $ and, for the canonical basis discussed in \S\ref{subsec amv}, $\tilde{\tau}'_{\text{ex,1}}(\boldsymbol{x}) = - \, \bar{\tilde{\tau}}(\boldsymbol{x}_{\text{ex,1}}) $. Given the transformation of the period matrix under the inversion discussed in Appendix \ref{app inv}, applying first (\ref{ABex v1}) and then the inversion, we get $\tau'_{\text{ex,1}}(\boldsymbol{x}) = \tau(\boldsymbol{x}_{\text{ex,1}})$ and  similarly for the tilded basis.

Another way to implement $A \leftrightarrow B$ is the following
\be
\label{ABex v2}
\left\{\begin{array}{l}
A_i \,\rightarrow\, B_{N-i \;\textrm{mod}\;N} \\
B_i \,\rightarrow\, A_{N-i\;\textrm{mod}\;N} 
\end{array},
\right.
\qquad
\left\{\begin{array}{l}
u_i \,\rightarrow\, u_{N-i+1} \\
v_i \,\rightarrow\, v_{N-i\;\textrm{mod}\;N} 
\end{array},
\right.
\ee
which is an involution for each component $A_i$ and $B_i$. This map inverts the orientation of all the intervals and it sends a twist field $\mathcal{T}_n$ into another field of the same kind, and similarly for $\bar{\mathcal{T}}_n$.
The change induced on $\boldsymbol{x}$ reads
\be
\label{ABex for x 2}
x_\zeta \;\rightarrow\; 
1-x_{2N-2-\zeta} \equiv (\boldsymbol{x}_{\text{ex,2}})_\zeta\,,
\qquad
\zeta = 1, \dots, 2N-3\,.
\ee
When $N=2$, both (\ref{ABex for x 1}) and (\ref{ABex for x 2}) give $x \rightarrow 1-x$. The transformation (\ref{ABex v2}) acts on the enlarged set of auxiliary cycles described above as $a^{\textrm{\tiny aux}}_{\alpha,j} \rightarrow b^{\textrm{\tiny aux}}_{N-\alpha,j}$ and $b^{\textrm{\tiny aux}}_{\alpha,j} \rightarrow a^{\textrm{\tiny aux}}_{N-\alpha,j+1}$. Through (\ref{eg cycles from aux}) and (\ref{amv cycles from aux}), this allows us to find respectively
\be
\Bigg\{\begin{array}{l}
a_{\alpha,j} \rightarrow \sum_{\gamma=N-\alpha}^{N-1} 
 (b_{\gamma,j-1} - b_{\gamma,j}  ) \\
 \rule{0pt}{.5cm}
  b_{\alpha,j} \rightarrow \sum_{k=j}^{n-1} (a_{N-\alpha,k} - a_{N-\alpha-1, k}  ) 
\end{array},
\ee
and
\be
\Bigg\{\begin{array}{l}
\tilde{a}_{\alpha,j} \rightarrow 
 \sum_{\gamma=N-\alpha}^{N-1} \sum_{k=1}^j \tilde{b}_{\gamma,k} \\
 \rule{0pt}{.5cm}
\tilde{b}_{\alpha,j} \rightarrow 
  \tilde{a}_{N-\alpha, j+1} - \tilde{a}_{N-\alpha, j} 
  - \tilde{a}_{N-\alpha-1, j+1} + \tilde{a}_{N-\alpha-1, j} 
\end{array},
\ee
whose expressions in matrix form read
\be
\label{Mex2}
 M_{\text{ex,2}} = 
\begin{pmatrix}
0_{g} & -( I_{n-1}^{\textrm{\tiny low}})^{-1} \otimes \check{I}_{N-1}^{\textrm{\tiny low}} \\ 
\,I_{n-1}^{\textrm{\tiny up}} \otimes ( \check{I}_{N-1}^{\textrm{\tiny up}})^{-1} & 0_{g} 
\end{pmatrix},
\ee
and
\be
\label{Mex2 tilde}
\tilde{M}_{\text{ex,2}} = 
\begin{pmatrix}
0_{g} &  I_{n-1}^{\textrm{\tiny low}} \otimes \check{I}_{N-1}^{\textrm{\tiny low}} \\ 
\,-(I_{n-1}^{\textrm{\tiny up}})^{-1} \otimes ( \check{I}_{N-1}^{\textrm{\tiny low}})^{-1} & 0_{g} 
\end{pmatrix},
\ee
where $(\check{I}_{N-1})_{\alpha\beta} \equiv  1 $ if $\alpha \geqslant N-\beta$ and $(\check{I}_{N-1})_{\alpha\beta} \equiv  0 $ otherwise. 
As for the change of the period matrix under (\ref{ABex for x 2}), applying the transformation rule (\ref{modular transf tau}) for (\ref{Mex2}) and (\ref{Mex2 tilde}), we find  $\tau'_{\text{ex,2}}(\boldsymbol{x}) = \tau(\boldsymbol{x}_{\text{ex,2}}) $ and $\tilde{\tau}'_{\text{ex,2}}(\boldsymbol{x}) = \tilde{\tau}(\boldsymbol{x}_{\text{ex,2}}) $ respectively.\\

We remark that, under the transformations considered in this subsection, the ratio within the absolute value in (\ref{FNn def general cft}) is left invariant. Indeed, the cyclic transformation and the inversion do not involve the endpoints of the intervals at all.
As for  $A \leftrightarrow B$, in the two cases shown  above,
either the sets $\{u_i , i=1, \dots, N\}$ and $\{v_i , i=1, \dots, N\}$ are exchanged or they are mapped into themselves.

\section {Some technical issues on the numerical analysis} 

\begin{figure}[t] 
\begin{center}
\vspace{0.3cm}
\hspace{-0.4cm}
\includegraphics[width=.49\textwidth]{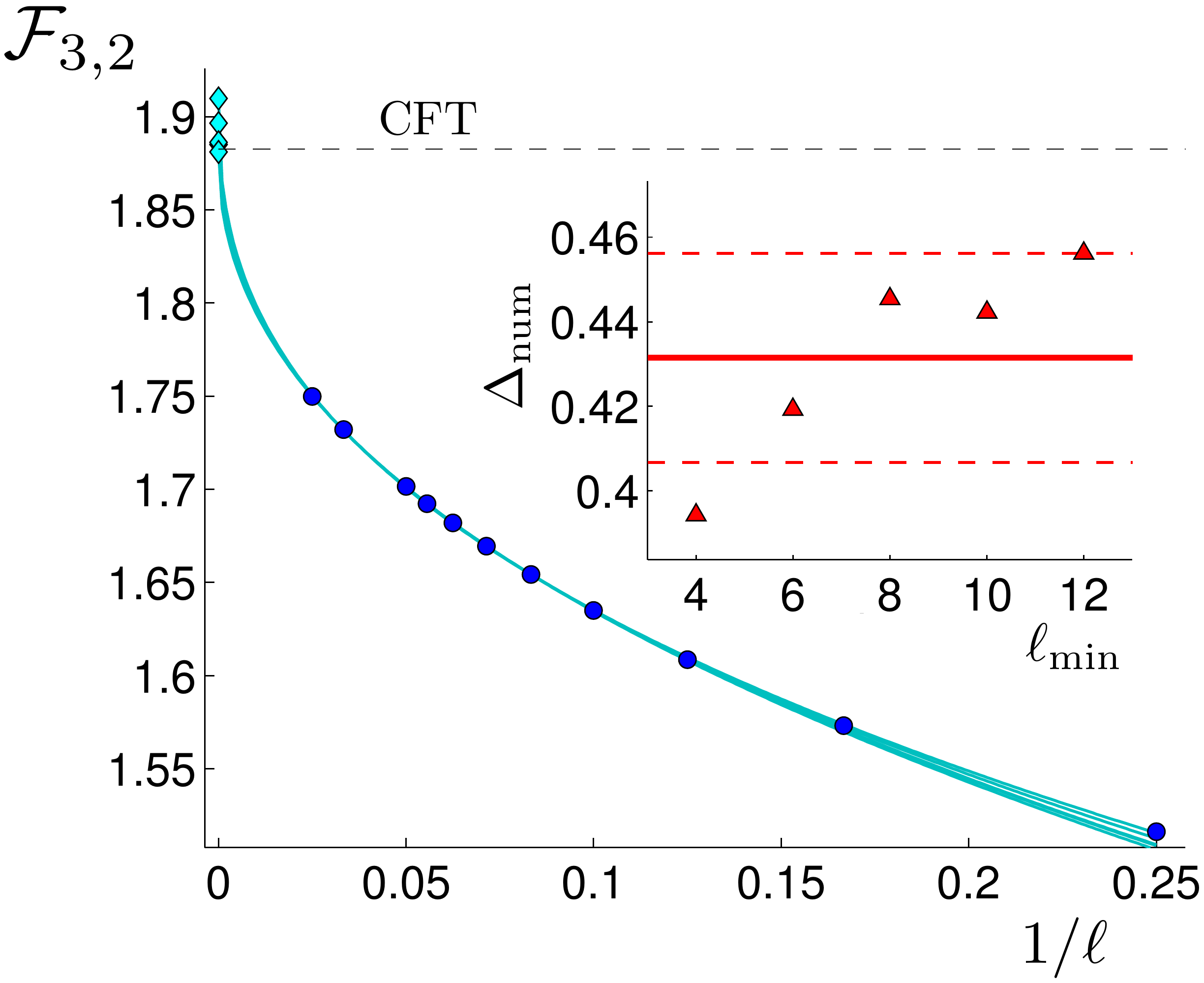}
\hspace{0.2cm}
\includegraphics[width=.49\textwidth]{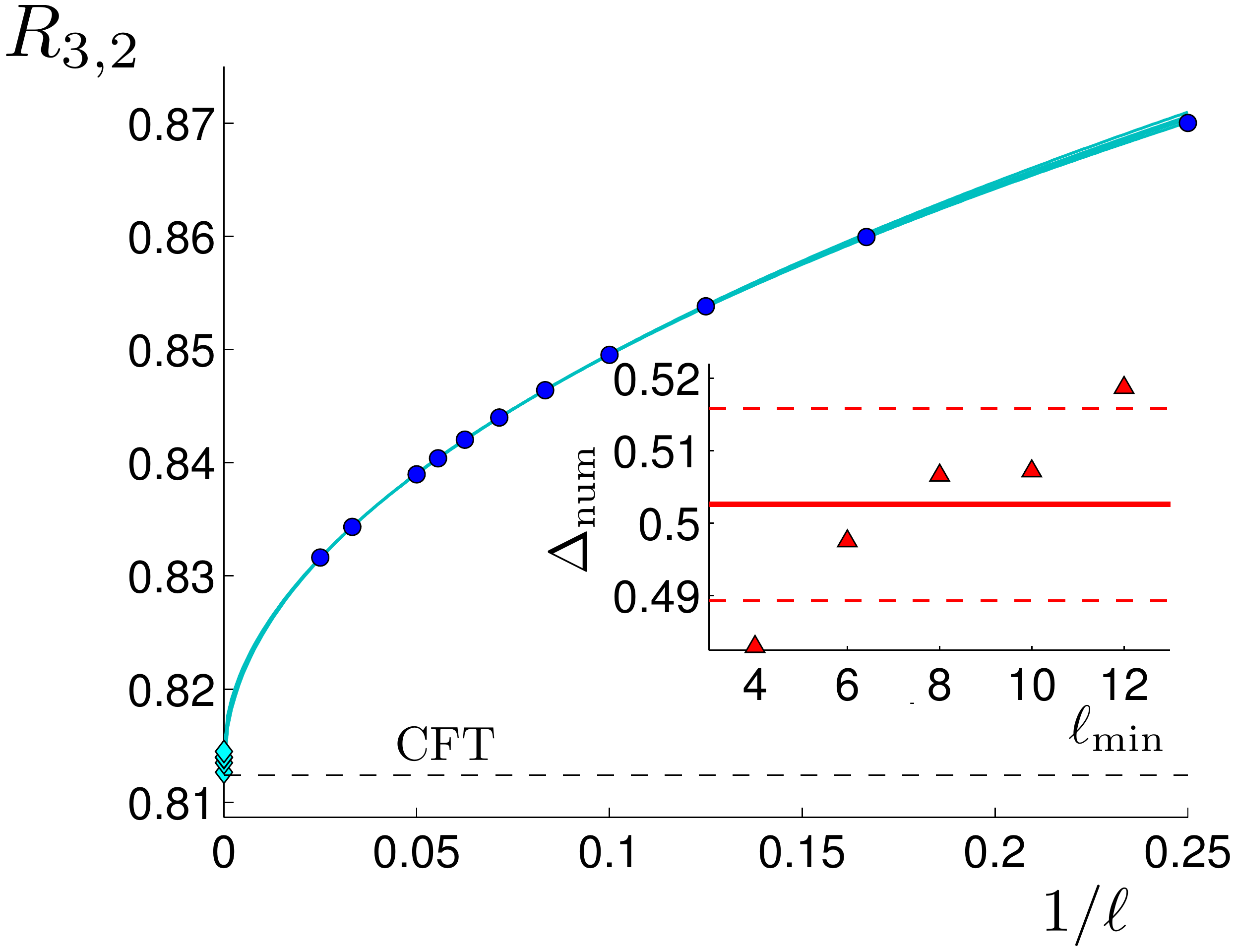}
\end{center}
\vspace{-.3cm}
\caption{\label{fig:isingDeltanum}
Leading corrections to the scaling of $\mathcal{F}_{3,2}$ (left) and $R_{3,2}$ (right) for the special case of $\alpha = 0.5$ and $x_1= 0.146$ (see top right of Figs. \ref{fig:isingN3aF} and \ref{fig:isingN3aR}), computed as explained in Appendix \ref{app:expo}.
In the inset we show the mean value of $\Delta_{\textrm{num}}$ and the error bars, obtained by fitting the data with the highest values of $\ell$, starting from $\ell_{\textrm{min}}$. Each fit provides a curve in the plot. The extrapolated values are show as cyan diamonds. 
}
\end{figure}

In this appendix we discuss some technical issues employed to extract the results of \S\ref{sec num ising}, performing also some additional analysis.
In Appendices \ref{app:expo} and \ref{app:fss3corr} we explain how the finite size scaling analysis has been performed by using either one correction or higher order ones, respectively. 
In Appendix \ref{app:finitechi} we briefly discuss some effects due to the finiteness of the bond dimension.

\subsection {The exponent in the first correction} 
\label{app:expo}

\begin{figure}[t] 
\begin{center}
\vspace{0.3cm}
\hspace{0cm}
\includegraphics[width=.48\textwidth]{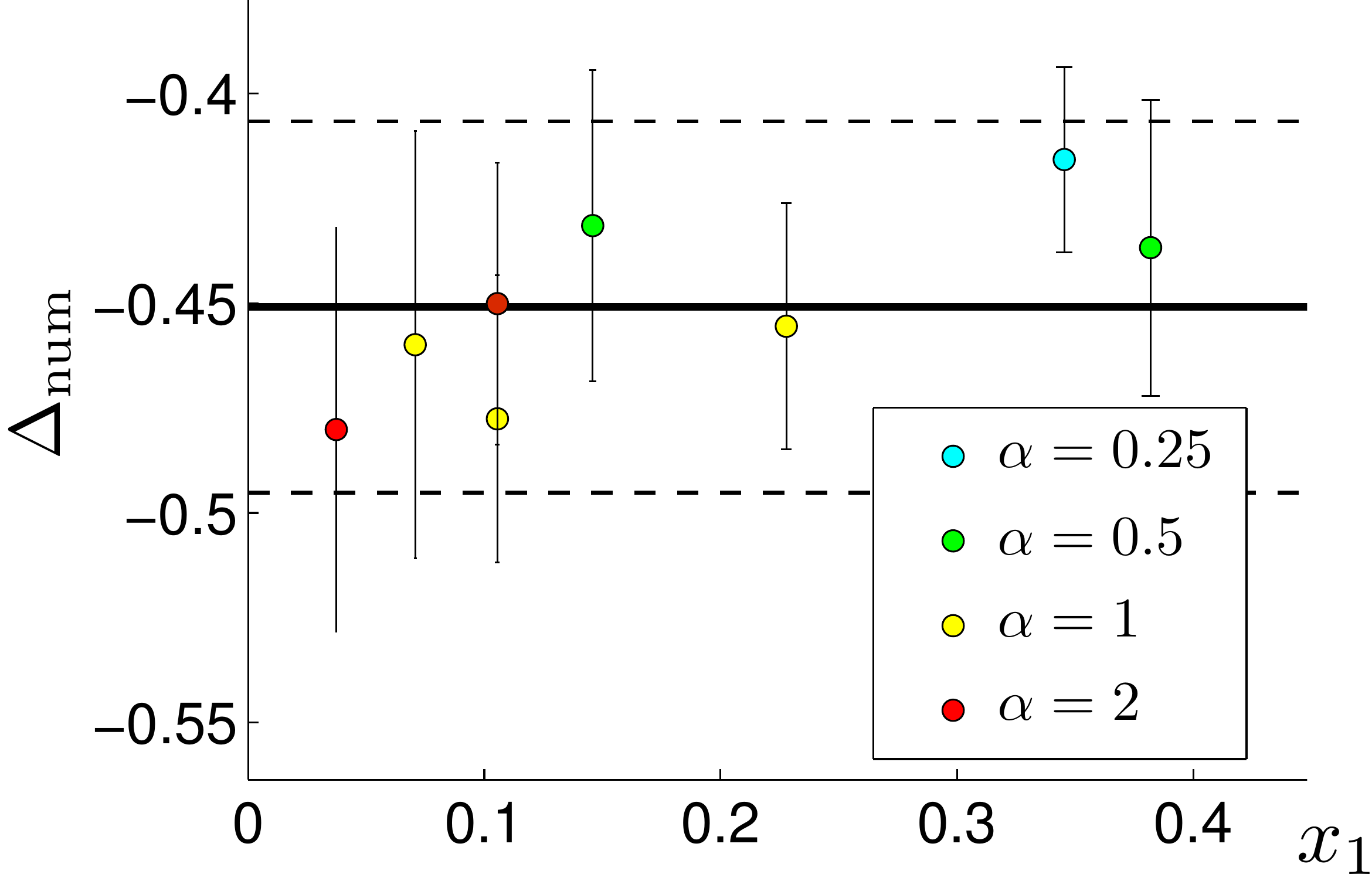}
\includegraphics[width=.48\textwidth]{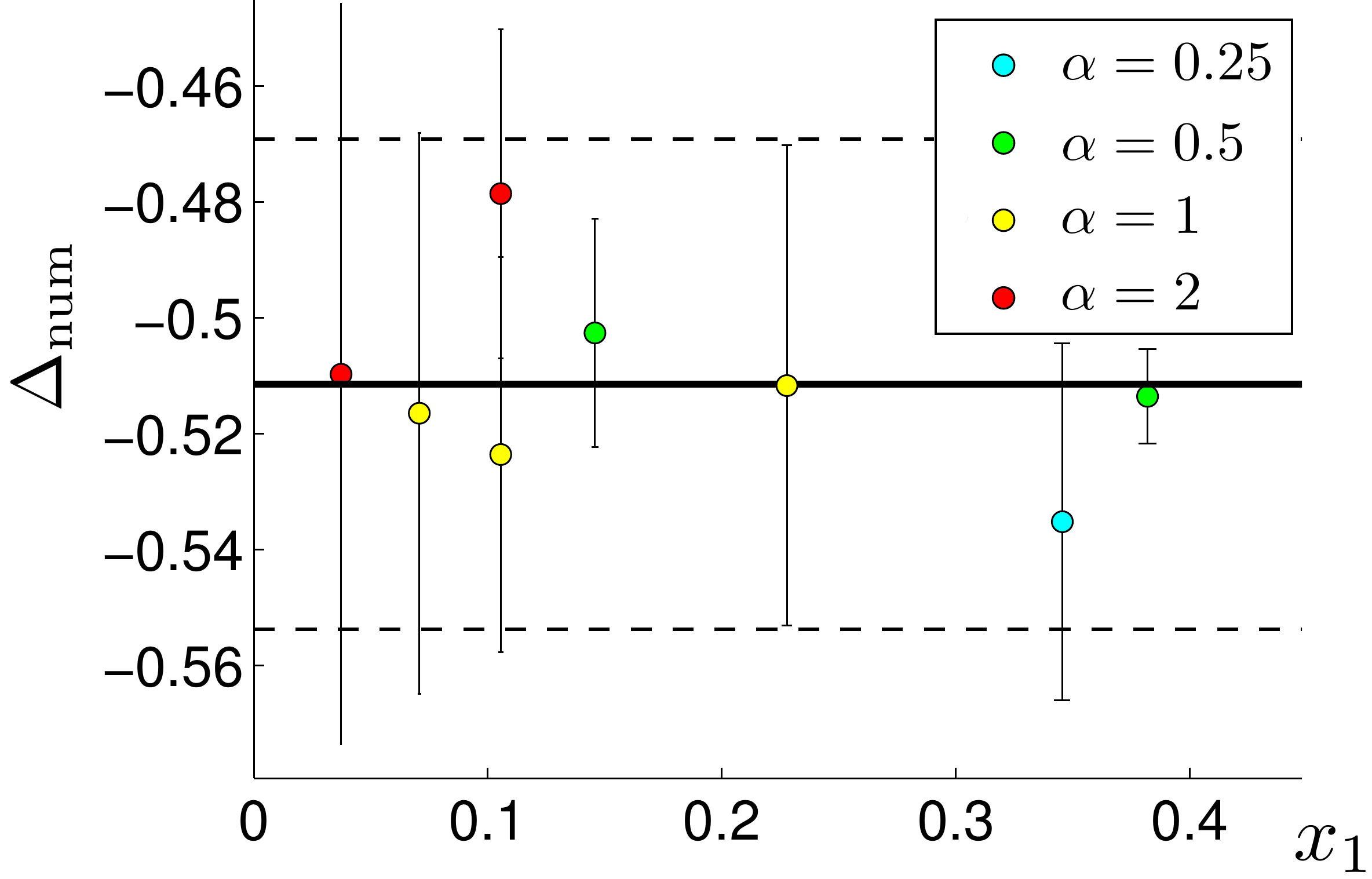}
\\
\vspace{.5cm}
\includegraphics[width=.48\textwidth]{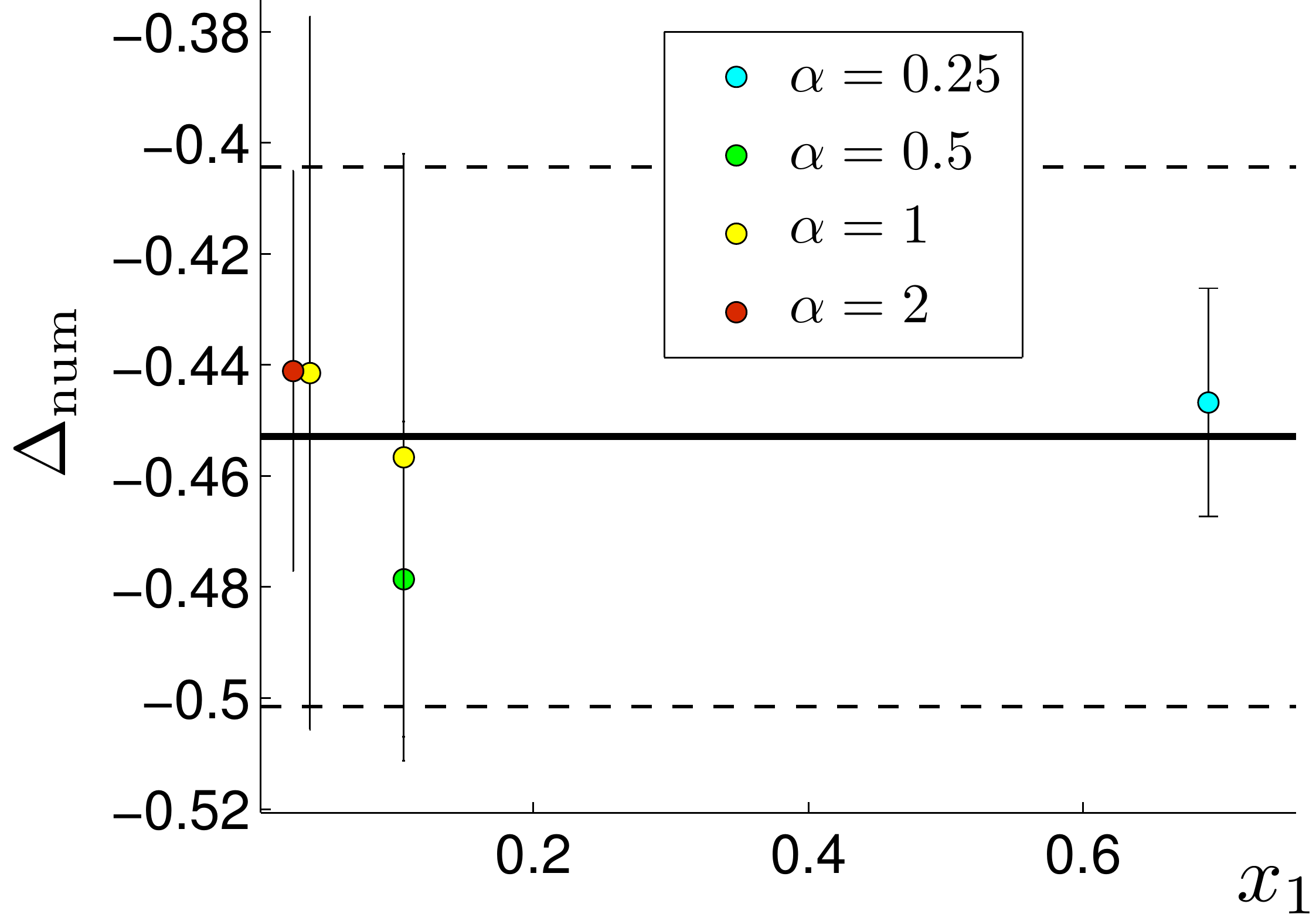}
\end{center}
\vspace{-.3cm}
\caption{\label{fig:isingdeltanum}
The value of the exponent $\Delta_{\textrm{num}}$ obtained from the numerical values of $\mathcal{F}_{3,2}$, $R_{3,2}$ and $\mathcal{F}_{4,2}$ (clockwise direction, starting from the top left). The values of $x_1$ correspond to the ones where several numerical points are available (see Figs. \ref{fig:isingN3aF}, \ref{fig:isingN3aR} and \ref{fig:isingN4} respectively).
The error bars are obtained by changing the number of numerical points in the fit (see Fig. \ref{fig:isingDeltanum}).}
\end{figure}

Given the large discrepancy between our numerical data for the Ising model and the corresponding CFT predictions, the finite size scaling analysis becomes crucial either to confirm or to discard them.  
As discussed in \S\ref{sec num ising}, we numerically study $\Tr \rho_A^2$ when $A$ is made by three or four disjoint intervals by considering ${\cal F}_{3,2}$,  $R_{3,2}$ and ${\cal F}_{4,2}$.

The first step in the finite size scaling analysis is the determination of the exponents of the corrections. To this aim, we start by taking only one correction into account.
Since we usually have only few numerical points for a fixed value of $\boldsymbol{x}$, let us focus on those $\boldsymbol{x}$'s with several of them coming from different values of $\chi$.
For these $\boldsymbol{x}$'s, which correspond to different $\alpha$'s, we fit the numerical data for ${\cal F}_{3,2}$,  $R_{3,2}$ and ${\cal F}_{4,2}$ by using the function $a_0+b_0/\ell^{\Delta_{\textrm{num}}}$, which has three parameters to determine. 
Changing the ranges of variation for $\ell$, we can check the stability of the results and also find an estimate of the error for the fitting process (see Fig. \ref{fig:isingDeltanum} for a typical example).
The results for $\Delta_{\textrm{num}}$ are shown in Fig. \ref{fig:isingdeltanum}: 
starting from the top left in clockwise direction, we find $\Delta_{\textrm{num}}=0.45(5)$, $\Delta_{\textrm{num}}=0.51(4)$ and $\Delta_{\textrm{num}}=0.45(5)$ for ${\cal F}_{3,2}$, $R_{3,2}$ and ${\cal F}_{4,2}$ respectively.
In this analysis the CFT formulas have not been used.
Notice that it is non trivial that $\Delta_{\textrm{num}}$ does not depend on $\boldsymbol{x}$. Our results are consistent with $\Delta_{\textrm{num}}=1/2$ found for $N=2$ \cite{atc-10, FagottiCalabrese10} and they show that it holds also for $N>2$. 

The values of $\Delta_{\textrm{num}}$ just given have been used in (\ref{1corr cft}) to find the extrapolated points in Figs. \ref{fig:isingN3aF}, \ref{fig:isingN3aR} and \ref{fig:isingN4}. 
Thus, for each $\boldsymbol{x}$, now there are two parameters to fit.
Notice that we have not employed the CFT formula yet.

\begin{figure}[t] 
\begin{center}
\vspace{0.3cm}
\hspace{0cm}
\includegraphics[width=.47\textwidth]{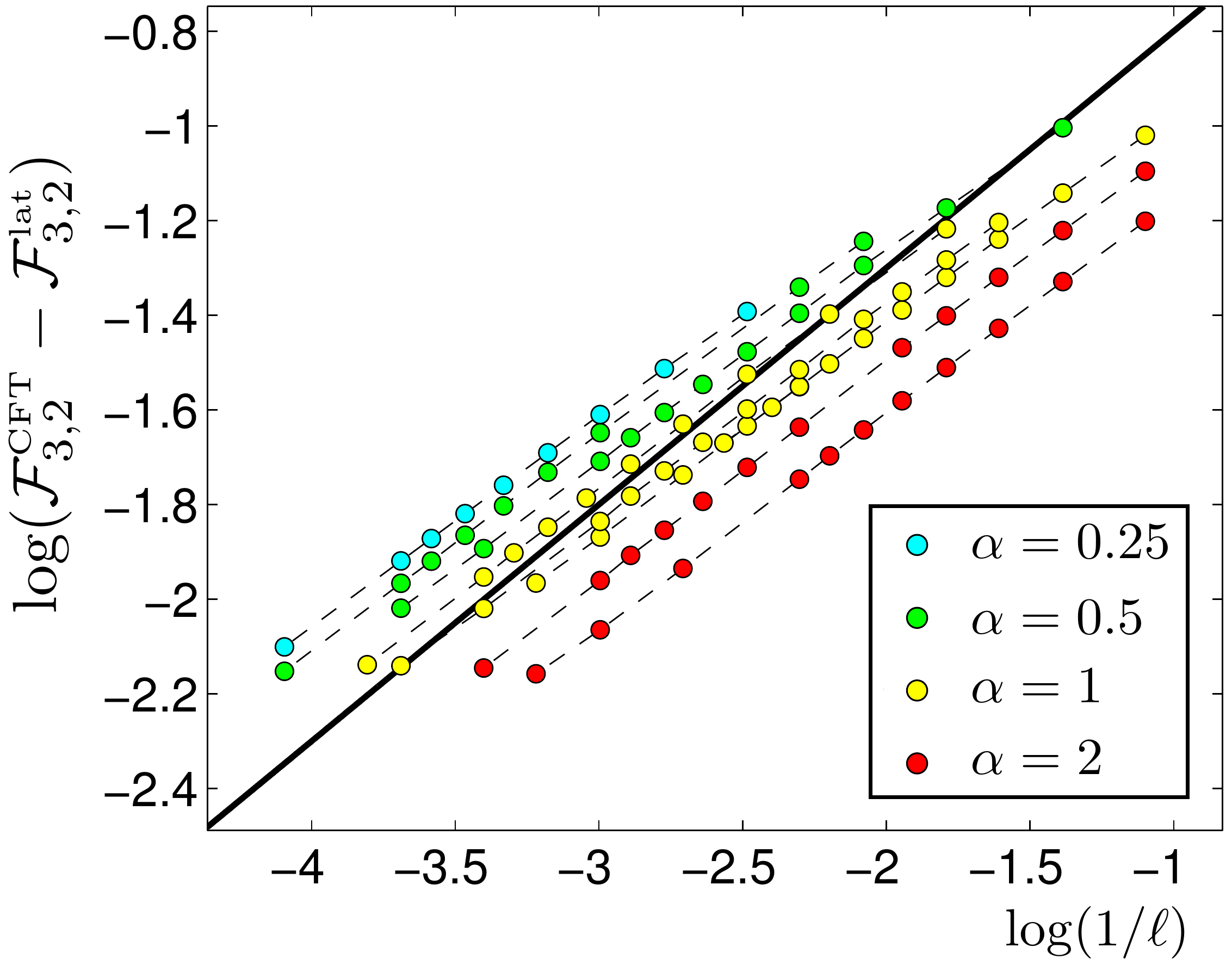}
\hspace{.4cm}
\includegraphics[width=.47\textwidth]{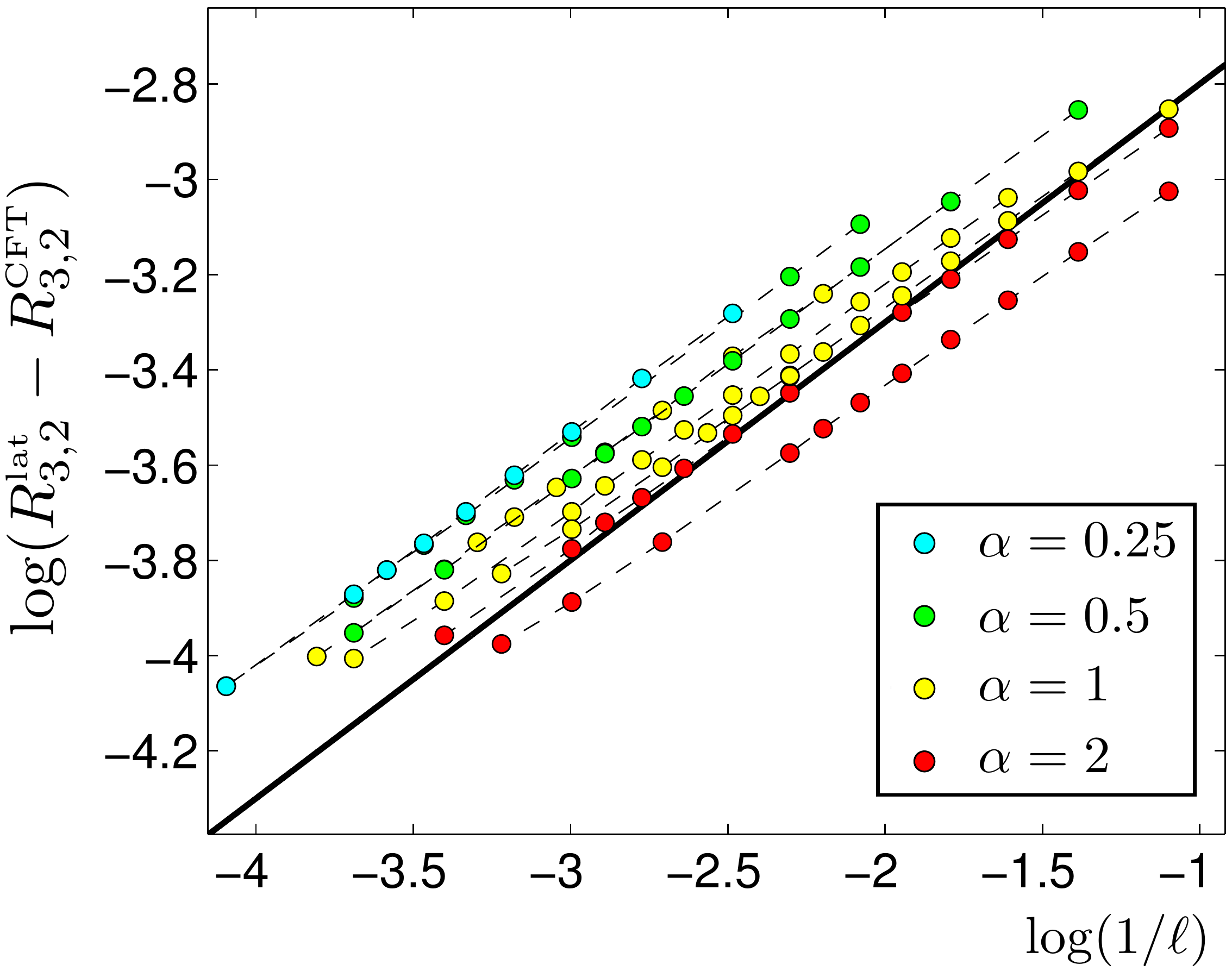}\\
\vspace{0.3cm}
\hspace{-.05cm}
\includegraphics[width=.47\textwidth]{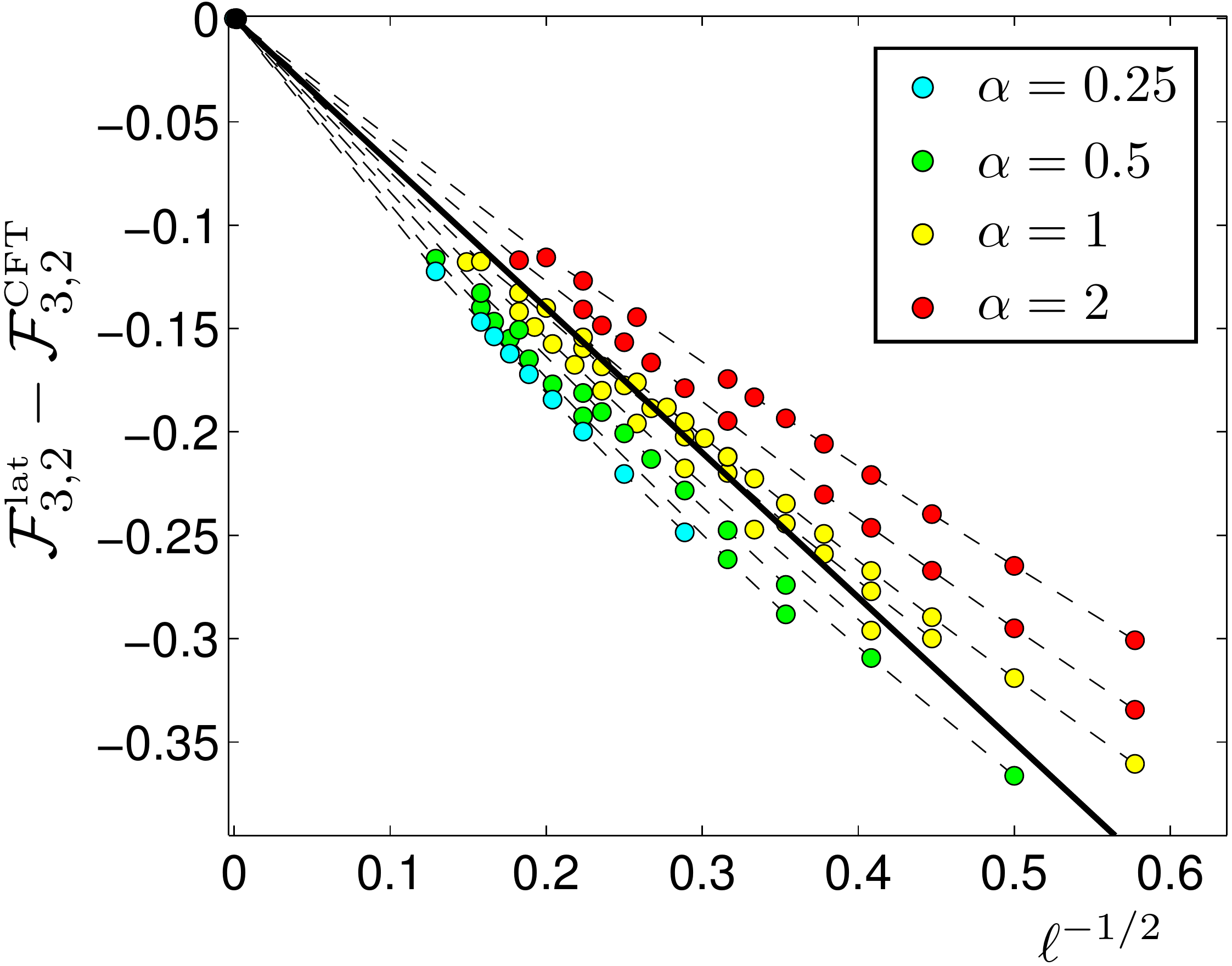}
\hspace{.4cm}
\includegraphics[width=.475\textwidth]{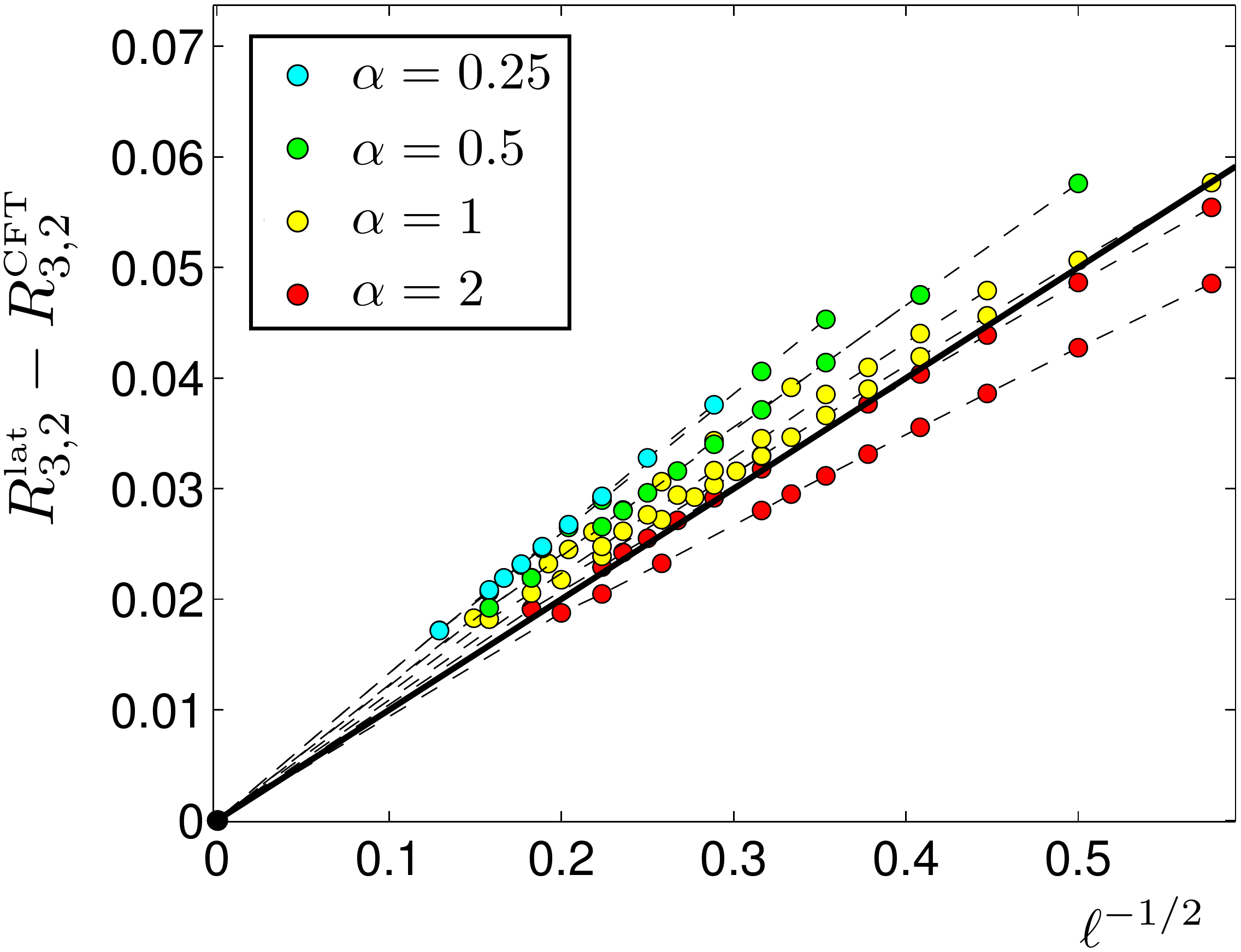}
\end{center}
\vspace{-.3cm}
\caption{\label{fig:isingLines}
Difference between the numerical data and the CFT prediction for $\mathcal{F}_{3,2}$ (left) and $R_{3,2}$ (right). The black solid line corresponds to $\Delta=1/2$ for the exponent of the leading correction, which is the value expected from CFT arguments. In the upper panels the results are shown in logarithmic scales in order to appreciate the fact that, joining the data having the same $\boldsymbol{x}$, we find almost straight lines having nearly the same slope.
}
\end{figure}

In Fig. \ref{fig:isingLines} we plot the difference between the numerical data and the CFT prediction in log-log scale, in order to visualize the leading correction. All the data lie on parallel lines whose slope is close to the one expected from the two intervals case.

\begin{figure}[t] 
\begin{center}
\vspace{0.3cm}
\hspace{-.25cm}
\includegraphics[width=.905\textwidth]{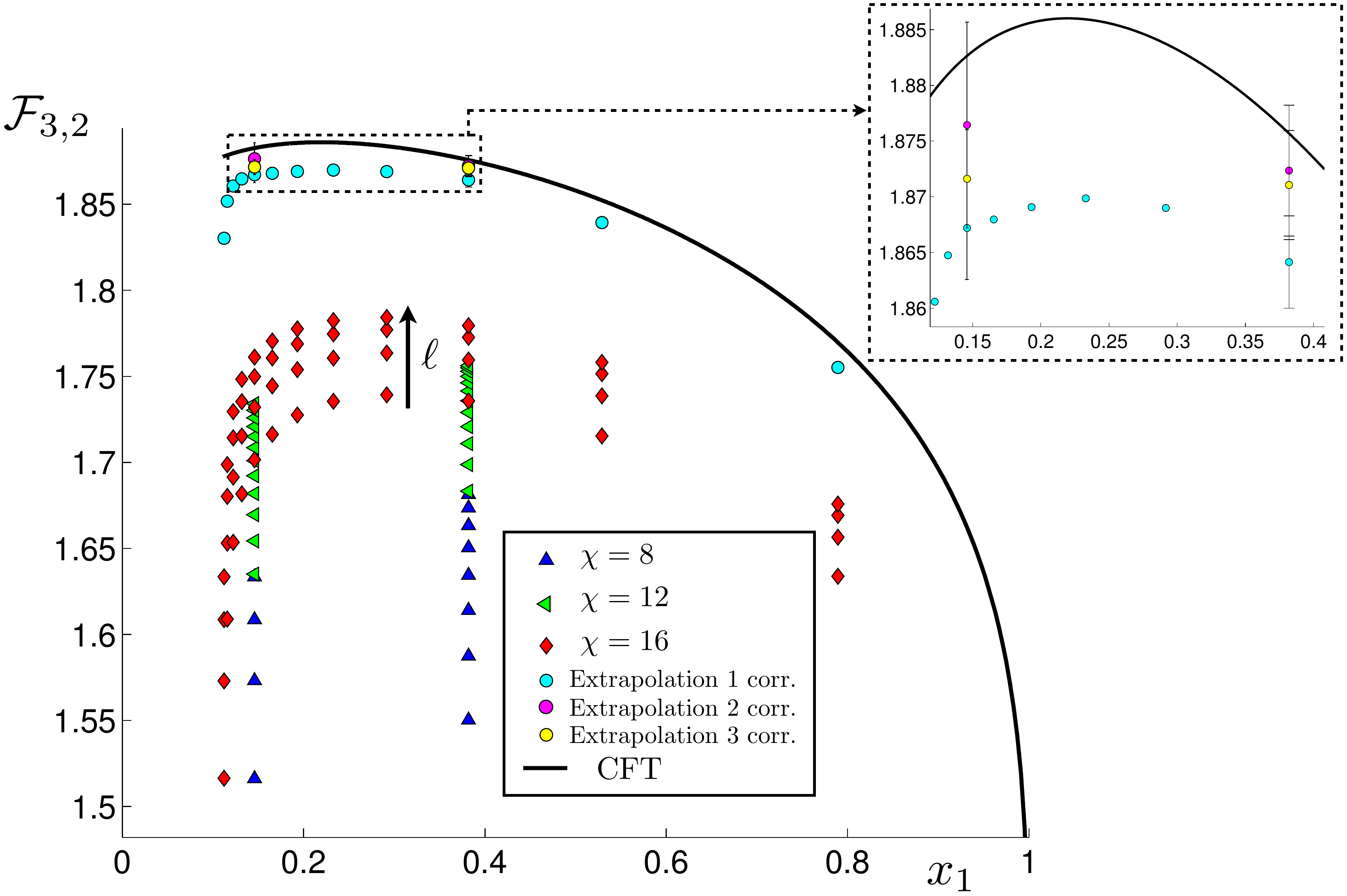}\\
\vspace{0.3cm}
\includegraphics[width=.9\textwidth]{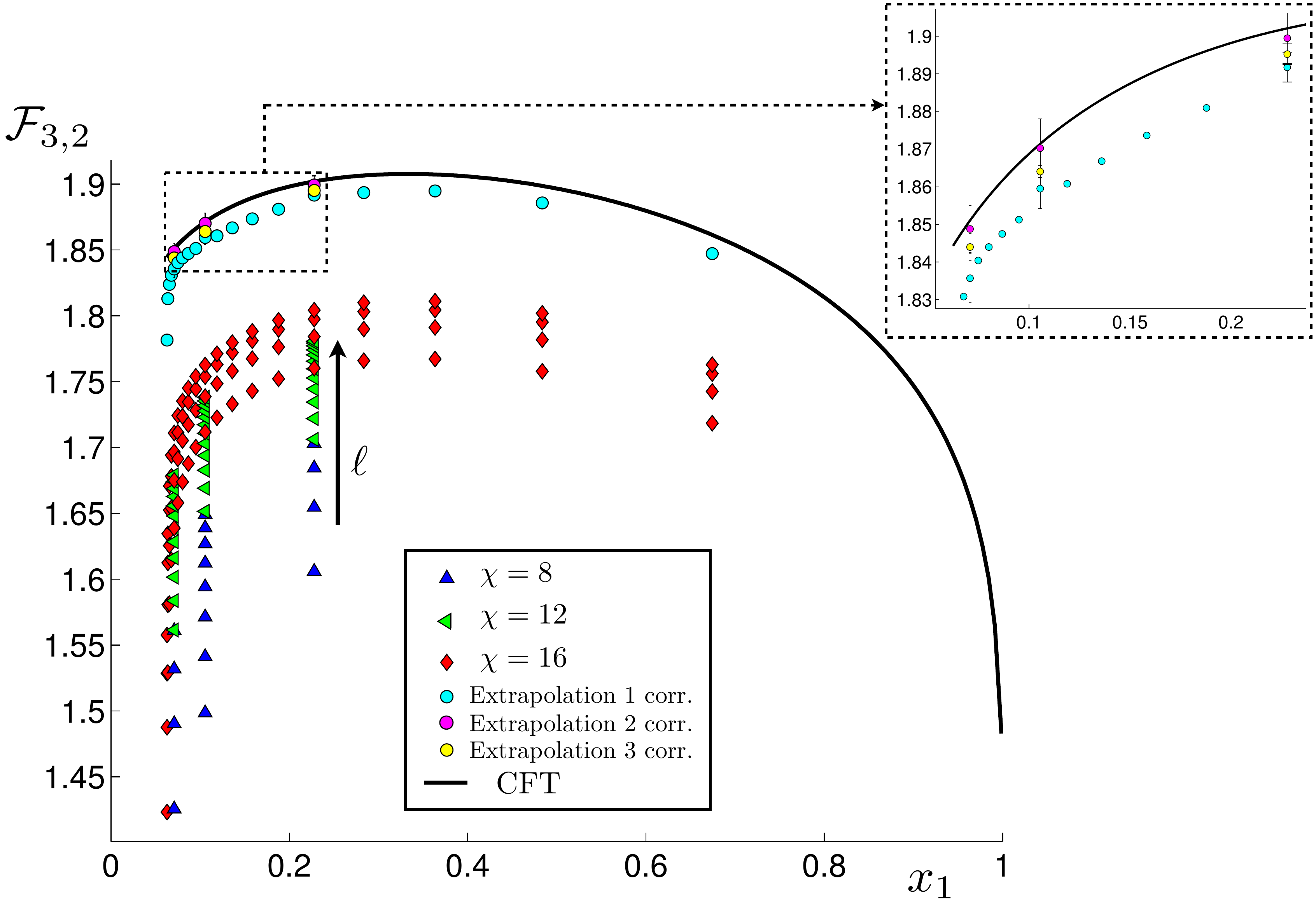}
\end{center}
\vspace{-.3cm}
\caption{\label{fig:isingN3fss3corrF}
Finite size scaling analysis with higher order corrections for $\mathcal{F}_{3,2}$ for the configurations characterized by $\alpha=0.5$ (top) and $\alpha=1$ (bottom).
The method is explained in Appendix \ref{app:fss3corr}. Three corrections can be taken into account only for those $\boldsymbol{x}$'s having several numerical points, as shown in the zoom. The third correction never improves the agreement with the CFT prediction.}
\end{figure}

\begin{figure}[t] 
\begin{center}
\vspace{0.3cm}
\hspace{0cm}
\includegraphics[width=.489\textwidth]{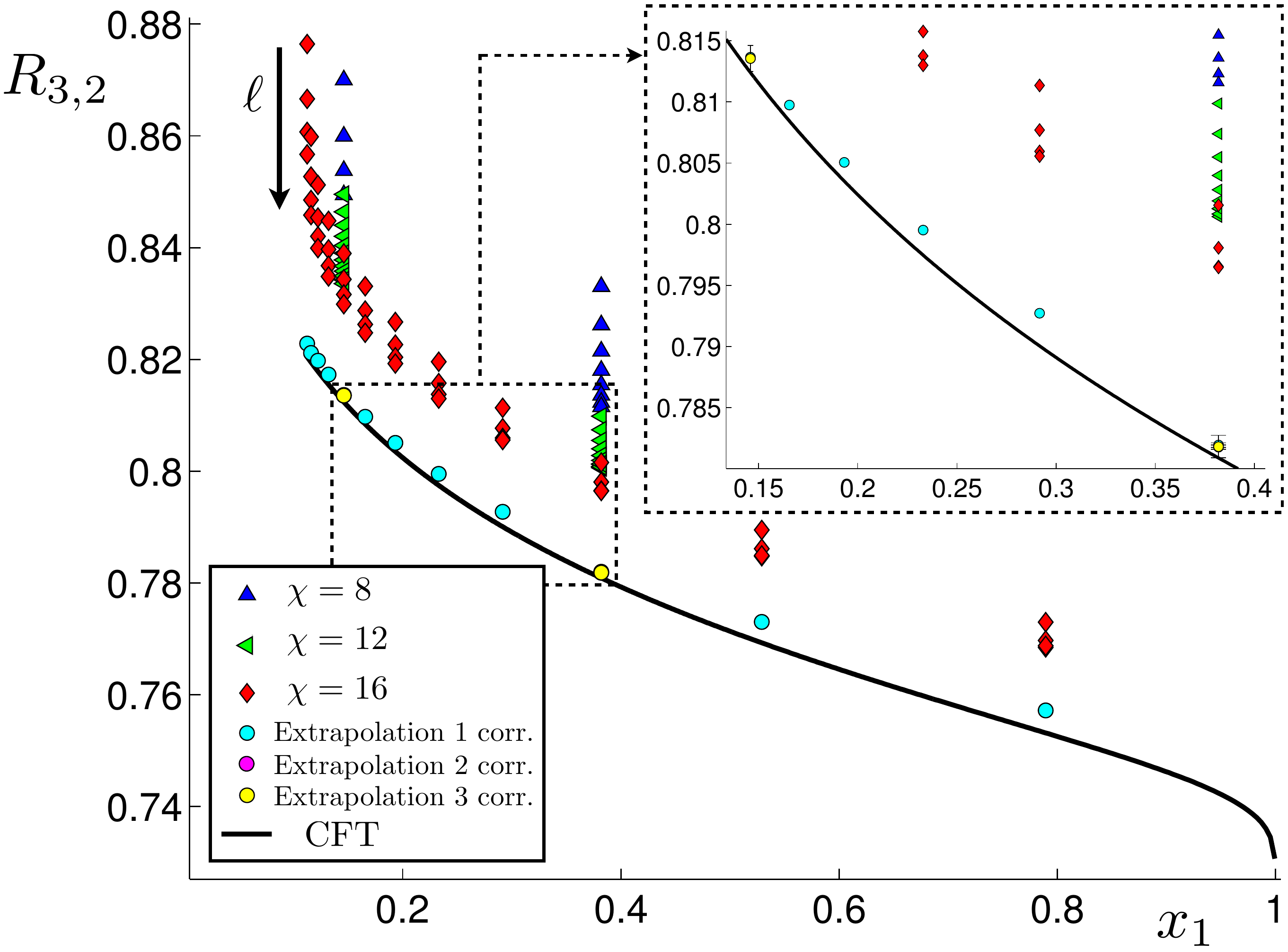}\,
\includegraphics[width=.489\textwidth]{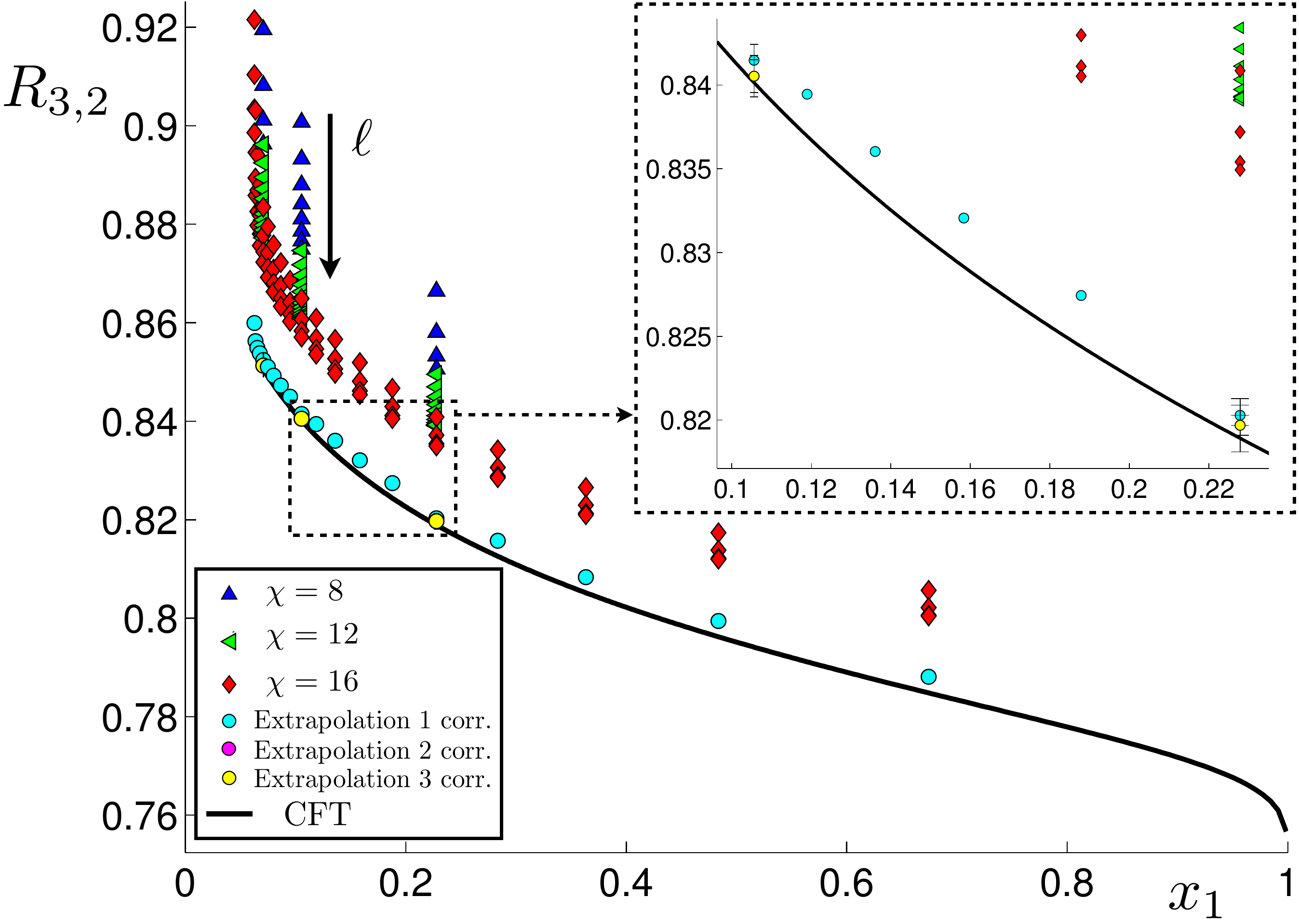}
\end{center}
\vspace{-.3cm}
\caption{\label{fig:isingN3fss3corrR}
Finite size scaling analysis with higher order corrections for $R_{3,2}$ for the configurations characterized by $\alpha=0.5$ (left) and $\alpha=1$ (right).
The method is explained in Appendix \ref{app:fss3corr}. Three corrections can be taken into account only for those $\boldsymbol{x}$'s having several numerical points, as shown in the zoom. The third correction never improves the agreement with the CFT prediction.}
\end{figure}

\subsection{A finite size scaling analysis with higher order corrections}
\label{app:fss3corr}

Instead of considering only one correction as discussed in \S\ref{sec num ising} and Appendix \ref{app:expo}, one can try to perform a finite size scaling analysis which includes more corrections
\cite{cc-unusual, CalabreseEssler-10, atc-10, FagottiCalabrese10, atc-11, ctt-neg-ising}.
In particular, we choose the following function
\be
\label{3corr formula}
a_0 + \frac{b_1}{\ell^{1/2}} + \frac{b_2}{\ell} + \frac{b_3}{\ell^{3/2}} \,.
\ee
The exponents are the ones giving agreement with the CFT predictions for $N=2$ \cite{ctt-neg-ising}.
Since in this case we have four parameters to fit, we can carry out this analysis only for few $\boldsymbol{x}$'s at fixed $\alpha$.
We have considered the same configurations of \S\ref{sec num ising}, namely $\alpha=p$ and $\alpha=1/p$ with $1\leqslant p \leqslant 8$ finding the same qualitative behavior.
Here we give only one representative example in Fig \ref{fig:isingN3fss3corrF} for $\mathcal{F}_{3,2}$ and in Fig \ref{fig:isingN3fss3corrR} for $R_{3,2}$. 
The error bars have been determined by choosing different minimum values for $\ell$ in the fitting procedure, as done for $\Delta_{\textrm{num}}$ in Appendix \ref{app:expo}.

It is instructive to analyze the contribution of the various corrections. Taking only the first correction into account (cyan circles in Figs. \ref{fig:isingN3fss3corrF} and \ref{fig:isingN3fss3corrR}), the extrapolated points are very close to the curves predicted by the CFT. Nevertheless, they do not coincide with it, staying systematically below for $\mathcal{F}_{N,2}$ or above for $R_{N,2}$. 
Adding the second correction, i.e. $b_1 \neq 0$ and $b_2 \neq 0$ in (\ref{3corr formula}), the extrapolations (green circles in Figs. \ref{fig:isingN3fss3corrF} and \ref{fig:isingN3fss3corrR}) usually improve, as expected, getting closer to the CFT prediction and, in some case, coinciding with it. 
As for the third correction, we notice that it does not improve the extrapolation in almost all the cases that we studied. This probably tells us that the range of $\ell$ available allows us to see at most two corrections to the scaling. 
As for the sign of the coefficients $b_1$, $b_2$ and $b_3$ in (\ref{3corr formula}), we find $(-,+,+)$ for $\mathcal{F}_{3,2}$ and $(+,-,+)$ for $R_{3,2}$. Notice that the sign of $b_1$ can be easily inferred from the position of the numerical points with respect to the CFT curve. For instance, since for $R_{3,2}$ they are all above the theoretical curve, we have that $b_1>0$ in this case.

 \subsection{On the finiteness of the bond dimension}
 \label{app:finitechi}
 
 \begin{figure}[t] 
\begin{center}
\vspace{0.3cm}
\hspace{-0.3cm}
\includegraphics[width=.487\textwidth]{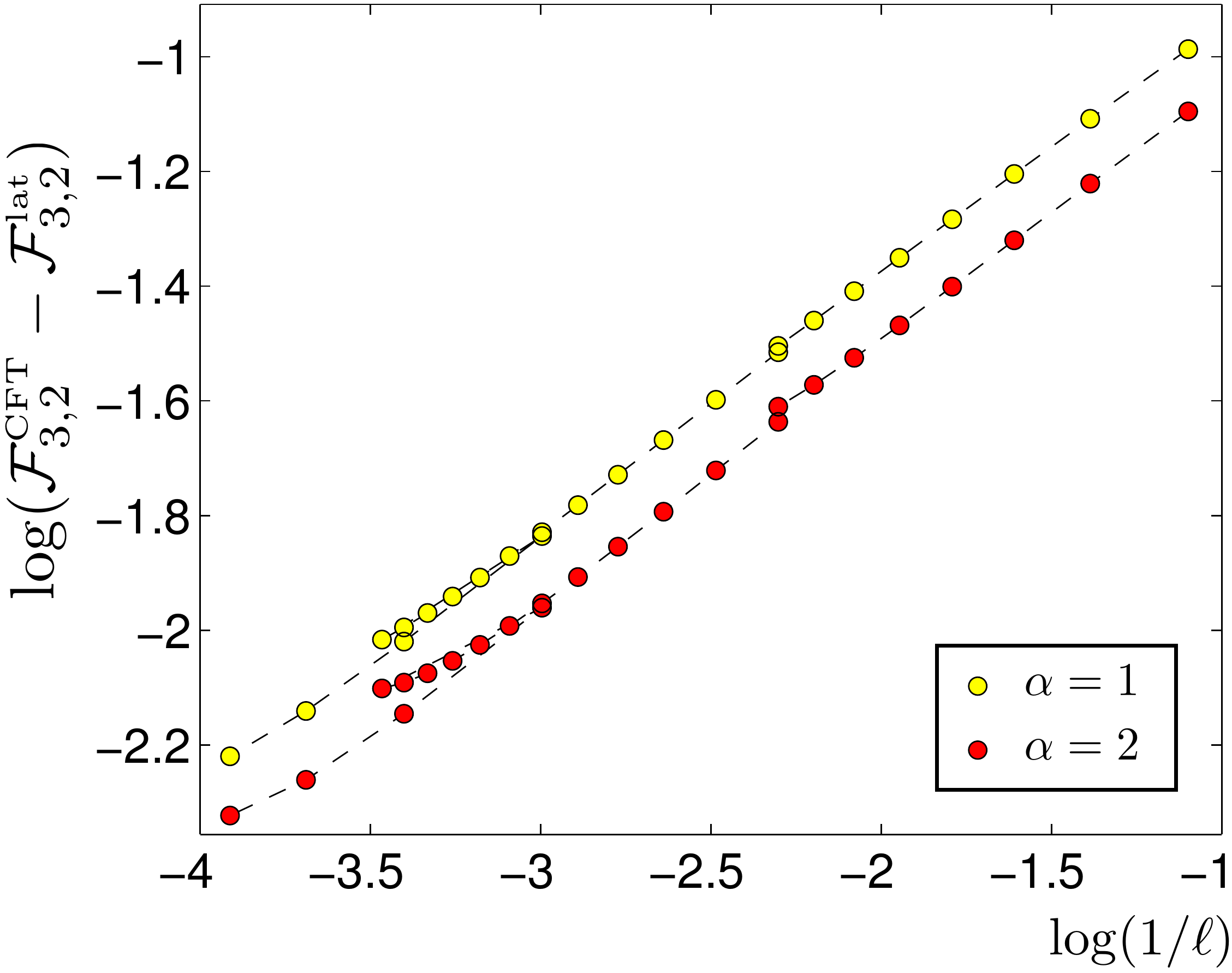}
\hspace{0.35cm}
\includegraphics[width=.486\textwidth]{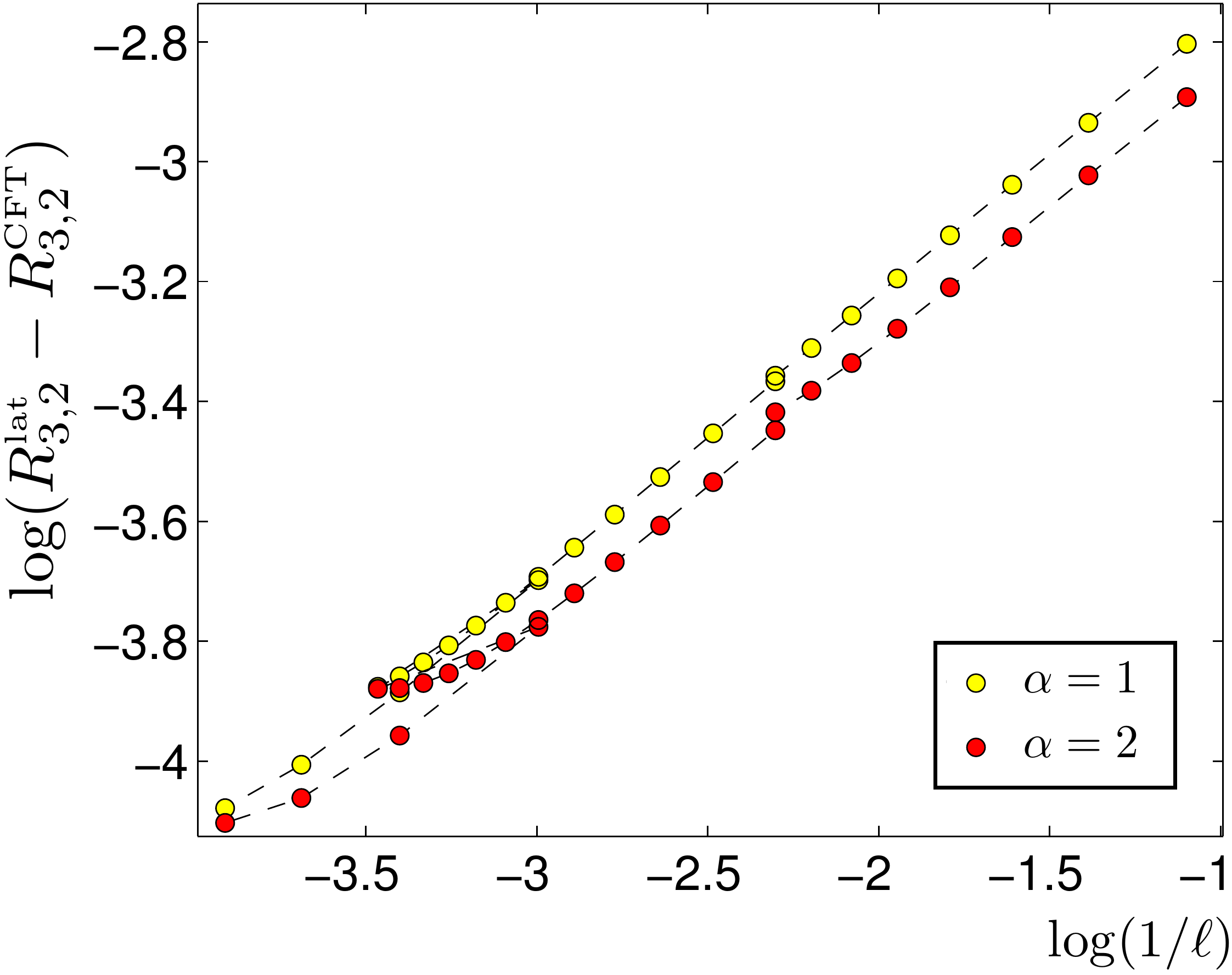}
\\
\vspace{0.5cm}
\includegraphics[width=.485\textwidth]{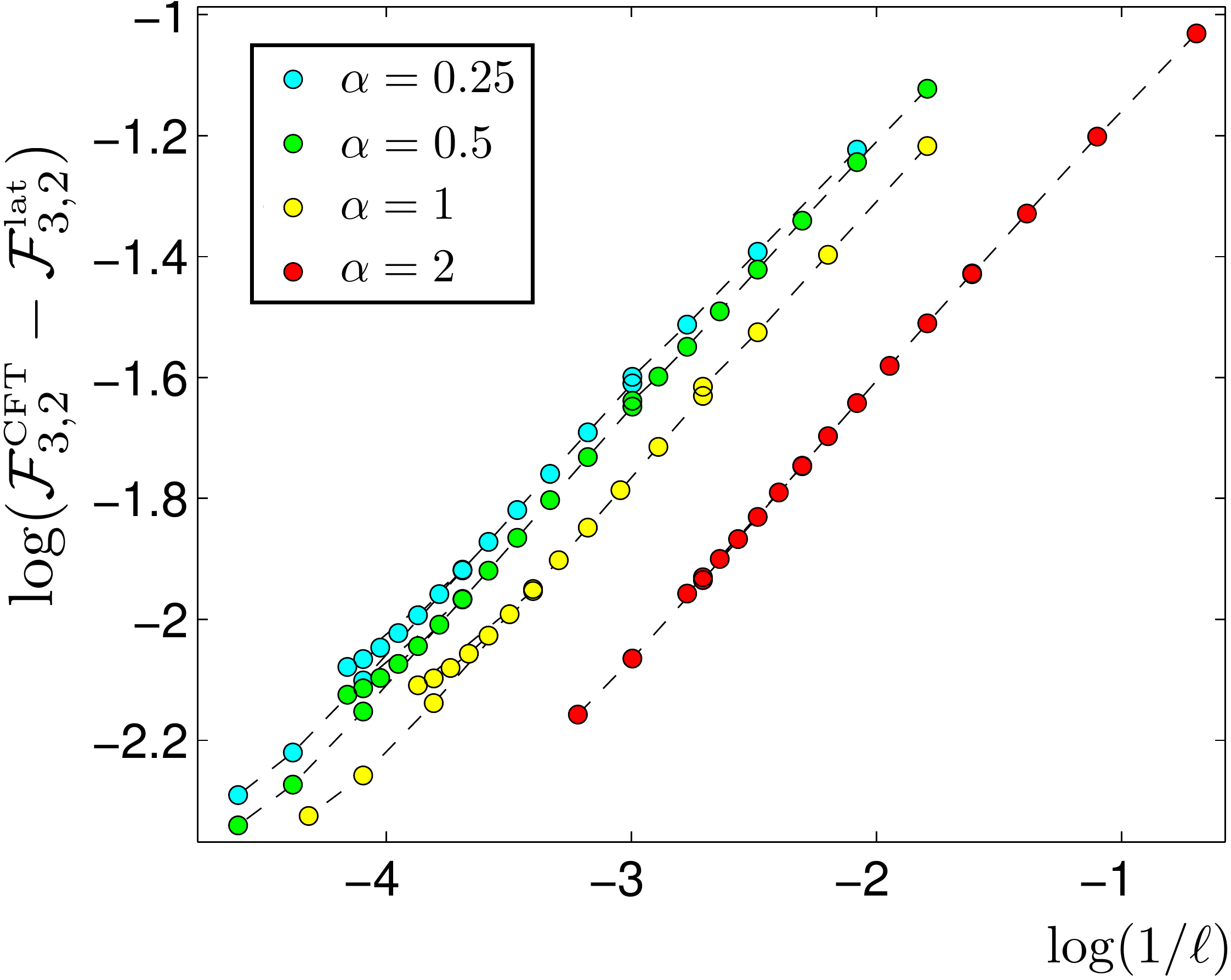}
\hspace{0.2cm}
\includegraphics[width=.485\textwidth]{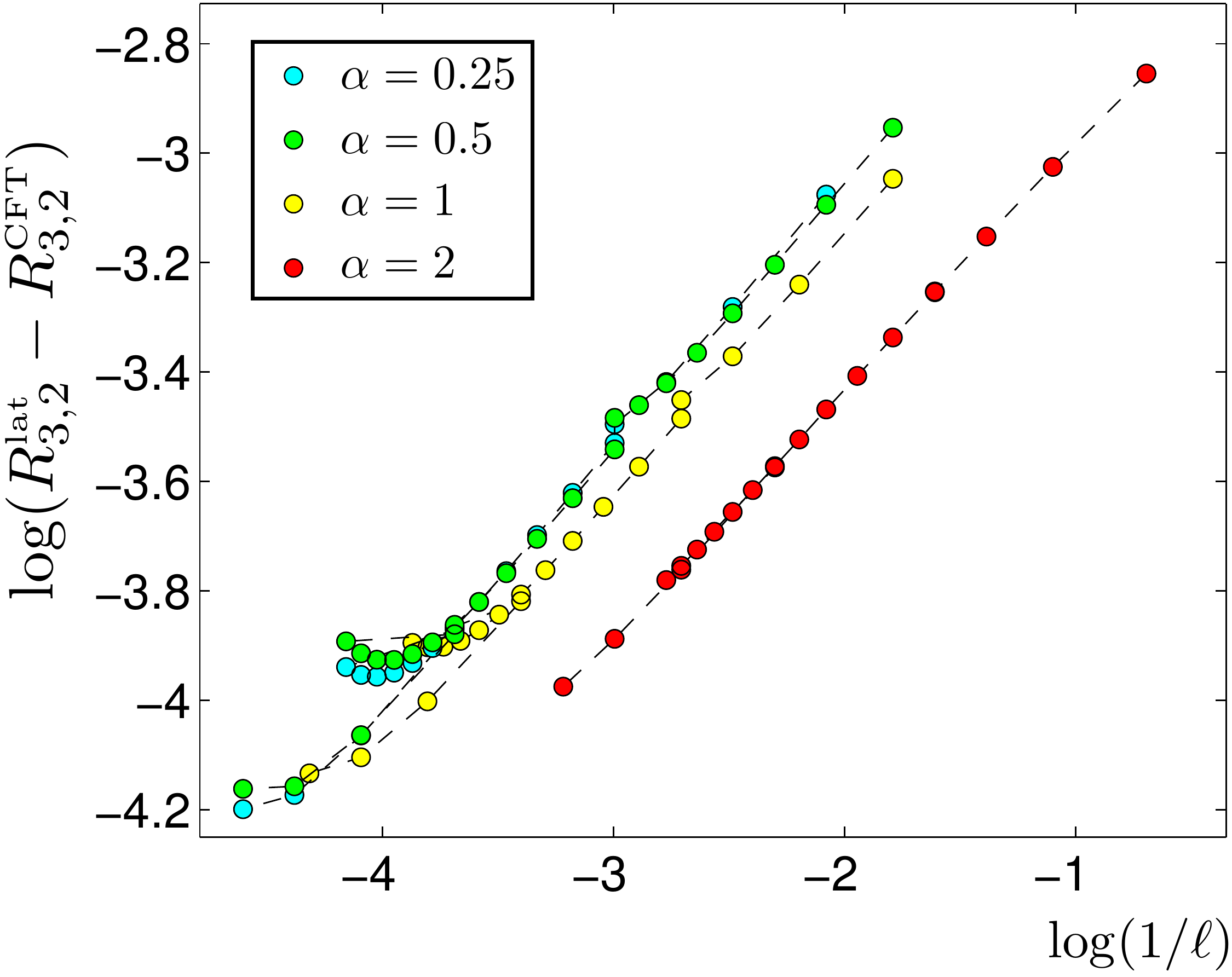}
\end{center}
\vspace{-.3cm}
\caption{\label{fig:isingfinitebond}
Effects of the finite bond dimension leading to deviations from the power law decays are shown for $\mathcal{F}_{3,2}$ (left) and $R_{3,2}$ (right). 
In the top panels $x_1=0.106$. For fixed $\ell$, the deviation from the straight line is more evident for points with larger $d$ ($\alpha=2$).
In the bottom panels we have:
$x_1=0.345$ ($\alpha=0.25$), $x_1=0.382$ ($\alpha=0.5$),
$x_1=0.228$ ($\alpha=1$) and $x_1=0.037$ ($\alpha=2$). 
For small values of $\alpha$, regimes of large $\ell$ can be considered, where deviations may also occur.
The points deviating from the straight line have been discarded from the numerical analysis.}
\end{figure}

Tensor networks, which include the MPS as a subclass, are variational approximations whose accuracy strongly depends on the bond dimension $\chi$.  
In principle, one would like to have access to the regime of $\chi \to \infty$ but, being the computational cost an increasing function of $\chi$, the results are always obtained for finite $\chi$.

The MPS are finitely correlated state, which means that they naturally describe systems where either the correlations do not decay or they decay exponentially at large distance \cite{Fannes92}. The two cases are distinguished by the ratio $e_2/e_1\leqslant 1$ between the  two largest eigenvalues  $e_1$ and $e_2$ of the MPS transfer matrix $E$.  
In particular, if $e_2 < e_1$, the finite correlation length of the MPS is $\xi_{\textrm{\tiny MPS}} \equiv 1/\log(e_1/e_2)$, while, when $e_2 = e_1$, the correlation function (\ref{eq:two_point}) is constant as a function of $r$ (long range order). 

The finite size of a critical system naturally induces a finite correlation length  $\xi_L  \propto L$. Thus, the MPS representation  can still be used to perform accurate finite size scaling analysis \cite{VerstraeteCirac06} and one would expect that a good MPS approximation has $\xi_{\textrm{\tiny MPS}} =\xi_{L} $.
However, it has been found that, when $\chi$ is too small, the best approximation of a critical system through a MPS with finite $\chi$ has a finite correlation length $\xi_{\textrm{\tiny MPS}} =\xi_{\chi} \propto \chi^ {\kappa}$ \cite{TagliacozzoOliveiraIblisdirLatorre}. 
In order to get $\xi_{\textrm{\tiny MPS}} = \xi_L $, one needs to increase $\chi$.
Since $\xi_L$ enters in the scaling of the two point correlation functions for critical systems, a useful criterion is obtained by considering \cite{PirvuEtAl, PirvuVidalVerstraeteTagliacozzo}
\be
\chi^\ast =\min\big\{\chi \, \big|\, \xi_{\chi}  > L/2 \big\}\,.
\label{eq:fchi}
\ee

However, notice that this result has been found by considering the two point functions of local operators, while in our problem both non local operators (whose support is of order $\xi_\chi$) and $2N \geqslant 4$ point functions are involved. 
In our numerical analysis we have adopted the criterion (\ref{eq:fchi}) and, indeed, we find that sometimes it fails. For instance, this happens in Fig. \ref{fig:isingfinitebond} whenever a deviation from the straight lines occurs.
We have taken this failure into account by discarding from the numerical analysis the points deviating from the straight lines.
Being (\ref{eq:fchi}) too optimistic for our computations, the criterion 
\be
\chi^\ast =\min\big\{\chi \, \big|\, \xi_{\chi}  > L \big\}
\label{eq:fchi2}
\ee
should be enough to avoid deviations from the expected power law decay and should be implemented in future studies.

\end{appendices}


\end{document}